\documentstyle[times,righttag,amsxtra,amssymb,euscript]{amsbook}
\newlength{\dinwidth}
\newlength{\dinmargin}

\newskip\AboveMaplePlot
\newskip\BelowMaplePlot
\newskip\AboveMapleSkip
\newskip\BelowMapleSkip
\newdimen\LeftMapleSkip
\newdimen\MaplePlotHeight
\newdimen\MaplePlotWidth
\newdimen\MapleSepLineWidth
\newdimen\MapleSepLineHeight \MapleSepLineHeight = 0.01in
\newif\ifMaple          \Maplefalse
\newif\ifMapleFirstLine
\newif\ifMaplePrompt
\newcount\MapleTab      \MapleTab = 8
\newcount\MapleParCount   \MapleParCount=0
\newtoks\MaplePromptString \MaplePromptString = {}
\MaplePromptString = {\raise 1pt \hbox{$\scriptstyle>$\space}}
\BelowMapleSkip = \AboveMapleSkip
\let\MapleSepLineWidth\linewidth  % \let so it will be redefined
\let\MapleFont\tt
\let\MapleSize\small
\MapleFirstLinefalse  % Ignore first \cr after \begin{mapleinput}
\MaplePrompttrue      % Include the prompt string for each input line.
\makeatletter
\def\MakeActive#1{\catcode `#1 = \active\relax }
\def\MakeTabActive{\MakeActive{\^^I}}
\def\MakeEolActive{\MakeActive{\^^M}}
\newif\if@IgnoreNewLine
\@IgnoreNewLinetrue
\def\@MaplePar{\nopagebreak[3]\par\@@par}%
\def\MapleSpace{\ }
\def\@ObeySpaces{\MakeActive{\ }\@@ObeySpaces\relax}
{%
\MakeActive{\ }\gdef\@@ObeySpaces{\edef {\MapleSpace}}%
}%
\newdimen\@MapleTabSize
\def\@ObeyTabs{\MakeTabActive\@@ObeyTabs\relax}
{%
\MakeTabActive\gdef\@@ObeyTabs{\def^^I{\@MapleTab}}
}%
\def\@MapleTab{%
  \leavevmode   % ensure that we are in horizontal mode
  \egroup      % Terminate box register 0
  \dimen0=\wd0 % store its length and round up to next stop
  \divide\dimen0 by \@MapleTabSize
  \advance\dimen0 by 1sp
  \multiply \dimen0 by \@MapleTabSize
  \wd0 = \dimen0
  \box0
  \setbox0 = \hbox\bgroup
}
\def\@ObeyEol{\MakeEolActive\@@ObeyEol\relax}%
{%
  \MakeEolActive %
  \gdef\@@ObeyEol{%
    \let^^M=\@MapleEol%
  }%
}
\def\@MapleEol{%
  \if@IgnoreNewLine
  \else
    \leavevmode%
    \egroup%
    \box 0%
    \@MaplePar%
  \fi
  \@IgnoreNewLinefalse
}
\def\@SetupMapleTty#1{%
\begingroup
  \setbox 0 = \hbox{\MapleSize\MapleFont X}
  \@MapleTabSize = \wd0
  \multiply\@MapleTabSize by \MapleTab
  \setbox0 = \hbox{\relax}  % Clearing \box 0
  \rightskip = 0pt
  \parindent=0pt
  \parskip = 0pt
  \leftskip=\LeftMapleSkip
  \parfillskip = 0pt plus 1fil
  \spaceskip = 0pt   \xspaceskip = 0pt
\ifnum #1 = 0 \@IgnoreNewLinetrue\else\@IgnoreNewLinefalse\fi%
\everypar = {\EveryParZ}%
\def\EveryParZ{%
  \ifMaplePrompt
    \the\MaplePromptString
  \fi
    \setbox0 = \hbox \bgroup
}%
\MapleSize\MapleFont%
\lineskiplimit=0\normalbaselineskip
\baselineskip=0.8\normalbaselineskip
\@noligs%
\let\do\@makeother \dospecials
\catcode ``=\active
\frenchspacing%
\@ObeySpaces%
\@ObeyTabs%
\@ObeyEol%
}
\def\@EndMapleInput{%
  \endgroup  %% matches the \begingroup inside of @SetupMapleInput
}
\let\@EndMapleTtyout\@EndMapleInput
\begingroup
  \catcode`| = 0  % becomes the escape character
  \catcode`( = 1  % becomes a begin group
  \catcode`) = 2  % becomes an end group
  \catcode`@ = 11 % becomes a regular character (for macro names)
  \catcode`\{ = 12 % becomes an other character
  \catcode`\} = 12 % becomes a other character
  \catcode`\\ = 12 % becomes a other character
  |gdef|@BeginMapleInput#1\end{mapleinput}(#1|end(mapleinput))%
  |gdef|@BeginMapleTtyout#1\end{maplettyout}(#1|end(maplettyout))%
|endgroup
\def\mapleinput{
\@MaplePar%
\if@minipage\removelastskip\vskip-1.3ex
\else\vskip\AboveMapleSkip\fi%
\@SetupMapleTty{0}% The 0 indicates ignore first new line
\@BeginMapleInput
}
\def\endmapleinput{
  \@EndMapleInput
  \nopagebreak[3]
}
\def\maplettyout{
  \MaplePromptfalse       % No prompt is required in this environment.
  \nopagebreak[3]\@MaplePar
  \removelastskip
  \@SetupMapleTty{0}   % A 1 would indicate "respect first new line"
  \@BeginMapleTtyout
}
\def\endmaplettyout{
  \@EndMapleTtyout
  \vskip\BelowMapleSkip
  \pagebreak[3]
  \par
}
\newenvironment{maplelatex}{
  \parindent=0pt
  \parskip = 0pt
  \parfillskip = 0pt plus 1fil
  \lineskiplimit=0\normalbaselineskip
  \baselineskip=0.6\normalbaselineskip
  \abovedisplayskip=.4\AboveMapleSkip
  \belowdisplayskip=.4\AboveMapleSkip
  
  \removelastskip
  \nopagebreak[3]\@@par
}{%
\baselineskip=\normalbaselineskip%
\pagebreak[3]\relax}
\newcommand{\maplesepline}{\vskip \parskip%
\hrule\@height\MapleSepLineHeight\@width\MapleSepLineWidth%
\vskip \parskip\relax}
\input psfig.sty
\psrotatefirst

\makeatother
\MaplePrompttrue      % generate a prompt at start of each line?
\MaplePromptString = {\raise 1pt \hbox{$\scriptstyle>$\space}}
\BelowMapleSkip = \AboveMapleSkip
\let\MapleSepLineWidth\linewidth  % \let so that it will be redefined
\let\MapleFont\tt     % font used for input and ttyout
\let\MapleSize\small  % font size for input and ttyout
\MapleFirstLinefalse  % hides first \cr of \mapleinput
\def\newblock{\hskip .11em plus.33em minus.07em}%
\newenvironment{enumeraterm}{%
\begin{enumerate}}{\end{enumerate}%
}
\newtheorem{Thm}{Theorem}[chapter]
\newtheorem{Lemma}[Thm]{Lemma}
\newtheorem{Cor}[Thm]{Corollary}
\newtheorem{Prop}[Thm]{Proposition}
\theoremstyle{definition}
\newtheorem{Dfn}[Thm]{Definition}
\newtheorem{exmp}[Thm]{Example}
\theoremstyle{remark}
\newtheorem{rem}{Remark} 
\newtheorem{obs}{Observation} 
\newtheorem{ack}{Acknowledgement} 
\newcommand{\script}{\EuScript}                                 % Script
\newcommand{\AO}{\mbox{${\mathfrak A}({\mathcal O})$}}          % Algebra
\newcommand{\calO}{\mbox{${\mathcal O}$}}                       % region O
\newcommand{\AUO}{\mbox{${\mathfrak A}_U({\mathcal O})$}}       % U-Algebra
\newcommand{\AUOH}{\mbox{$\hat{\mathfrak A}_U({\mathcal O})$}}  % HUT-Algebra
\newcommand{\notnabla}{\mbox{$\not \! \nabla$}}                 % nabla slash
\newcommand{\notn}{\mbox{$\not \! n$}}                          % n slash
\newcommand{\notD}{\mbox{$\not\!\! D$}}                         % D slash
\newcommand{\miDm}{(-i \notnabla -m)}                           % ()
\newcommand{\iDm}{(i \notnabla -m)}                             % ()
\newcommand{\CinfO}{C^\infty_0}                                 % ()
\newcommand{\Cinf}{C^\infty}                                    % ()
\newcommand{\Dsquare}{\left(\Box+(m^2+\frac{1}{4}R)\right)}     % Dirac
\newcommand{\rvec}[1]{\mbox{$\stackrel{\rightharpoonup}{#1}$}}
\newcommand{\lvec}[1]{\mbox{$\stackrel{\leftharpoonup}{#1}$}}
\newcommand{\AiB}{(A- i \gamma^5 B)}
\newcommand{\supp}{\operatorname{supp}}                         % Traeger
\newcommand{\identity}{\mbox{${\bf 1}$}}                        % Identity
\newcommand{\WF}{\mbox{\rm WF}}                                 % Wave-
\newcommand{\Rn}{{{\mathbb R}^n}}                               % R^n
\newcommand{\cO}{{\mathcal O}}                                  % \mathcal O
\newcommand{\cC}{{\mathcal C}}                                  % Cauchyfl.
\newcommand{\cCb}{{{\mathcal C}_b}}                             % Umgebung
\newcommand{\Tr}{\text{Tr}}                                     % Trace
\newcommand{\Gp}[1]{\sideset{^{\mathcal C}}{^+_{#1}}{G}}        % Supercharge
\newcommand{\bfm}[1]{\text{\boldmath $#1$}}                     % boldmath
\def\bbbR{{\mathbb R}}                                          % Reelle
\def\bbbZ{{\mathbb Z}}                                          % Ganze
\def\bbb1{{\bf 1}}                                              % Identitaet
\def\frakI{{\mathfrak 1}}                                       % Id fuer
\def\Phat{\mbox{$\hat{{\bfm{\mathcal P}}}_0$}}                  % Ueber-
\def\SL2C{\mbox{$SL(2,{\mathbb C})$}}                           % SL(2,C)
\def\Lor{{\mathcal L}_0^\uparrow}                               % Eigentliche
\def\bimply{\Leftrightarrow}
\begingroup\makeatletter
\def\x#1#2#3#4#5#6#7\relax{\def\x{#1#2#3#4#5#6}}%
\expandafter\x\fmtname xxxxxx\relax \def\y{splain}%
\ifx\x\y   % LaTeX or SliTeX?
\gdef\SetFigFont#1#2#3{%
  \ifnum #1<17\tiny\else \ifnum #1<20\small\else
  \ifnum #1<24\normalsize\else \ifnum #1<29\large\else
  \ifnum #1<34\Large\else \ifnum #1<41\LARGE\else
     \huge\fi\fi\fi\fi\fi\fi
  \csname #3\endcsname}%
\else
\gdef\SetFigFont#1#2#3{\begingroup
  \count@#1\relax \ifnum 25<\count@\count@25\fi
  \def\x{\endgroup\@setsize\SetFigFont{#2pt}}%
  \expandafter\x
    \csname \romannumeral\the\count@ pt\expandafter\endcsname
    \csname @\romannumeral\the\count@ pt\endcsname
  \csname #3\endcsname}%
\fi
\endgroup
\begin{document}
\title{The stress energy tensor of a locally
       supersymmetric quantum field on a curved spacetime}
\author{\vspace*{3cm}
        Dissertation\\
        zur Erlangung des Doktorgrades\\
        des Fachbereichs Physik\\
        der Universit\"at Hamburg\\[3.5cm]
        vorgelegt von\\
        Martin K\"ohler\\
        aus Berlin\\[3cm]
        Hamburg\\
        1995 \\
        {\tt DESY 95-080 \hfill hep-th/9505014}
        }
% Datum %%%%%%
%\date{\today}
%%%%%%%%%%%%%%
\maketitle
%%%%%%%%%%%%%%
\vspace*{\fill}
\begin{tabbing}
  Sprecher des Fachbereichs Physik \qquad \= Prof.~Dr.~B. Kramer      \kill{}
  Gutachter der Dissertation: \> Prof.~Dr.~K. Fredenhagen      \\
                              \> Prof.~Dr.~D. Buchholz         \\[0.2cm]
  Gutachter der Disputation:  \> Prof.~Dr.~K. Fredenhagen      \\
                              \> Prof.~Dr.~H. Nicolai          \\[0.2cm]
  Datum der Disputation:      \> 4.~April~1995                 \\[0.2cm]
  Sprecher des Fachbereichs Physik \>                          \\
  und Vorsitzender des        \>                               \\
  Promotionsausschusses:      \>  Prof.~Dr.~B. Kramer
\end{tabbing}
\newpage
\vspace*{\fill}
\begin{abstract}
  For an analogon of the free Wess-Zumino model on Ricci flat
  spacetimes, the relation between a conserved `supercurrent' and the
  point-separated improved energy momentum tensor is investigated and
  a similar relation as on Minkowski space is established. The
  expectation value of the latter in any globally Hadamard product
  state is found to be a priori finite in the coincidence limit if the
  theory is massive.
  On arbitrary globally hyperbolic spacetimes the `supercurrent' is
  shown to be a well defined operator valued distribution on the GNS
  Hilbertspace of any globally Hadamard product state. Viewed as a new
  field, all n-point distributions exist, giving a new example for a
  Wightman field on that manifold. Moreover, it is shown that this
  field satisfies a new wave front set spectrum condition in a non
  trivial way.\\[1cm]
  \language 1
  {\sc Zusammenfassung.}$\quad$ In dieser Arbeit wird ein Analogon zum
  freien supersymmetrischen Wess-Zumino Modell auf dem Hintergrund
  einer Ricci-flachen, global hyperbolischen Raumzeit betrachtet. Die
  zu diesem Modell geh\"orenden `Superstr\"ome' und deren Beziehungen
  zum erweiterten Energie Impuls Tensor werden untersucht. Es zeigt
  sich, da{\ss} sich in diesem Modell, wie im Minkowskiraum, der durch
  `Punktetrennung' regularisierte Energie Impuls Tensor mit Hilfe der
  `Superstr\"ome' erzeugen l\"a{\ss}t.  Ferner ergibt sich unter der
  Voraussetzung einer massiven Theorie, da{\ss} der Erwartungswert von
  letzterem auch bei Aufhebung der Regularisierung endlich bleibt.
  Ein weiteres Resultat dieser Arbeit ist, da{\ss} die `Superstr\"ome'
  auf beliebigen global hyperbolischen Raumzeiten operatorwertige
  Distributionen auf dem GNS-Hilbertraum eines jeden globalen
  Hadamardproduktzustandes definieren. Diese Distributionen erf\"ullen
  die \"ublichen Wightman Axiome und k\"onnen deshalb als Beispiele
  f\"ur neue Wightman Felder auf Mannigfaltigkeiten dienen.
  Insbesondere existieren alle n-Punkt Distributionen und diese
  erf\"ullen in nicht trivialer Weise eine neue
  Wellen\-fronten-Spek\-trums\-be\-din\-gung.
\end{abstract}
\vspace*{3cm}
\language 0
\tableofcontents
%
%%%%%%%%%%%%%%%%%%%%%%%%%%%%%%%%%%%%%%%%%%%%%%%%%%%%%%%%%%%%%%%%%%%%%%%%%
\chapter*{Introduction}
%%%%%%%%%%%%%%%%%%%%%%%%%%%%%%%%%%%%%%%%%%%%%%%%%%%%%%%%%%%%%%%%%%%%%%%%%
%
In the last decade there was remarkable progress in the construction
of unified theories for the fundamental forces of nature. The
unification of the electro-magnetic with the weak interaction summit
into the Salam-Weinberg theory, while attempts to incorporate strong
interaction, described by quantum chromodynamics, seem to achieve
success in the so-called `supersymmetric grand-unified theories'
(SUSY-GUTs). Ordinary GUTs, despite of their qualitative and
quantitative success, for instance equality of different coupling constants
at high energies, disparity of quark-lepton masses at low energy,
determination of the electro-weak mixing-angle, etc., unfortunately
contain baryon and lepton number violating interactions, which
contradict the presently observed stability of the proton (lifetime
$>10^{32}$~years). On the other hand the addition of supersymmetry
leaves the standard success of the ordinary GUTs more or less intact
and predicts $10^{32\pm{}2}$~years for the lifetime of the proton,
which is compatible with the lower experimental bound mentioned above.
However, an incorporation of gravity is still one of the greatest
challenges of modern physics. In the last years there were an
increasing number of approaches to this problem which look promising
and the question arose whether one can say anything about the
influence of gravity on quantum phenomena. In some sense this
situation is comparable to the one encountered just before the
discovery of quantum electrodynamics. In those days of quantum
mechanics, one treated the electro-magnetic field in the calculation
as an external classical potential, while the matter was already
described quantum mechanically. The resulting theory, which one would
call a semiclassical approximation nowadays, already gave some results
which were later shown to be in complete agreement with the fully
quantized theory, i.e., quantum electrodynamics. On the other hand
those calculations showed also that a semiclassical theory leads to
inconsistencies if it is stressed too far.
One may therefore hope that a similar semiclassical description is
also valid for gravity.  Quantum field theory on curved spacetimes
(QFT on CST) provides such a framework: The spacetime is described
classically by a manifold together with a metric field, while all the
matter is described as quantized fields, which propagate on that
spacetime. The attempt to translate the considerations on the domain
of validity from quantum electrodynamics (QED) to QFT on CST indicates
that the latter is expected to be huge, namely from the macroscopic
scale down to the order of the Planck length $(\approx
10^{-33}\,\text{cm})$ or Planck time $(\approx
10^{-44}\,\text{s})$ respectively. This huge domain should leave
enough room for a fruitful semiclassical theory. However, there is a
subtle problem with such naive comparison of QED and QFT on CST: Due
to the equivalence principle, gravitational energy couples equally
strong to geometry, as does all other matter. Hence gravity will enter
in a non trivial way at {\em all\/} scales.  Fortunately one may still
succeed in a semiclassical description of gravity.
To see the argument, recall that in classical general relativity
gravitational waves are described as small perturbations of the
background metric. These perturbations can be treated as a null fluid
and their contribution to the l.h.s.\ (i.e., to the geometrical side)
of the Einstein equation can be separated off from the (geometrical)
rest in such a way that they may be treated as part of the source,
i.e., as part of $T^{ab}$. It seems therefore reasonable to suppose
that the `graviton' field, representing linearized perturbations of
the background spacetime, can be included along with all other fields
as part of the matter rather than geometry. This approximation should
work so long as one stays clearly away from the Planck scale.
Due to the weakness of gravitational effects, possible direct
applications of this framework seemed to lie in the early cosmology or
in the description of microscopic black holes. This situation changed
when Hawking during his studies of quantum black holes in
1975~\cite{hawking:75} discovered the by now famous (macroscopic)
black hole radiation, opening a broad new area of physics. Although
this radiation is still too weak to be detected experimentally (e.g.,
the temperature of a black hole of a mass ten times that of
our sun is approximately $10^{-8}$~K, which is eight orders of
magnitude below the the temperature of the cosmic background
radiation), his result is remarkable for at least two reasons. First,
it seems to be universal in the sense that it can be established using
very general assumptions on the quantum fields
involved~\cite{fredenhagenHaag:90} and it has been reproduced in
various different ways. Second, it establishes and fixes the close
relation between black holes and thermodynamics, which had been
suggested already before the application of quantum field theory to
black hole physics.
Hawking's radiation corresponds to a continuous energy-flow from the
black hole to infinity and, at least due to energy conservation, should
be accompanied by a mass-loss of the former, i.e., the black hole
should evaporate.  For a precise mathematical formulation of this
back-reaction problem, a local stress energy tensor for the quantum
field is needed. This object, denoted by $T^{ab}$, is important for
another reason, too: Since the work of Unruh~\cite{unruh:76} it is
known that the traditional particle concept of Minkowski quantum field
theory is not suitable for quantum fields on curved spacetimes.
Instead the physical situation in the latter case should be described
in a more operational way, i.e., in terms of detectors and counting
rates. Such a description is provided by a local stress tensor.
Following Hawking's paper, in the mid seventies, there were undertaken
enormous efforts on the study of $T^{ab}$ and its expectation value in
some states $\omega$ ($<T^{ab}>_\omega$). Unfortunately the classical
expression for the stress energy tensor in the cases of the standard
fields (e.g., scalar, Dirac, photon, etc.) yields divergent results
after quantization, because products of field operators at the same
point occur. One therefore has to subtract out the divergences in some
way, i.e., to regularize and renormalize. The various renormalization
techniques which were developed in that time all gave a locally
conserved expectation value for the stress tensor, which therefore is
a good candidate for the r.h.s.\ of the semiclassical Einstein
equation, even if particle creation due to gravitational effects
occur.  Wald was even able to formulate a set of
axioms~\cite{wald:77,wald:78} which almost uniquely fixes these
renormalized expectation values for every state under consideration.
Despite of their success these methods are still burdened with some
fundamental difficulties. The first one concerns renormalization. In
Minkowski space one agrees on the fact that only energy {\em
  differences\/} are measurable. Hence any energy renormalization
should not change the physical content of the theory.  On the other
hand in classical general relativity the {\em absolute\/} value of the
energy curves spacetime. This means that energy renormalization could
change the physical content of the theory in that case. To avoid
renormalization, one could for example use a mathematical framework
which allows products of field operators at the same spacetime point
to be well defined. The new generalized functions of
Colombeau~\cite{colombeau:NGF}, which can be extended to
manifolds~\cite{koehler:91}, could be used to implement such a setting.
However, at the level of interpretation such a scheme is believed to be
in fact equivalent to the renormalization prescriptions mentioned
above. It follows that a model which has a finite stress energy tensor
a priori would be highly desirable. In fact such a model is already
discussed by Unruh~\cite{unruh:76} in 1+1 dimensions, but apparently
his ansatz was never extended to the four dimensional case.
Meanwhile one did discover on Minkowski space a whole class of models
with such a remarkable finiteness property. These models possess,
besides the usual Poincar{\'e} symmetry, an internal symmetry, the
generators of which anti-commute with those of the Poincar{\'e} group.
The resulting algebra for all generators turned out to be a graded
Lie-algebra, which is now called a supersymmetric extension of the
Poincar{\'e} algebra. Consequently, these models were summarized under
the name supersymmetric models. One of their fundamental properties is
the fact that they possess equally many bosonic and fermionic degrees
of freedom and all particles of one supermultiplett must have equal
masses\footnote{We do assume the supersymmetry to be unbroken.}.
Moreover, the stress energy tensor of these models can be expressed
locally in terms of spinor currents~\cite{ferraraZumino:75}, a
representation which manifestly shows a priori finiteness of the
energy. Wess and Zumino~\cite{wess_zumino:74} gave the first field
theoretical example of a global supersymmetric field theory in 1974
and the subject evolved rapidly from thereon. On the other hand it is
evident from experiments that (unbroken) supersymmetry is {\em not\/}
realized in nature. To solve this contradiction, the mechanism of
spontaneous symmetry breaking was introduced into the framework, but
it must be said that even today there are only hints on supersymmetry,
but no experimental verification.
It is well known that supersymmetry --together with general
covariance-- leads unavoidably to supergravity
theories~\cite{nieuwenhuizen:81}, hence leaves the frame of QFT on
CST. The only known non trivial example of a spacetime which admits a
supersymmetric extension of its isometry group is the anti-DeSitter
solution of Einstein's equation~\cite{breitenlohner_freedman:82}.
Unfortunately this manifold is not globally hyperbolic and this leads
to problems during quantization, namely boundary conditions at
infinity are needed. On the other hand if one is interested in a
cancellation of the divergences of the stress energy tensor only, a
model with the same {\em particle contents\/} as a `real'
supersymmetric model should suffice.
Coming back to the renormalization of the stress tensor on a curved
spacetime, a second difficulty is a direct implication of one of the
major, still unsolved problems of QFT in CST.  Namely the
characterization of physical admissible states. (Recall that the
expectation value of $T^{ab}$ must be calculated in some state
$\omega$.) The problem occurs, since field theories have infinite many
degrees of freedom and hence there exist inequivalent representations
of the canonical (anti) commutation relations. It follows that the
usual Hilbertspace approach, where all physical admissible states are
identified with positive trace class operators, is insufficient in
general. Instead one should use the algebraic approach to quantum
field theory, to which Haag's book~\cite{haag:alg} provides a
good introduction. The basic objects in this approach are the
`algebra of local observables' which, roughly speaking, describes
elementary events, together with states, which describe the
preparation of the physical system or equivalently the probability for
subsequent events. It is worth noting that every choice of a state on
the algebra of local observables uniquely fixes, via the GNS
reconstruction theorem, a Hilbertspace and a `vacuum' vector, on which
the observables act as bounded linear operators. For linear scalar and
spinor fields on globally hyperbolic spacetimes, the algebra of local
observables, satisfying the axioms of algebraic quantum field theory,
was successfully constructed by Dimock in the
1980's~\cite{dimock:80,dimock:82}.
In the exceptional case of Minkowski space, there exists an --usually
unique-- Poincar{\'e} invariant state, satisfying the spectrum
condition, i.e., the vacuum, which is used to construct all other
physical relevant states for particle physics. Unfortunately both,
Poincar{\'e} invariance and spectrum condition rely on the globally
flat structure of Minkowski space; thus they can not be extended
naturally to generic manifolds. On the other hand for free fields the
class of the so-called Hadamard states seems to be a suitable
replacement for the vacuum. Those states are characterized locally by
the short distance behavior of their two-point
distributions~\cite{fullingSweenyWald:78}.  It was shown
in~\cite{Verch:94} that they satisfy the condition of local
definiteness due to Haag, Narnhofer and Stein~\cite{haagnarnhofer:84}
and that their local von Neumann algebras are factors of type
$\text{III}_1$ (see also~\cite{wollenberg:92b}).  Finally Hadamard
states are precisely those states for which the expectation value of
the stress energy tensor can successfully be renormalized.  We should
also mention another class of states, which is believed to be
equivalent in the domain of application, namely the adiabatic vacuum
states, which were closely investigated by L{\"u}ders and
Roberts~\cite{luedersRoberts:90}
This work is organized as follows: Following this Introduction in
chapter~\ref{sec:SUSYinR4} the ideas to represent the stress energy
tensor in terms of supercurrents is motivated and some basic
properties of the free Wess-Zumino model, which is the simplest
supersymmetric model on Minkowski space, are reviewed. The classical
settings for this model are described and a simple argument is
presented, which shows that its energy operator is a priori normal
ordered after quantization. In the next chapter the basic notation for
the definition of spinors on Lorentz manifolds is established. We
expose the relation between topological properties of the spacetime
and the existence of classical spinor fields.
Chapter~\ref{sec:analogonWess} describes our analogon of the free
Wess-Zumino model. We begin with the classical settings and introduce
spinor currents which shall later generate the stress energy tensor.
The assumption that these currents should be locally conserved
restricts the underlying spacetime to have vanishing scalar curvature.
For the convenience of the reader this classical setting is followed
by a brief introduction to the algebraic approach to quantum field
theory.  In a next step the set of admissible states is fixed to
consist of Hadamard states only. We give a precise definition of these
states due to Kay and Wald~\cite{kay:91} and extend their definition
for scalar fields to fermions.  Unfortunately this condition is quite
complicated and difficult to check in actual calculations. Moreover,
various questions concerning multiplication of Hadamard distributions
and their restriction to hypersurfaces are almost impossible to deal
with in this setting. Fortunately H{\"o}rmander and Duistermaat
developed the mathematical theory of Fourier integral
operators~\cite{Hoermander:71,Hoermander:72}, in which the concept of
wave front sets occur. In terms of wave front sets the questions
mentioned above can be answered quite easily.  Radzikowski was the
first who gave a characterization of Hadamard states by their wave
front sets~\cite{Radzikowski:92}. However, his proof as it stands is
only valid in Minkowski space. The gap in his argument is filled in
section~\ref{sec:wfGlobHadam} of chapter~\ref{sec:analogonWess}, where
the wave front set of any Hadamard state on an arbitrary globally
hyperbolic Lorentz manifold is calculated. We close this chapter by
proposing a general wave front set spectrum condition, which might
serve in the future as a replacement for the usual spectrum condition
on manifolds.
In chapter~\ref{sec:emtCurved} the stress energy tensor of our model
is calculated. The quantized versions of the spinor currents defined
in chapter~\ref{sec:analogonWess} are shown to yield to new Wightman
fields on the spacetime, whose n-point distribution satisfy our new
wave front set spectrum condition. Mimicking the situation on
Minkowski space, we {\em define\/} a candidate for the energy momentum
tensor restricted to a spacelike hypersurface ${\mathcal C}$ in terms of
the spinor current. The result, which is a distribution {\em on ${\mathcal
    C}$\/} can be expressed in terms of the basic fields and equals
the point-separated `true' stress energy tensor if and only if the
spacetime is Ricci flat, i.e., if it is a vacuum solution of the
Einstein equation. Using the results of
Christensen~\cite{chris:76,chris:78}, it is shown next that the
expectation value of our point-separated stress energy tensor, with
respect to {\em every\/} Hadamard product state has finite coincidence
limit after elementary averaging, i.e., it defines a (locally
conserved) tensor field on the spacetime. As a consequence of this a
priori normal ordering feature, the stress energy tensor turns out to
be a second new Wightman field on the spacetime.  We close the work
with two examples: On the Schwarzschild and Kerr solutions of the
Einstein equation the finite terms of our energy momentum tensor in
the DeWitt-Schwinger pseudo-state are calculated.
%
%%%%%%%%%%%%%%%%%%%%%%%%%%%%%%%%%%%%%%%%%%%%%%%%%%%%%%%%%%%%%%%%%%%%%%%%%
\chapter{The Wess Zumino model on Minkowski spacetime}\label{sec:SUSYinR4}
%%%%%%%%%%%%%%%%%%%%%%%%%%%%%%%%%%%%%%%%%%%%%%%%%%%%%%%%%%%%%%%%%%%%%%%%%
%
A field theoretical realization of the super-Poincar\'e algebra
yielding its lowest dimensional representation was discovered and
studied by J. Wess and B. Zumino in 1974~\cite{wess_zumino:74}.  This
model describes two spin-0 fields ($A,B$), one of which is scalar
($A$), the other ($B$) is pseudo-scalar and a third ($\Psi$) which is a
Majorana spinor field. In this chapter we describe some features of
this model, which motivate our subsequent studies on curved
spacetimes. To establish the mathematical settings, we start with an
introduction to classical spinors on Minkowski space. The classical
Lagrangian of the Wess-Zumino model is given next and its internal
symmetries are investigated. We close the chapter by showing that the
energy operator in the quantized version of this supersymmetric theory
is a priori normal ordered and obtain a formula for the
energy-momentum tensor in terms of suitable spinor currents.
%
%%%%%%%%%
\section{Classical spinors in Minkowski spacetime}
%%%%%%%%%
%
Spinors arise most naturally in quantum field theory on Minkowski
spacetime. It is even possible to show that all classical (e.g.\
tensorial) objects can be constructed using spinors and some authors
even believe that spinors are the most fundamental objects in physics
(See~\cite{pen:twist1}). This statement clearly shows the importance
of the concept of spinors, even if one does not follow Penrose's and
Rindler's extreme point of view. This section starts with a
motivation, which is followed by the definition of classical spinor
fields on Minkowski space. The notion of Dirac- and charge conjugation
is introduced next and the definition of Majorana spinors closes the
section.
Consider a spacetime $(M,g_{ab})$, which we assume to be a four
dimensional $C^\infty$-manifold $M$ together with a Lorentz metric
$g_{ab}$ with signature $(+,-,-,-)$. Let $\mathcal S$ denote the set of
all physical states on this spacetime. At this stage, it is not
necessary to specify ${\mathcal S}$ more precisely. However, we assume the
elements of ${\mathcal S}$ to be characterized {\em locally\/}; the usual
tensor fields on $M$ are an example. The principle of `special
covariance', i.e., every physically admissible result of a measurement
which is performed by an observer ${\mathcal O}$ must also be a physically
admissible result of a measurement performed by a second observer
$\tilde{{\mathcal O}}$ who is related to ${\mathcal O}$ by an isometry,
induces for every isometry ${\varrho}$ of the spacetime a map on the
physical states. This map $\tilde{\varrho}:{\mathcal S} \rightarrow {\mathcal
  S}$ is fixed by the requirement that every state
$\tilde{\varrho}(\omega)$ $(\omega\in {\mathcal S})$ gives the same
results of measurement for the observer $\tilde{\mathcal O}$ as $\omega$
for ${\mathcal O}$.  The group of isometries for Minkowski space $({\mathbb
  R}^4, \eta_{ab})$ is the well known extended Poincar\'e group
(${\mathcal P}$). The experiments of Wu~(\cite{wu}) (violation of parity)
and Fitch~(\cite{fitch}) (violation of time-reversal if the
CPT-Theorem is valid) however indicate that the physical laws on
Minkowski space are specially covariant under the restricted
Poincar\'e group (${\mathcal P}_0$) only. We therefore restrict ourself to
this subgroup.
It was shown in 1939 by Wigner~\cite{wig:39} that the assumption that
$\tilde{\varrho}$ does not change the transition amplitudes for pure
states in quantum mechanics\footnote{It is assumed in the proof that
  pure states are represented as one-dimensional projectors on a
  separable Hilbertspace.} implies that $\tilde{\varrho}$ is a
projective representation of the proper Poincar\'e group.  Moreover
Wigner showed that adding the requirement that the representation
should depend continuously on the group elements fixes this
representation up to sign. As a next step it was proved by
Bargman~\cite{barg:54} in 1954 that those representations are in
one-to-one correspondence to the true representations of the covering
group \Phat{} of the proper Poincar\'e group.
Every unitary representation of \Phat{} can be decomposed into a
direct sum (or a direct integral) of irreducible representations. To
study the former it is therefore sufficient to classify all
irreducible representations. The latter can be classified partially by
the mass ($m^2$) and the sign of the energy ($\epsilon$).\footnote{Let
  $P^\mu$ and $M^{\lambda \nu}$ denote the generator of the
  representation.  The three invariants mass ($m^2$), sign of the
  energy ($\epsilon$) and spin ($s$) are given by: $m^2 = P_\mu P^\mu
  = P^2$; $\epsilon = \text{sign} (P^0)$; $-m^2(s^2 +s) = W^2$, where
  $W = \epsilon_{\mu \nu \lambda \rho}P^\nu M^{\lambda \rho}$ is the
  Pauli-Lubanski vector.} If $m^2 >0$ and $\epsilon=+1$ one further
invariant $s$, called spin, is needed to fix the representation up to
unitary equivalence. It can be shown (see e.g.~\cite{bog:axiom}) that
all irreducible physical representations of \Phat{} can be constructed
in the space of spinor wave functions (see below). We remark that this
space carries a representation of \Phat{} by construction. Moreover
there exists a positive definite \Phat{} invariant form such that the
equivalence classes of the spinor wave functions with respect to this
form span a pre-Hilbertspace (see \cite[p.\ 285]{bog:axiom} for the
details).  Completing this space yields a Hilbertspace which carries
the unitary representation of \Phat{} we are looking for.
\begin{rem}
  Note that the Hilbertspace mentioned above is a one-particle
  Hilbertspace. The quantum statistics will not be implemented until
  the introduction of a many-particle Hilbertspace (Fockspace).
\end{rem}
\subsection{\SL2C--spinors}
In this subsection the space of spinor tensor fields is constructed
using the two component formulation of Penrose
\&~Rindler~\cite{pen:twist1}. This formulation is more in the spirit
of the introduction than the usual four component Dirac approach which
will be used later on exclusively, since two component
\SL2C{}--spinors transform according to an {\em irreducible\/}
representation of the homogeneous part of \Phat{} and there is a close
relation between spinor tensors and spinor wave functions.
We note first that \Phat{} is isomorphic to $ISL(2,{\mathbb C})$, the
semidirect product of the group of translations with the group of
linear transformations with unit determinant acting on a two
dimensional complex Hilbertspace. Let $(W,\epsilon_{xy})$ be such a
Hilbertspace together with a volume form $\epsilon_{xy}$.
Then the homogeneous part of $ISL(2,{\mathbb C})$ in its defining
representation leaves the form $\epsilon_{xy}$ invariant. The pair
$(W,\epsilon_{xy})$ is called a {\em spinor space}, its elements are
called {\em{}spinors}.
Let $(W^*,\epsilon^{xy})$ be the dual space of $W$,
$(\bar{W},\bar{\epsilon}_{x'y'})$ the space of all anti-linear maps
form $W^*$ into $\mathbb C$. The map
\begin{equation*}
  \begin{split}
    \bar{\phantom{f}} : W & \rightarrow \bar{W} \\
         \xi^x           & \mapsto  \bar{\xi}^{x'} : ( \nu_x
                            \mapsto \overline{\xi^x\nu_x})
  \end{split}
\end{equation*}
is called {\em complex conjugation\/} and induces a mapping between
$\epsilon_{xy}$ and $\bar{\epsilon}_{x'y'}$.  To clarify the notation
we denote all elements of $\bar{W}$ with upper primed indices. The
dual space of $\bar{W}$ is denoted by
$(\bar{W}^*,\bar{\epsilon}^{x'y'})$. The elements of $W^*$
($\bar{W},\bar{W}^*$) are called dual (adjoint, adjoint dual) spinors.
The space of spinor tensors ${\bf T}$ is the tensor product of these
spaces. Note that the relative order of primed and unprimed indices is
irrelevant, i.e., $v^{xx'y}$ and $v^{xyx'}$ are equivalent, since $W
\otimes \bar{W} \otimes {W}$ and $W \otimes W \otimes \bar{W}$ are
naturally isomorphic.  However $v^{xyx'}$ and $v^{yxx'}$ are different
in general.  Smooth mappings from the Minkowski space into ${\bf T}$
are called {\em spinor tensor fields}.  An action of
$ISL(2,{\mathbb C})$ on these mappings can be defined by brute force as
follows: To any $g \in ISL(2,{\mathbb C})$ we associate the
transformation
\begin{equation}\label{eq:slTransform}
  \xi^x(q) \mapsto {L^x}_{y} \xi^y[{\mathcal P}^{-1}(q)],
\end{equation}
where ${L^x}_y \in SL(2,{\mathbb C})$ is the homogeneous part of $g$,
${\mathcal P}$ denotes the associated Poincar\'e element and $q$ is a
point in Minkowski space.
\begin{rem}
  $\phantom{~}$\par
  \begin{itemize}
  \item This transformation is a true representation of \Phat{}, while
    for the Poincar\'e group itself it is a representation up to sign
    only.
  \item The representation given by Eqn.~(\ref{eq:slTransform}) is
    {\em not\/} irreducible, since spinor tensors are defined on the
    configuration space. However it has the advantage that one can
    identify certain spinor tensors with ordinary tensors. To obtain
    all {\em irreducible\/} representations mentioned above, one has
    to introduce spinor wave functions, which are defined on momentum
    space together with an appropriate transformation law: For any non
    negative integer or half integer $j$, consider the symmetric
    tensor product of $2j$ copies of $W$ together with the self
    representation of $\SL2C$ and its complex conjugate. Mappings from
    the momentum space, which is isomorphic to ${\mathbb R}^4$, into one
    of these tensor products are spinor wave functions. A
    transformation law, which is identical to
    Eqn.~(\ref{eq:slTransform}), up to a factor containing the
    inhomogeneous part of the corresponding element $g\in ISL(2,{\mathbb
      C})$, finishes the construction. For more details, e.g., the
    explicit form of the transformation, for the proof of the claimed
    irreducibility as well as for the proof of the statement that all
    irreducible representation of \Phat{} can be obtained in this way,
    the reader is referred to the book of Bogolubov loc.\ cit.\ and
    the references therein.
  \end{itemize}
\end{rem}
We establish now the relation between so called real spinor tensor
fields and vector fields on Minkowski space. Let $(W,\epsilon_{xy})$
be a spinor space and let us denote the tensor product of $W$ with
$\bar{W}$ by $Y$. Complex conjugation maps $Y$ into itself. Those
elements in $Y$ which are invariant under complex conjugation are
called {\em real spinor tensors\/} and an analogous definition is
valid for spinor tensor fields.
\begin{rem}
  The term `real' does not make sense for spinors, i.e., for elements
  of $W$. To see this consider the case $W={\mathbb C}^2$, $z \in {\mathbb
    C}^2$. The definition $\text{Re}(z) := 1/2 (z + \bar{z})$
  obviously is {\em not\/} invariant under the action of $SL(2,{\mathbb
    C})$.
\end{rem}
The real elements of $Y$ span a (real) four dimensional subspace of
$Y$, which we denote by $V$. Let $\sigma^x$, $\iota^x$ be a basis in
$W$ with the property $\sigma^x \iota^y\epsilon_{xy}=1$. Then
\begin{eqnarray}   \label{eq:spinONB1}
  t^{xx'}  &  =  &  \frac{1}{\sqrt{2}}(\sigma^x\bar{\sigma}^{x'}
                                        + \iota^x\bar{\iota}^{x'}) \\
  x^{xx'}  &  =  &  \frac{1}{\sqrt{2}}(\sigma^x\bar{\iota}^{x'}
                                        + \iota^x\bar{\sigma}^{x'}) \\
  y^{xx'}  &  =  &  \frac{i}{\sqrt{2}}(\sigma^x\bar{\iota}^{x'}
                                        - \iota^x\bar{\sigma}^{x'}) \\
  z^{xx'}  &  =  &  \frac{1}{\sqrt{2}}(\sigma^x\bar{\sigma}^{x'}
                                        - \iota^x\bar{\iota}^{x'})
\label{eq:spinONB4}
\end{eqnarray}
are a basis in $V$. Moreover  $g_{xx'yy'}:=
\epsilon_{xy}\bar{\epsilon}_{x'y'}$ defines a Lorentz metric on $V$
(For the proof calculate explicitly that
(\ref{eq:spinONB1})--(\ref{eq:spinONB4}) are an orthogonal basis for
$g_{xx'yy'}$). Finally consider a tetrad field $(t^a,x^a,y^a,z^a)$ on
Minkowski space. The following hybrid vector spinor field
\[
\sigma^a_{xx'} := t^a t_{xx'} - x^a x_{xx'} -
y^a y_{xx'} -z^a z_{xx'}
\]
defines for every point $q\in {\mathbb R}^4$ a vector space isomorphism
between the tangential space $T_q{\mathbb R}^4$ at this point and $V$.
Note that we lowered the indices with the metric $g_{xx'yy'}$.  This
isomorphism respects the metric structure ($\eta^{ab} \equiv
{\sigma^a}_{xx'} {\sigma^b}_{yy'} g^{xx'yy'}$) and the action of the
Poincar\'e group (See~\cite{wald:gr}).  Thus having chosen the map
${\sigma^a}_{xx'}$ we can identify real spinor tensor fields with
vector fields.
\subsection{Dirac spinors}
While \SL2C{}--spinors simplify the calculations in most cases since
no Dirac $\gamma$--matrices occur, the study of uncharged fermions is best
done using the Dirac description. Consider the direct sum of two
spinor spaces $(W,\epsilon_{xy})$ and
$(\bar{W},\bar{\epsilon}_{x'y'})$. Denoting this space by $E$, $E$
carries a (reducible) representation of \SL2C{} inherited from $W$ and
$\bar{W}$ respectively:
\begin{equation}\label{eq:repr_SL2C}
  \pi : L \mapsto
                   \left(
                     \begin{matrix}
                     L & 0 \\
                     0 & \bar{L}
                   \end{matrix}
                   \right)
                   \equiv S \qquad \forall L \in \SL2C
\end{equation}
%
% Laut Streater Wightman muesste da noch ein \xi hin (Vorzeichen?)
% L^*-1 ist uebrigens die die dual complex conj Darstellung (von
% \bar{W}^*
%
Recall that $\bar{W}$ carries the complex conjugate representation of \SL2C.
$E$ is called {\em Dirac spinor space\/} and the image of $\SL2C$
under $\pi$ is naturally isomorphic to the group $\mbox{Spin}_0$ (See
below). Using the volume forms $\epsilon_{xy}$ and
$\bar{\epsilon}_{x'y'}$, the $\gamma$-matrices can be written in the
following $4 \times 4$ matrix representation
(See~\cite[p.~460]{pen:twist1}):
\[
{{\gamma_a}^A}_B = \sqrt{2}
                      \left(
                          \begin{matrix}
                            0 & {\sigma_a}^{zz'}
                            {\epsilon_z}^x \bar{\epsilon}_{z'y'}\\
                            {\sigma_a}^{zz'}
                            {\mbox{$\bar{\epsilon}$}_{z'}}^{x'}
                            \epsilon_{zy}&0 \\
                          \end{matrix}
                       \right),
\]
where $A= x \oplus x'$ and $B= y \oplus y'$, i.e., both indices $A$
and $B$ can take the four values $11'$, $12'$, $21'$ and $22'$
respectively. An elementary but lengthly calculation verifies that the
Clifford relations are satisfied:
\begin{equation}
  \gamma_a \gamma_b + \gamma_b \gamma_a = 2 \eta_{ab} \identity,
  \label{eq:dirac_clifford}
\end{equation}
where $\eta_{ab} = \mbox{diag} (+1, -1 ,-1,-1)$ is the Minkowski
metric and $\identity$ is the identity on $E$. We remark that
$\mbox{Spin}_0$ mentioned above is the connected component of the
identity in the group Spin which consists of all $4
\times 4$ matrices $S$ satisfying
\[
S \gamma^a S^{-1} = {\Lambda^a}_b \gamma^b
\]
for some real constant ${\Lambda^a}_b$. The elements of $E$ are called
{\em Dirac spinors}.  The elements of the dual vector-space of $E$,
which is called the Dirac cospinor space ($E^*$), are called {\em
  Dirac cospinors}. Finally smooth mappings from Minkowski space into
$E$ ($E^*$), denoted by $\Gamma(E)$ ($\Gamma(E^*)$), are called {\em
  Dirac spinor fields\/} ({\em Dirac cospinor fields\/}). By analogy
to the previous subsection, an action of $I\SL2C$ on these fields is
given by the transformation
\begin{align}
  \Psi^A(q) & \mapsto {\Psi'}^A(q) = {S^A}_B \Psi^B({\mathcal P}^{-1}(q))
 \qquad
  \forall g \in I\SL2C, \Psi^A \in \Gamma(E), \\
  \intertext{and}
  \bar{\Psi}_A(q) & \mapsto {\bar{\Psi}'}_A(q) = \bar{\Psi}_B (
  {\mathcal P}^{-1}(q)) {{S^{-1\,B}}_A} \qquad \forall g \in I\SL2C,
 \bar{\Psi}_A
  \in \Gamma(E^*),
\end{align}
where ${S^A}_B$ is the homogeneous part of $g$ in the representation
given by Eqn.~(\ref{eq:repr_SL2C}), ${\mathcal P}$ denotes the Poincar\'e
 element
associated to $g$ and $q$ is some point in Minkowski space. Since $E$
and $E^*$ are dual to each other, we can equip the vector spaces
$\Gamma(E)$ and $\Gamma(E^*)$ with an (antisymmetric) wedge product
$\wedge$. More precisely: For $\Psi^A \in \Gamma(E)$ and $\bar{\Psi}_A
\in \Gamma(E^*)$ we define
\begin{align}
  \bar{\Psi}_B \wedge \Psi^A :
  \left(\Gamma(E) \oplus \Gamma(E^*)\right) \otimes
    \left(\Gamma(E) \oplus \Gamma(E^*)\right)
  & \mapsto \Cinf({\mathbb R}^4) \nonumber\\
  \binom{{\phi_1}^B}{\bar{\phi}_{1A}} \otimes
  \binom{{\phi_2}^B}{\bar{\phi}_{2A}} \qquad
  &
  \mapsto
  \bar{\Psi}_B{\phi_1}^B \cdot \Psi^A \bar{\phi}_{2A} \nonumber\\
  & \phantom{\mapsto}
  -
  \bar{\Psi}_B{\phi_2}^B \cdot \Psi^A \bar{\phi}_{1A} \\
  \intertext{Moreover,}
  \bar{\Psi}_B(q) \wedge \Psi^{A}(q') :
  \left(\Gamma(E) \oplus \Gamma(E^*)\right) \otimes
    \left(\Gamma(E) \oplus \Gamma(E^*)\right)
  &  \mapsto {\mathbb C} \nonumber\\
  \binom{{\phi_1}^B}{\bar{\phi}_{1A}} \otimes
  \binom{{\phi_2}^B}{\bar{\phi}_{2A}} \qquad
  &
  \mapsto
  \bar{\Psi}_B{\phi_1}^B(q) \cdot \Psi^{A}\bar{\phi}_{2A}(q') \nonumber\\
  & \phantom{\mapsto}
  -
  \bar{\Psi}_B{\phi_2}^B(q) \cdot \Psi^{A} \bar{\phi}_{1A}(q')
\end{align}
is well defined. In the future we will omit the $\wedge$ and simply
write $\bar{\Psi}_B(q)\Psi^A(q')$ or even
$\bar{\Psi}_{B}\Psi^{A'}$.The prime reminds of the fact that $A'$ is a
Dirac spinor index referring to the point $q'$. We obviously have
\[
\{ \bar{\Psi}_{B},\Psi^{A'} \}_+=\{ \bar{\Psi}_{B},\bar{\Psi}_{B'} \}_+=
\{ \Psi^{A},\Psi^{A'} \}_+ = 0,
\]
where $\{\cdot,\cdot\}_+$ denotes the anticommutator. In this sense
classical Dirac spinor- and cospinor fields are {\em anti-commuting\/}
objects. Note that $\bar{\Psi}_{B}\Psi^{A'}$ transforms as a cospinor
at $p$ and as a spinor at $p'$. Such quantities are usually called
bi-spinors (See~\cite{DeWitt:60} for further details).
Classical Dirac spinors without further assumptions are suitable for
models describing charged fermionic fields. We are now going to define
neutral spinor fields which we need later for the definition of the
Wess-Zumino model.  In the literature neutral spinor fields often are
called Majorana spinor fields.
We begin with the implementation of a charge conjugation.  Majorana
spinor fields are the elements of $\Gamma(E)$ which are invariant
under this mapping.
%
%{Dirac conjugation}
%
Consider a fixed but arbitrary four dimensional matrix representation
of the Dirac $\gamma$-matrices. Since the adjoint representation also
satisfy Eqn.~(\ref{eq:dirac_clifford}) there exists a unique matrix
$\beta$ with unit determinant\footnote{In the literature one usually
  considers only those representations of the Dirac $\gamma$-matrices,
  where $\beta$ is identical $\gamma^0$} that intertwines between
these. The four maps
\begin{eqnarray*}
E_\Lambda : {\mathbb R}^4 &\rightarrow & E\\
                p      & \mapsto    & \omega_\Lambda,
\end{eqnarray*}
such that for for every point $p$ the four elements $\omega_\Lambda$,
$\Lambda=1,\ldots,4$ form a basis in $E$, define a {\em moving frame}.
The dual frame $(E^1,\ldots,E^4)$ consists of four maps $E^\Sigma$
from Minkowski space to $E^*$ satisfying $E^\Lambda(E_\Sigma)=
{\delta^\Lambda}_\Sigma$.  Every Dirac spinor field $\Psi$ (Dirac
cospinor field $\bar{\Psi}$) can be written as $\Psi=\Psi^\Lambda
E_\Lambda$ ($\bar{\Psi} = \bar{\Psi}_\Sigma E^\Sigma$), where
$\Psi^\Lambda$ ($\bar{\Psi}_\Sigma$) are complex valued functions on
Minkowski space called {\em components of\/} $\Psi$.  The Dirac
conjugation is given by the following definition:
\begin{Dfn}
  Let $\Psi^A\in \Gamma(E)$ be a Dirac spinor field with components
  $\Psi^\Lambda$.  We define the {\em Dirac conjugate spinor field\/}
  $\Psi^+_A\in \Gamma(E^*)$ by setting its components equal to
  \begin{equation}\label{eq:diracconjspinor}
    \Psi^+_\Sigma = \overline{\Psi^\Lambda} \beta_{\Lambda \Sigma}.
  \end{equation}
  Here $\overline{\phantom{f}}$ denotes complex conjugation and
  $\beta_{\Lambda \Gamma}$ are the matrix elements of $\beta$.
\end{Dfn}
\begin{Dfn}
  The Dirac adjoint of a Dirac cospinor field $\Psi^+ \in \Gamma(E^*)$ is
  defined by
  \begin{equation}\label{eq:diracconjcospinor}
    {(\Psi^+)}^+ :=
    {(\beta^*)}^{-1\Lambda\Sigma}\overline{\Psi^+_\Sigma} E_\Lambda
  \end{equation}
  where ${(\beta^*)}^{- 1\Lambda\Sigma}$ are the matrix elements of
  the adjoint matrix of $\beta^{- 1}$.
\end{Dfn}
For the proof that $\Psi^+$ and $(\Psi^+)^+$ defined by
Eqn.~(\ref{eq:diracconjspinor}) and (\ref{eq:diracconjcospinor})
respectively are well defined, observe first that
\begin{eqnarray*}
  {(\gamma_a)}^* & = & {(S^{-1} \gamma_b {\Lambda(S)^b}_a S )}^* \\
                 & = & S^* \beta \gamma_b \beta^{-1} {\Lambda(S)^b}_a
                       {(S^{-1})}^* \\
                 & = & S^* \beta S \gamma_a S^{-1} \beta^{-1} {(S^{-1})}^* \\
                 & \equiv & \beta \gamma_a \beta^{-1},
\end{eqnarray*}
since $\det(S^*\beta S) =+ 1$. Hence for all $S \in
\pi\left(SL(2,{\mathbb C})\right)$ we have $S^*\beta S=\beta$.
Under a change of frames which induces a change of
components $\Psi^\Lambda \mapsto {\Psi'}^\Lambda = {S^\Lambda}_\Sigma
\Psi^\Sigma$ one finds
\begin{eqnarray*}
  {(\Psi')^+}_\Sigma & := &
   \overline{ {\Psi'}^\Lambda } \beta_{\Lambda\Sigma}\\
  & = & {{\overline{S}^\Lambda}}_\Gamma
   \overline{\Psi^\Gamma} \beta_{\Lambda\Sigma} \\
  & = & \overline{\Psi^\Gamma}\; {{{\overline{S}}^T}_\Gamma}^\Lambda
   \beta_{\Lambda\Sigma}= \overline{\Psi^\Gamma} \beta_{\Gamma\Delta}
   {{S^{-1}}^\Delta}_\Sigma
   = {\Psi^+}_\Delta {{S^{-1}}^\Delta}_\Sigma
\end{eqnarray*}
since ${(\overline{S}^T)_\Gamma}^\Lambda = {(S^*)_\Gamma}^\Lambda$.
This shows that $\Psi^+:={\Psi^+}_\Sigma E^\Sigma$ is a frame
invariant object. For a cospinor field $\Psi^+$, locally given by
$\Psi^+ = \Psi^+_\Sigma E^\Sigma$, a similar calculation as above
shows that ${(\Psi^+)^+}$ transforms as a Dirac spinor field under a
change of frame, which coincides with $\Psi$.
The charge conjugation is defined next:
%
%[Charge conjugation]
%
The transposed Gamma matrices $\gamma^T$ satisfy the
anticommutation relation Eqn.~(\ref{eq:dirac_clifford}).
Hence there exists a unique $4 \times 4$ matrix $C$ with unit
determinant such that $C\gamma_a C^{-1} = - \gamma^T_a$. This matrix
is called {\em charge conjugation matrix}. Note that $C^T= -C =
C^{-1}$.
\begin{Dfn}
  For any Dirac spinor field
  $\Psi \in \Gamma(E)$ we define the {\em charge conjugate Dirac
    field\/} $\sideset{^c}{}\Psi$ by:
  \[
  \Gamma(E) \ni \sideset{^c}{}\Psi =
  {C}^{\Lambda\Sigma}\Psi^+_\Sigma E_\Lambda
  \]
  where ${C}^{\Lambda\Sigma}$ are the matrix elements of $C$
  and $\Psi^+_\Sigma$ are the components of the Dirac conjugate spinor
  field of $\Psi$.
\end{Dfn}
For the proof that this definition is sensible, use a similar argument
as above to show that for all $S \in \pi\left(SL(2,{\mathbb C})\right)$ we
 have
$CS={S^{-1}}^T C$ or equivalently $SC = C (S^{-1})^T$. Under a change
of frames
\begin{eqnarray*}
{({\sideset{^c}{}{\Psi'}}) }^\Lambda & = &
   {C}^{\Lambda\Delta} {\Psi^+}_\Gamma {{S^{-1}}^\Gamma}_\Delta \\
& = & {C}^{\Lambda\Delta} {{({S^{-1}}^T)}_\Delta}^\Gamma{\Psi^+}_\Gamma \\
& = & {S^\Lambda}_\Sigma {C}^{\Sigma\Delta} {\Psi^+}_\Delta \\
& = & {S^\Lambda}_\Sigma {(\sideset{^c}{}\Psi)}^\Sigma
\end{eqnarray*}
which shows that $\sideset{^c}{}\Psi:= {(\sideset{^c}{}\Psi)}^\Lambda
E_\Lambda \in \Gamma(E)$ is a frame invariant object.
\begin{Dfn}
  Dirac spinor fields which are invariant under charge conjugation are
  called {\em Majorana spinor fields}.
\end{Dfn}
In the remaining part of this chapter we will use Majorana spinor
fields exclusively. To shorten the notation we will omit the word
`Majorana' in the future whenever it appears unnecessary.
%
%%%%%%%%%
\section{The internal symmetries of the classical free Wess-Zumino
  Model}\label{sec:internalSymWZmodel}
%%%%%%%%%%
The Wess-Zumino model consists of three fields, two Spin-0
fields ($A,B$) one of which is scalar ($A$), the other ($B$) is
pseudo-scalar and the third ($\Psi$) is a Majorana spinor field.
The free Lagrangian density of this model is~\cite{wess_zumino:74}:
\begin{equation}
  \label{eq:WZ_Lagrange}
  {\mathcal L}=
   \frac{1}{2} (\nabla A)^2 - \frac{1}{2} m^2 A^2 +
   \frac{1}{2} (\nabla B)^2 - \frac{1}{2} m^2 B^2 +
   \frac{1}{2} \Psi^+ (i \rvec{\notnabla} - m ) \Psi,
\end{equation}
where $m>0$ is the mass of the fields and $\phantom{\Psi}^+$ denotes
Dirac conjugation. Note that all fields have {\em equal\/} masses. The
equations of motion resulting from (\ref{eq:WZ_Lagrange}) are the
Klein-Gordon and Dirac equations respectively.
\begin{align}\label{eq:eulerLagrange}
  ( \Box + m^2 ) A            & = 0 \nonumber \tag{\theequation a}\\
  ( \Box + m^2 ) B            & = 0  \nonumber \tag{\theequation b}\\
  ( i \rvec{\notnabla} -m ) \Psi  & = 0 \nonumber \tag{\theequation c}\\
  \Psi^+ (i \lvec{\notnabla} + m) & = 0 \nonumber \tag{\theequation d}
\end{align}
\addtocounter{equation}{1} % special construct to advance counter
Besides the usual Poincar\'e symmetry the model possesses an internal
symmetry, which is called {\em supersymmetry}. The latter transforms
Bose to Fermi fields and vice versa. To see this explicitly let us
consider the following transformations of the basic fields:
\begin{align}\label{eq:SUSYMink}
  A    & \mapsto A' = A + \delta A =
        A + \epsilon^+ \Psi \nonumber \tag{\theequation a}\\
  B    & \mapsto B' = B + \delta B =
        B - i \epsilon^+ \gamma^5 \Psi \nonumber \tag{\theequation b}\\
  \Psi & \mapsto \Psi' = \Psi + \delta \Psi = \Psi - ( i \rvec{\notnabla} +\
 m\
 ) \AiB \epsilon,
        \nonumber \tag{\theequation c}
\end{align}
\addtocounter{equation}{1} % special construct to advance counter
where $\epsilon$ is a $x$-independent spinor field. The resulting
transformation of ${\mathcal L}$ up to first order
\begin{equation}
  \label{eq:LagrangeVariation}
 \delta {\mathcal L} = {\mathcal L} (A',B',\Psi') - {\mathcal L}(A,B,\Psi) =
  \nabla^a\left( \nabla_a A \delta A + \nabla_a B  \delta B -
  \frac{1}{2} \Psi^+ i \gamma_a \delta \Psi \right) =: \nabla^a v_a
\end{equation}
is a pure divergence. This means that the action $S= \int d^4x {\mathcal
  L}$ is left invariant, hence
Eqn.~(\ref{eq:SUSYMink}a)--(\ref{eq:SUSYMink}c) is indeed a symmetry
transformation.
For the proof simply insert the definitions and use the
Fierz rearrangement formulas of Appendix~\ref{sec:usefulForm}:
  \begin{equation*}
    \begin{split}
      \delta {\mathcal L} & = {\mathcal L}(A',B',\Psi',) - {\mathcal\
 L}(A,B,\Psi)\\
      & =
      \left(
        \frac{1}{2}\left( \nabla(A+\delta A) \right)^2 -
        \frac{m^2}{2}(A+\delta A)^2
        +
        \frac{1}{2}\left( \nabla(B+\delta B) \right)^2 -
        \frac{m^2}{2}(B+\delta B)^2 \right.{}\\
      & \phantom{= \bigl( \frac{1}{2}\left( \nabla{}(A+\delta A) \right)^2}
        \left.{}
          +
          \frac{1}{2}(\Psi + \delta\Psi)^+ ( i \notnabla - m) ( \Psi +
          \delta\Psi)
        \right) \\
      & \phantom{=}
      -
      \left(
        \frac{1}{2}\left( \nabla{}A \right)^2 -
        \frac{m^2}{2} A^2
        +
        \frac{1}{2}\left( \nabla{}B \right)^2 -
        \frac{m^2}{2} B^2
        +
        \frac{1}{2}\Psi^+ ( i \notnabla - m)\Psi
      \right)
      \\
      & =
      \nabla_a A \nabla^a \delta A - m^2 A \delta A
      +
      \nabla_a B \nabla^a \delta B - m^2 B \delta B
      - \frac{1}{2} \Psi^+ ( i \notnabla -m) \delta\Psi
        + \frac{1}{2} (\delta \Psi)^+ ( i \notnabla -m) \Psi
        \\
      \intertext{(1. order)}
      & =
      \nabla^a ( \nabla_a A \delta A + \nabla_a B \delta B)\\
      & \phantom{=}
       -\Box A \delta A - m^2 A \delta A
       -\Box B \delta B - m^2 B \delta B
       + \frac{1}{2} \Psi^+ ( i \notnabla - m ) \delta\Psi
         - \frac{1}{2} \Psi^+ ( i \lvec{\notnabla} + m ) \delta\Psi\\
      \intertext{by (\ref{eq:fierz_indentity}d)}
      & = \nabla^a ( \nabla_a A \delta A + \nabla_a B \delta B
                       - \frac{i}{2} \Psi^+\gamma_a\delta\Psi)\\
      & \phantom{=}
      -   ( \Box + m^2 ) A \Psi^+ \epsilon
      + i ( \Box + m^2 ) B \Psi^+ \gamma^5 \epsilon
      +   ( \Box + m^2 ) A \Psi^+ \epsilon
      - i ( \Box + m^2 ) B \Psi^+ \gamma^5 \epsilon\\
      \intertext{using (\ref{eq:fierz_indentity}c) and since $(i \notnabla
        -m ) \delta\Psi = (\Box+m^2)\AiB\epsilon$}
      & = \nabla^a ( \nabla_a A \delta A + \nabla_a B \delta B
                       - \frac{i}{2} \Psi^+\gamma_a\delta\Psi)\\
      & =: \nabla^a v_a \\
    \end{split}
  \end{equation*}
\begin{rem}
  It is a generic feature of supersymmetry transformations, in
  contrast to other symmetries, e.g., Poincar\'e invariance, that they
  produce a nontrivial variation of the Lagrangian density.  The
  reader should note that Eqn.~(\ref{eq:LagrangeVariation}) is
  obtainable from Eqn.~(\ref{eq:WZ_Lagrange}) without using the field
  equations~(\ref{eq:eulerLagrange}a)--(\ref{eq:eulerLagrange}d).
\end{rem}
Using Noether's theorem one obtains a conserved current
associated with the
transformation~(\ref{eq:SUSYMink}a)--(\ref{eq:SUSYMink}c):
It is given by:
\[ i^a := v^a - \frac{\partial {\mathcal L}}{\partial(\partial_a \Phi)}
           \delta\Phi \qquad \Phi=(A,B,\Psi)
\]
Inserting
\[
\frac{\partial {\mathcal L}}{\partial(\partial_a A)} = \partial^a A,
\quad
\frac{\partial {\mathcal L}}{\partial(\partial_a B)} = \partial^a B,
\quad
\frac{\partial {\mathcal L}}{\partial(\partial_a \Psi)} = \frac{i}{2}
                                                      \Psi^+ \gamma^a
\]
one obtains --using the Fierz rearrangement identities again--:
\begin{equation}
  \label{eq:supercurrentMink}
  \begin{split}
    i^a & = -i \Psi^+ \gamma^a \delta\Psi\\
        & = i \Psi^+ \gamma^a ( i \rvec{\notnabla} +m ) \AiB \epsilon \\
        & =  \epsilon^+ i \AiB (i \lvec{\notnabla} - m) \gamma^a \Psi\\
        & =: {\epsilon^+}_A k^{aA}\\
  \end{split}
\end{equation}
The current $k^{aA}$ is usually called {\em supercurrent}.  Note that
it carries --besides the usual vector index-- an additional spinor
index, hence it is a mixed spinor
vector field. Its Dirac conjugate is
\begin{equation}
  \label{eq:con_supercurrent}
  {k^{+a}}_A = i \Psi^+ \gamma^a ( i \rvec{\notnabla}+m) \AiB
\end{equation}
\begin{obs}
  $\phantom{~}{}$\par{}
  \begin{enumerate}
  \item The charge $Q$ associated to the Noether current $i^a$ induces
    the symmetry transformation
    Eqn.~(\ref{eq:SUSYMink}a)--(\ref{eq:SUSYMink}c).
  \item The symmetry transformation is a realization of the
    supersymmetry algebra.
  \end{enumerate}
\end{obs}
\begin{pf}
  \begin{enumerate}
  \item  The classical charge $Q_\epsilon$ associated with $i^a$ is defined
    by
    \begin{equation}
      \label{eq:class_charge}
      Q_\epsilon := \int_\Sigma k^{+a} \epsilon d\sigma_a,
    \end{equation}
    where $\Sigma$ is the $t=0$ hypersurface. Recall that
    $Q_\epsilon$ is actually independent of $t$ and $\Sigma$, since
    $k^a$ is a conserved current. To see that $Q_\epsilon$ generates
    the symmetry transformations Eqn.~(\ref{eq:SUSYMink}a) -
    (\ref{eq:SUSYMink}c) we introduce the canonical momenta for the
    three fields: For $t=t_0$
    \begin{eqnarray*}
      \Pi_A & := & \frac{\partial{\mathcal L}}{\partial(\partial_0 A)} =
                 \partial^0 A\\
      \Pi_B & := & \frac{\partial{\mathcal L}}{\partial(\partial_0 B)} =
                 \partial^0 B\\
      \Pi_\Psi & = & i \Psi^+ \gamma^0
    \end{eqnarray*}
    Note that the last equation does {\em not\/} read $\Pi_\Psi=
    \frac{i}{2} \Psi^+ \gamma^0$ (Wrong~!), since $\Psi$ is a Majorana
    spinor field, i.e., $\Psi$ and $\Psi^+$ are {\em not\/}
    independent.  (See for example the discussion
    in~\cite[pp.~306]{roman:TheoryElemPart}).  Inserting these
    definitions in Eqn.~(\ref{eq:class_charge}) we obtain
    {\small
    \begin{equation}
      \begin{split}
        Q_\epsilon(A,\nabla A, B, \underline{\nabla} B , &
                   \Psi , \underline{\nabla} \Psi, \Pi_A,
                   \Pi_B, \Pi_\Psi)
        \\
        & =
        \int_\Sigma d^3\sigma \Pi_\Psi\left( i \gamma^0 \Pi_A + \gamma^0
        \gamma^5 \Pi_B + (i \underline{\notnabla} + m ) \AiB \right)\
 \epsilon,
      \end{split}
    \end{equation}
    }
    where the underscore denotes spatial (i.e.\ three dimensional)
    expressions.  Note that $Q_\epsilon$ is bilinear in it's
    variables. We now compute the Poisson brackets of $Q_\epsilon$
    with the fields\footnote{For simplicity we consider a
      representation of the Dirac $\gamma$-matrices, where $\beta
      \equiv \gamma^0$.}
    \begin{align}
      \begin{split}
        \{ Q_\epsilon , A \}
       & :=
       \int_\Sigma d^3\sigma \sum_\Phi
       \frac{\partial Q}{\partial \Phi} \frac{\partial A }{\partial
         \Pi_\Phi}
        -
        \frac{\partial Q}{\partial \Pi_\Phi} \frac{\partial A }{\partial
          \Phi}\\
       & = - \Pi_\Psi i \gamma^0 \epsilon = \Psi^+ \gamma^0 \gamma^0
       \epsilon = \epsilon^+ \Psi \equiv \delta A
     \end{split}\\
     \{ Q_\epsilon,B \}
     & = - \Pi_\Psi \gamma^0 \gamma^5 \epsilon = -i \Psi^+ \gamma^5
     \epsilon = -i \epsilon^+ \gamma^5 \Psi \equiv \delta B \\
     \{ Q_\epsilon,\Psi \}
     & = - ( \notnabla + m) \AiB \epsilon \equiv \delta \Psi \\
   \end{align}
   This shows that $Q_\epsilon$ indeed generates the supersymmetry
   transformation at the classical level.
 \item To establish the second statement, we investigate the algebra
   of the supersymmetry charge:
   \begin{equation}
     \left\{ \{Q_{\epsilon_1},Q_{\epsilon_2} \} , \cdot \right\},
     \end{equation}
     where the $\cdot$ denotes one of the basic fields. Let us
     evaluate these expressions explicitly using the Jacobi identity
     and the Fierz rearrangement formulas. We start with the basic
     field $A$:
     \begin{equation*}
       \begin{split}
         \left\{ \{Q_{\epsilon_1},Q_{\epsilon_2} \} , A \right\}
       &= \left\{ Q_{\epsilon_1}, \{Q_{\epsilon_2}, A \} \right\}
           - \left\{ Q_{\epsilon_2}, \{Q_{\epsilon_1}, A \} \right\}\\
       & = \{Q_{\epsilon_1}, \epsilon_2^+ \Psi \} -  \{Q_{\epsilon_2},
           \epsilon_1^+ \Psi \}\\
       & = - \epsilon_2^+ ( i \notnabla + m ) \AiB \epsilon_1
           + \epsilon_1^+ ( i \notnabla + m ) \AiB \epsilon_2     \\
       & = -2 i \epsilon_2^+ \notnabla \epsilon_1 A
      \end{split}
    \end{equation*}
    Similarly for $B$:
    \begin{equation*}
      \left\{ \{Q_{\epsilon_1},Q_{\epsilon_2} \}, B \right\} = - 2 i
        \epsilon_2^+ \notnabla \epsilon_1 B
    \end{equation*}
    Finally let us calculate
    $\left\{ \{Q_{\epsilon_1},Q_{\epsilon_2} \}, \Psi \right\}$:
    \begin{equation}
      \label{eq:charge_on_Psi}
    \begin{split}
      \left\{ \{Q_{\epsilon_1},Q_{\epsilon_2} \}, \Psi \right\}
      &= \left\{ Q_{\epsilon_1}, \{Q_{\epsilon_2}, \Psi \} \right\}
          - \left\{ Q_{\epsilon_2}, \{Q_{\epsilon_1}, \Psi \} \right\}\\
      & = \{ Q_{\epsilon_1}, (-1) ( i \notnabla+ m ) \AiB \epsilon_2\}\\
      & \phantom{=}
          - \{ Q_{\epsilon_2}, (-1) ( i \notnabla+ m ) \AiB \epsilon_1
          \}\\
      & = - (i\notnabla +m) ( \epsilon_1^+ \Psi \epsilon_2
                             - \gamma^5 \epsilon_1^+ \gamma^5 \Psi\
 \epsilon_2)
          - ( 1\leftrightarrow 2)\\
    \end{split}
  \end{equation}
  To evaluate the last term we use the following identity for products
  of spinors and arbitrary 4$\times$4 matrices $N$ and $M$
  \cite{west:SUSY_Intro}:
  \[
  \Psi^+_1 M \Psi_3 \Psi^+_4 N  \Psi_2 =
  - \frac{1}{4} \sum_R
  \sigma(R) \Psi^+_1 O_R \Psi_2 \Psi^+_4 N O_R M  \Psi_3,
  \]
  where
  \begin{alignat*}{2}
    \sigma(R) & =
    \begin{cases}
      +1 & \\
      -1 & \\
      +1 & {}
    \end{cases}
    & \qquad \text{for} \quad
    O_R & =
    \begin{cases}
      1, \gamma^5,\gamma^a & \\
      \gamma^5 \gamma^a & \\
      \frac{1}{4} ( \gamma^a \gamma^b - \gamma^b \gamma^a )
      \equiv G^{[ab]} & {}
    \end{cases}
  \end{alignat*}
  \begin{equation*}
    \begin{split}
      (\ref{eq:charge_on_Psi})
      & = - i \epsilon_1^+ \nabla_a \Psi \gamma^a \epsilon_2
          - m \epsilon_1^+ \Psi \epsilon_2
          + i \epsilon_1^+ \gamma^5 \nabla_a \Psi \gamma^a \gamma^5\
 \epsilon_2
          + m \epsilon_1^+ \gamma^5 \Psi \gamma^5 \epsilon_2
          - ( 1 \leftrightarrow 2 ) \\
      & = - \frac{1}{4} \sum_R
          - i \sigma(R) \epsilon_1^+ O_R \epsilon_2 \gamma^a
            ( O_R - \gamma^5 O_R \gamma^5 )
            \nabla_a \Psi \\
      & \phantom{= - \frac{1}{4} \sum_R}
          - m \sigma(R) \epsilon_1^+ O_R \epsilon_2
            ( O_R - \gamma^5 O_R \gamma^5 )
            \Psi
         - ( 1 \leftrightarrow 2 )\\ \nonumber
       \end{split}\nonumber
     \end{equation*}
     Note that the l.h.s.\ of Eqn.~(\ref{eq:charge_on_Psi}) is
     antisymmetric in $\epsilon_1$ and $\epsilon_2$ and the only
     antisymmetric terms on its r.h.s.\ involving $\epsilon_1$ and
     $\epsilon_2$ are $\epsilon_1^+ \gamma^a \epsilon_2$ and $\gamma^5
     G^{[ab]} \gamma^5$. But $\gamma^5 G^{[ab]} \gamma^5 =
     G^{[ab]}$  so that the brackets are zero in this case. We get
     \begin{equation*}
       \begin{split}
      (\ref{eq:charge_on_Psi})
      & = \sum_b
          i \epsilon_1^+ \gamma^b \epsilon_2 \gamma^a \gamma^b \nabla_a \Psi
           + m \epsilon_1^+ \gamma^b \epsilon_2 \gamma^b \Psi \\
      & = \sum_b
          \epsilon_1^+ \gamma^b \epsilon_2
            \left( - \gamma^b \left( (i \notnabla -m )\Psi \right) + 2i
              \eta^{ab} \nabla_a \Psi
            \right) \\
      & = -2 i \epsilon_2^+ \notnabla \epsilon_1  \Psi \\
    \end{split}
  \end{equation*}
  On the other hand the derivative $\nabla_a$ is the generator of the
  translations. Therefore the Poisson brackets of the supercharge
  generate the translations, i.e., $Q_\epsilon$ is the `square-root'
  of the 4-momentum $P_a = i \nabla_a$ and
  \begin{equation}\label{eq:poisson_charge}
    \{ Q_{\epsilon_1}, Q_{\epsilon_2} \}
    = -2 \epsilon_2^+ \gamma^a P_a \epsilon_1
  \end{equation}
\end{enumerate}
\end{pf}
%
%%%%%%%%%
\section{Normal Ordering}
%%%%%%%%%
We will show in this section that the momentum operator and the
Hamiltonian of the quantized Wess-Zumino model are a priori normal
ordered, i.e., that their vacuum expectation values vanish. Moreover
the stress energy tensor of the model can be written as an operator
valued distribution on the vacuum Hilbertspace in terms of the
supercurrent. We start with the supercurrent.  It is worth noting that
Eqn.~(\ref{eq:supercurrentMink}) is well defined after quantization,
since the spectrum condition ensures that products of {\em
  different\/} fields at the same spacetime point exist. For the
adjoint of the supercurrent one obtains:
\begin{equation*}
  \begin{split}
    {i^a_\epsilon}^*
    & =
    {( \epsilon^+ i \AiB (i \notnabla -m ) \gamma^a \Psi )}^* \\
    & = \Psi^+ \gamma^a ( i \notnabla + m ) \AiB \epsilon \\
    & = {k^+}^a \epsilon \\
    & = \epsilon^+ k^a = i_\epsilon^a,\\
  \end{split}
\end{equation*}
since $k^a$ is a Majorana spinor. Hence $j_\epsilon^a$ is a conserved
hermitian current.
Let $f_R \in \CinfO({\mathbb R}^3)$ and $f_T \in \CinfO({\mathbb R})$ be two
smooth function with compact support, such that
\[
f_R(x) =
\begin{cases}
   1        & | x | < R \\
   f_R(|x|)   & R \leq | x | \leq R + \Lambda \\
   0        & | x | >  R + \Lambda
\end{cases}
\]
for some $\Lambda \in {\mathbb R}$, and
\[
f_T : \quad
\begin{cases}
  \int f_T ( x_0 ) dx_0 = 1 & {} \\
  f_T (x_0) = 0             & | x_0 -t | > T \\
  f_T(x_0) = f_T(-x_0 + 2t) & {}
\end{cases}
\]
Then $Q_{\epsilon RT} = j_\epsilon^0 (f_R f_T)$ is a well defined
operator. We have the following two theorems
\begin{Thm}[Theorem~6.2.3 of~\cite{lopus:SUSYinQFT}]
  Let $A_L$ be a localized operator, then
  \[
  C_\epsilon(A_L) = [ Q_{\epsilon RT} , A_L ]_-
  \] applied to any state in the set
  \[
  D_\Omega=\{ u\in H | \quad \text{\rm $u$ is generated
    by polynomials of the smeared fields applied to the vacuum
    $\Omega\}$}
  \]
  is independent of both $f_R$ and $f_T$, whenever $j_\epsilon^a$ is a
  conserved current and $R$ is sufficiently large. $C_\epsilon(A_L)$
  is a localized operator.
\end{Thm}
\begin{Thm}[Theorem~6.2.8 and 6.2.9 of~\cite{lopus:SUSYinQFT}]
  Assume the existence of a mass gap and the local conservation of the
  current; then the limit $\lim_{R\rightarrow \infty} ( u, Q_{\epsilon
    RT} v) =: Q_\epsilon(u,v)$ exists for all quasi-local vectors $u$
  and $v$ and defines a linear operator $\epsilon^+G$ on the
  Hilbertspace:
  \[
  Q_\epsilon(u,v) = ( u, \epsilon^+G v ) \qquad
                     \forall u,v \in H, \text{\rm $v$ quasilocal}
  \]
\end{Thm}
\begin{rem}
  $\phantom{~}$\par
  \begin{itemize}
   \item The charge $\epsilon^+G$ is hermitian due to the fact that the
     corresponding current is. Moreover, since $i^a_\epsilon$ is
     translational covariant, it can be proved that $\epsilon^+G$ is
     self-adjoint with a dense invariant domain $D_\Omega^{\text{ex}}$,
     consisting of all states with finite number of particles and
     smooth momentum space wave-functions having compact support.
   \item By Theorem~6.2.12 of~\cite{lopus:SUSYinQFT} one finds $(
     \Omega, [ Q_{\epsilon RT} , A_L]_- \Omega) =0$ for all localized
     operators and sufficiently large $R$.  It follows that $\epsilon^+
    G$ annihilates the vacuum, i.e., supersymmetry is unbroken and on
     the vacuum $\lim_{R\rightarrow\infty} [Q_{\epsilon RT}, A_L]_-
     \Omega = [ \epsilon^+G,A_L]_- \Omega = \epsilon^+ G A_L \Omega$.
     The relation between currents and charges for models {\em with\/}
     spontaneously broken symmetries is closely investigated by
     Buchholz, Doplicher, Longo and Roberts~\cite{buchDopLonRob:92}.
  \end{itemize}
\end{rem}
Let us  now take a closer look at the charge operators: $Q_\epsilon$
is bilinear in the basic fields. Thus the results for the classical charges \
 \

obtained in the previous section translate to
the quantum field theoretical case:
\begin{alignat*}{2}
    [ \epsilon_1^+ G , \epsilon_2^+ G ]_- &
     = [ \epsilon_1^+ G , G^+ \epsilon_2]_-
    & \qquad & \text{since $G$ transforms as a Majorana spinor}\\
    & = \epsilon_1^+ [ G, G^+]_+ \epsilon_2
    & \qquad &
        \text{$\epsilon_1^+$ and $\epsilon_2$ anticommute}\\
    & = 2 \epsilon_1^+ \gamma^a P_a \epsilon_2
    & \qquad &
        \text{by Eqn.~(\ref{eq:poisson_charge})}
\end{alignat*}
It is therefore possible to write:
\begin{equation}
  \label{eq:charge_anticom}
  \{ G^A, G^+_B\}_+ = 2 {{\gamma^a}^A}_B P_a
\end{equation}
Now multiply Eqn.~(\ref{eq:charge_anticom}) by $\gamma^0$ and take the
trace over the spinor indices:
\begin{equation*}
  \begin{split}
    \text{Tr } ( \{G^A, G^+_B\}_+ {{\gamma^0}^B}_C )
    & = G^A {\bar{G}^T}_B {{\gamma^0}^B}_C {{\gamma^0}^C}_A
        + {\bar{G}^T}_B {{\gamma^0}^B}_C {{\gamma^0}^C}_A G^A \\
    & = G^A {G^{\dag}}_A + \text{h.c.} \\
    & = 8 P_0,\\
  \end{split}
\end{equation*}
where $\text{Tr } (\gamma^a \gamma^b) = 4 \eta^{ab}$ and ${\gamma^0}^2=1$
were used. As usual $\phantom{~}^{\dag}$ denotes hermitian
conjugation. Since $G^A {G^{\dag}}_A + \text{h.c.}$ is a positive
operator, we conclude that $P_0$ --which is the Hamiltonian-- is
positive too. A similar calculation yields
\[
0 = \text{Tr } <\Omega | \{G, G^+ \}_+ \gamma^a \Omega > =
    8 <\Omega|P^a \Omega>,
\]
which ensures that all components of $P^a$ are normal ordered. We
remind the reader that $P^a$ which is the generator of the
translations is a well defined self adjoint operator for the massive
Wess-Zumino model.
Heuristically this normal ordering feature is due to the fact that
bosonic and fermionic loops have opposite sign. Their infinite
contributions to the vacuum expectation values cancel each other in
models with equally many bosonic and fermionic degrees of freedom. In
view of Eqn.~(\ref{eq:charge_anticom}) the energy momentum tensor
$\theta^{ab}$ of the Wess-Zumino model can be written as (see
e.g.~\cite{ferraraZumino:75})
\begin{equation}
  \label{eq:SUSY_EMT}
  \theta^{ab}= + \frac{1}{16} \text{Tr}\left( \gamma^a\{k^b,G^+\}_+ +
  \gamma^b\{k^a,G^+\}_+ \right),
\end{equation}
where the spinor indices on $\gamma^a$, $k^a$ and $G^+$ were
suppressed\footnote{The explicit proof of Eqn.~(\ref{eq:SUSY_EMT})
  requires some lengthly calculations. Due to the fact that a similar
  identity on a manifold will be used and proved later in this thesis,
  the actual proof of (\ref{eq:SUSY_EMT}) is omitted.}. (Kraus and
Landau loc.cit.\ showed that the anti-commutator in
Eqn~(\ref{eq:SUSY_EMT}) is well defined on a suitable subset of
$D_\Omega^{\text{ex}}$). It follows that the vacuum expectation value of
$\theta^{ab}$ vanishes. Recall that the supersymmetry is not
spontaneously broken, i.e., the supercharge $G^+$ annihilates the
vacuum.
In remaining parts of this work the question whether this tameness of
the vacuum expectation value of the energy momentum tensor of
supersymmetric quantum field theories can be extended to curved space
times is investigated. Such a priori renormalization is a highly
desirable feature from a general relativistic viewpoint; contrary to
special relativity where only differences of energy are measurable, we
have measurable effects induced by the absolute value of the energy,
in the latter, namely energy curves spacetime.
It was already stated in the introduction, that supersymmetry on
curved spacetime, to be implemented consistently, leads to
supergravity theories in general~\cite{nieuwenhuizen:81}. Therefore we
will only mimic the {\em structure\/} of the Wess-Zumino model on a
curved spacetime.  If we restrict ourself to globally hyperbolic
manifolds, i.e., to spacetimes where a rigorous quantization of
various free fields has been
done~\cite{dimock:80,dimock:82,dimock:92}, there seems to be no known
nontrivial example of a manifold which possesses a true supersymmetric
extension of its isometry group. Moreover, in the general case
a Lorentz manifold does not possess an isometry group at all.
Coming back to the {\em structure\/} of a supersymmetric theory one
observes immediately that fermionic {\em and\/} bosonic fields are
needed. (Recall that a supersymmetric transformation changes half
integer spin fields to integer spin fields and vice versa). The latter
exist on all Lorentz manifolds, whereas for the existence of fermionic
fields special conditions on the topology of the manifold must be
satisfied~\cite{milnor:63}. On a globally hyperbolic manifold however
a Theorem due to Geroch~\cite{geroch:68,geroch:70} ensures the
existence of half integer spin fields in four dimensions (see below).
Nevertheless let us begin with a closer look on this topological
condition preventing the existence of spinor fields, because there are
serious attempts to extend the result of Dimock loc.\ cit.\ and others
to non globally hyperbolic spacetimes (See~\cite{kay:92}). Moreover,
there are some cosmologically interesting solutions of the Einstein
Equations --for example plain waves-- which fail to be globally
hyperbolic.
%
%%%%%%%%%%%%%%%%%%%%%%%%%%%%%%%%%%%%%%%%%%%%%%%%%%%%%%%%%%%%%%%%%%%%%%%%%
\chapter{Spinors on Lorentz manifolds}\label{sec:spinorsOnManifolds}
%%%%%%%%%%%%%%%%%%%%%%%%%%%%%%%%%%%%%%%%%%%%%%%%%%%%%%%%%%%%%%%%%%%%%%%%%
%
For the definition of spinors on Minkowski spacetime it was important that
the Poincar\'e group is the isometry group of this manifold. It was
already mentioned above that in the general case there might exist no
isometry group at all. Therefore the study of isometries does not seem
to be a good starting point for a definition\footnote{In view of
  this E. Cartan suggested in~\cite{cartan} that a sensible definition
  of spinors on manifolds is not possible.} of spinors, since we would\
 restrict
ourself to just a few very special cases. Let us therefore return to
the principle of `special covariance': Like on Minkowski space, on a
general manifold we can identify an observer with her measurement
device at a point $p$ with an orthonormal base system (vierbein) at
this point. Different observers at the {\em same\/} point will be
described by different vierbeins. Since a vierbein consists of four
linearly independent orthogonal tangent vectors at $p$, two different
vierbeins are connected by a Lorentz transformation\footnote{We
  continue to restrict ourself to {\em proper\/} Lorentz
  transformations, since there are experimental evidences (see
  Chapter~\ref{sec:SUSYinR4}) that our universe is time and space
  orientable. Moreover, we can distinguish these orientations by
  performing $T$ and $P$ asymmetry experiments.}. Thus to describe
spinors we will start with the proper Lorentz group, not with the
whole Poincar\'e group, as before. We are going to define spinors in
such a way that every transformation of a vierbein ($\hat{=}$ change
of observer) results in a corresponding spinor transformation up to
sign. The sign ambiguity evaporates --as in Minkowski space-- as soon
as a continuous curve of transformations starting from the identity is
specified. Mathematically this idea is described best in terms of
fiber bundles: Spinors are sections in a suitable associated fiber
bundle, called spinor bundle. The corresponding principal fiber bundle
is a special twofold covering of the principal fiber bundle of the
vierbeins. The existence of this covering is a topological property of
the underlying manifold.
In the first section of this chapter for the convenience of the reader
the definition of a principal fiber bundle is recalled and we define
the spinor bundle. Some facts on the preferred connection on the
latter are presented in the next section. For the proofs that were
omitted, as well as for further details, the reader is referred to the
book of Kobayashi and Nomizu~\cite{kob1:foundation}. We continue by
defining \v{C}ech-cohomology groups and Stiefel-Whitney classes
according to the book of Nakahara~\cite{nak:geometry}. The latter are
shown in section~\ref{sec:existenceSpinorStructure} to measure the
existence of a spin structure as Milnor already found out
in~1963~\cite{milnor:63}. We close this chapter by summarizing some
other, but also well known conditions on the existence of spin
structures which turn out to be more useful for applications.
%
%%%%%%%%%
\section{Bundle-formalism for spinors} \label{sec:spinorBundleFormalism}
%%%%%%%%%
%
Let ${\mathcal G}$ denote a Lie-group, $\bfm{P}$ a manifold and ${\bf L}\
 \colon
{\mathcal G} \times \bfm{P} \rightarrow \bfm{P}$ a smooth mapping.
\begin{Dfn}
  The mapping ${\bf L}$ is called {\em left action\/}$:\bimply$
  \begin{enumerate}
  \item $\forall g \in {\mathcal G}$ fixed, ${\bf L}_g \colon \bfm{P}
    \rightarrow \bfm{P}$ is a diffeomorphism.
  \item $\forall g_1,g_2 \in {\mathcal G}$, ${\bf L}_{g_1} \circ {\bf
      L}_{g_2}
    \equiv {\bf L}_{g_1g_2}$.
  \end{enumerate}
\end{Dfn}
\begin{Dfn}
  A left action is called {\em free\/}$:\bimply$ $\forall g\in {\mathcal
    G}$, $u\in \bfm{P}$, $g \neq e$ one has ${\bf L}_g(u) \neq{} u$, i.e.,
  ${\bf L}_g{}$ has no fix-points.
\end{Dfn}
\begin{Dfn}
  For all points $u \in \bfm{P}$ the set $O=\{ {\bf L}_g(u)| \quad g \in
  {\mathcal G}\}$ is called the {\em orbit\/} of $u$.
\end{Dfn}
\begin{rem}
  $\bfm{P}$ can be represented as a disjoint union of orbits.
\end{rem}
\begin{Dfn}\label{Dfn:principalfiber}
  A {\em principal fiber bundle\/} $(\bfm{P},{\mathcal G},M,{\bf L})$
  consists of a manifold $\bfm{P}$, the bundle manifold, a Lie group
  ${\mathcal G}$ (fiber group), a second manifold $M$ (base manifold) and
  a free left action ${\bf L}$, which fulfills the following
  conditions.
  \begin{enumerate}
  \item The orbits of ${\bf L}$ are in one-to-one correspondence to the
    points of $M$ and the projection
    \[ \pi \colon \bfm{P} \rightarrow M\]
    is surjective and smooth.
  \item For all $x \in M$ there exists an open neighborhood $U$ of $x$
    such that there is a diffeomorphism
    \[
    \Psi \colon \bfm{P} \supset \pi^{-1}(U) \rightarrow {\mathcal G} \times U
    \]
    and the left action of ${\mathcal G}$ on $\pi^{-1}(U)$ equals the left
    multiplication on ${\mathcal G}\times U$ (Let $\Psi(u) = (g,x)$, then
    $\Psi({\bf L}_{g'} u)= \Psi(g'g,x)$)
  \end{enumerate}
\end{Dfn}
Let $(\bfm{P},{\mathcal G},M,{\bf L})$ be a principal fiber bundle,
$\bfm{F}$ another manifold, and $\chi\colon {\mathcal G} \times \bfm{F}
\rightarrow \bfm{F}$ a second left action, which is not necessarily
assumed to be free. Then the associated fiber bundle $\bfm{B}$ is
defined by:
\begin{Dfn}\label{Dfn:principleFibre}
  Consider the product manifold $\bfm{P}\times \bfm{F}$. On this manifold
  \[
  \xi_g(u,f) = ({\bf L}_g(u),\chi_g(f))
  \]
  defines a left action. Let $\bfm{B}$ denote the set of all orbits of
  ${\mathcal G}$ on $\bfm{P}\times \bfm{F}$. By the construction there
  exists a natural projection $\tilde{\pi}$ from $E$ to $M$
  \begin{align*}
    \tilde{\pi} \colon E & \rightarrow M \\
    y=[u,f]_{\sim}  & \mapsto     \pi(u),
  \end{align*}
  where $[\cdot,\cdot]_{\sim}$ denotes the orbit of $(u,f)\in
  \bfm{P}\times \bfm{F}$. We equip $\bfm{B}$ with the structure of a
  topological manifold by demanding that for any neighborhood $U$
  satisfying condition~(2) of Definition~\ref{Dfn:principalfiber},
  $\tilde{\pi}^{-1}(U) \subset{} \bfm{B}$ is diffeomorphic to
  $\bfm{F}\times U$. $\bfm{B}=(\bfm{P}\times \bfm{F})_{/ {\mathcal G}}$ is
  called the {\em associated fiber bundle\/} to $(\bfm{P},{\mathcal
    G},M,{\bf L})$.
\end{Dfn}
\begin{rem}
  It follows that $\tilde{\pi}^{-1}(x)$ is diffeomorphic to $\bfm{F}$,
  that is, $\bfm{B}$ equals $M\times \bfm{F}$ locally.
\end{rem}
We have now available all tools for the definition of spinors on a
manifold. Consider the principal fiber bundle
$FM=(\bfm{P},\Lor{},M,{\bf L}){}$ which is the bundle consisting of
all oriented, time oriented orthonormal frames on $(M,g_{ab})$
together with the proper Lorentz group $\Lor{}$. Denoting its
universal covering group by $\text{Spin}_0$ we have
\begin{Dfn}
  An oriented, time oriented Lorentz manifold $(M,g_{ab})$ admits a
  {\em spin structure\/}$:\bimply{}$\\ There exists a
  $\text{Spin}_0$-principal fiber bundle
  $SM=(\hat{\bfm{P}},\text{Spin}_0,M,\hat{{\bf L}})$ over $M$, called
  the {\em spin frame bundle}, and a bundle homomorphism $\bfm{p} \colon SM
  \rightarrow FM$ preserving the base points which satisfies
  \[
  \bfm{p} \circ \hat{{\bf L}}_S = {\bf L}_{\Lambda(S)} \circ \bfm{p}
  \]
  Here $\Lambda(S)\in\Lor$ is the element of $\Lor$ associated to
  $S\in\text{Spin}_0$.
\end{Dfn}
\begin{rem}
  It was already mentioned above that not all manifolds admit a spin
  structure. Moreover, spin structures are not unique in general. The
  existence of a spin structure is a topological property of the
  manifold, i.e., it is independent of the Lorentz metric. We will
  come back to this point in
  section~\ref{sec:existenceSpinorStructure} of this chapter.
\end{rem}
\begin{Dfn}
  The {\em spinor bundle\/} is the associated vector bundle
  $DM=(SM\times {\mathbb C}^4)_{/\text{Spin}_0}$. Spinor fields are
  sections in $DM$, i.e., $\Cinf$-mappings $u\colon M \rightarrow DM$,
  such that $u(x) \in D_xM$, where $D_xM$ denotes the fiber over $x\in
  M$. The dual vector bundle $D^*M$ is called {\em cospinor bundle},
  its sections are called {\em cospinor fields}.
\end{Dfn}
In order to relate this bundle theoretical formulation with the
intuitive ideas of a spinor, consider a local cross section
$E\colon M \supset U \rightarrow SM$ in the spin structure. This
section  uniquely determines a `moving frame' $(E_1,\ldots,E_4)$ in the
spinor bundle $DM$: The mappings $E_\Lambda\colon U \rightarrow DM$
are local crossections in the spinor bundle given by
\[
E_\Lambda(x) = i(E(x),\omega_\Lambda),
\]
where $i\colon SM \times {\mathbb C}^4 \rightarrow DM$ is the natural
embedding of $(SM\times{\mathbb C}^4)$ into $DM$ and $\omega_\Lambda$,
$\Lambda=1,\cdots,4$ denote a basis in ${\mathbb C}^4$ . It follows that
every spinor field $u\in\Gamma(DM)$ can be written locally as
\[
 u = \sum_{\Lambda=1}^4 u^\Lambda E_\Lambda
\]
with  smooth functions $u^\Lambda$. Under a change of sections in
$SM$, given by $E \mapsto E' = \hat{{\bf L}}_S(E)$, where $\hat{{\bf L}}_S$
is the left-action of $S\in\text{Spin}_0$ on $SM$, the `moving frames'
transform as
\[
E_\Lambda \mapsto E'_\Lambda= E_\Gamma{(S^{-1})^\Gamma}_\Lambda.
\]
Consequently the components of a spinor field $u$ transform according
to
\begin{equation}\label{eq:spinorTransform}
u^\Lambda \mapsto {u'}^\Lambda= {(S)^\Lambda}_\Gamma u^\Gamma
\end{equation}
On the other hand by the bundle-homomorphism $p$ every change of section
in the spin structure induces a change of sections in the vierbein
bundle $FM$, i.e., the vierbeins in every fiber are undergoing a
Lorentz transformation. In this sense we found
\begin{quote}
  A spinor $u \in\Gamma(DM)$ transforms under a change of observers
  ($\hat{=}$ change of section in the spin structure) according to a
  (reducible) representation of the group $\text{Spin}_0$, the
  covering group of the Lorentz group.
\end{quote}
%
%%%%%%%%%
\section{The spinor connection}
\label{sec:nablaSpinor} \label{sec:spinorConnection}
%%%%%%%%%
%
It is well known that on the tangent bundle of a Lorentz manifold
there exists a preferred connection, called Levi-Civita connection. In
this section it is shown how this connection can be used to define a
preferred connection in the spinor bundle. It is worth noting that
this result depends crucial on the fact that the fiber groups of both
bundles, i.e., $\Lor{}$ and $\text{Spin}_0$, have the same
Lie-algebra.
Let us begin with two different characterizations of the notion of a
`connection' in a principal fiber bundle.  Let $(M,g_{ab})$ denote a
pseudo-Riemannian manifold of dimension $d$, ${\mathcal G}$ a Lie-group
and $(\bfm{P},M,{\mathcal G},{\bf R})$ a principal fiber bundle over $M$.
\begin{Dfn}
  Let $u$ be a point in $(\bfm{P},M,{\mathcal G},{\bf L})$ and  let\
 $T_u\bfm{P}$
  denote the tangent space of $\bfm{P}$ at $u$.
  A {\em connection\/} $\Xi$ in $\bfm{P}$ is an assignment
  of a subspace $Q_u$ of $T_u\bfm{P}$ for every point $u\in \bfm{P}$, such\
 that
  \begin{enumerate}
  \item $T_u \bfm{P}=Q_u+G_u$ (direct sum), where $G_u\subset T_u
    \bfm{P}$ is the subspace tangential to the fiber
    $\pi^{-1}(\pi(u))$; $\pi$ denotes the projection to the base
    point.
  \item $Q_{au}= ({\bf L}_a)_* Q_u$ (${\bf L}_a$ is the left action of
    ${\mathcal G}$)
  \item The mapping $u \mapsto Q_u$ is smooth
  \end{enumerate}
  $Q_u$ is called {\em horizontal}, whereas $G_u$ is called {\em
    vertical\/} subspace of $T_u\bfm{P}$, i.e., a connection uniquely
  assigns to every element $t\in T_u \bfm{P}$ its horizontal and vertical
  components, denoted by $H t$ and $V t$ respectively.
\end{Dfn}
Note that this Definition yields for every vector-field on $M$ a
unique lift, i.e., it allows the parallel transport of vectors along
curves in $M$.
\begin{Thm}
  Let $(\bfm{P},M,{\mathcal G},{\bf L})$ and $\Xi$ be given as above and
  let $X\in\Gamma(TM)$ be a vector-field on $M$, then there exists a
  unique ${\mathcal G}$ invariant lift $\tilde{X}\in \Gamma(T\bfm{P})$ of
  $X$.
\end{Thm}
\noindent
For the proof we refer the reader to~\cite[pp. 65]{kob1:foundation}.
As an immediate consequence one finds
\begin{Lemma}
  Let $\Xi$ be a connection in $(\bfm{P},M,{\mathcal G},{\bf L})$,\
 $\gamma\colon
  [0,1] \rightarrow M$ a curve in $M$, then there exists a
  diffeomorphism
  \[ T_\gamma : \pi^{-1}(\gamma(0)) \rightarrow \pi^{-1}(\gamma(1)) ,\]
  which is called {\em parallel transport along $\gamma$}.
\end{Lemma}
\begin{pf}
  Let $u\in\pi^{-1}(\gamma(0))$ be a point in the principal fiber
  bundle with base point $\gamma(0)$. We define $T_\gamma u$ to be the
  element in the fiber $\pi^{-1}(\gamma(1))$ for which $T_\gamma u =
  \tilde{\gamma}(1)$; $\tilde{\gamma}$ denotes the unique lift of
  $\gamma$ with $\tilde{\gamma}(0)=u$. We remark that the horizontal
  lift of a {\em curve\/} is the integral curve of the horizontal lift
  of its tangent vector field.
  On the other hand the Definition of a parallel transport uniquely
  fixes the connection: Let $\gamma$ be a curve in $M$,
  $u \in\pi^{-1}(\gamma(0))$ a point in the fiber bundle and $t\in T_u
  \bfm{P}$ be an element in the tangent apace of $u$, such that
   \[     d\pi(t)=\gamma_*(0) \equiv \dot{\gamma}(0). \]
   Then the horizontal part $Ht$ of $t$ is defined as follows: Let
   $\tilde{\gamma}$ be the lift of $\gamma$ with
   $\tilde{\gamma}(0)=u$. Then by assumption $\tilde{\gamma}$ denotes
   the parallel transport of $u$ along $\gamma$ and the horizontal
   part of $t$ is defined by $Ht := \tilde{\gamma}_*(0)$. This
   finishes the proof, since such a splitting of the tangent space of
   $\bfm{P}$ defines a connection.
\end{pf}
On every Lorentz manifold there exists a distinguished connection, the
Levi-Civita connection, with respect to which the metric is
covariantly constant.  In order to define the `pull-back' of this
connection to bundles associated to the tangential bundle, we need the
notion of the so called `connection form' $\Upsilon$.
\begin{Dfn}\label{Dfn:connectionForm}
  Let $(\bfm{P},M,{\mathcal G},{\bf L})$ be a principal fiber bundle and
  let Lie-${\mathcal G}$ denote the Lie algebra of ${\mathcal G}$, i.e.,
  Lie-${\mathcal G}$ is the Lie algebra of the left invariant vector
  fields over ${\mathcal G}$.  For every $u \in \bfm{P}$ let $\sigma_u$
  denote the smooth mapping
  \begin{eqnarray*}
    \sigma_u \colon \mathcal {G} & \rightarrow & \bfm{P} \\
    g     & \mapsto     & \sigma_u(g) = {\bf L}_g (u).
  \end{eqnarray*}
  A Lie algebra valued homomorphism $\lambda$ from Lie-${\mathcal G}$ into
  the Lie-algebra of the vertical\footnote{Note that the notion of a
    {\em vertical\/} vector field is independent of a connection.}
  vector fields over $\bfm{P}$ is given by:
  \begin{eqnarray*}
    \lambda \colon \text{Lie-}{\mathcal G}  & \rightarrow & \Gamma \
 (T\bfm{P}) \\
    X    & \mapsto     & \lambda X \colon
                              u \mapsto (\sigma_u)_* (X_e),
  \end{eqnarray*}
  where $X_e$ denotes the value of $X$ at the identity of ${\mathcal G}$.
  Recall that that Lie-${\mathcal G}$ is isomorphic to the tangent space
  $T_e{\mathcal G}$ of ${\mathcal G}$ at the identity. Given a connection
  $\Xi$ in $(\bfm{P},M,{\mathcal G},{\bf L})$, we define a 1-form
  $\Upsilon$ on $\bfm{P}$ with values in Lie-${\mathcal G}$ as follows.
  For every $t\in T_u \bfm{P}$ let $\Upsilon(t)$ to be the unique
  $X\in \text{Lie-}{\mathcal G}$, such that $\lambda X(u)$ is equal to the
  vertical component of $t$. The form $\Upsilon$ is called the {\em
    connection form\/} of the given connection $\Xi$.
\end{Dfn}
{\sloppy
\begin{Prop}[Proposition 1.1. of~{\cite[p.~64]{kob1:foundation}}]
  The Lie-algebra valued 1-form $\Upsilon$ is uniquely
  fixed by the following three requirements.
  \begin{enumerate}
  \item $\Upsilon(\lambda{X}(u))=X$ for all $u\in \bfm{P} $ and $
    X\in\text{Lie-}{\mathcal G}$.
  \item $\Upsilon$ is equivariant, i.e., $\Upsilon \circ d{\bf L}_g=
    \text{ad}(g^{-1}) \circ \Upsilon$ for all $g \in {\mathcal G}$, where
    $\text{ad}$ denotes the adjoint representation of ${\mathcal G}$ in
    $\text{Lie-}{\mathcal G}$.
  \item $\Upsilon$ is smooth.
  \end{enumerate}
\end{Prop}
}
For a given connection $\Xi$, $\Upsilon$ is given by
Definition~\ref{Dfn:connectionForm} above.  For the reverse relation
let $\Upsilon$ be a connection form. Then
\[ Q_u = \{ t \in T_u \bfm{P} |\; \Upsilon (t)=0 \}
 \qquad\text{and}\qquad
 G_u = \{ t \in T_u \bfm{P} | \; t = \lambda \, \Upsilon(t) \} \]
define the horizontal and vertical subspace of $u\in \bfm{P}$ respectively.
Any connection in a principal fiber bundle defines canonically a
connection in all associated fiber bundles: Let $(\bfm{P},M,{\mathcal
  G},{\bf L})$ be a principal fiber bundle with connection $\Xi$,
$\bfm{B}=(\bfm{P}\times \bfm{F})_{/{\mathcal G}}$ an associated fiber
bundle with fiber $\bfm{F}$. A parallel transport in $\bfm{B}$ along
some curve $\gamma$ in $M$ is defined by
\begin{Dfn}
  Let $\gamma\colon [0,1] \rightarrow M$ denote a smooth curve in $M$,
  $\tilde{\pi}$ be the base point projection in $\bfm{B}$ and $b \in
  \tilde{\pi}^{-1}(\gamma(0))$ be a point in $\bfm{B}$. Moreover, let
  $ f \in \bfm{F}$ and $u \in \bfm{P}$ be such that
  $\pi(u)=\gamma(0)$, $[u,f]_\sim = b$. By assumption there exists
  precisely one horizontal lift
  \[
  \tilde{\gamma}\colon [0,1]\rightarrow \bfm{P} \quad
  \mbox{with}
  \quad
  \tilde{\gamma}(0)=u.
  \]
  A lift of $\gamma$ to $\bfm{B}$ is defined by
  \[ \bar{\gamma}(s) := [ \tilde{\gamma}(s),f]_\sim.\qquad s\in [0,1] \]
  We define a {\em parallel transport\/} $\bar{T}_\gamma$ along
  $\gamma$ by
  \[ \bar{T}_\gamma = \hat u \circ T_\gamma(p) \circ \hat{u}^{-1}, \]
  where $\hat{u}$ denotes the mapping $\hat u \colon u \mapsto
  [u,f]_\sim$ from $\bfm{P}$ to $\bfm{B}$.
\end{Dfn}
For the details the reader is referred to the book by S. Kobayashi and
K. Nomizu loc.cit. We are now able to define the spinor connection:
\begin{Dfn}
  Let $\Xi$ denote the connection form of the Levi-Civita connection
  in the Lorentz manifold $(M,g_{ab})$ with spin structure $(SM,\bfm{p})$.
  The connection form $\sigma$ of the spinor connection is the
  pull-back of $\Xi$ to $(SM,\bfm{p})$
  \[
  \sigma := (\varpi)^{-1} \circ \bfm{p}^*(\Xi),
  \]
  where $\varpi$ denotes the Lie-algebra homomorphism between
  $\text{Lie-}\text{Spin}_0$ and $\text{Lie-}\Lor$.
\end{Dfn}
\begin{exmp}
  Let $(M,g_{ab})$ a four dimensional orientable, time orientable
  Lorentz manifold with spinor bundle $(SM,\bfm{p})$. Its Levi-Civita
  connection is characterized by the connection form $\Upsilon$.
  Alternatively it can be regarded as a covariant derivative $\nabla
  \colon \Gamma(TM) \rightarrow \Gamma(T^*M \otimes TM)$: If $e
  \in \Gamma(FM)$ denotes a local orthonormal frame, i.e., for $\mu =
  1,\cdots,4$ the four vector fields $e^{\mu a}\in\Gamma(TM)$ are
  everywhere linear independent and orthonormal, we have
  \begin{equation*}
    \begin{split}
      \nabla_a e^{\mu b}
      & = \sum_\nu ({\Upsilon^\mu}_\nu)_a \otimes e^{\nu b} \\
      & = \sum_{\lambda,\nu} \omega^{\lambda\mu\nu} e^\lambda_a
          \otimes e^{\nu b},
    \end{split}
  \end{equation*}
  where $({\Upsilon^\mu}_\nu)$ are the matrix elements of the
  Lie-$\Lor$ valued one form $e^*(\Upsilon)$ on $M$. Recall that
  $\Upsilon$ is a Lie-$\Lor$ valued one form on $FM$ and an
  orthonormal frame $e$ is a mapping $e \colon M \rightarrow FM$.
  Note that $\omega^{\lambda\mu\nu}$ is {\em anti-symmetric\/} in
  $\mu$ and $\nu$. The relation between $({\Upsilon^\mu}_\nu)$ and the
  usual Christoffel symbols is given by
 \[
  {\Gamma^\mu}_{\nu\rho} = {\Upsilon^\mu}_\nu \circ de (e_\rho),
 \]
 where $e_\rho$ denotes the $\rho$ `leg' of the frame $e$. In terms of
 the partial derivatives of the metric the Christoffel symbols are
 written as
 \[
 {\Gamma^c}_{ab} = \frac{1}{2} g^{cd}
        \left\{ \partial_a g_{bd} + \partial_b
          g_{ad} - \partial_d g_{ab} \right\}.
 \]
 Now let $\gamma_a$ be an arbitrary but fixed representation of the
 Dirac $\gamma$-matrices and let $E\in\Gamma(SM)$ be spin frame with
 $e= p \circ E$. Then one finds for the matrix elements of the
 Lie-$\mbox{Spin}_0$ valued one form $E^*(\sigma)$ on $M$
  \[
  {{\sigma_{a}}^A}_B =
      -\frac{1}{4} {\Gamma^b}_{ac} {{\gamma^c}^A}_C {{\gamma_b}^C}_B
 \]
 Combining these results, one finds that for any spinor tensor
 ${f^{aA}}_B \in \Gamma( TM \otimes DM \otimes D^*M)$ the covariant
 derivative has coordinate representation
 \begin{equation*}
   \begin{split}
     \nabla_a {f^{bA}}_B
     & =
     \partial_a  {f^{bA}}_B + {\Gamma^b}_{ac} {f^{cA}}_B
     + {{\sigma_b}^A}_D {f^{aD}}_B
     - {f^{aA}}_D {{\sigma_b}^D}_B.
   \end{split}
 \end{equation*}
 Further details and proofs can be found in the standard literature,
 for example~\cite{wald:gr,bishop:geom,misner:grav,kob1:foundation}.
\end{exmp}
Let us now investigate the conditions, which prevent the existence of
spin structures for a given manifold. We remind the reader that these
conditions are well known, but in order to formulate them, we have to
deviate.
%
%%%%%%%%%
\section{\v{C}ech-cohomology groups}
\label{sec:chech-chomology-groups}
%%%%%%%%%
%
Let $\bbbZ_2$ denote the multiplicative group $\{+1,-1\}$. Let
$U=(U_i)$ be an open covering of some manifold $M$.
\begin{Dfn}
  A mapping
  \[ f(i_0,\ldots,i_r) \colon \emptyset \neq U_{i_0} \cap
     U_{i_1} \cap \cdots \cap U_{i_r} \rightarrow \bbbZ_2 \]
  which is totally antisymmetric in the indices $(i_0,\ldots,i_r)$ is
  called a {\em r-\v{C}ech-cochain}.
\end{Dfn}
The multiplicative group of all r-cochains is denoted by
$S^r(U,\bbbZ_2)$.  We define a {\em boundary operator\/} $\delta$ by
\begin{align*}
  \delta\colon S^r(U,\bbbZ_2) & \rightarrow S^{r+1}(U,\bbbZ_2)\\
  f(i_0\ldots,i_r)            & \mapsto \prod_{j=0}^{r+1}
                                    f(i_0,\ldots,\hat{i_j},\ldots,i_r),
\end{align*}
where $\hat{\phantom{i}}$ denotes the suppression of the corresponding
index. It is easy to see that $\delta$ is nil potent\footnote{Since
  $\bbbZ_2$ is a multiplicative group, this means $\delta\delta f
  \equiv 1$.}. The set of all cocycles $Z^r(U,\bbbZ_2)$ and coboundaries
$B^r(U,\bbbZ_2)$ respectively are given by
\[
Z^r(U,\bbbZ_2) = \{ f \in S^r(U,\bbbZ_2) | \delta f=1\}
\]
and
\[
B^r(U,\bbbZ_2)=\{f\in S^r(U,\bbbZ_2) | f=\delta g \text{ for some } g
                  \in  S^{r-1}(U,\bbbZ_2)\}
\]
The r-\v{C}ech-cohomology group $H^r(U,\bbbZ_2)$ are the cocycles
modulo the coboundaries.
\[
H^r(U,\bbbZ_2) = Z^r(U,\bbbZ_2)/B^r(U,\bbbZ_2)=
                   \text{ker}\delta_r/\text{Im}\delta_{r-1}
\]
To achieve independence of the covering $U$ we use an inductive limes
process. The result
\[
H^r(M,\bbbZ_2)=\lim_U\text{ind} H^r(U,\bbbZ_2)
\]
is called {\em r-\v{C}ech-cohomology group\/} of $M$ (compare
Appendix~\ref{sec:algTopology}.
%
%%%%%%%%%
\section{Stiefel-Whitney classes}
     \label{sec:stiefel-Whitney-classes}
%%%%%%%%%
%

Stiefel-Whitney classes are characteristic classes with values in
$H^r(M,\bbbZ_2)$. In general these classes are defined for all fiber
bundles with an orthogonal group as fiber group. We are interested in
the special case of the tangent bundle of a four dimensional
Lorentz manifold only, i.e., a fiber bundle with $O(1,3)$ as fiber
group. Let $(M,g_{ab})$ be as usual a four dimensional Lorentz
manifold and $U=(U_i)$ a simple covering of $M$, that is a covering
such that $U_i \cap U_j$ is either empty or simply connected for
all pairs $(i,j)$. Let $\{e_{i\alpha}\}_{(\alpha=1,2,3,4)}$ denote a
vierbein field over $U_i$. Let $U_j$ denote another element of $U$
then
\[
e_{i\alpha}=t_{ij}e_{j\alpha},
\]
where $t_{ij}\colon U_i \cap U_j \rightarrow O(3,1)$ is the transition
function for these two vierbein fields. The mapping
\begin{align*}
  f\colon U_i \cap U_j & \rightarrow \bbbZ_2 \\
                t_{ij} & \mapsto \det(t_{ij}) = \pm 1
\end{align*}
defines a 1-cochain, since $f(i,j)=f(j,i)$. Moreover, we have
\[
(\delta f)(i,j,k)= \det(t_{jk}) \det(t_{ik}) \det(t_{ij}) =
\det(t_{jk}t_{ik}t_{ij})= +1,
\]
since for all transition functions $t_{ij}$ in a fiber bundle the
consistency condition
\begin{equation}
  \label{eq:consistency}
  t_{ij}(p) t_{jk}(p) = t_{ik}(p)
\end{equation}
must be satisfied. It follows that $f\in Z^1(M,\bbbZ_2)$ is a cocycle
defining an element
\[
[f] \in H^1(M,\bbbZ_2)
\]
in the first cohomology group.
\begin{Lemma}
  $[f] \in H^1(M,\bbbZ_2)$ is independent of the vierbein field
  $\{e_{i\alpha}\}$.
\end{Lemma}
\enlargethispage{-3em}
\begin{pf}
  Let $\{\tilde{e}_{i\alpha}\}$ denote another vierbein field over
  $U_i$ with $\tilde{e}_{i\alpha}=h_i e_{i\alpha}$, where $h_i$ is a
  mapping from $U_i$ into $O(1,3)$. Assume we are given the following
  transition functions $\tilde{t}_{ij}$, such that $
  \tilde{e}_{i\alpha}= \tilde{t}_{ij}\tilde{e}_{j\alpha}$ is
  satisfied. Between the original transition functions $t_{ij}$ and
  $\tilde{t}_{ij}$ there is the following relation:
  \[
  \tilde{t}_{ij} = h_i t_{ij} {h^{-1}}_j.
  \]
  Therefore
  \[
  \tilde{f}(i,j) := \det(h_i t_{ij} {h^{-1}}_j) = f(i,j) \cdot \delta(f_0)
  \]
  where $f_0(i) := \det (h_i)$. Note that
  $\det(h_i)\equiv\det({h^{-1}_i})$ since $f_0 \equiv \pm 1$. This
  shows that $f$ and $\tilde{f}$ differ by a coboundary only, i.e.,
  they define the same element $[f] \in H^1(M,\bbbZ_2)$.
\end{pf}
\begin{Dfn}
  The set of all elements $H^1(M,\bbbZ_2)\ni [f] = w_1(M)$ is called
  {\em first Stiefel-Whitney class of $M$}.
\end{Dfn}
\begin{Lemma}[\cite{nak:geometry}]
  $M$ is orientable if and only if $w_1(M)$ is trivial.
\end{Lemma}
\begin{pf}
  Assume $M$ to be orientable. The structure group of the tangential
  bundle is reducible to $\text{SO}(1,3)$ in this case and
  $\det{(t_{ij})}\equiv{}+1$. This means that $w_1(M)$ is trivial. On
  the other hand assume the triviality of $w_1(M)$. This implies that
  $f(i,j)$ is a coboundary, i.e., $f(i,j)=(\delta f_0)(i,j)$. Since
  $f_0(i)=\pm 1$ for all $i$, we can find suitable elements
  $h_i\colon U_i\rightarrow O(1,3)$, such that $f_0(i)=\det{(h_i)}$
  for all $i$. Suppose we are given vierbein fields $\{e_{i\alpha}
  \}$ on $U_i$. Use the functions $h_{i}$ to define new vierbein
  fields via $\tilde{e}_{i\alpha} = h_i e_{i\alpha}$. The
  corresponding transition functions $\tilde{t}_{ij}$ satisfy the
  relation $\det(\tilde{t}_{ij})= +1$ for all pairs of indices $(i,j)$
  such that $U_i \cap U_j \neq \emptyset$:\\
  Suppose we had $f(i,j) := \det(t_{ij}) = -1$ for some indices
  $(i,j)$. Choosing $h_i$ and $h_j$ such that $  \det(h_i)=+1$ and
  $\det(h_j)=-1$ respectively, we find for the new functions
  $\tilde{f}(i,j)$:
  \[
  \tilde{f}(i,j) = \det(h_i t_{ij} {(h^{-1})}_j) = +1
  \]
  \samepage
\end{pf}
For the Definition of the second Stiefel-Whitney class consider a
four dimensional {\em oriented\/} Lorentz manifold $(M,g_{ab})$. Let
$TM$ denote its tangent bundle. Note that $TM$ has $SO(1,3)$ as
fiber group. Let $U=\{U_i\}$ denote a simple covering of $M$
and $t_{ij}\colon M \rightarrow SO(1,3)$ are the local
transition functions. We will denote the universal (twofold) covering
group of $SO(1,3)$ by $\text{Spin}$. Let $\phi\colon \text{Spin}
\rightarrow SO(1,3)$ be the corresponding 2:1 group homomorphism. A
family $\{\tilde{t}_{ij}\}$, $\tilde{t}_{ij} \colon M \rightarrow\
 \text{Spin}$
is called a {\em lift\/} of $t_{ij}$, if and only if for all pairs $(i,j)$\
 such
that $U_i \cap U_j \neq\emptyset$
\begin{enumerate}
  \item $\phi(\tilde{t}_{ij})= t_{ij}$
  \item $\tilde{t}_{ji} = \tilde{t}^{-1}_{ij}$
\end{enumerate}
are satisfied. Such families always exist locally. Since we have the
consistency condition Eqn.~(\ref{eq:consistency}) it follows that
\[
\phi(\tilde{t}_{ij} \tilde{t}_{jk} \tilde{t}_{ki} ) =\bbb1
\]
which implies
\[
(\tilde{t}_{ij} \tilde{t}_{jk} \tilde{t}_{ki} ) =\pm \bbb1.
\]
On the other hand the transition functions $\tilde{t}_{ij} \in
\text{Spin}$ are defining a spin structure if and only if they
fulfill the consistency condition
\begin{equation}
  \label{eq:consistencySpin}
  \tilde{t}_{ij} \tilde{t}_{jk}= \tilde{t}_{ik} \quad \bimply \quad
  (\tilde{t}_{ij} \tilde{t}_{jk} \tilde{t}_{ki}) =  + \bbb1
\end{equation}
We define a 2-\v{C}ech-cochain $f\colon U_i \cap U_j \cap U_k
\rightarrow \bbbZ_2$ by
\[
  (\tilde{t}_{ij} \tilde{t}_{jk} \tilde{t}_{ki}) = f(i,j,k) \cdot \bbb1
\]
By Eqn.~(\ref{eq:consistencySpin}) we have
$f(i,j,k)\in S^2(M,\bbbZ_2)$.\footnote{It is enough to show the
  symmetry: By Definition we have $f(i,j,k)=f(i,k,j)^{-1}$. It remains
  to show that $f(i,j,k)=f(j,k,i)$.
  \[
    f(i,j,k)= \tilde{t}_{ij} \tilde{t}_{jk} \tilde{t}_{ki} =
    \tilde{t}_{ij} \tilde{t}_{jk}
    \underbrace{ \tilde{t}_{ki} \tilde{t}_{ij} }_{=f(j,k,i)\tilde{t}_{kj} }
    \tilde{t}_{jk} \tilde{t}_{ki} f(i,j,k)
    = f(i,j,k)f(j,k,i)f(i,j,k)
  \]
  It follows that $f(i,j,k)=f(j,k,i)^{-1}=f(j,k,i)$.
  } We also find
\begin{equation*}
  \begin{split}
     (\delta f)(i,j,k,l) & =  f(j,k,l)f(i,k,l)f(i,j,l)f(i,j,k) \\
     & =  (\pm 1)^{2\cdot 6} \; t_{ij}t_{jk}t_{ki} \; t_{ij}t_{jl}t_{li} \;
                         t_{ik}t_{kl}t_{li} \; t_{jk}t_{kl}t_{li}  \\
     &  = + \bbb1.
  \end{split}
\end{equation*}
Note that the second equality follows from the fact that we have a
freedom of sign in every transition function and all transition
function occur twice. Finally, using the consistency condition
Eqn.~(\ref{eq:consistency}) for the functions $t_{ij}$ gives the
result. It follows that $f$ defines an element in
$H^2(M,\bbbZ_2)$. The set of all elements in
$H^2(M,\bbbZ_2) \ni [f]= w_2(M)$ is called {\em 2. Stiefel-Whitney
  class\/} of $M$. It is easy to show that $[f]$ is independent of the
vierbein family $\{e_{i\alpha}\}$ and the `lift'.
%
%%%%%%%%%
\section{An existence theorem}
\label{sec:preventSpinorStructure}\label{sec:existenceSpinorStructure}
%%%%%%%%%
%
In this section we prove the following Theorem which ensures the existence\
 of\
 a
spin structure for a given manifold. A particular property which prevents
the existence of spin structures is described next.
\begin{Thm}[\cite{nak:geometry}]\label{thm:existenceSpinorStructure}
  Let $TM$ denote the tangent bundle of an orientable, time orientable,
  four dimensional Lorentz manifold $M$. There exists a spinor bundle
  over $M$ precisely when the second Stiefel-Whitney class
  $w_2(TM)$ of $TM$ is trivial.
\end{Thm}
It is worth noting that this Theorem involves {\em global\/} properties of\
 the
manifold and is independent of the metric structure, i.e., the
existence of a spin structure is a {\em topological\/} property of the
manifold.
\begin{pf}
Suppose there exists a spin structure over $M$, then all transition
functions $\tilde{t}_{ij}$ satisfy the compatibility condition
Eqn.~(\ref{eq:consistencySpin}) for all overlapping regions $U_i$,
$U_j$, $U_k$. It follows that $w_2(M)$ is trivial. On the other hand
assume the triviality of $w_2(M)$, i.e.,
\[
f(i,j,k)= (\delta f_1)(i,j,k) = f_1(j,k) f_1(i,k) f_1(i,j),
\]
where $f_1$ is a 1-cochain. Let $\tilde{t}_{ij}$ be some given
transition functions. The transition functions
\[
\tilde{t}'_{ij} = f_1(i,j) \tilde{t}_{ij}
\]
satisfy
 \[ (\tilde{t'}_{ij}\tilde{t'}_{jk}\tilde{t'}_{ki})=
 f_1(i,j)f_1(j,k)f_1(k,i) \cdot
 \tilde{t}_{ij}\tilde{t}_{jk}\tilde{t}_{ki} =\delta f_1 \delta f_1 =
 + \bbb1, \]
which ensures that $\{\tilde{t'}_{ij}\}$  defines a spin structure over $M$.
\end{pf}
We do now consider a `counterexample': Let $(M,g_{ab})$ be a
4-dimensional orientable and time orientable Lorentz manifold.
Consider a family $\{l_s\}_{s\in[0,1]}$ of closed curves in $M$.
Suppose further that $l_0 \equiv l_1 \equiv p$, i.e., $l$ is a special
embedding of the two-sphere into $M$. Let $\{e_{i\alpha}\}$ denote an
oriented, time oriented vierbein in the tangential space of the point
$p$. The parallel transport of this vierbein along $l$ yields a curve
in the proper Lorentz group since both, the starting and the ending
point of $l$ coincide:
\begin{align*}
  h_s \colon [0,1] & \rightarrow \Lor \\
  s                & \mapsto     h_s
\end{align*}
Moreover, since $h_0 \equiv h_1 \equiv \bbb1$ this curve is closed.
\begin{Lemma}\label{Lemma:counterExmp}
  If $h_1$ is not homotop to to the trivial curve then there exists no
  notion of a parallel transport for spinors, i.e., $M$ does not admit
  a spin structure\footnote{The parallel transport on the manifold
    induces --by a pull back of its connection form (see
    section~\ref{sec:nablaSpinor})-- a parallel transport in the
    spin structure.}.
\end{Lemma}
\begin{pf*}{Proof by contradiction}
  Assume we have a spin structure and $h_1$ is {\em not\/} null
  homotop. Consider the parallel transport of some spinor along $l_1$.
  On one hand $l_1$ is a constant curve, such that the spinor remains
  unchanged during its transport. On the other hand, since $h_1$ is
  homotop to a $2\pi$ rotation by assumption we conclude using the
  continuity of the parallel transport, that the spinor must change
  sign during the transport. Contradiction.
\end{pf*}
Other properties of Lorentz manifolds {\em without\/} spin structure
are investigated by
Ohlmeyer~\cite{ohlmeyer:94}, to which we refer the interested reader.
To summarize the results of this section, the following statements are
equivalent: Let $(M,g_{ab})$ denote a 4-dimensional orientable, time
orientable Lorentz manifold, then
\begin{description}
\item[  I] $M$ admits a spin structure
\item[ II] The second Stiefel-Whitney class of $M$ is trivial
\item[III] Let $\bfm{B}$ denote the bundle of all orthonormal bases of $M$.
  Its fundamental group can be written as $\pi_1(\bfm{B}) = \pi_1(M) \times
  \bbbZ_2$, where $\pi_1(M)$ denote the fundamental group of the
  manifold $M$, $\bbbZ_2=\{-1,+1 \}$ and $\times$ denotes the direct
  product for groups.
\item[ IV] On every two sphere in $M$ one can find three everywhere
  linearly independent vector fields.
\end{description}
\noindent
If $M$ is in addition non compact
\begin{description}
\item[  V] $M$ is parallelizable~\cite{geroch:70}
\end{description}
{\sloppy
Note that `{\bf V}' implies that all 4-dimensional globally hyperbolic
manifolds admit a spin structure (See \cite{geroch:70}): As is pointed
out by Geroch, an orientable or time orientable manifold does admit a
spin structure precisely when every of its covering manifolds does. Without
loss of generality we may therefore restrict ourselves to the
case of an orientable, time orientable, globally hyperbolic
manifold. These manifolds can be written as $M=\bbbR \times \Sigma$,
where $\Sigma$ is a smooth spacelike hypersurface. $\Sigma$ is three
dimensional and by assumption orientable; hence parallelizable (see
e.g.~\cite[page 204]{steenrod:topology}). Due to the fact that $M$ is
a Cartesian product of ${\mathbb R}$ and $\Sigma$ this parallelizability can\
 be
extended to the whole of $M$. Using `{\bf V}' this implies the existence
of a spin structure.
}
\begin{pf*}{On the Proof of the equivalences}
  \begin{description}
  $\phantom{~}{}$\par
  \item[I$\bimply$II ] Theorem~\ref{thm:existenceSpinorStructure}.
  \item[I$\Leftarrow$III] Consider the following covering space
    $\tilde{\bfm{B}}$ of $\bfm{B}$. Let $p_0\in \bfm{B}$ be fixed. Let $p$\
 denote a
    curve in $\bfm{B}$  from the point $p_0$ to $p\in \bfm{B}$. We introduce\
 the
    following equivalence relation:
    \begin{eqnarray*}
       p \sim \tilde p & :\bimply & \mbox{i) $p$ and $\tilde p$ have
                                       the same end points}\\
                       &          & \mbox{ii) $p \circ \tilde p$
                                      belongs to the homotopy class
                                    $(g_1,\bbb1) \in \pi_1(\bfm{B})$ in
                                     $\bfm{B}$,}                      \\
                       &          & \mbox{where $g_1 \in \pi_1(M)$
                                    is a suitable element of the
                                    homotopy group of $M$}
      \end{eqnarray*}
      It follows that $\tilde{\bfm{B}}$ is a $\SL2C$ bundle over $M$. i.e.,
      a spin structure.
    \item[I$\Rightarrow$III] This implication follows immediately from
      the Definition of the spin structure: A spin structure is
      the two-fold covering of the bundle of all orthonormal
      bases. Thus any simply closed curve\footnote{A curve $\gamma$ is
        called {\em simple}, if and only if there exists no closed curve
        $\tilde{\gamma}$ such that $\tilde{\gamma}\circ\tilde{\gamma}
        = \gamma$.} admits precisely two lifts --corresponding to the
      two elements in $\bbbZ_2$-- in $SM$. Moreover, traveling along
      $\gamma$ twice, yields lifts which are homotop.
    \item[I$\bimply$IV ] We refer to the book of
      Penrose~\cite{pen:twist1} for the proof of this
      equivalence\footnote{N.b.\ Actually Penrose shows: `{\bf IV}'
        implies that no situation as described in the `counterexample'
        Lemma~\ref{Lemma:counterExmp} above can occur.}.
    \item[I$\bimply$V ] This equivalence is shown in~\cite{geroch:70}.
      We refer the reader to the original work for the proof. Note
      however that `{\bf V} $\Rightarrow$ {\bf I}' is trivial and
      follows form the Definition of a spin structure or from: `{\bf V}
      $\Rightarrow$  {\bf IV} \text{and} {\bf IV} $\bimply$ {\bf I})'.
  \end{description}
\end{pf*}
%
%%%%%%%%%%%%%%%%%%%%%%%%%%%%%%%%%%%%%%%%%%%%%%%%%%%%%%%%%%%%%%%%%%%%
\chapter{An analogon of the free Wess-Zumino model}
\label{sec:analogonWess}
%%%%%%%%%%%%%%%%%%%%%%%%%%%%%%%%%%%%%%%%%%%%%%%%%%%%%%%%%%%%%%%%%%%%
%
In this chapter we define an analogon of the free Wess-Zumino model.
It was already mentioned above that supersymmetric models do not exist
on a general spacetime. Thus we can only mimic the structure of
supersymmetric models defined on Minkowski space.  Since the
Wess-Zumino model is the simplest of those, we mimic its structure, by
considering a model which has the same field content. Namely one that
consists of two bosonic and one fermionic (Majorana) field. We will
refer to this kind of models as {\em locally supersymmetric\/} models.
It is emphasized that this notion has no relation to supergravity,
where local, i.e., spacetime dependent, supersymmetry transformations
are considered. In our setting the word `local' should remind of the
fact that our model `looks' like a supersymmetric model at every {\em
  point\/} of the spacetime.  We begin with the description of the
classical settings for the bosonic and fermionic parts, followed by a
brief introduction to the algebraic approach to quantum field theory
on curved spacetimes. In section~\ref{sec:hadamardStates} we review
the global Hadamard condition of Kay and Wald and extend their
definition to Fermi fields. In order to describe Hadamard states
locally we introduce wave front sets in the next section. The chapter
is closed with a proposal of a general wave front set spectrum
condition, which might generalize the usual Minkowski spectrum
condition to Lorentz manifolds and should be applicable even to models
containing interaction.
Let $M$ be a four dimensional, smooth, second countable, paracompact,
Hausdorff manifold equip\-ped with a metric field $g_{ab}$ with
signature $(+,-,-,-)$. The pair $(M,g_{ab})$ is called a {\em (curved)
  spacetime}. In the following we will use the timelike sign
convention of Landau and Lifshitz which differs from the Misner, Thorne
and Wheeler convention (\cite{misner:grav}) in the sign of the metric.
In particular we define the Riemann tensor by
\[ (\nabla_a \nabla_b - \nabla_b \nabla_a) \omega_c =: {R_{abc}}^{d}
\omega_d \hspace{4em} \mbox{$\forall$ dual vector fields $\omega_c$} \]
and the Ricci tensor by $R_{ab}:={R_{acb}}^c$. Let us assume further
that $(M,g_{ab})$ is orientable, time orientable and globally
hyperbolic.  Recall that a manifold is called globally hyperbolic if
and only if it is of the form $M=\Sigma \times {\mathbb R}$, where
$\Sigma$ is a (spacelike) Cauchysurface. The last assumption
guarantees the existence of a spin-structure, which is necessary for
the definition of spinor fields (see \cite{geroch:68} or
Theorem~\ref{thm:existenceSpinorStructure} above).
\begin{rem}
  It should be noted that at least for the metric both sign conventions
  are used in the literature. This makes the comparison of results
  sometimes quite difficult, especially if the authors do not
  explicitly state their conventions. For a recipe to change the
  formulas between different conventions, the reader is referred to
  the book of Birrell and Davis~\cite{birrell:qft}.
\end{rem}
\vspace{0.5cm} \noindent
{\em Spin $0$\/}: A classical neutral massive scalar field $\Phi \in
C^\infty (M)$ propagating in a curved spacetime $(M,g_{ab})$ is
described by the action functional
\begin{equation} \label{scalaraction}
S[\Phi ] =  + \frac{1}{2} \int
           \left( \Phi_{;a} \Phi^{;a} -
                   (m^2 + \xi R ) \Phi^2 \right) d\mu,
\end{equation}
where $d\mu$ is the invariant volume measure, $m \geq 0$ is the mass,
$\xi$ describes the coupling ($\xi = 0$ minimal, $\xi =
\frac{1}{6}$ conformal and $\xi = \frac{1}{4}$ SUSY) and $R \equiv
{R_a}^a$ is the scalar curvature. As usual a semicolon denotes
covariant differentiation. The field equation is the Klein
Gordon equation
\[ 0 = \frac{\delta S}{\delta \Phi} =
       \left( \Box + ( m^2 + \xi R ) \right) \Phi (x), \]
where $\Box= \nabla_\mu \nabla^\mu$ is the D'Alembert operator.
The canonical energy momentum tensor is
\begin{equation*}
  \begin{split}
    \Theta^{ab} & =
         \frac{\delta {\mathcal L}}{\delta (\frac{\partial
             \Phi_i}{\partial x^{(a}})}
         \frac{\partial \Phi_i}{\partial x_{b)}} - g^{ab} {\mathcal L} \\
         & =
          \Phi^{;(a}\Phi^{;b)}-\frac{1}{2} g^{ab} \left( \Phi_{;c}\Phi^{;c} -
          (m^2 + \xi R) \Phi^2 \right),
  \end{split}\notag
\end{equation*}
where ${\mathcal L}$ denotes the Lagrangian density of $\Phi$ and $(ab)$
means symmetrization. However, as potential source term for the
Einstein equation the improved energy momentum tensor $T^{ab}$
introduced by Callan et al.~\cite{callanEtAl:70} seems more
appropriate. This improved energy momentum tensor is obtained by
functionally differentiating the action with respect to the metric
field. This treats the metric as a dynamical variable which seems
natural in view of general relativity. (See the discussion in the
paper by DeWitt and Brehme~\cite{DeWitt:60} or Fulling's
book~\cite{Fulling:aspects_of_qft} for further details.)  We find
that in the SUSY coupled case
\begin{eqnarray} \label{eq:scalaremt}
T^{ab} & := & 2 g^{-1/2} \frac{\delta S}{\delta g_{ab}}  \\
& \equiv &
\left\{
  \frac{1}{4} \{\Phi^{;a},\Phi^{;b} \}_+
- \frac{1}{4} \{\Phi^{;ab},\Phi \}_+
+ \frac{1}{16} g^{ab} \{{\Phi_{; c}}^{c},\Phi \}_+
- \frac{1}{8} ( R^{ab} + \frac{1}{8} R g^{ab}) \{\Phi,\Phi \}_+
\right. \nonumber \\
& & \left.
\phantom{g^{1/2} \biggl\{ }
+ \frac{1}{16} m^2 g^{ab} \{\Phi,\Phi \}_+
    \right\} \nonumber
\end{eqnarray}
where $\{ \cdot,\cdot \}_+$ denotes the anticommutator and $g^{1/2}$ is
the square root of the determinant of the metric. We now do
point splitting: Write each bracket in Eqn.~(\ref{eq:scalaremt}) in the form
\begin{eqnarray*}
{\{ \Phi^{;a},\Phi^{;b}\}}_+ & \rightarrow &
      \lim_{x' \rightarrow x} \frac{1}{2} \left(
       {g^a}_{c '} {\{\Phi,\Phi '\}_+}^{; c ' b} +
       {g^b}_{c '} {\{\Phi,\Phi '\}_+}^{; ac '} \right) \\
{\{ \Phi^{;ab},\Phi\}}_+ & \rightarrow &
      \lim_{x' \rightarrow x} \frac{1}{2} \left(
       {\{\Phi,\Phi '\}_+}^{;ab} +
       {g^a}_{c '}{g^b}_{d '}
       {\{\Phi,\Phi '\}_+}^{; c ' d '} \right) \\
{\{ \Phi , \Phi \}}_+ & \rightarrow &
      \lim_{x' \rightarrow x} \left(
       {\{\Phi,\Phi '\}_+} \right)
\end{eqnarray*}
where $x'$ is a point near $x$, $\Phi ' = \Phi(x')$ and ${\{\cdot ,
  \cdot\}_+}^{;a '}$ represents the covariant derivative at $x'$.
${g^a}_{c '}$ is the bi-tensor of parallel transport defined by the
equation $ \sigma_{;a} {g^a}_{c '} = 0 $ together with the boundary
condition ${g^a}_c(x,x)={\delta^a}_c(x,x)$. The bi-scalar
$\sigma(x,x')$ is half the square of the geodesic distance between $x$
and $x'$. Recall that $\sigma$ is well defined whenever $x$ and $x'$
are sufficiently close to each other.  We end up with
\begin{equation}
  \label{eq:scalaremtsplit}
  \begin{split}
    T^{ab} & = \lim_{x'\rightarrow x}
       \biggl\{
       \frac{1}{8}
       \left(
         {g^a}_{a '} {\{\Phi,\Phi '\}_+}^{;a ' b} +
         {g^b}_{b '} {\{\Phi,\Phi '\}_+}^{; a b '}
       \right)
       - \frac{1}{8} \left(
       {\{\Phi,\Phi '\}_+}^{;ab} +
       {g^a}_{a '}{g^b}_{b '}
                {\{\Phi,\Phi '\}_+}^{a ' b '}
              \right) \\
      &  \phantom{\lim_{x' \rightarrow x} g^{1/2} \biggl\{}
      + \frac{1}{32} g^{ab} \left(
       {\{\Phi,\Phi '\}_{+;c} }^{c} +
       g_{cd} {g^c}_{c '}{g^d}_{d '}
                {\{ \Phi,\Phi ' \}_+}^{c ' d '} \right)\\
      &  \phantom{\lim_{x' \rightarrow x} g^{1/2} \biggl\{}
      - \frac{1}{8} ( R^{ab} + \frac{1}{8} R g^{ab}) \{\Phi,\Phi'\}_+
      + \frac{1}{16} m^2 g^{ab} \{\Phi,\Phi'\}_+
      \biggr\}
  \end{split}
\end{equation}
\vspace{0.5cm} \noindent
{\em Spin $1/2$\/}: A classical (anti-commuting!) massive Majorana
spinor field $\Psi$  propagating in a spacetime $(M,g_{ab})$ with
spin structure $(SM,p)$ has the action functional (see
\cite{DeWitt:group,birrell:qft})
\[
S[\Psi ] =
  \frac{1}{2} \int  \Psi^+(i \notnabla \Psi - m \Psi) d\mu.
\]
Note the factor $1/2$ which is necessary since we assume a Majorana
spinor field.  The field equation is the Dirac equation
\[ ( i \notnabla - m) \Psi = 0 \]
and for the improved energy momentum tensor one finds~(\cite{birrell:qft})
\[
T^{ab} = -
  \frac{i}{8}  \left(
    {\Psi^+}^{;a} \gamma^b \Psi
  + {\Psi^+}^{;b} \gamma^a \Psi
  - {\Psi^+} \gamma^a \Psi^{; b}
  - {\Psi^+} \gamma^b \Psi^{; a}
                       \right)
\]
Since $\Psi$ is a Majorana spinor, we use the identities of
Appendix~\ref{sec:usefulForm} and write:
\[
T^{ab} = \frac{i}{8} \sum_{A,B} {\gamma^{(a A}}_B \left(
{[\Psi^{B},{{\Psi^+}_A}^{;b )}]}_- - {[\Psi^{;b ) B},{\Psi^+}_A]}_-
\right)\] where $(~)$ means symmetrization, $[\cdot,\cdot]_-$ is the
commutator and a local spin frame $E_A$ was chosen. Note that $T^{ab}$
transforms as a second rank tensor under changes of frames. Since it
is already symmetric we do point splitting by writing
\[
T^{ab} = \lim_{x'\rightarrow x} \frac{i}{8}
\sum_{A,B}
  {\gamma^{(a A}}_B
   \left(
     {g^b}_{b '} {{\mathcal J}_A}^{A '}
       {[ \Psi^{B},{ \Psi^{'+} }_{A '} ] }_-^{;b ' )}
   - {{\mathcal J}_A}^{A '} {[ \Psi^B,{\Psi^{'+} }_{A '} ] }_-^{;b )}
   \right)
\]
Here ${{\mathcal J}_A}^{A '}$ is the bi-spinor of parallel transport which
is defined by the equation $ \sigma^{a} {{{\mathcal J}_A}^{B '}}_{;a} = 0
$ together with the boundary condition ${{\mathcal J}_A}^{B}(x,x)=
{\frakI_A}^B(x,x)$.
\vspace{0.5cm} \noindent {\em Local supersymmetry}: A classical
locally supersymmetric field theoretical model is a free field model
with two SUSY coupled neutral scalar fields $(A,B)$ together with one
(anti-commuting) Majorana spinor field $(\Psi)$.  The action
functional of this model is
\[
S[A,B,\Psi]=
 \frac{1}{2} \int
           \left({ A}_{;a} {A}^{;a} -
                   ( m^2 - \xi R)  {A}^2
               + { B}_{;a} {B}^{;a} -
                   (m^2- \xi R)  {B}^2
               + \Psi^+(i \notnabla - m ) \Psi) \right) d\mu.
\]
The equation of motion for the three fields obviously decouples,
giving the Klein Gordon equation for both $A$ and $B$
and the Dirac equation for $\Psi$. The already point-separated
improved energy momentum tensor is
{\small
\begin{equation}
  \label{eq:SUSYemtsplit}
  \begin{split}
T^{ab}  & =  \lim_{x' \rightarrow x}\\
        & \biggl\{
        \frac{1}{8}
        \left(
          {g^a}_{a '} {\{A,A '\}_+}^{;a ' b} +
          {g^b}_{b '} {\{A,A '\}_+}^{; a b '}
        \right)
        - \frac{1}{8} \left(
        {\{A,A '\}_+}^{;ab} +
        {g^a}_{a '}{g^b}_{b '}
        {\{A,A '\}_+}^{a ' b '}
        \right) \\
        &  \phantom{=\biggl\{}
        + \frac{1}{32} g^{ab} \left(
        {\{A,A '\}_{+;c}}^{c} +
       g_{cd} {g^c}_{c '}{g^d}_{d '}
                {\{A,A '\}_+}^{c ' d '} \right)\\
      &  \phantom{= \biggl\{}
      - \frac{1}{8} ( R^{ab} + \frac{1}{8} R g^{ab}) \{A,A'\}_+
      + \frac{1}{16} m^2 g^{ab} \{A,A'\}_+
      + \frac{1}{8}
       \left(
         {g^a}_{a '} {\{B,B '\}_+}^{;a ' b} +
         {g^b}_{b '} {\{B,B '\}_+}^{; a b '}
       \right)\\
       &  \phantom{= \biggl\{}
       - \frac{1}{8} \left(
       {\{\phi,\phi '\}_+}^{;ab} +
       {g^a}_{a '}{g^b}_{b '}
                {\{B,B'\}_+}^{a ' b '}
              \right)
       + \frac{1}{32} g^{ab} \left(
       {\{B,B '\}_{+;c} }^{c} +
       g_{cd} {g^c}_{c '}{g^d}_{d '}
                {\{B,B '\}_+}^{c ' d '} \right)\\
      &  \phantom{= \biggl\{}
      - \frac{1}{8} ( R^{ab} + \frac{1}{8} R g^{ab}) \{B,B'\}_+
      + \frac{1}{16} m^2 g^{ab} \{B,B'\}_+\biggr\} \\
      &  \phantom{= \biggl\{}
      +
      \frac{i}{8}
      \sum_{A,B}
      {\gamma^{(a A}}_B
      \left(
        {g^b}_{b '} {{\mathcal J}_A}^{A '}
        {[ \Psi^{B},{\Psi^{'+} }_{A '} ] }_-^{;b ' )}
        - {{\mathcal J}_A}^{A '} {[ \Psi^B,{\Psi^{'+} }_{A '} ] }_-^{;b )}
      \right)
    \end{split}
  \end{equation}
}
On a Ricci flat spacetime ($R^{ab} \equiv{}0$) we obtain,
\begin{equation}\label{eq:SUSY_EMTR_O}
  \begin{split}
    T^{ab} & = \lim_{x'\rightarrow x}\\
           & \phantom{=}
           \biggl\{ \frac{1}{8}
           \left(
             {g^a}_{c '} {\{{A},{A} '\}_+}^{; c ' b} +
             {g^b}_{c '} {\{{A},{A} '\}_+}^{; a c '}
           \right)
           - \frac{1}{8}
           \left(
             {\{{A},{A} '\}_+}^{;ab} +
             {g^a}_{c '}{g^b}_{d '}
             {\{{A},{A} '\}_+}^{; c ' d '}
           \right)\\
           & \phantom{=\biggl\{}{}
           + \frac{1}{8}
           \left(
             {g^a}_{c '} {\{{B},{B} '\}_+}^{; c ' b} +
             {g^b}_{c '} {\{{B},{B} '\}_+}^{; a c '}
           \right)
           - \frac{1}{8}
           \left(
             {\{{B},{B} '\}_+}^{;ab} +
             {g^a}_{c '}{g^b}_{d '}
             {\{{B},{B} '\}_+}^{c ' d '}
           \right)\\
           & \phantom{=\biggl\{}{}
           + \frac{i}{8}
           \sum_{A,B}
           {\gamma^{(a A}}_B
           \Bigl(
             {g^b}_{b '} {{\mathcal J}_A}^{A '}
             {[\Psi^{B},{\Psi^{'+}}_{A '} ] }_-^{;b ' )}
             -  {{\mathcal J}_A}^{A '} {[ \Psi^B,{\Psi^{'+}}_{A '} ] }_-^{;b\
 )}
           \Bigr)
           \biggr\} \qquad \text{iff $R^{ab}\equiv{}0$}.
  \end{split}\notag
\end{equation}
Note that we used the field equations for $A$ and $B$.
\begin{rem}
  For classical fields Eqn.~(\ref{eq:SUSYemtsplit}) is well defined.
  However in the quantized version of this model the basic fields are
  operator valued distributions on some GNS Hilbertspace (see
  section~\ref{sec:algApproach} this chapter below). In this case
  Eqn.~(\ref{eq:SUSYemtsplit}) contains --in the limit-- the product
  of the fields in the sense of a product of distributions. As is well
  known from Minkowski space such products of operator valued
  distributions do not make sense in general. Thus at this stage
  Eqn.~(\ref{eq:SUSYemtsplit}) is only formal for the quantized model.
  On the other hand Christensen describes in~\cite{chris:76,chris:78}
  a method for renormalizing the expectation values of $T^{ab}$ given
  by Eqn.~(\ref{eq:SUSYemtsplit}) for a special class of states,
  namely for Hadamard states.  We will use his results in
  chapter~\ref{sec:emtCurved} to show that
  Eqn.~(\ref{eq:SUSY_EMTR_O}) without renormalization remains well
  defined after quantization.
\end{rem}
%
%%%%%%%%%
\section{A locally conserved `supercurrent'}
\label{sec:supercurrent}
%%%%%%%%%
%
It is amusing to observe that on a special class of curved spacetimes
a locally supersymmetric model possesses due its Bose-Fermi symmetry a
locally conserved `supercurrent'. In contrast to the Minkowski space,
however, this current can {\em not\/} be obtained by a Noether method,
since on generic spacetimes there exist no covariant constant
(i.e.~parallel) spinor fields. We remark that there is a tight
connection between such parallel spinors and the notion of a K\"ahler
manifold, for which we refer the interested reader to chapter 10.8
in~\cite{benn:spinors}.
To obtain this current we consider the `variable' part of
Eqn.~(\ref{eq:supercurrentMink}).
\begin{Prop}\label{prop:supercurConsCurv}
  The current
  \begin{equation}
    \label{eq:supercurrentCurved}
    k^{aA} = i\left( \AiB ( i \lvec{\notnabla} - m) \gamma^a \Psi \right)^A
  \end{equation}
  is locally conserved on spacetimes with vanishing scalar curvature ($R=0$).
\end{Prop}
\begin{pf}
  By straightforward calculation
  \begin{equation*}
    \begin{split}
      \nabla_a k^{aA}
      & = - \left( \AiB ( \lvec{\notnabla} \lvec{\notnabla} + m^2)
         \Psi\right)^A
         -i \left( \AiB \lvec{\notnabla} (i \rvec{\notnabla} - m) \Psi
       \right)^A\\
      & = 0 \qquad \text{iff $R=0$}.
    \end{split}
  \end{equation*}
We used the field equations and the fact that $\notnabla
\notnabla$ applied to a scalar is the D'Alembert operator. Note that the
assumption on the scalar curvature of the spacetime enters via the
scalar field equations.
\end{pf}
%
%%%%%%%%%
\section{The improved `supercurrent'}
\label{sec:imprsupercurCurved}
%%%%%%%%%
%
In chapter~\ref{sec:SUSYinR4} we saw that the (canonical) energy
momentum tensor of the free Wess Zumino model can be expressed in
terms of the supercurrent $k^{a}$
(Eqn.~(\ref{eq:SUSY_EMT})). To obtain the {\em improved\/} energy
momentum tensor in this way it is necessary to use the following
`improved' supercurrent:
\begin{equation}
  \label{eq:impsupercurCurved}
  \begin{split}
    j^{aA}
    & = k^{aA}  - 2 \xi {(\gamma^a \gamma^b - \gamma^b \gamma^a)^A}_B
    \nabla_b \left( \AiB \Psi \right)^B
  \end{split}
\end{equation}
\begin{rem}
  Ferrara and Zumino introduced the `improved' supercurrent in
  1975~\cite{ferraraZumino:75} to simplify their studies of the
  transformation properties of various supercurrents in Minkowski
  space. They show that the `improved' supercurrent can be used to
  construct all generators of the supersymmetry transformations and
  emphasize that this step is perfectly analogous to the introduction
  of the `improved' energy momentum tensor~\cite{callanEtAl:70} from
  which the generators of the Minkowski space symmetries can be
  constructed.
\end{rem}
It is obvious that $\nabla_a j^{aA}= 0$ on Minkowski space since
$\nabla_a \nabla_b$ is symmetric in this case. Moreover, since
$j^{aA}$ differs from $k^{aA}$ by a pure divergence only, it follows
that both of these currents generate the same charge on Minkowski
space. On the other hand on a generic spacetime the improved
`supercurrent' is {\em not\/} covariantly conserved.  Only on
spacetimes with vanishing scalar curvature we find in analogy to
Proposition~\ref{prop:supercurConsCurv}:
\begin{Prop}\label{prop:superCurCurvWeakCons}
  Let $(M,g_{ab})$ denote a Lorentz manifold with spinor
  structure. The current
  \begin{equation*}
    \begin{split}
      j^{aA}
      & = k^{aA}  - 2 \xi {(\gamma^a \gamma^b - \gamma^b \gamma^a)^A}_B
      \nabla_b \left( \AiB \Psi \right)^B
    \end{split}
  \end{equation*}
  is covariantly constant if and only if
  $(M,g_{ab})$ has vanishing scalar curvature, i.e., $R\equiv 0$.
\end{Prop}
\begin{pf}
  Let us abbreviate $(\gamma^a\gamma^b - \gamma^b \gamma^a)$ by
  $4G^{[ab]}$, where the square brackets reminds of the fact that
  $G^{[ab]}$ is antisymmetric in $a$ and $b$. We have the following
  useful identity
  \[
  G^{[ab]}G_{[cd]} {R^{cd}}_{ab} = + \frac{1}{8} R,
  \]
  which is valid since
  \begin{equation*}
    \begin{split}
      G^{[ab]}G_{[cd]} {R^{cd}}_{ab}
      & = \frac{1}{4} \gamma^a \gamma^b \gamma^c \gamma^d R_{abcd},\\
      \intertext{since $R_{abcd} = R_{[ab][cd]}$}\\
      & = \frac{1}{4} \gamma^{(a} \gamma^b \gamma^{c)} \gamma^d
      R_{cdab},\\
      \intertext{since $R_{[abc]d}=0$ by the Bianchi identities}\\
      & = \frac{1}{24} \left( g^{ab} \gamma^c + g^{bc} \gamma^a +
      \frac{\gamma^a \gamma^b \gamma^c + \gamma^c \gamma^b
        \gamma^a}{2}\right)
      \gamma^d R_{cdab}\\
      & = \frac{1}{24}\left( 2 g^{bc}\gamma^a - g^{ac}\gamma^b  \right)
      \gamma^d R_{cdab},\\
      \intertext{since $R_{abcd}$ is antisymmetric in $ab$}\\
      & = - \frac{1}{8} \gamma^a \gamma^d {R_{bad}}^b\\
      & = + \frac{1}{8} R,\\
      \intertext{since $R_{ab}={R^c}_{acb}$ is symmetric in $ab$}
    \end{split}
  \end{equation*}
  We obtain
  \begin{equation}\label{eq:divImprSuperCur}
    \begin{split}
      \nabla_a j^{aA}
      & = \underbrace{\nabla_a k^{aA}}_{=0} - 8 \xi{}  G^{[ab]}
      \nabla_a\nabla_b \left( \AiB\Psi \right)\\
      & =
      - 8 \xi{} G^{[ab]} \biggl(  (\nabla_{(a} \nabla_{b)} \AiB) \Psi
        + \left[ 2 (\nabla_{(a} \AiB) (\nabla_{b)} \Psi) \right]
        + \AiB \nabla_a \nabla_b \Psi
        \biggr),\\
       \intertext{where round brackets denote symmetrization}\\
       & =
       -2 \xi{} \AiB{} G^{[ab]}G_{[cd]} {R^{cd}}_{ab} \Psi\\
       & = - \frac{1}{4} \xi{} \AiB{} R \Psi
    \end{split}\notag
  \end{equation}
  Here we used the fact that the commutator of two covariant
  derivations on scalars vanishes whereas on spinors it is given by
  \[
  [\nabla_a,\nabla_b]_- \Psi
  = ( \nabla_a \nabla_b -  \nabla_b \nabla_a ) \Psi
  = 2 \nabla_{[a} \nabla_{b]} \Psi
  =  \frac{1}{2} G_{[cd]} {R^{cd}}_{ab} \Psi.
  \]
  It follows that the expression Eqn.~(\ref{eq:divImprSuperCur}) vanishes
  precisely when $R=0$, i.e., if and only if the spacetime has
  vanishing scalar curvature.
\end{pf}
\begin{rem}
  In particular the improved `supercurrent' is conserved on all
  spacetimes which are {\em vacuum solutions\/} of the Einstein field
  equations, provided these solutions admit the existence of spinors.
\end{rem}
%
%%%%%%%%%%%%%%%%%%%%%%%%%%%%%%%%%%%%%%%%%%%%%%%%%%%%%
\section{Quantization: The algebraic approach to quantum field theory}
\label{sec:algApproach}
%%%%%%%%%%%%%%%%%%%%%%%%%%%%%%%%%%%%%%%%%%%%%%%%%%%%%
This section is written for the convenience of the reader and gives a
brief introduction into the terminology and notions of the algebraic
approach to quantum field theory in curved spacetimes. Fur further
information the reader is referred to the books of
Haag~\cite{haag:alg}, Baumg\"artel \& Wollenberg \cite{baum:causal} or
Horuzhi~\cite{hor:intro}.
Recall that in quantum physics a ``basic field'' $\Phi$ is an operator valued
distribution on some Hilbertspace ${\EuScript H}$ satisfying certain
conditions which will not be recalled here~(see Haag's
book~\cite{haag:alg} for details and references). These fields can be
used to associate to each open region ${\mathcal O}$ of the spacetime
an algebra $\AO$ of (unbounded) operators on ${\EuScript
  H}$, namely the algebra generated by $\Phi (f)$ --the basic field
``smeared out'' with test functions $f$ with support in ${\mathcal O}$.
The elements of \AO\ can be interpreted as representing physical
operations performable in ${\mathcal O}$.  Furthermore, once the
correspondence
\begin{equation} \label{net}
{\mathcal O} \rightarrow \AO
\end{equation}
is known, one can (in principle) compute all relevant physical
quantities, e.g., cross-sections etc. This suggests that all relevant
physical information is encoded in the relation~(\ref{net}). For
mathematical convenience one restricts to local algebras generated by
bounded operators, i.e., to algebras generated by $e^{i\Phi(f)}$ in
the notation used above. The corresponding net is called {\em the net
  of local algebras\/} if and only if a generalization of the Haag
Kastler axioms are fulfilled (see~\cite{dimock:80}).  However for the
purpose of this text it is more convenient to consider a larger net of
algebras, namely the net of Borchers Uhlmann algebras.  This net,
first mentioned by Borchers and Uhlmann
(\cite{borchers:62,uhlmann:62}), contains kinematic information only,
meaning for example that our model consists of two scalar and one
Majorana spinor field. We will denote the net of Borchers Uhlmann
algebras by \AUO. A remark concerning the relation between these nets
is given at the end of this section. For a scalar field and a given
region ${\mathcal O} \subset M$, \AUO\ is the tensor algebra of smooth
functions with support in ${\mathcal O}$. To be more precise, let
$D_m({\mathcal O})$ denote $\bigotimes^m C_0^\infty ({\mathcal O})$, $m \geq\
 1$,
equipped with the usual topology and define $D_0({\mathcal O}) \equiv
{\mathbb C}$. $D'_m(\calO)$ denotes $\bigotimes^m \left[ D'(\calO)
\right]$, the dual space of $D_m(\calO)$. An element $\lambda^m \in
D'_m(\calO)$ is called a {\em m-point distribution}. For a family of
functions $\left\{ f_m \right\}_{m \in {\mathbb N}_0}$, where $f_m \in
D_m(\calO)$ with only finitely many $f_m$ non-vanishing, define
$f:=\bigoplus^m f_m$. Define an involution as $f^*:=\bigoplus^m
f^*_m$, where $f^*_m(x_1,\ldots,x_m):= \bar{f}_m(x_m,\ldots,x_1)$, a
non-commutative product for $f$ and $g=\bigoplus^m g_m$ by $f \times g
= \bigoplus^m(f \times g)_m$, where $(f \times g)_m := \sum_{i=0}^m
f_i(x_1,\ldots,x_i)g_{m-i}(x_{i+1},\ldots,x_m)$.  Together with
grade-wise addition the set of such functions $f$ becomes the
involutive algebra \AUO. We equip \AUO\ with the direct sum topology
coming from $D_m(\calO)$. The dual space of this algebra --denoted by
${\mathfrak A}'(\calO)$-- consists of a hierarchy of m-point
distributions:
\[{\mathfrak A}'(\calO) \ni \lambda = (\lambda^1,\ldots,
\lambda^m,\ldots)\]
with $\lambda(f)=\sum_{m=0}^\infty \lambda^m(f_m)$.
A state $\omega$ on the Borchers Uhlmann algebra \AUO\ is a
normalized linear form on ${\mathfrak A}_U = \bigcup_{\calO} \AUO $,
i.e.,
\begin{eqnarray*}
  \omega(f^*f) & \geq & 0 \hspace{4em}
                           \forall f \in {\mathfrak A}_U \\
  \omega(\identity) & \equiv & \omega_0(1)=1
\end{eqnarray*}
Such a state is said to satisfy {\em local commutativity\/} if and only if\
 for
all $m \geq 2$ and all $ 1 \leq i \leq m-1 $ we have
\[ \omega^m(f_1 \otimes \cdots \otimes f_i \otimes f_{i+1}
            \otimes \cdots \otimes f_m) =
   \omega^m(f_1 \otimes \cdots \otimes f_{i+1} \otimes f_{i}
            \otimes \cdots \otimes f_m) \]
for all $f_1,\ldots,f_m \in C^\infty_0(\calO)$ such that supp$\;f_i$
and supp$\;f_{i+1}$ are spacelike separated. The state is called {\em
  quasifree\/} if and only if its one point distribution vanishes and
all higher n-point distributions $\omega^n$ are completely determined
by its two point distribution $\omega^2$.
The relation between ``quantum fields'' mentioned at the beginning of
this section and the algebraic approach via the Borchers Uhlmann
algebra is the GNS-reconstruction theorem, Theorem~\ref{scalarRec}
below for which we need the following definition (Wightman axioms).
\begin{Dfn}[See, e.g.~\cite{Radzikowski:92}] \label{scalarQFTM}
  A quadruple $({\EuScript H},D,\Phi,\Omega)$ is called a neutral
  scalar quantum field model on a spacetime $(M,g_{ab})$, if the
  following properties are satisfied:
  \begin{enumerate}
  \item \label{Hs} ${\EuScript H}$ is a separable Hilbertspace with
    positive definite inner product $< \cdot , \cdot >$.
  \item $D$ is a dense subspace of ${\EuScript H}$.
  \item $\Phi$ is an ${\EuScript H}$ operator valued distribution on
    $C^\infty_0(M)$.
  \item For each $f \in C^\infty_0(M)$ the domain of $\Phi(f)$
    contains $D$, and
    \[ \Phi(f) D \subset D \]
  \item \label{cyclicity} The vector $\Omega \in {\EuScript H}$ is
    cyclic for $\Phi$ and $D$ is the linear span of the set
    \[ \{ \Phi(f_1) \cdots \Phi(f_m) \Omega \in {\EuScript H} |
    f_1,\ldots,f_m\in C^\infty_0(M) \} \]
  \item Hermicity of the field. For each $f \in C^\infty_0(M)$, the
    domain of the adjoint of $\Phi(f)$ (denoted by $\Phi(f)^*$)
    contains $D$ and \[ \Phi(f)^* = \Phi(\bar{f}) \qquad \text{on
      $D$},\] where $\overline{\phantom{f}}$ means complex conjugation.
  \item Local commutativity. For two test-functions $f_1,f_2 \in
    C^\infty_0(M)$ with spacelike separated support, the commutator
    ${[\Phi(f_1),\Phi(f_2)]}_-$ vanishes on $D$.
  \end{enumerate}
\end{Dfn}
\begin{Thm}[GNS-Reconstruction] \label{scalarRec}
{\sloppy
  For every state $\omega$ on the Borchers Uhlmann algebra \AUO\
  satisfying local commutativity there is a scalar quantum field
  theoretical model $({\EuScript H},D,\Phi,\Omega)$, unique up to
  unitary equivalence, such that for $f_1,\ldots,f_m \in C^\infty_0(
  {\calO} )$ we have
}
  \begin{equation}\label{eq:GNS-Bose}
    \omega^m(f_1 \otimes \cdots \otimes f_m) =
    <\Omega,\Phi(f_1),\cdots,\Phi(f_m) \Omega >.
  \end{equation}
  Conversely by Eqn.~(\ref{eq:GNS-Bose}) every scalar quantum field
  theoretical model $({\EuScript H},D,\Phi,\Omega)$ defines a state
  $\omega$ on \AUO\ satisfying local commutativity.
\end{Thm}
For a proof of this Theorem see the standard literature~(for
example~\cite{baum:causal,bog:axiom}).

So far we have only considered models with spin zero. For models
with higher spin the Borchers Uhlmann algebra has to be modified
slightly. For the spin~$1/2$ case, i.e., fermions, the corresponding
definitions are given in the next paragraph.
In analogy to the scalar case we define the Borchers Uhlmann algebra
for Dirac spinor fields to be the tensor algebra generated by sections
with compact support in the spinor and cospinor bundle respectively.
To be more precise, let $\hat{D}_0(\calO) \equiv {\mathbb C}$ for all open
subsets $\calO \subset M$ with compact closure. Denote by
$\hat{D}_1(\calO) = \Gamma \left( {\mathcal D}M(\calO) \right)$ the vector
space of all sections in ${\mathcal D}M := DM \oplus D^*M$ with support in
$\calO$ equipped with the topology induced by $D_1(\calO)$.
$\hat{D}_m(\calO)$ is the m-fold tensor product of the spaces
$\hat{D}_1(\calO)$\footnote{Recall that the tensor product of two
  vector bundles with typical fiber $E$ over some base manifold $M$ is
  the vector bundle over $M$ with typical fiber $E\otimes E$. Its
  Fr\'echet topology is defined analogously to the one of $D_1({\mathcal
    O})$, i.e., by the usual family of semi-norms
  (See, e.g.~\cite{dieudonne:analysisIII}~17.2).}
Note that this product is neither commutative nor anti-commutative.
\begin{rem}
  Let $\hat{D}'_1(\calO)$ denote the dual space of $\hat{D}_1(\calO)$.
  We sometimes denote by $( f \otimes g)(x,x')$ the
  integral kernel of the map
  $f \otimes g : \hat{D}'_1(\calO) \times \hat{D}'_1(\calO) \mapsto {\mathbb\
 C}$
  defined by
  \[ (f \otimes g) (f^*,g^*) \equiv
  \int_M \! \int_M (f \otimes g) (x,x') f^*(x) g^*(x')
  d\mu \; d\mu'
  \]
  The transformation properties of the components of $(f \otimes
  g) (x,x')$ under a change of frames show that $(f \otimes g)$ is a
  {\em bi-spinor}. If $f,\hat{f} \in \Gamma(DM)
  \stackrel{i}{\hookrightarrow} \hat{D}_1(\calO)$ are two spinors with
  compact support we abbreviate $(f \otimes \hat{f}) (x,x')$ by $f^{A}
  {\hat{f}}^{A'}$. The prime on the second index emphasizes the bi-spinor
  character.
\end{rem}
We consider now a family of functions $\left\{ f_m \right\}_{m
\in{\mathbb N}_0}$, where $f_m \in \hat{D}_m(\calO)$ and with only finite\
 many
$f_m$ non-vanishing and  define $f:=\bigoplus^m f_m$.
For those elements $f$, such that all $f_m$ can be written in the form $f_m =
g_1 \otimes \cdots \otimes g_m$ an involution is
defined by $f^*:=\bigoplus^m f^*_m$ where $f^*_m \equiv {\left( g_1
\otimes \cdots \otimes g_m \right)}^* := g_m^+\otimes \cdots \otimes g_1^+$.
$(\cdot)^+$ means Dirac conjugation.
For the other elements we define the involution by continuous extension.
A product of $f$ and $g=\bigoplus^m g_m$ is defined as $f \times g
= \bigoplus^m {(f \times g )}_m $ where ${(f \times g )}_m =
\sum_{n=1}^m f_n \otimes g_{m-n}$. With grade-wise addition
the set of such $f$'s becomes the involutive algebra \AUOH, the
Borchers algebra for Dirac Spinors. The topology on \AUOH\ is the
one inherited from $\hat{D}_m(\calO)$. We do not give the definition of
the dual space and of states since it is identical to the scalar
case~(see above).
\begin{Dfn}
  A Dirac quantum field theoretical model is a
  quadruple $({\EuScript H},D,\Psi,\Psi^+,\Omega)$ such that the
  properties~\ref{Hs}-\ref{cyclicity} of
  Definition~\ref{scalarQFTM} --replacing $\Phi$ by $\Psi$ and
  $C^\infty_0(M)$ by $\Gamma ({\mathcal D}M)$-- are fulfilled. The
  remaining two properties are substituted by
  \begin{enumerate}
  \item For each $f \in \Gamma ({\mathcal D}M)$, the
    domain of the adjoint of $\Psi(f)$ --denoted by
    $\Psi(f)^*$-- and of the Dirac adjoint --denoted by
    $\Psi^+(f)$
    both contain $D$ and
    \[ \Psi(f)^* = \Psi^+(f^+) \qquad\text{on $D$},\]
    where $f^+$ denotes the Dirac conjugate of $f$.
  \item Local anti-commutativity. For two test-functions $f_1,f_2 \in
    \Gamma ({\mathcal D}M)$ with spacelike separated support, the
    anticommutator $[\Psi(f_1),\Psi^+(f_2)]_+$ vanishes on $D$.
  \end{enumerate}
\end{Dfn}
The GNS-reconstruction theorem for Dirac fields is {\sloppy
\begin{Thm}[GNS-reconstruction] \label{fermiRec}
  For every state $\omega$ on the Borchers Uhlmann algebra \AUOH\
  for the Dirac field satisfying local anti-commutativity there is a
  Dirac quantum field theoretical model $({\mathcal
    H},D,\Psi,\Psi^+,\Omega)$, unique up to unitary equivalence, such
  that for $f_1,\ldots,f_m \in \Gamma ({\mathcal D}M)$ we have
  \begin{equation}{}\label{eq:GNS-Fermi}
    \omega^m(f_1 \otimes \cdots \otimes f_m) =
  <\Omega,\Psi(f_1),\cdots,\Psi(f_m) \Omega >.
  \end{equation}
  Conversely by Eqn.~(\ref{eq:GNS-Fermi}) every Dirac quantum field
  theoretical model $({\mathcal H},D,\Psi,\Psi^+,\Omega)$ defines a state
  $\omega$ on \AUOH\ satisfying local anti-commutativity.
\end{Thm}}
\begin{rem}
  $\Psi(f_1) = \Psi \left( \left( \begin{array}{c} g \\ h^+
  \end{array} \right) \right)$ is identical to
  $\Psi(g) + \Psi^+(h^+)$ in usual notation.
\end{rem}
We get the Borchers Uhlmann algebra for Majorana Spinors by
restricting ourself to the Majorana subbundle of ${\mathcal D}M$ and
repeating the construction above.
To clarify the relation between the net of local observables mentioned
in the introduction and the net of Borchers Uhlmann algebras we follow
Dimock's construction of the former~\cite{dimock:80,dimock:82}. Recall
that states on the Borchers Uhlmann algebra satisfying local
(anti) commutativity define a (uni\-que) field theoretical model. Let us
denote the corresponding Bose and Fermi fields with $\Phi$ and $\Psi$
respectively.  Dimock constructs the net of local algebras with these
field operators, showing that the result is independent of the state
chosen before. Roughly speaking for scalar fields the local algebra
\AO\ is the algebra generated by the exponentiated field operators
($e^{i\Phi(f)}$) smeared with testfunctions with support in $\mathcal O$.
On the other hand for Fermi fields the corresponding algebra of local
observables is generated by $\Psi(h)\Psi^+(h)$, with $h$ having
support in $\mathcal O$.
For a description of models with more than one field things are
greatly simplified by the fact that the Borchers Uhlmann algebras are
tensor algebras over some spaces of test-function.
\begin{Dfn}
{\sloppy
  Let ${\mathfrak A}_1, \cdots, {\mathfrak A}_N$ be a finite number of
  Borchers Uhlmann algebras over spaces of test-functions ${\mathcal A}_1,
  \cdots, {\mathcal A}_N$ respectively. The Borchers Uhlmann algebra over
  the space ${\mathcal A}=\bigoplus_{i=1}^{N} {\mathcal A}_i$ is called the
  {\em product\/} of these algebras and is denoted by ${\mathfrak A} :=
  {\mathfrak A}_1 \ast \cdots \ast {\mathfrak A}_N$, i.e., $\mathfrak A$ is\
 the
  {\em free\/} tensor product of the algebras ${\mathfrak
    A}_1,\cdots,{\mathfrak A}_N$.
  }
\end{Dfn}
\begin{rem}
  The monomial parts of $\mathfrak A$ are $\mathbb C$, ${\mathfrak A}_1\
 \oplus
  \cdots \oplus {\mathfrak A}_N$, $({\mathfrak A}_1 \oplus \cdots \oplus
  {\mathfrak A}_N) \otimes ( {\mathfrak A}_1 \oplus \cdots \oplus {\mathfrak
    A}_N)$, $\cdots$, the unit element is given by $\identity =
  (1,0,\cdots)$ and multiplication is the usual tensor product. This
  product --in difference to the usual tensor product of algebras-- is
  not commutative. (Recall that for two algebras $\mathfrak A$,$\mathfrak B$,
  $\mathfrak A \otimes \mathfrak B$ is well defined. The embeddings are
  ${\mathfrak A} \stackrel{i}{\hookrightarrow} {\mathfrak A} \otimes
  \identity$, ${\mathfrak B} \stackrel{i}{\hookrightarrow} \identity
  \otimes {\mathfrak B}$.  It follows that for $a \in {\mathfrak A}$ and $b
  \in {\mathfrak B}$ their product is given by $a \cdot b := (a \otimes
  \identity) ( \identity \otimes b) \equiv (a \otimes b) \equiv b
  \cdot a$).
\end{rem}
We take the opportunity to introduce product states on these algebras.
These states are used to implement the idea of `fields not
interacting with each other'.
\begin{Dfn}
  Let $\omega$ and $\omega_i$ be states on ${\mathfrak A} = {\mathfrak A}_1
  \ast \cdots \ast {\mathfrak A}_N$ and ${\mathfrak A}_i$ respectively.
  $\omega$ is called a {\em product state\/} if and only if all its
  $m$-point distributions $\omega^m$ can be written as products of the
  $\omega_i$, e.g., let ${\mathfrak A} = {\mathfrak A}_1 \ast {\mathfrak\
 A}_2$ and
  $\omega = \omega_1 \cdot \omega_2$ be given, then for $f_1 \otimes f_2
  \otimes g \in {\mathcal A}_1 \otimes {\mathcal A}_1 \otimes {\mathcal A}_2\
 $
  the corresponding  three point distribution $\omega^3$ may read
  \[
  \omega^3(f\otimes g_1 \otimes g_2) = \omega^2_1(f_1 \otimes f_2)
  \cdot \omega^1_2(g).
  \]
\end{Dfn}
%
%%%%%%%%%
\section{Hadamard states}
\label{sec:hadamardStates}
%%%%%%%%%
%
The Hadamard condition for quasifree states of quantum fields on a
manifold is believed to be a necessary condition for physical states
since the work of DeWitt and Brehme in 1960~\cite{DeWitt:60} and was
intensively studied since this time by various authors (See the
references in Fulling's book~\cite{Fulling:aspects_of_qft}).
Unfortunately it is a global condition which moreover, can be
formulated for free fields only. The first mathematical rigorous
formulation of this condition for free scalar fields is due to Kay and
Wald~\cite{kay:91}. We state their definition in this section for the
convenience of the reader and extend it, using the ansatz of Najmi and
Ottewill~\cite{Najmi:84}, to the fermionic case.
\vspace{0.5cm}\noindent
{\em Scalar case\/}: Let $\omega$ be a quasifree state on the Borchers
Uhlmann algebra for a scalar field on a globally hyperbolic manifold
$(M,g_{ab})$ satisfying the Klein-Gordon equation and local
commutativity.  Assume that a preferred time orientation has been
chosen on $(M,g_{ab})$ and let $T$ be a global time function
increasing towards the future.  Recall the definition of a convex
normal neighborhood: An open subset $U \subset M$ is called a convex
normal neighborhood if for all points $x_1,x_2 \in U$ there exists a
unique geodesic {\em contained\/} in $U$ connecting $x_1$ and $x_2$.
Let ${\cO} \subset M \times M$ be an open neighborhood of the set of
causally related points $(x_1,x_2)\in M \times M$, such that $J^+(x_1)
\cap J^-(x_2)$ and $J^+(x_2) \cap J^-(x_1)$ are contained within a
convex normal neighborhood. As usual $J^{\pm}(x)$ denotes the causal
future (past) of the point $x$. Then the square of the geodesic distance,
$\sigma$, is well defined and smooth on $\cO$. For each integer $p$
and $\epsilon \geq 0$ define for $(x,x') \in \cO$ the complex valued
function
\begin{equation} \label{eq:kernel}
G_\epsilon^{T,p} (x,x') = \frac{1}{{(2 \pi)}^2}
  \left( \frac{\Delta^{1/2} (x,x')}
         {\sigma (x,x') + 2 i \epsilon t + \epsilon^2}
        + v^{(p)}(x,x') \ln [\sigma(x,x')+2 i \epsilon t +
                         \epsilon^2]
  \right)
\end{equation}
where $t \equiv T(x) - T(x')$ and $\Delta^{1/2}$ and $v^{(p)}$ are
smooth functions uniquely determined by the geometry on $M$
($\Delta^{1/2}$ is the van Vleck Morette
determinant~\cite{DeWitt:group} and $v^{(p)}$ is given by the Hadamard
recursion relation up to order $p$ (See
Appendix~\ref{sec:hadamRecRel}).  The branch cut of the logarithm is
taken to lie on the negative real axis). Let $\cC$ be a Cauchy surface
and let $N$ be a neighborhood of $\cC$ with the property: For each
pair of points $x_1,x_2 \in N$ such that $x_1$ can be reached by a
causal curve emerging from $x_2$ ($x_1 \in J^+(x_2)$) one can find a
convex normal neighborhood in $M$ containing $J^-(x_1) \cap J^+(x_2)$.
$N$ is called a causal normal neighborhood of $\cC$.  Now let $\cO '$
be a neighborhood in $N \times N$ such that its closure is contained
in $\cO$ and choose a smooth real valued function $\chi$ on $M \times
M$ with the properties $\chi (x,y) = 0$ if $(x,y) \not\in \cO$ and
$\chi(x,y) = 1$ if $(x,y) \in \cO '$.
\begin{Dfn}[Globally Hadamard states]\label{dfn:glob_hadam}
  We say the state $\omega$ mentioned above is {\em globally
    Ha\-da\-mard\/} if and only if its two-point
  distribution\footnote{Note  that this Definition discusses the
    Hadamard condition in terms of the (not symmetrized) two point
    distribution of the state. The symmetric two-point distribution,
    which is discussed usually in the literature, is the real part of
    $\omega^2$.} $\omega^2$ is such that for each integer $p$ there
  exists a $C^p$- function $H^{(p)}(x,y)$ on $N \times N$ such that
  for all $F_1,F_2
\in C_0^\infty(N)$ we have
\begin{equation} \label{ScalarHadam}
\omega^2(F_1 \otimes F_2) =
                       \lim_{\epsilon \rightarrow 0}
                       \left( \quad
   \int_{N \times N} \left( \chi (x,y) G_\epsilon^{T,p}(x,y) +
        H^{(p)}(x,y) \right) F_1(x) F_2(y) d\mu_x d\mu_y
                        \right)
\end{equation}
\end{Dfn}
\begin{rem}
  Kay and Wald prove in~\cite{kay:91} that this Definition is
  independent of the Cauchy surface $\cC$ and the time function $T$.
  This in turn means that the Cauchy evolution preserves the Hadamard
  structure, i.e., $\omega^2$ restricted to a causal normal
  neighborhood $N$ of a Cauchy surface $\cC$ already fixes $\omega^2$
  throughout the whole spacetime.
\end{rem}
For a detailed discussion of this definition the reader is referred to
the work of Kay and Wald~\cite{kay:91}.
\vspace{0.5cm}\noindent
{\em Fermi case\/}: Let $\omega$ be a quasifree state on the Borchers
Uhlmann algebra for a Majorana spinor field satisfying the Dirac
equation on $(M,g_{ab})$. In analogy to the scalar case the Hadamard
condition is formulated in terms of its two point distribution
$\omega^2$. Since the Dirac
operator itself is not a strictly hyperbolic differential operator in
the sense of Leray~\cite{leray:diffeqn} we introduce an auxiliary two
point distribution $\tilde{\omega}^2$ such that for all $F_1 \in
\Gamma(D^*M(N)) \stackrel{i}{\hookrightarrow} \hat{D}_1(N))$, $F_2 \in
\Gamma(DM(N)) \stackrel{i}{\hookrightarrow}
  \hat{D}_1(N))$
\[ \omega^2 (F_1 \otimes F_2) = ( i \notnabla_x + m )
                       \tilde{\omega}^2(F_1 \otimes F_2)
                       \equiv \tilde{\omega}^2
                          \left( ((  i \notnabla + m ) F_1 )
                                            \otimes F_2 \right) \]
where $\notnabla$ is the Dirac operator (minus adjoint Dirac
operator) on the spinor (cospinor) part of ${\mathcal D}M$. Since
$\omega^2$ is a bi-solution of the Dirac equation it follows
that $\tilde{\omega}^2$ is a bi-solution of the SUSY coupled Klein
Gordon equation
\[
\tilde{\omega}^2
\left( \left( \left[ \Box + \left\{ m^2 + \frac{1}{4} R \right\}
\right] \frakI F_1 \right) \otimes F_2 \right)
= \tilde{\omega}^2\left( F_1
\otimes \left( \left[ \Box + \left\{ m^2 + \frac{1}{4} R \right\}
\right] \frakI F_2 \right) \right) =
   0. \]
   Having chosen a moving frame $E_A$ we write the components of the
   kernel of $\tilde{\omega}^2$ as\footnote{For simplicity only the
     case $F_1 \in \Gamma(D^*M)$ and  $F_2 \in \Gamma (DM)$
     respectively is considered.}
\begin{equation*}
 \omega^2(h \otimes j) = {( i \notnabla_x + m \frakI )^B}_C
                         {{\mbox{$\tilde{\omega}$}^2}^C}_{A'}(x,x'){F_1}_B(x)
                            F_2^{A'}(x').
\end{equation*}
The SUSY coupled Klein Gordon operator is strictly hyperbolic and we
define in analogy to the scalar case: The spinor state $\omega$
mentioned above is a {\em (spinor) Hadamard state\/} if and only if it's
auxiliary two point distribution $\tilde{\omega}^2$ is a Hadamard
distribution.  More precisely: Let
${{({\mathcal G}_\epsilon^{T,p})}^B}_{A'}(x,x')$ be the bi-spinor function
\[
{{({\mathcal G}_\epsilon^{T,p})}^B}_{A'}(x,x')=
\frac{1}{{(2 \pi)}^2}
  \left( \frac{ {{\Delta^{1/2}}^B}_{A'} (x,x')}
         {\sigma (x,x') + 2 i \epsilon t + \epsilon^2}
        + {{v^{(p)}}^B}_{A'}(x,x')
                 \ln [\sigma(x,x')+2 i \epsilon t + \epsilon^2]
  \right)
\]
where ${{\Delta^{1/2}}^B}_{A'}$ is the spinor van Vleck Morette
determinant and ${{v^{(p)}}^B}_{A'}$ is given by the spinor
version of the Hadamard recursion relation (See
Appendix~\ref{sec:hadamRecRel}).
\begin{Dfn}[Spinor Hadamard states]
  $\omega$ is a {\em spinor Hadamard state\/} if and only if it's
  auxiliary two point distribution $\tilde{\omega}^2$ is such that for
  each integer $p$ there exist a $C^p$-function $H^{(p)}(\cdot,\cdot)$
  on $\hat{D}_2(N)$ such that for all $F_1 \in \Gamma(D^*M(N))
  \stackrel{i}{\hookrightarrow} \hat{D}_1(N))$, $F_2 \in \Gamma(DM(N))
  \stackrel{i}{\hookrightarrow}
  \hat{D}_1(N))$ (in a local frame $E_A$) we have
\[
\tilde{\omega}^2 ( F_1 \otimes F_2 ) = \lim_{ \epsilon \rightarrow 0 }
\left(
\int_{N \times N}
    \left(
           \chi (x,y)
         { {{\mathcal G}_\epsilon^{T,p}}^B }_{A'} (x,x')
    \right)
{F_1}_B(x) {F_2}^{A'}(x')
d\mu_x d\mu_{x'} + H^{(p)}(F_1,F_2)
\right)
\]
\end{Dfn}
\begin{rem}
  The reader might wonder why the operator in front of the auxiliary
  two-point distribution $\tilde{\omega}^2$ is $(i\notnabla +m)$ and
  not $-(i \notnabla +m)$; the second being  the correct choice for
  the fundamental solution (See Proposition~\ref{Prop:Dimock2_2.1}
  below). To see the reason consider a free massive Majorana spinor
  field $\Psi$ on Minkowski space. Let $\Psi^+$ denote its Dirac
  adjoint and $\Psi\dag$ its hermitian conjugate. Then one
  demands~\cite{itzykson:qft}:
  \begin{equation}
    \label{eq:anticommFermi}
    \{\Psi(x),\Psi^+(y)\}_+ = -i S (x,y) \equiv i (i \not\partial + m)
    E(x,y),
  \end{equation}
  where $S$ is the fundamental solution of the Dirac operator
  $(i \not\partial +m)$, and $E$ is the fundamental solution of the
  Klein Gordon operator. The sign setting is required to ensure the
  equal time anti-commutation relations
  \[
  \{\Psi(t,\boldsymbol{x}),\Psi^+(t,\boldsymbol{y})\}_+ = \gamma^0
  \delta^{(3)} (\boldsymbol{x},\boldsymbol{y}),
  \]
  which means that $i \Psi\dag$ is conjugate to $\Psi$. Recall that
  this result is suggested by the corresponding Lagrangian (compare
  chapter~\ref{sec:analogonWess}) and that the l.h.s.\ of
  Eqn.~(\ref{eq:anticommFermi}) equals the imaginary part of
  $\omega^2$. Comparing Eqn.~(\ref{eq:anticommFermi}) with the
  corresponding relation for the scalar field, where one demands the
  commutator to be $+i$ times the fundamental solution $E$, one sees
  that $(i \notnabla +m)$ must be used above for consistency.
\end{rem}
The Hadamard condition in the formulation of Kay and Wald is a global
condition, for example it requires continuity of $H^{(p)}$ on the {\em
  whole\/} spacetime. In his thesis Radzikowski~\cite{Radzikowski:92}
recently gave a {\em local\/} characterization of the Hadamard
condition in terms of wave front sets. However the proof of his claim
is --as it stands-- only valid for the rather trivial case of a flat
spacetime. The proof presented below (See
subsection~\ref{sec:wfGlobHadam}) verifies his results in the general
case. This wave front set characterization will be extremely useful
for all following calculations in this thesis. It also motivates the wave
front set spectrum condition, which is presented in the last section
of this chapter.
%
%%%%%%%%%%%%%%%%%%%%%%%%%%%%%%%%%%%%%%%%%%%%%%%%%%%%%
\section{Wave front sets}
\label{sec:wfSets}
%%%%%%%%%%%%%%%%%%%%%%%%%%%%%%%%%%%%%%%%%%%%%%%%%%%%%
In this section some properties of the wave front set of a
distribution are collected. For further details and the proofs which
are omitted, we refer the reader to the original literature
(\cite{Hoermander:71,Hoermander:72}) or to the monographs of
H\"ormander, Taylor and  Reed \&
Simon~\cite{hoermander:analysisI,Taylor:81,reed:MethodsII}.
The theory of wave front sets was developed in the seventies by
H\"ormander together with Duistermaat for their studies of
pseudodifferential operators and differential equations on
manifolds~\cite{Hoermander:71,Hoermander:72}.  Wave front sets ($\WF$)
are refinements of the notion of the singular support (sing supp) of a
distribution. One main reason for using them in favor of the sing supp
is that they provide a simple characterization for the existence of
products of distributions and eliminate the difference between local
and global results. It is interesting that Duistermaat and H\"ormander
even mention relations between their `micro-local analysis' which they
used to study solutions of the Klein-Gordon equation on a manifold and
quantum field theory.
\begin{Dfn} \label{def:sigma_z}
  Let $v \in D' (\Rn)$ be a distribution on ${\mathbb R}^n$.  The
  set $\Sigma_z (v)$ is the complement in ${\mathbb R}^n \setminus \{ 0
  \}$ of the set of all nonzero $\xi \in \Rn $ for which there is a
  smooth function $\phi$ with compact support not vanishing at $z$ and
  a conic neighborhood $C_\xi$ of $\xi$ such that for all $N \in \mathbb
  Z$ there exists a constant $C_N$ such that for all $\xi' \in
  C_{\xi}$
   \[ (1 + \left| \xi' \right| )^N \left| \widehat{\phi u}(\xi') \right|
                              \leq C_N \]
The hat $\hat{~}$ denotes Fourier transformation.
\end{Dfn}
\begin{Dfn}\label{def:wavefrontset}
The wave front set of $v$ is defined by
\begin{equation}
\WF(v) = \{ (z,\theta) \in T^*\Rn \setminus \{ 0 \}| \quad
                  \theta \in \Sigma_z (v)\}
\label{wavefront}
\end{equation}
\end{Dfn}
All points $<x,\xi> \in T^*\Rn$ which are {\em not\/} in the
wave front set are called {\em regular directed points}.
\begin{rem} \label{remark}
\samepage $\phantom{~}$\par
  \begin{enumerate}
  \item $\WF{}(v)$ is a closed subset of $T^*\Rn \setminus \{0\}$
    since each regular directed point $<x,\xi> \not\in \WF(v)$ has by
    definition an open neighborhood in $T^*\Rn \setminus \{0\}$
    consisting of regular directed points, too.
  \item\label{rem:wfsmooth}
    The wave front set of a smooth function is the empty set
  \item For $v \in D'(\Rn)$ with wave front set
    \WF$(v)$, the projection of the wave front set to the base point
    gives the singular support of $v$.
  \item\label{rem:wfinklusion} For all smooth functions $\tilde{\phi}$
    with compact support $\WF(\tilde{\phi}v) \subset \WF(v)$.
  \item\label{rem:addition}
    For two distributions $v,w \in D'(\Rn)$ with wave front sets
    \WF$(v)$ and \WF$(w)$ respectively, the wave front set of $(v + w)
    \in D' (\Rn)$ is contained in $\WF(v) \cup \WF(w)$
  \end{enumerate}
\end{rem}
In order to extend this definition to manifolds we use the behavior of
$\WF(v)$ under diffeomorphisms of $\Rn$ to $\Rn$
(Theorem~\ref{thm:TransWF} below). Although
Definition~\ref{def:sigma_z} contains a Fourier transform, it turns
out that the elements of $\WF(v)$ transform as elements of $T^*\Rn$,
the cotangential bundle of $\Rn$. It is therefore possible to define
the wave front set of a distribution on a manifold via charts.
\begin{Thm}[Theorem~IX.44 and Problem~75 in~\cite{reed:MethodsII}]
  \label{thm:TransWF}
  Let $v \in D'(\Rn)$ be a distribution on $\Rn$, $\chi$ be a
  diffeomorphism of $\Rn$ to $\Rn$ and let $v \circ \chi$ be the
  distribution\footnote{Note the misprint in~\cite{reed:MethodsII}.}
  \[ v \circ \chi (f) := v( g^{-1} ( f \circ \chi^{-1})) \]
  where $g$ is the determinant of the Jacobian matrix $d\chi$. Define
  $\chi_* : \Rn \times (\Rn \setminus \{0\}) \rightarrow
  \Rn \times (\Rn \setminus \{0\}) $ by
  \[ \chi_* <x,\xi> \; = \; <\chi(x), d\chi^*(\xi)> \]
  where $d\chi^*$ is the adjoint of $d\chi$ with respect to the
  Euclidean inner product on $\Rn$. Then
  \[ \WF(v \circ \chi^{-1}) = \chi_* \left( \WF(v) \right) \]
\end{Thm}
For the proof of this Theorem, which is only sketched
in~\cite{reed:MethodsII}, we refer the reader to the monograph of
H\"ormander~\cite[p.~265]{hoermander:analysisI}.
\begin{Dfn}\label{dfn:wfmanifold}
  Let $ u \in D'(M) $ be a distribution over some manifold $M$.  Let
  $\{X_\lambda\}_{\lambda \in {\mathbb N}}$ denote an open covering of
  $M$.  Choose a compatible partition of unity
  $\{\Phi_\lambda\}_{\lambda \in {\mathbb N}}$. Without loss of
  generality we may assume that $\supp \Phi_\lambda$ is contained in a
  single coordinate patch for every $\lambda \in {\mathbb N}$. The
  corresponding charts are denoted by $ \chi_\lambda : \supp
  \Phi_\lambda \subset X_\lambda \rightarrow \Rn$.  Using the same
  notation as in Theorem~\ref{thm:TransWF} above we find as a
  Corollary
\begin{equation}\label{wfcovering}
  {\chi_\lambda}_* \left( \WF(\Phi_\lambda u) \right) \equiv
  \WF(\Phi_\lambda u \circ \chi_\lambda^{-1})
\end{equation}
  where $\Phi_\lambda u \circ \chi_\lambda^{-1}$ obviously is a
  distribution with compact support over $\Rn$.  We define the {\em wave
    front set\/} of $u$ by
  \[  \WF(u) = \bigcup_\lambda \WF(\Phi_\lambda u) \]
\end{Dfn}
\begin{rem}
  This is a {\em local\/} definition. It follows that the
  following results which are stated in~\cite{reed:MethodsII} for
  distributions on $\Rn$, are valid for distributions defined on
  manifolds, too.
\end{rem}
A useful application of wave front sets is the definition of products
of distributions.  Wave front sets provide a simple characterization
for the existence of such products, which furthermore can be extended
to manifolds.  The following definition of a product and its relation
to wave front sets can be found for example in Reed \&
Simon~\cite[p.90--97]{reed:MethodsII}.
\begin{Dfn}
  Let $v,w \in D'(\Rn)$. The distribution $T\in D'(\Rn)$
  is the {\em product\/} of $v$ and $w$ if and only if for all $x \in
  \Rn$ there exists a smooth function $f$ not vanishing at $x$ such
  that for all $k \in \Rn$
  \begin{equation}
    \widehat{f^2T}(k) = (2\pi)^{-n/2} \int_{\Rn} \widehat{fv}(l)
    \widehat{fw}(k-l) d^nl
    \label{eq:productRn}
  \end{equation}
  where the integral is absolutely convergent.
\end{Dfn}
\begin{rem}
  \samepage$\phantom{~}$\par
  \begin{itemize}
  \item The product is well defined, since if such a $T$ exists, it is
    unique: \\
    Let $g \in D(\Rn)$ then
    \[
    \widehat{gf^2T} = (2\pi)^{-n/2} \widehat{gfv} * \widehat{fw} =
    (2\pi)^{-n/2} \widehat{fv} * \widehat{gfw},
    \]
    since the change of variables is legitimate due to the assumption
    of absolute convergence of (\ref{eq:productRn}). Now suppose $T_1$
    and $T_2$ both fulfill (\ref{eq:productRn}), e.g.\ for all $x \in
    \Rn$ there exist functions $f$ and $g$ not vanishing at $x$ such
    that $\widehat{f^2T_1}=\widehat{fv} * \widehat{fw}$ and
    $\widehat{g^2T_2}=\widehat{gv} * \widehat{gw}$.
    We conclude $\widehat{f^2g^2T_1} = \widehat{f^2g^2T_2}$, so
    $(T_1-T_2)$ vanishes near $x$ for all $x$, that is it is zero.
  \item The product of two distributions with disjoint support is
    zero.
  \item Since this is a local definition, we can extend it immediately
    to manifolds, using charts (See \cite{Hoermander:71}).
  \end{itemize}
\end{rem}
The following Theorem gives the relation between the wave front sets
of two distributions and the existence of their product.
\begin{Thm}[Theorem IX.54 of \cite{reed:MethodsII}]\label{Thm:IX.54}
  Let $v,w$ be two distributions on $M$ such that
  \[
  \WF(v) \oplus \WF(w) \equiv \{ <x,k_1 + k_2> | \quad <x,k_1> \in \WF(v),
  <x,k_2> \in \WF(w)  \}
  \]
  does not contain any element of the form $<x,0>$, then the product
  $v \cdot w$ exists and has wave front set
  \[
  \WF(v\cdot w) \subset \WF(v) \cup \WF(w) \cup (\WF(v) \oplus \WF(w))
  \]
\end{Thm}
Another interesting result, which will be used frequently in this work
is the following Theorem of
H\"or\-man\-der~\cite[Theorem~2.5.11']{Hoermander:71}.
\begin{Thm}\label{thm:hoermander2.5.11}
  Let ${\mathcal C}$ be a sub-manifold of $M$ with embedding $\varphi$ and\
 let
  \[
  N_\varphi = \{ (\varphi(y),\xi) \in T^*(M) | <\xi,d\varphi(y) \eta> = 0\
 \quad
  \forall \eta \in T(M) \}
  \]
  be the set of all normals of the map $ \varphi$ ($<\cdot,\cdot>$
  denotes dual pairing). If $u\in D'(M)$ and $\WF(u)\cap N_\varphi =
  \emptyset$, we can define the restriction (pullback)
  $u\restriction_{\mathcal C}\equiv{} \varphi^*u$ of $u$ in one and only
  one way so that it is equal to the composition $u\circ \varphi$ when
  $u$ is a continuous function and it is a sequentially
  continuous\footnote{We refer the reader to H\"ormanders original
    work for the definition of the pseudo topology $D'_\Gamma$ is
    equipped with.} function from $D'_\Gamma(M)=\{ u \in D'(M) |
  \WF(u) \subseteq \Gamma \}$ to $D'({\mathcal C})$ for
  any closed cone $\Gamma\subset T^*(M) \setminus \{0\}$ with $\Gamma
  \cap N_\varphi = \emptyset$.  Moreover,
  \[
  \WF(\varphi^*(u)) \subset \varphi^*\WF(u) = \{ (y,
  \sideset{^t}{}{d\varphi}(y) \xi | (\varphi(y),\xi) \in \WF(u) \},
  \]
  where $\sideset{^t}{}{d\varphi}(y)$ denotes the transposition of
  $d\varphi(y)$ with respect to the dual pairing.
\end{Thm}
\begin{rem}
  $\phantom{~}$\par
  \begin{itemize}
  \item The smooth functions of $M$ are sequentially dense in
    $D'_\Gamma$ for all admissible $\Gamma \in T^*M\setminus\{0\}$
    (See \cite[p.~125]{Hoermander:71}).
  \item Theorem~\ref{thm:TransWF} above may be viewed as a Corollary
    to this Theorem setting ${\mathcal C} \equiv M = \Rn$
    (Compare~\cite[p.~265]{hoermander:analysisI}).
  \end{itemize}
\end{rem}
It is worth noting that the restriction mapping $\varphi{}^*$,due to
its continuity, can be approximated by a suitable family of test
densities:
\begin{Cor}\label{Cor:restrApprox}
  Let ${\mathcal C},M,\varphi$ as in Theorem~\ref{thm:hoermander2.5.11}
  above be given and let $u\in D'(M)$ denote a distribution, such that
  $\varphi^*u$ exists. Then for all smooth densities $f$ on ${\mathcal C}$
  there exists a family of smooth densities $f_\rho$ on $M$, such that
  $u(f_\rho) \rightarrow \varphi^*u(f)$ for $\rho \rightarrow 0$.
\end{Cor}
\begin{pf}
  Let $\{f_\rho\}$ be a family of smooth densities on $M$, such that
  for all smooth functions $u_l$ on $M$ $\lim_{\rho\rightarrow 0}
  u_l(f_\rho) = u \circ \varphi (f) \equiv \varphi^* u_l(f)$. Such
  families always exist due to the fact that the Dirac measure on $M$
  can be approximated by smooth densities (see
  e.g.~\cite[p.~242]{dieudonne:analysisIII}). Now let $\{u_l\}$ denote
  a Cauchy sequence of smooth functions converging to $u$, then
  \begin{equation*}
    \begin{split}
      \| \varphi^* u (f) - u (f_\rho) \|
      & \leq \| \varphi^* u (f) - \varphi^* u_l(f) \|
         +   \| \varphi^* u_l(f) - u_l(f_\rho) \|
         +   \| u_l(f_\rho) - u(f_\rho) \| \\
      & \leq \epsilon,
    \end{split}
  \end{equation*}
  for $\rho \leq \delta$ and $l\geq N_\rho$, since $\varphi^*u_l$
  converges to $\varphi^* u$ by Theorem~\ref{thm:hoermander2.5.11}, as
  well as $u_l(f_\rho)$ and $u_l$ converge to $\varphi^* u_l(f)$ and
  $u$ respectively.
\end{pf}
In the sequel of this work, Theorem~\ref{thm:hoermander2.5.11} above
will be used for instance to define restrictions of distributions on
$M\times M$ to the diagonal. As a second Corollary one finds:
\begin{Cor}\label{Cor:restrictionDiag}
  Let $f \in \CinfO{}(M \times M)$ be a smooth function on $M\times M$
  and let $u\in D'(M)$ be an arbitrary distribution.  Then $f\cdot u$
  exists as a distribution on $M\times M$ and can be restricted to the
  diagonal.
\end{Cor}
\begin{pf}
  Let $\WF{}(u)$ denote the wave front set of $u$. Obviously $f \cdot
  u$ exists as a distribution on $M\times M$ with wave front set
  \[
  \WF(f\cdot u) = \{ (x,0;y,\xi) \in T^*(M \times M) \setminus \{0\} |\quad
                      x\in M, (y,\xi)\in \WF(u)\}.
  \]
  Let $\varphi : M \ni x \mapsto (x,x) \in M \times M$ denote the
  embedding of $M$ onto the diagonal in $M \times M$. $\varphi$ is smooth
  and for the set $N_\varphi$ introduced above one finds:
  \[
  N_\varphi = \{(x,\eta;x,-\eta)\in T^*(M\times M)| \quad (x,\eta)\in
  T^*(M)\}.
  \]
  Thus the assumption of Theorem~\ref{thm:hoermander2.5.11} are
  satisfied, i.e.,
  \[
  \WF(f\cdot u) \cap N_\varphi = \emptyset,
  \]
  which finishes the proof.
\end{pf}
\begin{rem}
  We will denote the kernel of this special restriction $\varphi^*(f\cdot
  u)$ by $\varphi^*(f\cdot u)(y)$, $f(y,y)u(y)$ or $\lim_{x\rightarrow
    y} f(x,y) u(y)$ repectively.
\end{rem}
\subsection{The wave front set of a globally Hadamard state}
\label{sec:wfGlobHadam}
The result of this subsection (Corollary~\ref{Cor:wavefrontscalar}
below) can already be found in the dissertation of Radzikowski.
However the proof he gave is --as it stands-- only valid for the
rather trivial case of a flat spacetime. The proof presented here
verifies his claim in the general case.
Let $\omega^2$ w.l.o.g.\ denote the two point distribution of a
globally Hadamard state for a free scalar Klein Gordon field on
$(M,g_{ab})$. Following Radzikowski's line of argument, consider first
the free massive Klein-Gordon field on Minkowski space in the vacuum
state.  The vacuum is known to fulfil the global Hadamard condition
with $\Delta^{1/2} \equiv 1$ and $v^{(p)} \equiv \sum_{n=0}^p 2
\fracwithdelims(){m^2}{2}^{n+1} \bigl/ n!\,
(n+1)!$~(\cite{castagnino_harari:84}).  The wave front set of the
corresponding two-point distribution
\mbox{$\sideset{^{\text{Mk}}}{^2}{\omega}$} is also well known (see
for instance~\cite{reed:MethodsII}):
\begin{Thm}[Theorem IX.48 of~\cite{reed:MethodsII}]\label{Thm:WfMinkScal}
  The two-point distribution \mbox{$\sideset{^{\text{{\rm
          Mk}}}}{^2}{\omega}$} of
  the free massive scalar field on Minkowski space in the vacuum state
  has wave front set
%  \tracingcommands=1
  \begin{equation}\label{eq:WFscalarMink}
    \begin{split}
      & \WF(\sideset{^{\text{\rm Mk}}}{^2}{\omega})\\
      & = \{ (x_1,k_1),(x_2,k_2) \in {\mathbb R}^4 \times ( {\mathbb R}^4
      \setminus \{0\}) | \quad x_1 \neq x_2; \; (x_1 -x_2)^2 =0;\\
      & \phantom{= \{ (x_1,k_1),(x_2,k_2) \in {\mathbb R}^4 \times (\
 {\mathbb\
 R}^4
      \setminus \{0\}) | \quad}
      k_1 \| (x_1 -x_2); k_1+k_2 =0;\\
      & \phantom{= \{ (x_1,k_1),(x_2,k_2) \in {\mathbb R}^4 \times (\
 {\mathbb\
 R}^4
      \setminus \{0\}) | \quad }
                         k_1^0 \geq 0 \} \\
      & \phantom{=\quad}
      \bigcup
      \{ (x,k_1),(x,k_2) \in {\mathbb R}^4 \times ({\mathbb R}^4\setminus
      \{0\}) | \quad k_1 + k_2 = 0; \; k_1^2=0 ; \; k_1^0 \geq 0 \}\\
      & \mbox
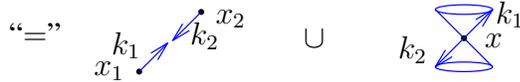
{
\begin{picture}(0,0)%
\includegraphics{figures/wellenfront.pstex}%
\end{picture}%
\setlength{\unitlength}{0.012500in}%
\begin{picture}(219,37)(65,753)
\put(148,766){\makebox(0,0)[lb]{\smash{\SetFigFont{12}{14.4}{rm}$k_2$}}}
\put(115,762){\makebox(0,0)[lb]{\smash{\SetFigFont{12}{14.4}{rm}$k_1$}}}
\put( 83,766){\makebox(0,0)[b]{\smash{\SetFigFont{12}{14.4}{rm}``$=$''}}}
\put(114,753){\makebox(0,0)[b]{\smash{\SetFigFont{12}{14.4}{rm}$x_1$}}}
\put(276,775){\makebox(0,0)[lb]{\smash{\SetFigFont{12}{14.4}{rm}$k_1$}}}
\put(235,760){\makebox(0,0)[lb]{\smash{\SetFigFont{12}{14.4}{rm}$k_2$}}}
\put(200,765){\makebox(0,0)[b]{\smash{\SetFigFont{12}{14.4}{rm}$\cup$}}}
\put(165,775){\makebox(0,0)[b]{\smash{\SetFigFont{12}{14.4}{rm}$x_2$}}}
\put(275,765){\makebox(0,0)[b]{\smash{\SetFigFont{12}{14.4}{rm}$x$}}}
\end{picture}
        }
    \label{eq:wfscalarMink}
  \end{split}
\end{equation}
\end{Thm}
For a proof see~\cite{reed:MethodsII} or use the following
representation of the Fourier transform of
\mbox{$\sideset{^{\text{\rm Mk}}}{^2}{\omega}$}:
\[
  \widehat{\sideset{^{\text{\rm Mk}}}{^2}{\omega}} = (2\pi)^{-1}
  \delta(k_1 + k_2) \Theta(k_1^0) \delta(k_1^2 - m^2)
  \in {\mathcal S}' ({\mathbb R}^4 \times {\mathbb R}^4)
\]
Using this representation and Definition~\ref{def:wavefrontset} it is an
easy, but lengthly calculation to verify Eqn.~(\ref{eq:WFscalarMink})
explicitly.
In order to extend this result to arbitrary (globally hyperbolic)
manifolds, let $x \in M$ be a point on a Cauchy surface $\cC$ and let
$U_x$ be a convex normal neighborhood of $x$. Denote by $\cCb \subset
\cC$ an arbitrary (non void) open subset in $\cC$, containing $x$,
such that the domain of causal dependence, $\Diamond(\cCb)$, is
contained in $U_x$. We are going to calculate the wavefront set of
$\omega^2$ for all base points $x_1,x_2 \in D^+(\cCb) :=
\Diamond(\cCb) \cap {J}^+ (\cCb)$ using some properties of Hadamard
states on a special smooth deformation $(\hat{M},\hat{g}_{ab})$ of our
original spacetime\footnote{The author thanks R. Verch for calling his
  attention to the deformation argument of Fulling, Narcowich and
  Wald~\cite{FullingNarcowichWald:81}. Note that the metric $g_{ab}$
  restricted to $D^+(\cCb)$ is {\em not\/} flat in general; the latter
  was assumed implicitly in Radzikowski's argument at some point,
  making the following modification of his proof necessary.}.  Finally
the propagation of singularities theorem
(Theorem~\ref{Thm:prop_of_sing} below) and our knowledge that
$\omega^2$ has no singularities for spacelike separated points allows
us to extend this result to the whole spacetime $(M,g_{ab})$.
Given $\cC$, $x \in \cC$ and $\cCb$ it will be shown below that
there exists a globally hyperbolic spacetime
$(\tilde{M},\tilde{g}_{ab})$ with the following properties:
\begin{enumeraterm}
  \item A neighborhood $U$ of $\cC$ in $M$ is isometrically isomorphic
    to a neighborhood $\tilde{U}$ in $\tilde{M}$ and the isometry $\varrho$
    is also an isometry between $\cC$ and $\tilde{\cC}$; e.\ g.\
    $(\tilde{M},\tilde{g}_{ab})$ is a smooth deformation of
    $(M,g_{ab})$.
  \item For all points $x_1,x_2 \in D^+(\cCb)$ there exists a Cauchy
    surface $\tilde{S}$ with neighborhood $\hat{U}$ in $\tilde{M}$,
    such that the metric $\tilde{g}_{ab}$ restricted to $\hat{U}$ is
    flat (i.e.\ Minkowskian) and $\Diamond(\varrho(\cCb)) \subset\
 \Diamond(\hat{U})$.
\end{enumeraterm}
Now $\omega^2$ induces  canonically a Hadamard distribution
$\tilde{\omega}^2$ on $(\tilde{M},\tilde{g}_{ab})$, since it does so
on $\tilde{U}$: Being a Hadamard distribution on $\tilde{U}$ implies,
by the Remark following Definition~\ref{dfn:glob_hadam}, that
$\tilde{\omega}^2$ is a Hadamard distribution throughout
$(\tilde{M},\tilde{g}_{ab})$.  Moreover, the wave front set of
$\tilde{\omega}^2$ on $\tilde{U}$ determines that of $\omega$ at $U$
and vice versa by Theorem~\ref{thm:TransWF}. On the other hand, using
the propagation of singularities theorem and the smoothness of
$\tilde{\omega}^2$ for spacelike separated points, the wave front set
of $\tilde{\omega}^2$ at $\tilde{x}_1 = \varrho(x_1)$ and $\tilde{x}_2 =
\varrho(x_2)$ is already fixed by the wave front set of that distribution
at all points in $\hat{U}$ (See below and note that $\tilde{x}_1$ and
$\tilde{x}_2 \in \Diamond(\hat{U})$). To obtain the wave front set of
$\omega^2$ at $x_1$ and $x_2$ it is therefore sufficient to calculate
the wave front set of
$\tilde{\omega}^2$ for all points in $\hat{U}$.
%%%%%%%%%%%%%%%%%%%%%%%%%%%%%%%%%%%%%%%%%%
The following computations are performed entirely in the flat part of
$\tilde{M}$ and were partly sketched in~\cite{Radzikowski:92}. Note
first that Eqn.~(\ref{ScalarHadam}) is valid in
$(\tilde{M},\tilde{g}_{ab})$, too.  To keep the notation simple we
omit the $\tilde{\phantom{x}}$ in what follows. Choose a point $x \in
S \equiv \hat{\cC}(t_1)$ and a convex normal neighborhood $U_x \subset
\hat{U}$ of $x$.  Using normal coordinates and an adapted time
function $T$ the distribution $G^{T,p}_\epsilon$, when restricted to
$U_x \times U_x$ is
\begin{equation*}
  \begin{split}
    G^p & :=
    \lim_{\epsilon\rightarrow 0}\quad \chi G^p_\epsilon\\
    & =
    \lim_{\epsilon\rightarrow 0}\quad
    \frac{\Delta^{1/2}(x_1,x_2)}{-(x_1-x_2)^2 + 2i\epsilon
      (x_1^0-x_2^0) +\epsilon^2}
    + v^{(p)}(x_1,x_2) \ln(-(x_1-x_2)^2 + 2i\epsilon
      (x_1^0-x_2^0) +\epsilon^2)
  \end{split}
\end{equation*}
The metric $g_{ab}$ is flat on $U_x$ and hence $2 \sigma(x,y) \equiv
-(x_1-x_2)^2$ in these coordinates. Moreover the van Vleck Morette
determinant $\Delta^{1/2}$ and $v^{(p)}$ are identical to the
corresponding functions on Minkowski space, since only {\em local\/}
properties of the underlying geometry enter into the Hadamard
recursion relations.  We conclude that $\WF(\sideset{^{\text{\rm
      Mk}}}{^2_{m=0}}{\omega})=\WF(G^p)$ in normal coordinates on $U'_x
\times U'_x$. Using Definition~\ref{dfn:wfmanifold} this result can be
pulled back to the
manifold:
\begin{equation*}
 \begin{split}
  \WF(G^p)
  = \{ (x_1,k_1),(x_2,k_2) & \in (T^*M \times
  T^*M) \setminus \{0\} | \\
   &
  (x_1,k_1) \sim (x_2,-k_2); \quad k_1^0 \geq 0 \}
  \quad \text{on $U'_x \times U'_x$}
 \end{split}
\end{equation*}
where $(x_1,k_1) \sim (x_2,k_2)$ means (i) $x_1$ and $x_2$ can be
joined by a null geodesic $\gamma$, (ii) $k_1 (= k_{1\nu})$ is a
cotangent vector such that $k_1^\mu=k_{1\nu} g^{\mu\nu}$ is tangential
to $\gamma$, (iii) The parallel transport of $k_1$ along $\gamma$
yields $k_2$ {\em or\/} (i) $x_1=x_2$, (ii) $k_1^2=0$ and (iii)
$k_1=k_2$. Now $H^{(p)} \in C^p(M\times M)$ which implies for
$\phi \in \CinfO(M)$ and $\phi(x) \neq 0$
\[
(( {\phi\otimes\phi)H^{(p)})}\sphat (k_1,k_2) \leq C_p(1+|k|)^{-p}
\]
for a constant $C_p$ and all $k=(k_1,k_2)$. Since $\omega^2 = G^p +
H^{(p)}$ for all $p$ we conclude that the wave front set of $\omega^2$
restricted to $U_x' \times U_x'$ is
\begin{multline} \label{eq:wfscalarUx}
\WF(\omega^2) = \{ (x_1,k_1),(x_2,k_2) \in (T^*M \times T^*M)
\setminus \{0\} | \\
(x_1,k_1) \sim (x_2,- k_2); \;\; k_1^0 \geq 0 \} \quad
           \text{on $U_x' \times U_x'$}
\end{multline}
As a Corollary to the following Theorem of H\"ormander this extends to
all of $\tilde{M}$ (Compare section~2.2 in~\cite{Radzikowski:92}).
\begin{Thm}[Corollary to Theorem 6.1.1 of \cite{Hoermander:72}]
  \label{Thm:prop_of_sing}
  Let $P= \identity \otimes (\Box + \nabla_\mu V^\mu \linebreak[2] + b)$ be a
  pseudo\-differential
  operator on a globally hyperbolic manifold $(M,g_{ab})$. $\Box$
  denotes the D'Alembert operator and $V^\mu$ and $b$ are a smooth
  vector field and a smooth function on $M$ respectively. The principal
  symbol of $P$ is denoted by $p$ and is given here by
  \begin{align*}
    p: (T^*M  \times T^*M) \setminus \{ 0\}
    & \rightarrow {\mathbb R} \\
    (x_1,k_1;x_2,k_2) & \mapsto g^{\mu\nu}(x_2) k_{2\mu} k_{2\nu}
  \end{align*}
  If $u \in D'(M\times M)$ is a weak solution of $Pu=0$ with wave front
  set $\WF(u)$, it follows that \\
  a) $\WF(u) \subseteq p^{-1}(0)$ and\\
  b) $\WF(u)$ is invariant under the (Hamiltonian) vector field $H_p$
  given by
  \[
   H_p
   = \sum_{i=1}^{2n} \frac{\partial p(x,k)}{\partial x_i}
  \frac{\partial}{\partial k_i} - \frac{\partial p(x,k)}{\partial k_i}
  \frac{\partial}{\partial x_i}
  \]
  in local coordinates.
\end{Thm}
\begin{Dfn}
  The bi-characteristic strips of $P$ are the curves on the sub-manifold
  $p^{-1}(0) \subset (T^*M \times T^*M) \setminus \{ 0
  \}$
  which are generated by $H_p$.
\end{Dfn}
Note that b) means: If $(x_1,k_1;x_2,k_2)$ and $(x'_1,k'_1;x'_2,k'_2)$ are on
the same bi-characteristic strip, denoted by $(x_1,k_1;x_2,k_2) \approx
(x'_1,k'_1;x'_2,k'_2)$ and $(x_1,k_1,x_2,k_2) \in \WF(u)$, then
$(x'_1,k'_1;x'_2,k'_2) \in \WF(u)$.
For $P=\identity \otimes (\Box + \nabla_\mu V^\mu + b)$ one finds (See
Proposition~2.8 in~\cite{Radzikowski:92})
\[
(x_1,k_1;x_2,k_2) \approx (x'_1,k'_1;x'_2,k'_2)
       \Leftrightarrow
(x_1,k_1) = (x'_1,k'_1) \quad \text{and}\quad (x_2,k_2) \sim (x'_2,k'_2)
\]
\begin{Cor}[See Theorem~2.6 in~\cite{Radzikowski:92}]
  \label{Cor:wavefrontscalar}
  The two-point distribution $\omega^2$ of a free massive Klein-Gordon
  field on a globally hyperbolic spacetime in a globally Ha\-da\-mard
  state has wave front set
\begin{equation}\label{eq:wfhadamscalar}
  \begin{split}
  \WF(\omega^2) & = \{ (x_1,k_1;x_2,-k_2) \in (T^*M \times
  T^*M ) \setminus \{0\} |
  \quad
  (x_1,k_1) \sim (x_2,k_2); \quad k_1^0 \geq 0 \}\\
  & \mbox{{
\begin{picture}(0,0)%
\includegraphics{figures/wellenfront.pstex}%
\end{picture}%
\setlength{\unitlength}{0.012500in}%
\begin{picture}(219,37)(65,753)
\put(148,766){\makebox(0,0)[lb]{\smash{\SetFigFont{12}{14.4}{rm}$k_2$}}}
\put(115,762){\makebox(0,0)[lb]{\smash{\SetFigFont{12}{14.4}{rm}$k_1$}}}
\put( 83,766){\makebox(0,0)[b]{\smash{\SetFigFont{12}{14.4}{rm}``$=$''}}}
\put(114,753){\makebox(0,0)[b]{\smash{\SetFigFont{12}{14.4}{rm}$x_1$}}}
\put(276,775){\makebox(0,0)[lb]{\smash{\SetFigFont{12}{14.4}{rm}$k_1$}}}
\put(235,760){\makebox(0,0)[lb]{\smash{\SetFigFont{12}{14.4}{rm}$k_2$}}}
\put(200,765){\makebox(0,0)[b]{\smash{\SetFigFont{12}{14.4}{rm}$\cup$}}}
\put(165,775){\makebox(0,0)[b]{\smash{\SetFigFont{12}{14.4}{rm}$x_2$}}}
\put(275,765){\makebox(0,0)[b]{\smash{\SetFigFont{12}{14.4}{rm}$x$}}}
\end{picture}
}}
  \end{split}
\end{equation}
\end{Cor}
\begin{pf}
  Since $\omega^2$ is a bisolution of the Klein-Gordon operator, we
  can apply Theorem~\ref{Thm:prop_of_sing} with $P_1= \identity{}
  \otimes (\Box + m^2)$ or $P_2 = (\Box + m^2) \otimes \identity$.
  In the first case we conclude by a) that
  \[
  \WF(\omega^2) \subseteq \{ (x_1,k_1;x_2,k_2) \in (T^*M \times  T^*M)
  \setminus \{ 0 \} | \quad k_2^2 =0 \}
  \]
  and --using b)--
  \begin{equation}\label{eq:propk_2}
  \left( (x_1,k_1; x_2,k_2) \in \WF(\omega^2)
  \wedge (x_2,k_2) \sim (x'_2,k'_2) \right)
  \Rightarrow
   \left( (x_1,k_1; x'_2,k'_2) \in \WF(\omega^2)\right)
  \end{equation}
  Obviously the second case leads to
  \[
  \WF(\omega^2) \subseteq \{ (x_1,k_1;x_2,k_2) \in (T^*M \times  T^*M)
  \setminus \{ 0 \} | \quad k_1^2 =0 \}
  \]
  and
  \begin{equation}\label{eq:propk_1}
  \left( (x_1,k_1; x_2,k_2) \in \WF(\omega^2)
  \wedge  (x_1,k_1) \sim (x'_1,k'_1) \right)
  \Rightarrow
   \left( (x'_1,k'_1; x_2,k_2) \in \WF(\omega^2) \right)
   \end{equation}
  To decide whether $(x_1,k_1;x_2,k_2)$ with $k_1^2 =k_2^2 =0$ is in
  the wave front set of $\omega^2$, one checks first whether $x_1$ and
  $x_2$ are spacelike separated. If this is true, then
  $(x_1,k_1;x_2,k_2) \not \in \WF(\omega^2)$, since the projection of
  the wave front set of a distribution to the base point gives the
  singular support of that distribution, but $\omega^2$ is assumed to
  have no singularities at spacelike separated points. If they are
  causally related, we use Eqn.~(\ref{eq:propk_2})
  and~(\ref{eq:propk_1}) to propagate $(x_i,k_i)$ along the null
  geodesic $\gamma_i$ with tangent vector $k_i$ at $x_i$ to the Cauchy
  surface $\cC$. Note that due to the global hyperbolicity of $M$
  this is always possible. If the points on $\mathcal C$ do not coincide,
  we have $(x_1,k_1;x_2,k_2) \not \in \WF(\omega^2)$, since different
  points on a Cauchy surface are spacelike separated. Should they
  coincide at say $x \in \mathcal C$, we check whether this particular
  combination $(x,k_1;x,k_2)$ is in the wave front set using
  Eqn.~(\ref{eq:wfscalarUx}).
\end{pf}
%%%%%%%%%%%%%%%%%%%%%%%%%%%%%%%%%%%%%%%%%%%%
\begin{figure}[hbtp]
  \begin{center}
    \leavevmode
\begin{picture}(0,0)%
\includegraphics{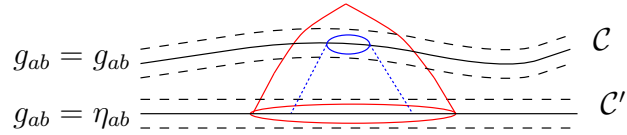}%
\end{picture}%
\setlength{\unitlength}{0.012500in}%
\begin{picture}(246,52)(117,777)
 \
 \put(117,804){\makebox(0,0)[lb]{\smash{\SetFigFont{12}{14.4}{rm}
$g_{ab}=g_{ab}$
}}}
 \
 \put(117,780){\makebox(0,0)[lb]{\smash{\SetFigFont{12}{14.4}{rm}
\mbox{$g_{ab}=\eta_{ab}$}
}}}
 \
 \put(363,783){\makebox(0,0)[lb]{\smash{\SetFigFont{12}{14.4}{rm}
${\mathcal C'}$
}}}
 \
 \put(360,810){\makebox(0,0)[lb]{\smash{\SetFigFont{12}{14.4}{rm}
${\mathcal C}$
}}}
\end{picture}
  \end{center}
  \caption{Spacetime Deformation}
  \label{fig:deformation}
\end{figure}
To finish the argument, we have to construct a spacetime
$(\tilde{M},\tilde{g}_{ab})$ that satisfies the properties (i) and
(ii).  This can be done by using the methods of Appendix~C
in~\cite{FullingNarcowichWald:81} (Compare
Figure~\ref{fig:deformation} and see also~\cite{Verch:93}). Let
$\Diamond(\cCb) \subset V \subset N$ be a causal normal neighborhood
of $\cC$, such that $\cC$ is also a Cauchy surface for $V$. $V$ is, by
the normal exponential map $\varrho$ of $\cC$, diffeomorphic to an
open neighborhood $\tilde{V} \subset {\mathbb R} \times \cC$ of
$\tilde{\cC} = \{ 0 \} \times \cC = \varrho(\cC)$. Using normal
coordinates $(t,\mbox{{\bf x}})$ around some point in $\tilde{\cC}$,
the metric $\varrho_*g$ on $\tilde{V}$ takes the form
\[ dt^2 - h_{ij}(t,\mbox{{\bf x}}) d\mbox{x}^i d\mbox{x}^j \]
Choose $t_1 < t_2 <t_3 <0$ such that
\[ \overline{ \hat{\cC}(t_i) \cap J^-(\tilde{\cCb})} \subset
     \varrho(U_x) \qquad i= 1,2,3,
\]
where $\hat{\cC}(t) := \{ t \} \times \cC$, $t \in {\mathbb R}$ and
$\tilde{\cCb} := \varrho(\cCb)$ (As a consequence of the global
hyperbolicity and due to the assumption $\Diamond(\cCb) \subset U_x$ such a
choice is always possible). Recall that $\varrho(U_x)$ can be covered
with a single coordinate patch by assumption. Next choose a
neighborhood $\tilde{U}$ of $\tilde{\cC}$ such that
$\overline{\tilde{U}} \subset \tilde{V} \cap \mbox{{\rm int}} \left(
J^+ (\hat{\cC}(t_3)) \right)$. Due to the fact that $t_1 < t_2$ there
exists a neighborhood $\hat{U}$ of $\overline{ J^- ( \tilde{\cCb} )
  \cap \hat{\cC}(t_1) }$ in $\tilde{M}$ such that (i) $\overline{
  \hat{U}} \subset \varrho(U_x)$ and (ii) $\overline{\hat{U}} \cap J^+
(\hat{\cC}(t_2)) = \emptyset $. Let $f\in \Cinf({\mathbb R}\times \cC,
{\mathbb R})$ be a smooth function $0 \leq f \leq 1$, $f\equiv 0$ on
$\tilde{U}$ and $f\equiv 1$ outside the closure of $\tilde{V}$ or in
the past of $J^-(\hat{\cC}(t_2))$. Let $\tilde{h}$ be a complete
Riemannian metric on $\cC$ being flat on the spatial component of
$\hat{U} \cap (\hat{\cC}(t_3))$. Note that the existence of such a
$\tilde{h}$ follows from the fact that $\overline{ \hat{U}}$ can be
covered by a single coordinate patch (i.e.\ $\varrho(U_x)$). Let
$\beta\in\Cinf({\mathbb R}\times \cC, {\mathbb R}^+)$ be a function equal to
one on $\tilde{U}$ and on $(-\infty,t_2)$. Define a Lorentzian metric
$\tilde{g}_{ab}$ on ${\mathbb R}\times \cC$ by setting in the coordinates
above the coordinate expression of $\tilde{g}_{ab}$ equal to
\[
\beta(t,\mbox{{\bf x}}) dt^2 - (( 1 -f(t,\mbox{{\bf
    x}}))h_{ij}(t,\mbox{{\bf x}}) + f(t,\mbox{{\bf x}})
\tilde{h}_{ij}(t,\mbox{{\bf x}}) ) d\mbox{x}^i d\mbox{x}^j
\]
By choosing $\beta$ sufficiently small outside the region where it is
demanded to be one\footnote{%
I.e.\ we ``close'' up the light cone.},
we can ensure that $(\tilde{M}:={\mathbb R}\times \cC,\tilde{g}_{ab})$
is globally hyperbolic. Setting $U:= \varrho^{-1}(\tilde{U})$ and
$\tilde{S} := \varrho^{-1}(\hat{\cC}(t_1))$ finishes the\
 construction.\\[0.5cm]
%%%%%%%%%%%%%%%%%%%%%%%%%%%%%%%%%%
%
A first consequence of Corollary~\ref{Cor:wavefrontscalar} is
\begin{Cor}\label{Cor:scalar_power}
  Let $\omega^n$ be the n-point distribution arising from a quasifree
  Hadamard state of a massive Klein-Gordon field propagating on a
  globally hyperbolic spacetime $(M,g_{ab})$. Then all finite powers
  of $\omega^n$ exist as products of distributions.
\end{Cor}
\begin{pf}
  Assuming all odd n-point distributions to vanish, we have, since
  $\omega^n$ arises from a quasifree state
  \[
  \omega^n(x_1,\ldots,x_n) = \sum_P \prod_r \omega^2(x_{(r,1)},x_{(r,2)}),
  \]
  where $P$ denotes a partition of the set of points $\{ x_i\}$ into
  subsets which are pairings of points, labeled by $r$. Note that the
  ordering of the points in $\omega^2$ is preserved, e.g.\ $(r,1) <
  (r,2)$ and no two arguments are identical. The latter fact ensures
  the existence of the product $\prod_r$ whenever $\omega^2(x_i,x_j)$
  are distributions. For the wave front set of $\omega^n$ one finds
  using Theorem~\ref{Thm:IX.54}
  \begin{equation}
    \label{eq:wf_omega_n}
    \begin{split}
      &\WF(\omega^n) \\
      & = \bigcup_{(x_1,\ldots,x_n)\in M^n}
                        \WF\left(\omega^n(x_1,\ldots,x_n)\right) \\
                    & \subseteq \bigcup_{(x_1,\ldots,x_n)\in M^n}
                    \left(
                      \bigcup_P \bigcup_p \bigoplus_{r_p}
                      \left[ M^{n-2} \times
                        \WF\left( \omega^2(x_{(r_p,1)},x_{(r_p,2)})
                            \right)
                      \right]
                    \right),
    \end{split}
  \end{equation}
  where $p$ denotes a subset of $P$, $r_p$ labels the elements of $p$
  and
  {\small
  \begin{equation*}
  \begin{split}
  & \left[
    M^{n-2} \times \WF\left( \omega^2(x_{(r_p,1)},x_{(r_p,2)})\right)
  \right]\\
  & = \{ (x_1,0; \cdots; x_{(r_p,1)}, k_{(r_p,1)}; \cdots; x_{(r_p,2)},
  k_{(r_p,2)}; \cdots; x_n,0) \in (T^*M)^n \setminus \{0\} | \\
  & \phantom{=\{}{}\quad
   (x_{(r_p,1)}, k_{(r_p,1)};x_{(r_p,2)}, k_{(r_p,2)}) \in
  \WF\left( \omega^2(x_{(r_p,1)},x_{(r_p,2)}) \right)\}.
  \end{split}
  \end{equation*}
  }
  (Compare Example~\ref{exmp:fourpointDist} below, which demonstrates
  the notation on the four point function).  To prove the Corollary it
  is --by Theorem~\ref{Thm:IX.54}-- necessary and sufficient to show
  that finite sums of $\WF(\omega_n)$ do not contain zero. We know the
  wave front sets of all distributions in Eqn.~\ref{eq:wf_omega_n}
  explicitly (See Corollary~\ref{Cor:wavefrontscalar}). Consider the
  following two cases:
  \begin{enumerate}
  \item The direction in $\WF(\omega_n)$ associated to the first
    variable is not zero. Then its time component is strictly
    positive by Corollary~\ref{Cor:wavefrontscalar}. Since the time
    components of all directions associated to the first variable of
    $\WF(\omega_n)$ are always greater or equal to zero, the sum of
    all these directions can not vanish and the Corollary is proved.
  \item The direction in $\WF(\omega_n)$ associated to the first
    variable is zero for all summands. Then all directions associated
    to the second variable must have time components greater or equal
    to zero and we apply the same argumentation as for the first
    variable. Note that it is excluded by the definition of the wave
    front set that all directions of all variables vanish
    simultaneously. Therefore there exists a variable $(x_i)$ $i<n$
    such that the analogon of (1) holds.
  \end{enumerate}
\end{pf}
\begin{exmp}\label{exmp:fourpointDist}
  Consider the four-point distribution of a
    quasifree state:
    \[
      \omega^4(x_1,x_2,x_3,x_4) =
     \omega^2(x_1,x_2)\omega^2(x_3,x_4) +
     \omega^2(x_1,x_3)\omega^2(x_2,x_4) +
     \omega^2(x_1,x_4)\omega^2(x_2,x_3)
   \]
   We have
    \[P \in \left\{ \{(x_1,x_2),(x_3,x_4) \},
    \{(x_1,x_3),(x_2,x_4) \}, \{(x_1,x_4),(x_2,x_3) \} \right\}
    \]
    Let
    $P= \{ (x_1,x_3), (x_2,x_4) \}$, then
    $p \in \left\{ \{(x_1,x_3)\}, \{(x_2,x_4)\}, \{
    (x_1,x_3),(x_2,x_4) \} \right\}$. Assume $p = \{ (x_1,x_3)\}$, then
    $r_p=(x_1,x_3)$, i.e.\ $x_{(r_p,1)}=x_1$ and
    $x_{(r_p,2)}=x_3$. Therefore
    \begin{equation*}
      \begin{split}
         &\WF(\omega^4)\\
        & \subseteq \bigcup_{(x_1,\ldots,x_4) \in M^4} \biggl(
        \left[ M^2 \times \WF\bigl(\omega^2(x_1,x_2)\bigr) \right] \cup
        \left[ M^2 \times \WF\bigl(\omega^2(x_3,x_4)\bigr) \right]\\
        & \phantom{\subseteq \bigcup_{(x_1,\ldots,x_4) \in M^4} \biggl(
        \left[ M^2 \times ) \right] \cup }
        \cup
        \left[
          \left( M^2 \times\WF\bigl(\omega^2(x_1,x_2)\bigr)
          \right)
          \oplus
          \left( M^2 \times\WF\bigl(\omega^2(x_3,x_4)\bigr)
          \right)
        \right] \\
        & \phantom{\subseteq \bigcup_{(x_1,\ldots,x_4) \in
            M^4}\biggl(}
        \cup 2 \leftrightarrow 3 \\
        & \phantom{\subseteq \bigcup_{(x_1,\ldots,x_4) \in
            M^4}\biggl(}
        \cup 2 \leftrightarrow 4 \biggr)\\
        & \equiv
        \left\{ (x_1,k_1;x_2,k_2;x_3,0;x_4,0) | \quad
          (x_1,k_1;x_2,k_2) \in \WF(\omega^2) \right\} \\
        & \phantom{\equiv} \cup
        \left\{ (x_1,0;x_2,0;x_3,k_3;x_4,k_4) | \quad
          (x_3,k_3;x_4,k_4) \in \WF(\omega^2) \right\} \\
        & \phantom{\equiv} \cup
        \left\{ (x_1,k_1;x_2,k_2;x_3,k_3;x_4,k_4) | \quad
          (x_1,k_1;x_2,k_2) \in \WF(\omega^2);
          (x_3,k_3;x_4,k_4) \in \WF(\omega^2) \right\} \\
        & \phantom{\equiv\quad} \cup \cdots
      \end{split}
    \end{equation*}
\end{exmp}
We can extend Corollary~\ref{Cor:wavefrontscalar} and
Corollary~\ref{Cor:scalar_power} immediately to the fermionic case,
since both, the spinor van Vleck Morette determinant
${{\Delta^{1/2}}^B}_{A'}$ and ${{v^{(p)}}^B}_{A'}$ are smooth near the
diagonal for all spinor indices $A',B$ and the Dirac operator, which
acts as a differential operator, does not enlarge the wave front set.
\begin{Cor}\label{Cor:wfhadamfermi}
  The two point distribution ${{\omega^2}_A}^{A'}$ of a free Majorana
  spinor field in a globally Hadamard state has wave front set
  contained in
  \begin{equation}\label{eq:wfhadamfermi}
    \WF({{\omega^2}_A}^{A'})
    \subseteq \{ (x_1,k_1),(x_2,- k_2) \in T^*M\setminus \{0\} \times
    T^*M\setminus \{0\} |
    (x_1,k_1) \sim (x_2,k_2); \quad k_1^0 \geq 0 \}
  \end{equation}
\end{Cor}
The extension of Corollary~\ref{Cor:scalar_power} above to the fermionic case
reads:
\begin{Cor}\label{Cor:fermi_power}
  Let ${{\omega^n}_A}^{A'}$ be the n-point distribution arising from a
  quasifree Hadamard of a massive Majorana spinor field propagating on a
  globally hyperbolic spacetime $(M,g_{ab})$. Then all finite powers
  of ${{\omega^n}_A}^{A'}$ exist as product of distributions.
\end{Cor}
\begin{pf}
  For fixed Spinor indices the wave front set of the fermionic
  two-point distribution is contained in the wave front set for the
  scalar field (See Eqn.~(\ref{eq:wfhadamfermi})). Applying
  the same argumentation as in the proof of
  Corollary~\ref{Cor:scalar_power} finishes the proof.
\end{pf}
\begin{rem}
  It is worth noting that for the  wave front set of all finite powers of
  all Hadamard two point distributions $\omega^2$ satisfies
  \begin{equation}
    \label{eq:WFpower2pointdistr}
    \WF(\omega^2 \cdot \cdots \cdot \omega^2) \subseteq (\WF(\omega^2)
    \oplus \WF(\omega^2) ) \cup \WF(\omega^2).
  \end{equation}
  This relation follows by Theorem~\ref{Thm:IX.54} together with the
  observation that the first summand in
  Eqn.~(\ref{eq:WFpower2pointdistr}) is invariant under further
  additions of $\WF(\omega^2)$, i.e.,
  \[
  \WF(\omega^2) \oplus \WF(\omega^2) \oplus\WF(\omega^2) =
  \WF(\omega^2) \oplus \WF(\omega^2)
  \]

\end{rem}
\subsection{A local characterization of globally Hadamard states}
One of the main motivations for using wave front sets in quantum field
theory is the fact that they allow the specification of global
properties locally. The following Theorem, which is one of
the main results of Radzikowski's dissertation, gives a local
characterization of globally Hadamard states.
\begin{Thm}[Theorem 2.6 of \cite{Radzikowski:92}]
  \label{Thm:locHadam}
  Let $\omega^2$ be the two-point distribution arising from a state of
  a massive Klein-Gordon field propagating on a globally hyperbolic
  spacetime. If $\omega^2$ has wave front set as in
  Corollary~\ref{Cor:wavefrontscalar} above then $\omega^2$ is
  globally Hadamard.
\end{Thm}
\begin{rem}
  Our assumption implies that $\omega^2$ fulfils the Klein-Gordon
  equation and has $i$ times the commutator distribution of $\Box +
  m^2$ as its antisymmetric part. The
  Corollary~\ref{Cor:wavefrontscalar} above may be viewed as the
  converse of this Theorem.
\end{rem}
The proof of this Theorem is rather long; we refer the reader to the
dissertation of Radzikowski for the details.  Radzikowski's
Theorem~2.6 states the equivalence of the wave front set assumption
and the globally Hadamard condition $\mod \Cinf$. To prove it, he
introduces the Feynman two-point distribution $\omega_F$ defined by
$\omega_F:= i \omega^2 + E^+$, where $E^+$ is the advanced fundamental
solution of $\Box + m^2$. The assumption on the wave front set of
$\omega^2$ uniquely fixes this distribution (up to $\Cinf$) to be the
distinguished Feynman parametrix defined in~\cite{Hoermander:72}.
This in turn implies that $\omega^2$ is globally Hadamard $\mod
\Cinf$.
%
%%%%%%%%%
\section{The wave front set spectrum condition}\label{sec:WFSSC}
%%%%%%%%%
%
{\sloppy
In this section we propose a condition on the wave front sets of
states of a quantum field on a manifold $M$, which might generalize
the usual Minkowski spectrum condition to Lorentz manifolds.  The
idea to use wave front sets for a formulation of some kind of spectrum
condition is due to Rad\-zi\-kows\-ki~\cite{Radzikowski:92}.  For a
motivation of this idea recall that the wave front set of a state can
be used to characterize this state globally. However
Equation~(\ref{eq:wfhadamscalar}) restricts the singular support of
$\omega_2(x_1,x_2)$ to points $x_1$ and $x_2$ which are null related;
hence $\omega_2$ is smooth for spacelike related points.
This smoothness is known to be true for reasonable quantum field
theories on Minkowski space satisfying the true spectrum condition by
the Bargmann-Hall-Wightman Theorem. For timelike related points
however a similar general prediction on the smoothness does not exist.
In order to include possible singularities at timelike related
points, Radzikowski extended in~\cite{Radzikowski:92} the right hand
side of Eqn.~(\ref{eq:wfhadamscalar}) to all causally related points.
He proposed that the wave front set of the two-point distributions of
any physically reasonable state should be contained in this extended
set and called this proposal the `wave front set spectrum condition'
(WFSSC).  He also proposed a WFSSC for higher n-point distributions
and showed that both of his Definitions are compatible to the usual
spectrum condition: Each $\omega^n$ fulfilling his WFSSC satisfies the
true spectrum condition on Minkowski space $\mod \Cinf$ (Theorem~4.10
of~\cite{Radzikowski:92}) and vice versa. He gives further evidence on
the legitimacy of his Definitions by linking them to the scaling limit
condition of Fredenhagen and Haag~\cite{fredenhagen:87}: Both
Definitions imply the true spectrum condition in the scaling limit if
this limit exists (Theorem~4.11 of~\cite{Radzikowski:92}).
Unfortunately it can be shown that the n-point distributions for $n >
2$ associated to a quasifree Hadamard state of a scalar field on a
globally hyperbolic spacetime do not satisfy his WFSSC in general.
Moreover, the Wick products defined below give rise to counterexamples
even for his two point WFSSC.  Thus his original WFSSC needs to be
modified.
}
The counterexample mentioned above and the wish to include fields
which are Wick products of `simple' fields, leads us to propose the
following `conic' WFSSC\footnote{I.e.,
  $(x,k_1;\ldots;y,l_1),(x,k_2;\ldots;y,l_2) \in \WF(\omega_n)
  \Rightarrow (x,\lambda k_1 + \mu k_2; \ldots;y, \lambda l_1 + \mu
  l_2) \in \WF(\omega_n) \quad \forall \lambda,\mu > 0$.}.
\begin{figure}[hbtp]
  \begin{center}
    \leavevmode
    \begin{picture}(0,0)%
      \includegraphics{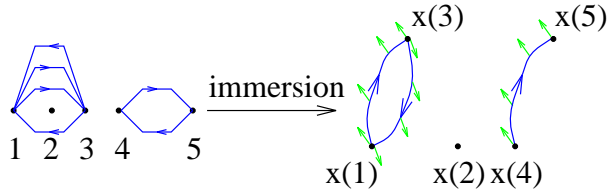}%
    \end{picture}%
    \setlength{\unitlength}{0.012500in}%
    \begin{picture}(251,81)(27,735)
      \put( 45,745){\makebox(0,0)[b]{\smash{\SetFigFont{12}{14.4}{rm}2}}}
      \put( 60,745){\makebox(0,0)[b]{\smash{\SetFigFont{12}{14.4}{rm}3}}}
      \put( 75,745){\makebox(0,0)[b]{\smash{\SetFigFont{12}{14.4}{rm}4}}}
      \put(105,745){\makebox(0,0)[b]{\smash{\SetFigFont{12}{14.4}{rm}5}}}
      \put(138,771){\makebox(0,0)[b]{
          \smash{\SetFigFont{12}{14.4}{rm}immersion}}}
      \put( 30,745){\makebox(0,0)[b]{\smash{\SetFigFont{12}{14.4}{rm}1}}}
      \put(171,735){\makebox(0,0)[b]{\smash{\SetFigFont{12}{14.4}{rm}x(1)}}}
      \put(216,735){\makebox(0,0)[b]{\smash{\SetFigFont{12}{14.4}{rm}x(2)}}}
      \put(240,735){\makebox(0,0)[b]{\smash{\SetFigFont{12}{14.4}{rm}x(4)}}}
      \put(207,801){\makebox(0,0)[b]{\smash{\SetFigFont{12}{14.4}{rm}x(3)}}}
      \put(267,801){\makebox(0,0)[b]{\smash{\SetFigFont{12}{14.4}{rm}x(5)}}}
    \end{picture}
  \end{center}
  \caption{An immersion of a graph}
  \label{fig:immersion}
\end{figure}
Let ${\mathcal G}_n$ denote the set of all undirected finite
graphs\footnote{We allow each two vertices of a graph to be connected
  by multiple edges.} with vertices $\{1,\ldots,n\}$, such that for
every element $G \in {\mathcal G}$ all edges occur in both admissible
directions. An {\em immersion\/} of a graph $G\in {\mathcal G}_n$ into
some manifold $M$ is an assignment of the vertices of $G$ to points in
$M$, $\nu\rightarrow x(\nu)$, and of the edges of $G$ to causal curves
in $M$, $e\rightarrow \gamma(e)$, together with a covariantly constant
causal covector field $k_e$ on $\gamma$, such that
\begin{enumerate}
\item If $e$ is an edge from $\nu$ to $\nu'$ then $\gamma(e)$
  connects $x(\nu)$ and $x(\nu')$,
\item If $e^{-1}$ denotes the edge with opposite direction as
  $e$, then the corresponding curve $\gamma(e^{-1})$ is the inverse of
  $\gamma(e)$,
\item For every edge $e$ from $\nu$ to $\nu'$, $k_e$ is
  directed towards the future whenever $\nu<\nu'$,
\item $k_{e^{-1}}=-k_e$.
\end{enumerate}
(Compare Figure~\ref{fig:immersion})
\begin{Dfn}[GWFSSC]\label{Dfn:GWFSSC}
  A state $\omega$ with n-point distributions $\omega^n$ is said to
  satisfy the {\em general wave front set spectrum condition\/}
  (GWFSSC) if and only if
  \begin{equation*}
  \begin{split}
    & \WF(\omega^n)\\
    & \subseteq \biggl\{(x_1,k_1;\ldots;x_n,k_n)\in
  T^*M^n\setminus\{0\}| \quad \exists G\in {\mathcal G}_n
  \text{~and an immersion $(x,\gamma,k)$ of $G$,}\\
  & \phantom{\subseteq \biggl\{(x_1)}
   \text{such that}
  \quad
  \text{~(i)~}  x_i = x(i) \quad \forall i=1,\ldots, n\\
  & \phantom{\subseteq \biggl\{(x_1)
   \text{such that}
  \quad }
  \text{(ii)~}  k_i =\sum_{\begin{Sb}
                             e\\
                             i \text{~is starting}\\
                             \text{point of~}e
                           \end{Sb}
                           } k_e (x_i)
  \biggr\}\\
\end{split}\nonumber
\end{equation*}
\end{Dfn}
\begin{exmp}
    \begin{center}
      $m=3$\\
      \begin{picture}(0,0)%
        \includegraphics{figures/muSC.pstex}%
      \end{picture}%
      \setlength{\unitlength}{0.012500in}%
      \begin{picture}(198,129)(40,690)
        \put(166,746){\makebox(0,0)[b]{\smash{\SetFigFont{12}{14.4}{rm}2}}}
        \put(126,802){\makebox(0,0)[b]{\smash{\SetFigFont{12}{14.4}{rm}3}}}
        \put(111,690){\makebox(0,0)[b]{\smash{\SetFigFont{12}{14.4}{rm}1}}}
        \put(180,692){\makebox(0,0)[b]{\smash{\SetFigFont{12}{14.4}{rm}1}}}
        \put(235,748){\makebox(0,0)[b]{\smash{\SetFigFont{12}{14.4}{rm}2}}}
        \put(195,804){\makebox(0,0)[b]{\smash{\SetFigFont{12}{14.4}{rm}3}}}
        \put( 45,690){\makebox(0,0)[b]{\smash{\SetFigFont{12}{14.4}{rm}1}}}
        \put(100,746){\makebox(0,0)[b]{\smash{\SetFigFont{12}{14.4}{rm}2}}}
        \put( 60,802){\makebox(0,0)[b]{\smash{\SetFigFont{12}{14.4}{rm}3}}}
      \end{picture}
    \end{center}
\end{exmp}
\begin{rem}
  For every set of base points $(x_1,\ldots{},x_n)\in
  \text{sing~supp~}(\omega^n)$ the first non zero direction $k_l$ in
  the wave front set is future directed.
\end{rem}
As an important consequence of this Definition we have the following
\begin{Prop}
  Let $\omega$ be a state which satisfies the GWFSSC, then all finite
  powers of its n-point distributions exist.
\end{Prop}
\begin{pf}
  To prove the Proposition it is --by Theorem~\ref{Thm:IX.54}--
  sufficient to show that finite sums of $\WF(\omega^n)$ do not
  contain zero. Consider the following two cases:
  \begin{enumerate}
  \item The direction in $\WF(\omega^n)$ associated to the first
    variable is not zero. Then its time component is strictly positive
    since the sum of future directed covectors is again future
    directed. Thus the sum of {\em all\/} directions associated to the
    first variable can not vanish and the Proposition is proved.
  \item The direction in $\WF(\omega^n)$ associated to the first
    variable is zero for all summands. Then all directions associated
    to the second variable must have time components greater or equal
    to zero and we apply the same argumentation as for the first
    variable. Note that it is excluded by the definition of the wave
    front set that all directions of all variables vanish
    simultaneously. Therefore there exists a variable $(x_i)$ $i<n$
    such that the analogon of (1) holds.
  \end{enumerate}
\end{pf}
To show that there exists non trivial states which verify our
Definition we prove the following
\begin{Lemma}\label{Lemma:GWFSSCKleinGord}
  Let $\omega$ denote a quasifree Hadamard state for the Klein-Gordon
  field on a globally hyperbolic manifold $(M,g_{ab})$, then $\omega$
  satisfies the general wave front set spectrum condition.
\end{Lemma}
{\sloppy
\begin{pf}
  Note first that all odd n-point distributions vanish by assumption,
  hence $\omega^n$ satisfies the GWFSSC trivially for odd n.  Consider
  the two point distribution of $\omega$. Its wave front set is
  explicitly known (see Proposition~\ref{Cor:wavefrontscalar}) and
  obviously satisfies the GWFSSC.  For a general n-point distribution,
  with n even, consider a point $p=(x_1,k_1;\ldots;x_n,k_n)\in
  \WF(\omega_n)$. Using the same notation as in
  Definition~\ref{Cor:scalar_power} and Eqn.~(\ref{eq:wf_omega_n})
  this point can be written as
  \[
  p=(x_1,\sum_{i=1}^{n/2} k_{1_i};\ldots;x_n,\sum_{j=1}^{n/2} k_{n_j}),
  \]
  where the directions $k_{l_m}$ are suitable chosen with the
  constraint $(x_l,k_{l_i};x_m,k_{m_j}) \in \WF(\omega^2)$ or
  $(x_l,k_{l_i};x_m,k_{m_j}) = (x_l,0;x_m,0)$.  We remark that this
  representation follows trivially from Eqn.~(\ref{eq:wf_omega_n}).
  Now by Eqn.~(\ref{eq:wfhadamscalar}), whenever $k_{l_i},k_{m_j} \neq
  0$, $(x_l,k_{l_i})\sim (x_m,-k_{m_j})$, i.e., $x_l$ and $x_m$ are
  immersed vertices $\nu_l$ and $\nu_m$ of some graph $G$, such that
  the corresponding edge $e$ connecting $\nu_l$ and $\nu_m$ has an
  immersion $\gamma(e)$, which is the lightlike geodesic connecting
  $x_l$ and $x_m$. Moreover, there exists a covector field $k_e$ such
  that $k_e({x_l})=k_{l_i}$ and $k_{e^{-1}}({x_m})=- k_{m_j}$. Note
  that $k_{l_i}$ is directed towards the future whenever $l < m$. Thus
  $G$ with this immersion $(x,\gamma,k)$ satisfy (i) and (ii) in the
  Definition of the GWFSSC.
\end{pf}
}
We close this chapter by giving  our definition of the quantized version of
our locally supersymmetric model using the algebraic approach:
A scalar quantum field theoretical model $({\EuScript
  H},D,\Phi,\Omega)$ on a spacetime $(M,g_{ab})$ which we assume to be
orientable, time orientable and globally hyperbolic, is said to
satisfy the Klein Gordon equation (with coupling $\xi$) if and only if
\[ \mbox{For all $f \in C_0^\infty(M)$: }\hspace{4em} \Phi(K_\xi f) = 0,\]
where $K_\xi := \left( \Box + ( m^2 + \xi R) \right)$ is the generic
Klein Gordon operator associated to $g_{ab}$.
A Majorana quantum field theoretical model $({\EuScript
  H},D,\Psi,\Psi^+,\Omega)$ on the same spacetime $(M,g_{ab})$ with
spin structure $(SM,\bfm{p})$ is said to satisfy the Dirac equation if
and only if
\[ \mbox{For all $f \in \Gamma(DM(\calO)) \stackrel{i}{\hookrightarrow}
\hat{D}_1(\calO)$: }\hspace{4em} \Psi(\notD f) = 0\]
and
\[ \mbox{For all $f^+ \in \Gamma(D^*M(\calO))
  \stackrel{i}{\hookrightarrow} \hat{D}_1(\calO)$: }\hspace{4em}
\Psi^+(\notD^+ f^+)
= 0\]
where $\notD := (i \notnabla - m)$ ($\notD^+ := (-i \notnabla - m)$)
is the generic Dirac (adjoint Dirac) operator on the spinor (cospinor)\
 bundle.
The states of these models fulfill the same equation of motion as the
corresponding fields. Let now ${\mathfrak A}_1,{\mathfrak A}_2$ be two scalar
Borchers Uhlmann algebras and ${\hat{\mathfrak A}}_3$ be a Majorana
Borchers Uhlmann algebra. Denote by ${\mathfrak A}$ the product ${\mathfrak
  A}_1 \ast {\mathfrak A}_2 \ast {\hat{\mathfrak A}}_3$ of these three
algebras. ${\mathfrak A}$ is called a {\em locally supersymmetric\/}
Borchers Uhlmann algebra. A product state $\omega = \omega_1 \cdot
\omega_2 \cdot \omega_3$ on ${\mathfrak A}$ such that each $\omega_{1,2}$
defines a scalar quantum field theoretical model satisfying the Klein
Gordon equation with $\xi= 1/4$, whereas $\omega_3$ gives a Majorana
quantum field theoretical model satisfying the Dirac equation, defines
a {\em locally supersymmetric quantum field theoretical model\/} on
the spacetime $(M,g_{ab})$.  For convenience such a state will be
called a local SUSY product state (on the local SUSY Borchers Uhlmann
algebra ${\mathfrak A}$).
%
%%%%%%%%%%%%%%%%%%%%%%%%%%%%%%%%%%%%%%%%%%%%%%%%%%%%%
\chapter{The energy momentum tensor}
\label{sec:emtCurved}
%%%%%%%%%%%%%%%%%%%%%%%%%%%%%%%%%%%%%%%%%%%%%%%%%%%%%
%
It was shown in chapter~\ref{sec:SUSYinR4} that there is a close
relation between the (conserved) supercurrent and the energy momentum
tensor in the free Wess-Zumino model on Minkowski space. We expect a
similar relation on a curved spacetime, too. It seems therefore
reasonable to {\em define\/} a candidate for the energy momentum
tensor of a locally supersymmetric model on a curved spacetime by an
analogon of Eqn.~(\ref{eq:SUSY_EMT}) using the improved
`supercurrent', i.e., by inserting the improved `supercurrent'
(Eqn.~\ref{eq:impsupercurCurved}) into Eqn.~(\ref{eq:SUSY_EMT}).  It is
worth noting that this substitution is necessary in order to obtain
the {\em improved\/} stress energy tensor instead of the canonical
one.  Unfortunately there are some conceptual and mathematical
difficulties with this approach which will be discussed in the sequel:
\begin{enumerate}
\item In classical general relativity the energy momentum tensor
  $T^{ab}$ of the matter fields has to be locally conserved. Recall
  that this assumption reflects the well established fundamental law
  of energy conservation. Moreover, it is required as a consistency
  condition in the Einstein equations, where $T^{ab}$ acts as a source
  term\footnote{For a discussion of this assumption see
    e.g.~\cite{hawking:largeScale}.}. It is easy to see --at least in
  Minkowski space--, that only a {\em locally conserved\/}
  supercurrent leads to a conserved energy momentum tensor. The fact
  that $k^{aA}$ is locally conserved on Lorentz manifolds with
  vanishing scalar curvature only, indicates that our Definition is
  applicable at most to those special spacetimes.
\item  A mathematical difficulty comes from the fact that on a general
  Lorentz manifold the integration of a spinor vector current, such as
  $j^{aA}$, over some spacelike hypersurface ${\mathcal C}$ with normal
  covector field $n_a$ does not make sense, since the parallel
  transport of spinors is path dependent. Recall that such an
  integration was necessary on Minkowski space for the definition of
  the spinor supercharge $Q^A$. However locally, i.e., for a
  sufficiently small region ${\mathcal O}_{\mathcal C}$ of ${\mathcal C}$\
 there exists
  an obvious solution to this problem: The expression $j^{aA}(x){{\mathcal
      J}_A}^{A'}(x,x')$, where ${{\mathcal J}_A}^{A'}(x,x')$ denotes the
  bi-spinor of parallel transport introduced in
  chapter~\ref{sec:analogonWess} transforms as a cospinor at $x$ and as
  a spinor at $x'$, hence the integral
  \begin{equation}
    \label{eq:supercurSigmaExamp}
    \int_{{\mathcal O}_{\mathcal C}} j^{aA}{{\mathcal J}_A}^{A'} n_a\
 d{\mathcal C}
  \end{equation}
  does make sense (at least at the classical level). Moreover, since
  $\lim_{x\rightarrow x'} ( \nabla_a {{\mathcal J}_A}^{A'} ) = 0 $, we
  have ``asymptotic conservation'' of $j^{aA}{{\mathcal J}_A}^{A'}$, i.e.,
  the coincidence limit $\lim_{x\rightarrow x'}\bigl(\nabla_a
  (j^{aA}{{\mathcal J}_A}^{A'})\bigr)$ vanishes whenever $j^{aA}$ is
  locally conserved. Note however that the
  integral~(\ref{eq:supercurSigmaExamp}) obviously depends on $x'$.
\end{enumerate}
In the remaining part of this work we will roughly speaking
investigate the expression
\begin{equation}
  \label{eq:formalTraceCurved}
  + \frac{1}{8} \left( \gamma^{(a} \{j^{b)}, Q^+ \}_+ \right),
\end{equation}
for a quantized version of our locally supersymmetric model.  $Q^+$ is
the `supercharge' associated to $k^{aA}$ and the point separated
version of Eqn.~(\ref{eq:formalTraceCurved}) is regarded as an
operator valued distribution on a suitable Hilbertspace (See below
for the actual definitions).
\begin{rem}
  The use of $k^{aA}$ instead of $j^{aA}$ for the definition of the
  `supercharge' $Q^+$ is motivated by the observation that in the
  Wess-Zumino model the divergence, by which $j^{aA}$ and $k^{aA}$
  differ, does not contribute to the supercharge $G$.
\end{rem}
%
%%%%%%%%%
\section{New Wightman fields: The `supercurrents' after quantization}
\label{sec:superCurCurv}
%%%%%%%%%
%
Consider the locally supersymmetric Borchers Uhlmann algebra ${\mathfrak
  A}$ on $(M,g_{ab})$ together with a locally SUSY Hadamard product
state $\omega$, i.e., $\omega=\omega_1\cdot\omega_2\cdot\omega_3$ is a
local SUSY product state, such that $\omega_1$, $\omega_2$ and
$\omega_3$ are three Hadamard states. Let $(\script
H,D,A,B,\Psi,\Psi^+,\Omega)$ denote the corresponding quantum field
theoretical model, where $A$ and $B$ are scalar and pseudo-scalar
fields respectively and $\Psi$ denotes a Majorana spinor field.
Recall that ${\script H}$ is the GNS Hilbertspace of $\omega$ and that
$\Omega$ denotes the corresponding cyclic vector. Let us study the
improved `supercurrent'
\begin{equation}
  \label{eq:impsupercurCurvedQuant}
  \begin{split}
    j^{aA}
    & = i \AiB ( i \lvec{\notnabla} - m) \gamma^a \Psi \\
    & \phantom{=} - 2 \xi (\gamma^a \gamma^b - \gamma^b \gamma^a)
    \nabla_b \left( \AiB \Psi \right),
  \end{split}
\end{equation}
which involves the pointwise product of the basic fields. We find
{\sloppy
\begin{Prop}\label{prop:wightmanCur}
  $j^{aA}$ is well defined and gives a new Wightman field on the GNS
  Hil\-bert\-space $({\script H},A,B,\Psi,\Omega)$ of every quasifree\
 globally
  Hadamard product state $\omega$.
\end{Prop}
}
\begin{pf}
  We show that $j^{aA}$ is well defined and fulfils the following
  four Wightman axioms:
\begin{enumerate}
\item For all $f\in \Gamma_0(D^*M \times T^*M)$, $j^{a A}$ is a well
  defined operator valued distribution on the GNS Hilbertspace
  $(H,\Omega,A,B,\Psi)$
  with domain
  \begin{equation*}
    \begin{split}
      D & = \mbox{Span} \{ \Phi(f_1) \cdots \Phi(f_n) \Omega |\\
        & \phantom{=\mbox{Span} \{ }
        \Phi \in (A,B,\Psi,j^{a A}),
        f_1,\ldots,f_n \in (\CinfO(M),\Gamma_0(D^*M),\Gamma_0(D^*M
        \otimes T^*M))
        \}
      \end{split}
    \end{equation*}
  By $\Gamma_0(\cdot)$ we denote sections with compact support.
\item $D$ is dense in $H$.
\item $j^{a A}$ leaves $D$ invariant.
\item $j^{a A}$ is anti-local (e.g.\ fermionic).
\end{enumerate}
To simplify the notation, these four Wightman axioms are actually shown
for the $k^{aA}$ part of $j^{aA}$ only. However the proof below is
obviously equally valid for $j^{aA}$ itself.
\begin{enumerate}
\item Let $D'= \mbox{Span} \{ \Psi = k(f_1) \cdots k(f_n) \Omega | f_i
  \in \Gamma_0(D^*M \otimes T^*M) \}$. We remark that the definition
  of $D'$ is only formal at this stage. To show that $k^{a A}$ is well
  defined on $D'$, it is sufficient to show that
  \begin{alignat}{2}
    \left\| k(f) \Psi \right\| & < \infty && \qquad \forall f \in
      \Gamma_0(D^*M \times T^*M), \quad \Psi = k(f_1) \cdots k(f_n)
      \Omega \in D' \label{norm}
  \end{alignat}
  Choose a moving
  frame in the spinor bundle ($E_A$) which induces moving frames in the
  cospinor bundle ($E^A$), in the tangent bundle ($e_\mu$) and in the
  cotangent bundle ($e^\mu$).
  Inserting the definitions of $k$ and $\Psi$ in Equation~(\ref{norm})
  we obtain in these local coordinates:
  \begin{equation}  \label{eq:normkf}
    \begin{split}
      & {\| k(f) \Psi \|}^2\\
      & = < k(f)k(f_1)\cdots k(f_n) \Omega | k(f)k(f_1)\cdots k(f_n) \Omega>\
 \\
      & = <\Omega |
        k^+(f^+_n)\cdots k^+(f^+_1)k^+(f^+)k(f)k(f_1)\cdots
        k(f_n)\Omega> \\
      & = \idotsint d\mu_{x_n} \cdots d\mu_{x_1} d\mu_x d\mu_y
                  d\mu_{y_1} \cdots d\mu_{y_n} \\
      & \phantom{=}
        {{\sideset{^\Psi}{_{2n+2}}{\omega}}_{A_n\cdots A_1
          A}}^{A'A'_1\cdots A'_n}
           (x_n,\ldots,x_1,x,y,y_1,\ldots,y_n)
        {{\gamma^{a_n}}^{A_n}}_{B_n} \cdots
        {{\gamma^{a_1}}^{A_1}}_{B_1}
        {{\gamma^{a}}^{A}}_{B}\\
      & \phantom{=}
        \biggl(
        ( -i \rvec{\underset{x_n}{\notnabla}} -m {)^{B_n}}_{C_n}
        \cdots
        ( -i \rvec{\underset{x}{\notnabla}} -m {)^{B}}_{C}
        ( i \rvec{\underset{y}{\notnabla}} -m {)^{C'}}_{B'}
        \cdots
        ( i \rvec{\underset{y_n}{\notnabla}} -m {)^{C'_n}}_{B'_n}\\
      & \phantom{\biggl( (-i\rvec{\underset{x_n}{\notnabla}}
                 -m {)^{B_n}}_{C_n}}
        \sideset{^A}{_{2n+2}}{\omega}
              (x_n,\ldots,x_1,x,y,y_1,\ldots,y_n)
        \\
      & \phantom{= \biggl(}
        -
        {\bigl( ( -i \rvec{\underset{x_n}{\notnabla}} -m )\gamma^5
         \bigr)^{B_n}}_{C_n}
       \cdots
       {\bigl(( -i \rvec{\underset{x}{\notnabla}} -m )\gamma^5
        \bigr)^{B}}_{C}
       {\bigl(\gamma^5
         (i\rvec{\underset{y}{\notnabla}}-m)
       \bigr)^{B'}}_{C'}
       \cdots
       {\bigl(\gamma^5
         (i\rvec{\underset{y_n}{\notnabla}}-m)
       \bigr)^{B'_n}}_{C'_n}
      \\
      &  \phantom{= \biggl(
                  -
                  {\bigl( ( -i \rvec{\underset{x_n}{\notnabla}} -m )\gamma^5
                  \bigr)^{B_n}}_{C_n}}
        \sideset{^B}{_{2n+2}}{\omega}
                (x_n,\ldots,x_1,x,y,y_1,\ldots,y_n)
        \biggr)\\
      & \phantom{=}
        {{\gamma^{a'}}^{B'}}_{A'}
        {{\gamma^{a'_1}}^{B'_1}}_{A'_1} \cdots
        {{\gamma^{a'_n}}^{B'_n}}_{A'_n}
        {f^+_{a_n}}^{C_n}(x_n) \cdots {f^+_{a_1}}^{C_1}(x_1)
        {f^+_{a}}^{C}(x)
        {f_{a' C'}}(y) {f_{a'_1 C'_1}}(y_1) \cdots
        {f_{a'_n C'_n}}(y_n)\\
    \end{split}
  \end{equation}
  The cross $\phantom{f}^+$ denotes Dirac conjugation and the equation
  separates since we are dealing with a product
  state. $\sideset{^\Phi}{_{2n+2}}{\omega}$ denotes the n-point
  distribution of the basic field $\Phi$ and the r.h.s\ of
  Eqn.~(\ref{eq:normkf}) {\em defines} $\| k(f) \Psi \|^2$.
  All n-point distributions above
  arise from quasifree states. Hence they decay into the sum of products
  of two-point distributions. Using
  Corollary~\ref{Cor:scalar_power} and~\ref{Cor:fermi_power} one sees
  that arbitrary (finite) products of these distributions with each
  other exist. This ensures the finiteness of Eqn.~(\ref{eq:normkf}). Now
  suppose we replace one or more $k(f_i)$ in Eqn.~(\ref{eq:normkf}) by
  $\Phi(g_1,\ldots,g_k,\tilde{g}_1,\ldots,\tilde{g_l},h_1,\ldots,h_m)
  \equiv A(g_1) \cdots A(g_k) B(\tilde{g}_1)  \cdots B(\tilde{g}_l) \Psi(h_1)
  \cdots \Psi(h_m) $ with $g_1,\ldots,\tilde{g_l} \in \CinfO (M)$ and $h_1,
  \ldots h_m \in \Gamma_0(D^*M )$. Redoing the calculation
  in Eqn.~(\ref{eq:normkf}) one ends up again with various products of
  two-point distributions. However all two-point distributions which
  include parts of the ``$\Phi$'s'' are in fact smooth functions (for
  example $\sideset{^A}{_2}{\omega}(g_i,x) \in \Cinf(M);
  \sideset{^A}{_2}{\omega}(g_i,g_j) \in {\mathbb C} $). Thus replacing an
  operator $k(f_i)$ by $\Phi(g_1,\ldots,h_m)$ does not change the
  finiteness of the whole equation. Moreover since the products in
  Eqn.~(\ref{eq:normkf}) are again {\em distributions}, it follows
  that $k_{\Psi\Psi'} \colon \Gamma(D^*M \otimes T^*M) \ni f \mapsto
  <\Psi | k(f) \Psi'>$ is well defined and continuous for all $\Psi,
  \Psi'\in D$. This proves the assertion. Note that the relation
  between $k$ and its adjoint $k^*$, i.e. $k^*(f) = k^+(f^+)$, follows
  by Corollary~\ref{Cor:restrApprox} from the corresponding properties
  of the basic fields.
\item is obvious since ${\EuScript H}$ is the GNS Hilbertspace of the
  basic fields,i.e., the ``vacuum'' $\Omega$ is cyclic for $A,B$ and $\Psi$.
  \item follows directly from our definition of $D$.
\item Note that $A,B$ and $\Psi$ mutually commute,
  \[ [A(f),A(g)]_- = [B(f),B(g)]_- = 0 \qquad
                     \text{if $\supp(f) \subset \supp(g)^c$},
  \]
  whereas two $\Psi$'s anti-commute for causally disjoint arguments.
  $\supp(g)^c$ denotes the causal complement of the support of $g$. Therefore
  \[ \{ k(f),k(g)\}_+ = 0 \qquad
                    \text{if $\supp(f) \subset \supp(g)^c$},
  \]
\end{enumerate}
\end{pf}
\begin{rem}
  $\phantom{~}{}$\par{}
  \begin{itemize}
  \item $D'$ is not dense in $H$, since for example $\Psi = A(f)
    |\Omega>$ can not be approximated by elements in $D'$: $\forall
    \Phi\in D'$ we have $\| \Psi - \Phi \|^2 = \| \Psi\|^2 + \|\Phi\|^2$
    since $<\Psi,\Phi> = 0$.)
  \item $|\Omega>$ is not cyclic for the fields $j^{aA}$.
  \end{itemize}
\end{rem}
It is worth noting that all n-point distributions of $k^{aA}$ do
satisfy our new GWFSSC in a non trivial way. The proof of this
statement which is completely analogous to the considerations of
Lemma~\ref{Lemma:GWFSSCKleinGord} is omitted.
\begin{Cor}\label{Cor:wfsupercurrent}
  Let a spacelike hypersurface ${\mathcal C}$ and $f,g \in \Gamma(D^*M
  \otimes T^*M)$ be given. It follows that the wave front set of the
  distribution $(f,g) \mapsto <\Omega | j^+(f^+) j(g) \Omega>$ does
  contain directions normal to ${\mathcal C}\times {\mathcal C}$, i.e., it
  violates the assumption of Theorem~\ref{thm:hoermander2.5.11}.
\end{Cor}
\begin{pf}
  Consider $< \Omega | j^+(g^+) j(f) \Omega >$. By Eqn.~(\ref{eq:normkf})\
 this
  expression contains sums of the product of two 2-point
  distributions, hence the wave front set of this distribution is
  contained (and in fact equals) the set
  \begin{equation*}
    \begin{split}
      W & \equiv
      (\WF(\omega^2) \oplus \WF(\omega^2) ) \cup \WF(\omega^2)\}.
    \end{split}
  \end{equation*}
  However, due to the fact that in the first term the wave front sets
  are added, $W$ contains --associated to the first entrance--
  directions in the whole forward lightcone, that is it contains all
  timelike direction. On the other hand the normals of ${\mathcal C}$
  are timelike by definition, which proves the Corollary.
\end{pf}
\begin{rem}
  Unfortunately the assumptions of Theorem~\ref{thm:hoermander2.5.11}
  are only sufficient for the existence of a restriction. Therefore we
  may not use the results of Corollary~\ref{Cor:wfsupercurrent} to
  conclude that the restriction of $j^{aA}$ to a spacelike
  hypersurface does not exist. On the other hand it is well known that
  on Minkowski space the product of free fields can not be restricted
  to spacelike hypersurfaces. Thus one might conjecture that a similar
  statement holds on a manifold.
\end{rem}
%
%%%%%%%%%
\section{The `supercharge'}
\label{sec:superCharCurv}
%%%%%%%%%
We saw above that we can not apply Theorem~\ref{thm:hoermander2.5.11}
to restrict the `supercurrent' to a spacelike hypersurface. However,
we will show below that the {\em (anti) commutator\/} of the
`supercurrent' with a basic field gives an operator valued
distribution which can be restricted to any spacelike hypersurface
${\mathcal C}$. Hence it is possible to define terms like $ {[{Q}^+_A,A(x)
  ]}_- \restriction_{\mathcal C}$ rigorously. However, for this definition
we need some more notation which is related to the splitting of the
spacetime into space and time (See~\cite[pp.\ 255]{wald:gr}).
%
%%%%%%%%%%%%
\subsection{Splitting spacetime into space and time}
\label{sec:splittingSpacetime}
%%%%%%%%%%%%
Let $(M,g_{ab})$ be a Lorentz manifold and let ${\mathcal C}$ denote a
smooth spacelike hypersurface in $M$ with unit normal covector field
$\tilde{n}_a$. Then there exists a covector field $n_a$ near ${\mathcal
  C}$ on $M$, such that $n_a$ restricted to ${\mathcal C}$ is
$\tilde{n}_a$ {\em and\/} such that the integral curves of
$n^a=g^{ab}n_b$ are (timelike) geodesics.
\begin{Dfn}\label{Dfn:inducedMetric}
  {\sloppy
  The spacetime metric $g_{ab}$ induces a spatial,i.e., a three
  dimensional, Rie\-mannian\footnote{Note that the signature of $h_{ab}$
    is $(-,-,-)$.} metric on ${\mathcal C}$ via the formula
  \[
  h_{ab} := g_{ab} - n_a n_b.
  \]
  }
\end{Dfn}
The Levi-Civita connection $D_a$ associated to this spatial metric is
given by
\begin{Lemma}[Lemma 10.2.1 of~\cite{wald:gr}]\label{Lemma:inducedLeviCiv}
  The Levi-Civita connection of $h_{ab}$ is given by
  \begin{equation}\label{eq:inducedLeviCiv}
  D_c {T^{a_1 \cdots a_l}}_{b_1 \cdots b_m} = {h^{a_1}}_{d_1} \cdots
                 {h_{b_m}}^{e_m} {h_c}^f \nabla_f {T^{d_1\cdots
                     d_l}}_{e_1\cdots e_m},
  \end{equation}
  where $\nabla_a$ is the Levi-Civita connection of $g_{ab}$.
\end{Lemma}
\begin{Dfn}
  The {\em extrinsic curvature\/} of ${\mathcal C}$ in $M$, $K_{ab}$, is
  defined by
  \[
  K_{ab}= {h_a}^c \nabla_c n_b.
  \]
\end{Dfn}
Note that $K_{ab}$ is {\em symmetric\/} in in its indices due to the
fact that it is half the Lie derivative of $h_{ab}$ with respect to
the vector field $n$ (See~\cite{wald:gr}).  Moreover, since we assumed
$n_a$ to be a geodesic covector field, we have
\begin{equation}
  \label{eq:Kab}
  K_{ab} = \nabla_a n_b \qquad\text{if $n_a$ is geodesic},
\end{equation}
since $n_a n^c \nabla_c n_b$ vanishes by assumption. Another useful
identity is $\nabla_a(v^a) = D_a v^a + n^a n_b \nabla_a v^b$ which is
valid for all vector-fields tangential to ${\mathcal C}$.
\begin{Dfn}
  The trace of $K_{ab}$, ${K^a}_a$, is denoted by $K$ --the extrinsic
  curvature scalar-- and a hypersurface with $K=0$ is called {\em maximal}.
\end{Dfn}
\begin{rem}
  $\phantom{~}$ \par
  \begin{itemize}
  \item It is not known whether an arbitrary spacetime admits maximal
    hypersurfaces {\em globally}, since there are various constraints.
    For example the extrinsic curvature and the full Ricci tensor are
    connected by the Gauss-Codazzi relations. For a special class of
    spacetimes this difficult problem is investigated
    in~\cite{chrusandWald:94}.  Fortunately the situation in this work
    is far less complicated, since we need maximal spacelike
    hypersurfaces {\em locally\/} only, i.e., for every point $p$ in
    the spacetime there should exist a neighborhood $U$ and a
    spacelike hypersurface ${\mathcal C}$ such that $U \cap {\mathcal C}$ has
    extrinsic curvature zero.  Using the following line of argument
    together with a result of Bartnik~\cite{bart:84}, it should be
    possible to show that {\em all\/} spacetimes admit maximal
    hypersurfaces locally through every point.
    Bartnik shows in~\cite{bart:84} (Theorem~5.4) that all
    asymptotically flat spacetimes satisfying an interior condition
    possess maximal spacelike hypersurfaces globally. It is worth
    noting that his interior condition demands certain quantities
    which vanish on Minkowski space (equipped with the usual time
    function) to be uniformly bounded throughout the spacetime, hence
    a spacetime which differ only slightly from Minkowski space, e.g.,
    on a compact set, does satisfy his interior condition
    automatically. Consider now a point $p$ in our spacetime with
    three sufficiently small open neighborhoods $U$, $U'$ and $U''$,
    such that the closure $\bar{U}$ of $U$ is contained in $U'$,
    $\bar{U'} \subset U''$ and $U''$ can be covered by a single chart.
    It follows that there exists a deformed spacetime $(M',g'_{ab})$
    together with a global time function, such that $M'$ is contained
    in $U''\subset M$, $g'$ equals $g$ on $U$ and $g'$ is flat outside
    of $U'$. Note that $(M',g'_{ab})$ can be constructed using a chart
    $\varrho$ to pull back the metric $g$ to ${\mathbb R}^4$. Then one
    deforms this metric in the region $\varrho(U' \setminus \bar{U}$
    to the Minkowski metric.  The time function $t$ is chosen, such
    that it is the usual time Minkowski time function outside of
    $\varrho(U'')$ and some arbitrary time function inside
    $\varrho(U)$. We interpolate $t$ between $\varrho(U)$ and
    $\varrho(U')$ such that $\nabla_a t$ is nonzero and everywhere
    timelike. Finally, using a conformal mapping, which does not
    change the metric inside $\varrho(U)$, the resulting spacetime is
    mapped into $\varrho(U'')$.  Then $(M',g'_{ab})$ together with $t$
    satisfy the interior condition of Bartnik's Theorem~5.4, which
    ensures that a maximal spacelike hypersurface ${\mathcal C}$ in
    $(M',g'_{ab}) \stackrel{i}{\hookrightarrow} (M,g_{ab})$ passing
    through $p$ exists. Thus $U \cap {\mathcal C}$ is a spacelike
    hypersurface with extrinsic curvature zero in $(M,g_{ab})$.
  \item It is worth noting that for a geodesic covector field $n_a$,
    which is in addition a Killing covector field, $K_{ab}= 0$
    automatically.
  \end{itemize}
\end{rem}
%
%%%%%%%%%%%%
\subsection{\mbox{(Anti-)} commutators for the `supercharge'}
\label{sec:anticommSupercharge}
%%%%%%%%%%%%
%
In this subsection we will explicitly calculate the \mbox{(anti-)}
commutators of the `supercharge' with the basic fields and their
derivatives. The former is obtained by smearing the distributional
`supercurrent' with a suitable testdensity. We remind the reader that
we define distributions according to the notation of
H\"ormander~loc.cit., i.e., distributions are extensions of smooth
functions. It follows that the former correspond to the continuous
linear forms on the smooth {\em densities\/} with compact support
(See~\cite[p.~145]{hoermander:analysisI}). It is worth noting that the
latter can be identified on every spacetime with smooth functions with
compact support by using the natural volume element on this Lorentz
manifold.
\begin{Dfn}\label{Dfn:regSupercharge}
  Let $k^{a A}$ denote the `supercurrent', ${\mathcal C}$ a spacelike
  hypersurface, $n^a \in \Gamma(TM)$ a unit timelike geodesic vector
  field which, when restricted to ${\mathcal C}$ is normal to this
  surface.  Fix a point $x_0 \in {\mathcal C} \subset M$ and let $\chi$ be
  a smooth function with compact support in a normal neighborhood of
  $x_0$, which can be identified with the density $\chi d\mu$, where
  $d\mu$ denotes the natural volume element on $M$.  The {\em regularized
  `supercharge'} $\sideset{^{\chi d\mu }}{^{A'}}Q (x_0)$ is defined by
  \begin{equation}
    \sideset{^{\chi d\mu}}{^{A'}}Q (x_0) := \int_M k^{a A}(x)
                        {{\mathcal J}_{A}}^{A'}(x,x_0) n_a (x) \chi(x)
                           d\mu_x.
    \label{reg_supercharge}
  \end{equation}
\end{Dfn}
Note that $\sideset{^{\chi d\mu }}{^{A'}}Q (x_0)$ as well as it's Dirac
adjoint $\sideset{^{\chi d\mu }}{^+_{A'}}{{Q}}(x_0)$ are well defined
operators on the GNS Hilbertspace which transform as a spinor
(cospinors respectively) at the point $x_0$.
\begin{Prop}\label{prop:commutQCurv}
  Let ${\mathcal C}$, $n^a$ and $x_0$ be the same as in
  Definition~\ref{Dfn:regSupercharge}. Then the six operator valued
  distributions
  \begin{align*}
     D'(M \times M) \ni \sideset{^A}{_{x_0}}V \colon (\chi d\mu ,h )
     & \mapsto {[\sideset{^{\chi d\mu}}{^+_{A'}}{{Q}}(x_0), A (h)]}_-\\
     D'(M \times M) \ni \sideset{^B}{_{x_0}}V \colon (\chi d\mu ,h )
     & \mapsto {[\sideset{^{\chi d\mu}}{^+_{A'}}{{Q}}(x_0), B (h)]}_-\\
     D'(M \times DM) \ni \sideset{^\Psi}{_{x_0}}V \colon (\chi d\mu ,h_B )
     & \mapsto {\{ \Psi^B(h_B) , \sideset{^{\chi d\mu
        f}}{^+_{A'}}{{Q}}(x_0) \}}_+\\
     D'(M \times T^*M) \ni \nabla_a\sideset{^A}{_{x_0}}{V} \colon
          (\chi d\mu ,g^a )
     & \mapsto {[\sideset{^{\chi d\mu}}{^+_{A'}}{{Q}}(x_0),
              \nabla_a A (g^a)]}_-\\
     D'(M \times T^*M) \ni \nabla_a\sideset{^B}{_{x_0}}{V} \colon
          (\chi d\mu ,g^a )
     & \mapsto {[\sideset{^{\chi d\mu}}{^+_{A'}}{{Q}}(x_0),
              \nabla_a B (g^a)]}_-\\
     D'(M \times DM \otimes T^*M ) \ni
            \nabla_a\sideset{^\Psi}{_{x_0}}{V} \colon (\chi d\mu ,g^a_B )
     & \mapsto {\{ \nabla_a \Psi^B(g^a_B) ,
              \sideset{^{\chi d\mu}}{^+_{A'}}{{Q}}(x_0) \}}_+
 \end{align*}
 have unique restrictions to the spacelike hypersurface ${\mathcal C}$.
\end{Prop}
\begin{pf}
  Let $(H,\Omega,A,B,\Psi)$ denote the GNS-Hilbertspace of a globally
  Hadamard product-state $\omega$ and let $D$ be the usual dense
  invariant domain for the three fields.  We prove first of all that
  the distribution
  \[
  < \sideset{^A}{_{x_0}} V (\cdot,\cdot) \Omega | \sideset{^A}{_{x_0}} V
  (\cdot,\cdot) \Omega > \in D'(M^4)
  \]
  on $M^4$ can be restricted to ${\mathcal C}^4$.  Inserting the
  definitions we obtain:
  \begin{equation}\label{eq:normV}
    \begin{split}{}
    &   {\| \sideset{^A}{_{x_0}} V (\chi d\mu ,hd\mu) \Omega \|}^2\\
    & = \int_{M^4} d\mu_x d\mu_y d\mu_z d\mu_v\\
    & \phantom{= \int{M^4}}
      <\Omega |
      -i{[{{k^+}^a}_A(x)
        {{\mathcal J }^A}_Z(x,x_0)n_a(x)\chi (x), A(y)]}_- \cdot \\
    & \phantom{= \int{M^4}}
      i{[{k^{a''}}^{B''}(z){{\mathcal J}_{B''}}^Y(z,x_0) n_{a''}(z) \chi
        (z), A(v) ]}_- h(v)
      \Omega >\\
    & = \int_{M^4} d\mu_x d\mu_y d\mu_z d\mu_v
    <\Omega | {\Psi^+}_A(x) \Psi^{A''}(z) \Omega >
    {{\gamma^a}^A}_B \\
    & \phantom{= \int_{M^4}}
    \Biggl(
   {{\left( -i \underset{x}{\notnabla} - m \right)}^B}_C
        <\Omega|{[A(x),A(y)]}_-\Omega>
   {{\left( i \underset{z}{\notnabla} -m \right)}^{B''}}_{C''}
        <\Omega|{[A(z),A(v)]}_-\Omega>\\
    & \phantom{= \int_{M^4}}
    -  {{\left((-i \underset{x}{\notnabla} -m)\gamma^5\right)}^B}_C
    <\Omega|{[B(x),B(y)]}_-\Omega>\\
    & \phantom{= \int_{M^4}\Biggl(}
    {{\left( (i \underset{z}{\notnabla} -m)\gamma^5\right)}^{B''}}_{C''}
    <\Omega|{[B(z),B(v)]}_-\Omega>
    \Biggr)\\
    & \phantom{= \int_{M^4}}
    {{\mathcal J}^C}_Z(x,x_0) n_a \chi(x) {{\gamma^{a''}}^{C''}}_{A''}
    {{\mathcal J}_{B''}}^Y(z,x_0) n_{a''}(z) \chi(v) \\
    &= \int_{M^4} d\mu_x d\mu_y d\mu_z d\mu_v
    (-1)  {{\sideset{^\Psi}{^2}{\omega}}_A}^{A'}(x,z) {{\gamma^a}^{A}}_{B} \\
    & \phantom{= \int_{M^4} }
   \Biggl(
   {{\left( -i \underset{x}{\notnabla} - m \right)}^B}_C
        E(x,y)
   {{\left( i \underset{z}{\notnabla} -m \right)}^{B''}}_{C''}
        E(z,v)\\
    &\phantom{= \int_{M^4} \Biggl( }
    -  {{\left((-i\underset{x}{\notnabla}-m)\gamma^5\right)}^B}_C
    E(x,y)
    {{\left( (i\underset{z}{\notnabla}-m)\gamma^5\right)}^{B''}}_{C''}
    E (z,v)
    \Biggr) \\
    & \phantom{= \int_{M^4} }
    {{\mathcal J}^C}_Z(x,x_0) n_a \chi(x) {{\gamma^{a''}}^{C''}}_{A''}
    {{\mathcal J}_{B''}}^Y(z,x_0) n_{a''}(z) \chi(v)
  \end{split}
\end{equation}
We used the fact that $\Omega$ is a quasifree product state and
introduced the following  abbreviations for the vacuum expectation
values:
\begin{align*}
  i E(x,y) & := <\Omega|{[A(x),A(y)]}_-\Omega> \\
  i E(x,y) & := <\Omega|{[B(x),B(y)]}_-\Omega> \\
  {{\sideset{^\Psi}{^2}{\omega}}_A}^{A'}(x,y)
           & :=  <\Omega|\Psi^+_A(x) \Psi^{A'}(y) \Omega>
\end{align*}
Note that $ < \Omega | [ A(x),A(y)]_- [A(z),A(v)]_- \Omega > =
   i E(x,y) i E (z,v)$, since the we are dealing with free fields.
For our claim it is sufficient to show that the following product of
distributions
\[ \tilde{V} : \equiv
  ( E(y,x) \otimes \identity \otimes \identity ) \cdot
   ( \identity \otimes {{\sideset{^\Psi}{^2}{\omega}}^A}_{A''}(x,z)
   \otimes \identity ) \cdot
   ( \identity \otimes \identity \otimes E(z,v) )
\]
can be restricted to ${\mathcal C}$, since (\ref{eq:normV}) consists of
terms of this type together with derivatives of it, multiplied by
smooth functions, only.  For the verification of this statement we
calculate the wave front set of the distribution $\tilde{V}$.  Note
that for a Distribution $u \in D'(M)$ with wave front set $\WF(u)$ the
wave front set of $\identity \otimes u$ is contained in the set
$(M,0)\times \WF (u)$ (See for example~\cite{Hoermander:71} p.\ 130)
and recall that differentiation does not enlarge the wave front set.
For the wave front sets of $E$, which is the difference of the
advanced and retarded fundamental solutions of the Klein Gordon
equation (see below) one finds (see for example~\cite{Hoermander:71}):
\begin{align}
\WF(E) & =
    \{ (x,\xi_1,y,\xi_2) \in T^*M \times T^*M
         \backslash \{0\} | (x,\xi_1) \sim (y,\xi_2) ; {\xi_1}^2 = 0
         \},
         \label{eq:wavefrontscalar}\\
\intertext{whereas for the wave front set of
  ${{\sideset{^\Psi}{^2}{\omega}}^A}_{A'}$ we already found in
  Corollary~\ref{Cor:wfhadamfermi}}
\WF({{\sideset{^\Psi}{^2}{\omega}}^A}_{A'}) &  \subseteq
    \{ (x,\xi_1,y,\xi_2) \in T^*M \times T^*M
         \backslash \{0\} | (x,\xi_1) \sim (y,\xi_2) ; {\xi_1}^2 = 0 ;
         \xi_1^0 > 0 \}\label{eq:wavefrontfermi}.
\end{align}
By Theorem~\ref{Thm:IX.54} the wavefront set of $\tilde{V}$ and all
its derivatives is contained in the following set:
\begin{equation*}
  \begin{split}
  \WF(\tilde{V})
  & \subseteq S \\
  & : =
    \bigl(
     \underbrace{\WF(E \otimes \identity \otimes \identity)}_{=:\Gamma_1}
   + \underbrace{\WF(\identity \otimes E \otimes \identity)}_{=:\Gamma_2}
   + \underbrace{\WF(\identity \otimes \identity \otimes E)}_{=:\Gamma_3}
    \bigr) \\
  & \phantom{\subseteq}
  \cup \Gamma_1 \cup \Gamma_2 \cup \Gamma_3 \cup
   (\Gamma_1 \oplus \Gamma_2) \cup (\Gamma_1 \oplus \Gamma_3)
   \cup (\Gamma_2 \oplus \Gamma_3)
   \\
  & \subseteq
  \{ (y,\xi_1,x,-\xi_1',z,0,v,0) \in {T^*M}^4 \backslash \{0\} |
     (y,\xi_1) \sim (x,\xi_1') ; {\xi_1}^2=0 \} \\
  & \phantom{\subseteq}
  \bigcup
  \{ (y,0,x,\xi_2,z,-\xi_2',v,0) \in {T^*M}^4 \backslash \{0\} |
  (x,\xi_2) \sim (z,\xi_2') ; {\xi_2}^2=0 ; \xi_2^0 > 0 \}\\
  & \phantom{\subseteq}
  \bigcup
  \{ (y,0,x,0,z,\xi_3,v,-\xi_3') \in {T^*M}^4 \backslash \{0\} |
  (z,\xi_3) \sim (v,\xi_3') ; {\xi_1}^2=0 \} \\
  & \phantom{\subseteq}
  \bigcup
  \{(y,\xi_1,x,\xi_2-\xi_1',z,-\xi_2',v,0) \in {T^*M}^4 \backslash\{0\} |
  (y,\xi_1) \sim (x,\xi_1');(x,\xi_2) \sim (z,\xi_2');\\
  & \phantom{\subseteq \bigcup \{(y,\xi_1,x,\xi_2-\xi_1',z,-\xi_2',v,0)}
    {\xi_1}^2={\xi_2}^2=0; \xi_2^0 > 0 \} \\
  & \phantom{\subseteq}
  \bigcup
  \{(y,\xi_1,x,-\xi_1',z,\xi_3',v,-\xi_3) \in {T^*M}^4 \backslash\{0\} |
  (y,\xi_1) \sim (x,\xi_1');(z,\xi_3) \sim (v,\xi_3');\\
  & \phantom{\subseteq \bigcup \{(y,\xi_1,x,\xi_2-\xi_1',z,-\xi_2',v,0)}
  {\xi_1}^2={\xi_3}^2=0 \} \\
  & \phantom{\subseteq}
  \bigcup
  \{(y,0,x,\xi_2,z,\xi_3-\xi_2',v,-\xi_3') \in {T^*M}^4 \backslash\{0\} |
  (x,\xi_2) \sim (z,\xi_2');(z,\xi_3) \sim (v,\xi_3');\\
  & \phantom{\subseteq \bigcup \{(y,\xi_1,x,\xi_2-\xi_1',z,-\xi_2',v,0)}
     {\xi_2}^2={\xi_3}^2=0; \xi_2^0 > 0 \} \\
& \phantom{\subseteq}
  \bigcup
  \{(y,\xi_1,x,\xi_2-\xi_1',z,\xi_3-\xi_2',v,-\xi_3')
            \in {T^*M}^4 \backslash\{0\} |\\
& \phantom{\subseteq \bigcup \{ }
   (y,\xi_1) \sim (x,\xi_1');(x,\xi_2) \sim (z,\xi_2');
    (z,\xi_3) \sim (v,\xi_3');\\
& \phantom{\subseteq \bigcup \{ }
    {\xi_1}^2={\xi_2}^2= {\xi_3}^2=0; \xi_2^0 > 0 \},
  \end{split}
\end{equation*}
where we used the fact that the wave front set of
${{\sideset{^\Psi}{^2}{\omega}}^A}_{A'}$ is contained in that of $E$.
Now let $\Tilde{\Tilde{\varphi}}$ denote the natural embedding of
${\mathcal C}^4$ into $M^4$ and $N_{\Tilde{\Tilde{\varphi}}}$ the
associated set of normals of $\Tilde{\Tilde{\varphi}}$. Then the two
sets $S$ and $N_{\Tilde{\Tilde{\varphi}}}$ do not intersect, hence the
restriction of $\tilde{V}$ as well as that of $\|\sideset{^A}{_{x_0}}V
( \cdot,\cdot) \Omega \|$ to ${\mathcal C}^4$ exists by
Theorem~\ref{thm:hoermander2.5.11}.  Now replace the vector $\Omega$
in Eqn.~(\ref{eq:normV}) by an arbitrary element $\Phi\in D\subset H$.
Since $\Omega$ is quasifree one ends up with a sum of terms which are
tensor products of $\tilde{V}$ and its derivatives with two-point
distributions of the basic fields or fundamental solutions.  Since the
latter can be restricted\footnote{Recall that their wave front sets
  contain lightlike singular directions only.} to ${\mathcal C}\times
{\mathcal C}$, it follows that the tensor products mentioned above can be
restricted to ${\mathcal C}\times{\mathcal C}$, too. Especially the
restriction to ${\mathcal C}$ for the internal variables, i.e., the
variables associated to $\sideset{^A}{_{x_0}}V$, exists, which proves
the claim for the first operator valued distribution
$\sideset{^A}{_{x_0}}V$. The five remaining cases follow completely
analogous. The corresponding proofs are omitted in order to avoid
repeating the argument verbatim.
\end{pf}
We have the following Theorem.
\begin{Thm}\label{Thm:commutQCurvedCx}
  Using the same notation as in Proposition~\ref{prop:commutQCurv}:
  For all smooth testdensities on ${\mathcal C}$ with support sufficiently
  close to $x_0$, the resulting six distributions on ${\mathcal
    C}\times{\mathcal C}$ are independent of the density $\bfm{\chi}
  d{\mathcal C}$, provided $\bfm{\chi}\equiv 1$ in a neighborhood of the
  support of the second testdensity. $d{\mathcal C}$ denotes the induced
  volume form on ${\mathcal C}$. More precisely: Let $\tilde{\varphi}$ and
  $\varphi$ denote the embedding of ${\mathcal C} \times {\mathcal C}$ and
  ${\mathcal C}$ into $M \times M$ and $M$ respectively.  Then
{\small
   \begin{align*}\allowdisplaybreaks{}
     D'({\mathcal C}) \ni \tilde{\varphi}^* (\sideset{^A}{_{x_0}}{V})
     (\bfm{\chi} d{\mathcal C},\cdot) \colon
       \bfm{h} \mapsto &
            {[ \sideset{^{\mathcal C}}{^+_{A''}}{{G}}(x_0), A (\bfm{h})
              ]}_-
            \\
     & := {[ \sideset{^{\bfm{\chi} d{\mathcal C}}}{^+_{A''}}{{Q}}(x_0),
            A (\bfm{h}) ]}_-\\
     & \phantom{:}
        =
     \varphi^*\biggl(
     + i {\Psi}^+_{A} {{\mathcal J}^{A}}_{A''}(x_0)
     \biggr)(\bfm{h})\\
     D'({\mathcal C}) \ni \tilde{\varphi}^* (\sideset{^B}{_{x_0}}{V})
     (\bfm{\chi} d{\mathcal C}, \cdot) \colon
     \bfm{h} \mapsto &
     {[ \sideset{^{\mathcal C}}{^+_{A''}}{{G}}(x_0), B (\bfm{h}) ]}_-\\
     & := {[ \sideset{^{\bfm{\chi} d{\mathcal C}}}{^+_{A''}}{{Q}}(x_0),
            B (\bfm{h}) ]}_-\\
     & \phantom{:}
        =
     \varphi^*\biggl(
     + {({\Psi}^+\gamma^5)}_{A} {{\mathcal J}^{A}}_{A''}(x_0)
     \biggr) (\bfm{h})\\
     D'(D_{\mathcal C}M) \ni \tilde{\varphi}^*
     (\sideset{^\Psi}{_{x_0}}{V}) (\bfm{\chi} d{\mathcal C},\cdot)
     \colon
     \bfm{h}_B \mapsto &
     {\{ \Psi^B(\bfm{h}_B), \sideset{^{\mathcal C}}{^+_{A''}}{{G}}(x_0)\
 \}}_+\\
     & := {\{\Psi^B(h_B),
           \sideset{^{\bfm{\chi} d{\mathcal C}}}{^+_{A''}}{{Q}}(x_0)\}}_+\\
     & \phantom{:}
        =
     \varphi^*\biggl(
     + i \biggl( {\bigl( ( i {\notnabla} + m )
           \AiB \bigr)} {\mathcal J}(x_0){\biggr)^{B}}_{A''}
         \biggr) (\bfm{h}_B)\\
     D'(T^*{\mathcal C}) \ni \tilde{\varphi}^* (\nabla_a
     \sideset{^A}{_{x_0}}{V})
     (\bfm{\chi} d{\mathcal C},\cdot)
     \colon
       \bfm{g}^a \mapsto &
       {[\sideset{^{\mathcal C}}{^+_{A''}}{{G}}(x_0),\nabla_a A (g^a)]}_-\\
       & := {[ \sideset{^{\bfm{\chi} d{\mathcal C}}}{^+_{A''}}{{Q}}(x_0),
         \nabla_a A (\bfm{g}^a)]}_-\\
       & \phantom{:}
       =
       \varphi^*\biggl(
       i D_a {{( {\Psi}^+ {\mathcal J} )}}_{A''}
       + i  n_a D_b {( {\Psi}^+ \notn \gamma^b
         {\mathcal J} )}_{A''}\\
       & \phantom{:= \varphi^*\biggl(}
           - m  n_a {({\Psi}^+ \notn {\mathcal J})}_{A''}
           \biggr) (\bfm{g}^a)
       \\
     D'(T^*{\mathcal C}) \ni \tilde{\varphi}^*
       (\nabla_a \sideset{^B}{_{x_0}}{V})
       (\bfm{\chi} d{\mathcal C},\cdot)
       \colon
     \bfm{g}^a \mapsto &
       {[\sideset{^{\mathcal C}}{^+_{A''}}{{G}}(x_0),\nabla_a B(g^a) ]}_- \\
       & := {[ \sideset{^{\bfm{\chi} d{\mathcal C}}}{^+_{A''}}{{Q}}(x_0),
         \nabla_a B(\bfm{g}^a) ]}_-\\
       & \phantom{:}
       =
          -i \varphi^*\biggl(
           i  D_a {( {\Psi}^+ \gamma^5 {\mathcal J} )}_{A''}
           + i n_a D_b {( {\Psi}^+ \notn \gamma^b
             \gamma^5 {\mathcal J} )}_{A''}\\
       & \phantom{:= -i \varphi^*\biggl(}
       - m n_a {({\Psi}^+ \notn \gamma^5 {\mathcal J})}_{A''}
       \biggr) (\bfm{g}^a)
       \\
     D'(D_{\mathcal C}M \otimes T^*{\mathcal C}) \ni
     \tilde{\varphi}^* (\nabla_a \sideset{^\Psi}{_{x_0}}{V})
     (\bfm{\chi} d{\mathcal C},\cdot)
     \colon
       \bfm{g}_B^a \mapsto  &
       {\{\nabla_a\Psi^B(\bfm{g}_B^a),
         \sideset{^{\mathcal C}}{^+_{A''}}{{G}}(x_0) \}}_+ \\
       & : = {\{\nabla_a\Psi^B(\bfm{g}_B^a),
         \sideset{^{\bfm{\chi} d{\mathcal C}}}{^+_{A''}}{{Q}}(x_0) \}}_+\\
       &\phantom{:}
       =    \varphi^*
    \biggl(
% II:
    - i {h_a}^c \nabla_c
    \biggl(
     \bigl( (-i {\notnabla} -m ) \AiB \bigr)
    {\mathcal J}
    {{\biggr)}^{B}}_{A''}\\
    & \phantom{:= \varphi^*\biggl(}
% III:
    -i n_a \gamma^b h{_b}^c \nabla_c
    \bigl(
    \notn \bigl( (-i {\notnabla} -m ) \AiB \bigr)
    {\mathcal J}
    {{\bigr)}^{B}}_{A''}\\
    & \phantom{:= \varphi^*\biggl(}
% VI:
    - m n_a
    \bigl(
     \notn
     \bigl(
      ( - i \notnabla -m ) \AiB
     \bigr)
     {\mathcal J}
    {\bigr)^{B}}_{A''}
    \biggr)
    (\bfm{g}_B^a),
\end{align*}
}
provided $\bfm{\chi} \equiv 1 $ o the support of the testdensities.
Here $D_{\mathcal C}M$ denotes all elements in $DM$ with basepoint in
${\mathcal C}$.
\end{Thm}
Note that we used boldface type, e.g., $\bfm{\chi}$ and $\bfm{h}$, to
indicate smooth functions and densities respectively on ${\mathcal C}$.
The notion $\sideset{^{\mathcal C}}{^+_{A''}}{G}(x_0)$ was introduced to
emphasize that the result does not depend on $\bfm{\chi}$.
It is obvious that the six operator valued distributions in
Theorem~\ref{Thm:commutQCurvedCx} smoothly depend on $x_0$, hence
they have a natural extension to operator valued distributions on a
neighborhood of $U\subset {\mathcal C} \times {\mathcal C}$ of the diagonal\
 in
${\mathcal C}\times{\mathcal C}$.
\begin{Cor}\label{Cor:commutQCurvedC}
  The six distributions of Theorem~\ref{Thm:commutQCurvedCx}, when
  extended to $U \subset {\mathcal C}\times {\mathcal C}$ can be restricted\
 to
  the diagonal in ${\mathcal C}\times{\mathcal C}$, i.e., they uniquely\
 define
  six operator valued distributions on ${\mathcal C}$. For the
  corresponding kernels one finds:
     \begin{align*}
     [\sideset{^{\mathcal C}}{^+_A}{G}, A(x)]_-
     & =
     \lim_{x_0 \rightarrow x}
     {[ \sideset{^{\bfm{\chi}}}{^+_{A'}}{{Q}}(x_0), A (x) ]}_-
     =
     + i {\Psi}^+_{A} (x) \\
     [\sideset{^{\mathcal C}}{^+_A}{G}, B(x)]_-
     & =
     \lim_{x_0 \rightarrow x}
     {[ \sideset{^{\bfm{\chi}}}{^+_{A'}}{{Q}}(x_0), B (x) ]}_-
     =
     {({\Psi}^+\gamma^5)}_{A} (x)\\
     \{\Psi^B(x),\sideset{^{\mathcal C}}{^+_A}{G}\}_+
     & =
     \lim_{x_0 \rightarrow x}
     {\{ \Psi^B(x), \sideset{^{\bfm{\chi}}}{^+_{A'}}{{Q}}(x_0) \}
           }_+
     =
     + i {\left( ( i {\notnabla} + m )
           \AiB \right)^{B}}_{A} \\
     \begin{split}
       [\sideset{^{\mathcal C}}{^+_A}{G},\nabla_a A(x)]_-
       & =
       \lim_{x_0 \rightarrow x}
       {[ \sideset{^{\bfm{\chi}}}{^+_{A'}}{{Q}}(x_0), \nabla_a A(x)]}_-\\
       & =
       i {h_a}^b \nabla_b {\Psi}^+_{A}
       + i n_a {h_b}^c \nabla_c {( {\Psi}^+ \notn \gamma^b )}_{A}\\
       & \phantom{= \int}
           - m n_a {({\Psi}^+ \notn )}_{A}
     \end{split}\\
     \begin{split}
       [\sideset{^{\mathcal C}}{^+_A}{G},\nabla_a B]_-
       & =
       \lim_{x_0 \rightarrow x}
       {[ \sideset{^{\bfm{\chi}}}{^+_{A'}}{{Q}}(x_0),\nabla_a B(x) ]}_-\\
       & =
       {h_a}^b \nabla_b {( {\Psi}^+ \gamma^5 )}_{A}
       + n_a {h_b}^c \nabla_c {( {\Psi}^+ \notn \gamma^b
            \gamma^5 )}_{A}\\
       & \phantom{= -i \int}
       - m n_a {({\Psi}^+ \notn \gamma^5 )}_{A}
     \end{split}
     \\
     \begin{split}
       \{\nabla_a\Psi^B(x),\sideset{^{\mathcal C}}{^+_A}{G}\}_+
       & =
       \lim_{x_0 \rightarrow x}
       {\{ \nabla_a \Psi^B(x), \sideset{^{\bfm{\chi}
               f}}{^+_{A'}}{{Q}}(x_0) \}}_+\\
       & =
       -i {h_a}^b \nabla_b {{\left(
            ( - i \notnabla - m ) \AiB  \right)}^B}_{A}\\
       & \phantom{= \int}
       -i  n_a {{\left\{ \gamma^d {h_d}^c
            \nabla_c  \left( \notn ( -i \notnabla - m ) \AiB
            \right)\right\} }^B}_{A} \\
         & \phantom{= \int}
             - m  n_a {{\left( \notn ( -i \notnabla - m )
             \AiB \right)}^B}_{A}.
       \end{split}
   \end{align*}
\end{Cor}
Corollary~\ref{Cor:commutQCurvedC} is obvious, since we may use the
fact that the pointwise product of a smooth function $g\in \Cinf(M)$
with a section in the spinorbundle $h^A\in\Gamma(DM)$, $g(x)h^A(x)$,
is again a section in the spinorbundle. Similarly the pointwise
product of $g_A\in\Gamma(D^*M)$ with $h^A \in \Gamma(DM)$, $g_A(x)
h^A(x)$, defines a smooth function on $M$ by dual pairing and the
claim follows immediately using Corollary~\ref{Cor:restrictionDiag}.
\begin{rem}
  The kernel of $\sideset{^{\cdot}}{^+_{A'}}{Q}$ is not covariantly
  constant. Thus the results of Proposition~\ref{prop:commutQCurv} and
  Theorem~\ref{Thm:commutQCurvedCx} depend on the hypersurface ${\mathcal
    C}$. On the other hand the results of
  Corollary~\ref{Cor:commutQCurvedC} --as long as no derivatives
  occur-- have a natural extension to all of $M$ independent of ${\mathcal
    C}$.
\end{rem}
For the proof of the Theorem we need some properties of the
fundamental solutions of the Klein-Gordon and Dirac equation
respectively. We begin with the scalar case.
%
%%%%%%%%%%%%
\subsection{The fundamental solution of the Klein Gordon equation}
\label{sec:fundamSolKleinG}
%%%%%%%%%%%%
%
In this subsection some useful identities for the fundamental solution
of the Klein Gordon equation are established.  Let $(M,g_{ab})$ denote
a globally hyperbolic Lorentz manifold with Cauchy surface ${\mathcal C}$.
For the sake of simplicity we identify in the sequel of this work test
densities and test functions using the natural measure $d\mu=
\sqrt{|g_{ab}|}d^4x$ on $M$ and denote the space of testdensities on
$M$ with compact support by $D(M)$.  Let $P$ be a pseudodifferential
operator on $M$ which has the form $P= \Box +a$, where $\Box$ denotes
the D'Alembert operator associated to $g_{ab}$ and $a$ is some smooth
function on $M$. The `squared' Dirac operator and the Klein Gordon
operator are examples for $P$. As a consequence of the global
hyperbolicity, there exist global fundamental solutions for $P$
uniquely determined by their support properties (see
Leray~\cite{leray:diffeqn}, Choquet-Bruhat~\cite{choquet:77}). That is
there exist unique distributions ${E}^\pm \in D'(M\times M)$, viewed
as linear mappings $\Delta^\pm \colon \CinfO(M) \rightarrow \Cinf(M)$
by the identification
\[
\int_M f_1 \Delta^\pm(f_2) d\mu = E^\pm(f_1 d\mu \otimes f_2d\mu) \quad
\text{for all } f_1,f_2 \in \CinfO(M),
\]
which are called advanced and retarded fundamental solutions
respectively. They are uniquely determined by the following two
properties:
\begin{enumerate}
\item \begin{equation}\label{eq:fundSolScalar}
      P{\Delta}^\pm = {\Delta}^\pm P = \identity,
      \end{equation}
      where $\identity$ denotes the identity on $\CinfO(M)$.
\item If $f\in \CinfO(M)$ then $\supp \Delta^\pm(f)$ is compact
  with respect to the future~($+$) or past~($-$).
\end{enumerate}
Here, a set $U\subset M$ is called compact with respect to the future
(past) if and only if for any point $x\in M$, $J^+(x) \cap J^-(U)$
($J^-(x) \cap J^+(U)$ ) is compact. Recall that $J^+(x)$ ($J^-(x)$) denotes\
 the
set of all points $y\in M$ which can be connected to $x$ with a future
(past) directed causal curve from $x$ to $y$.
We define the fundamental solutions $E = {E}^+ - {E}^-$ and $\Delta =
\Delta^+ - \Delta^-$. Given $\Delta$ it is easy to solve the Cauchy
problem for the differential equation $Pu=0$, $u \in \CinfO(M)$.  The
result is:
\begin{Prop}[Huygens principle (smooth form)]\label{Prop:Huygenssmooth}
  Let $\rho_0$ ($\rho_1$) denote the restriction (normal restriction)
  of any smooth function on $M$ to the Cauchy surface ${\mathcal C}$. Then
  for any smooth solution $u$ of $Pu=0$ on $M$ and any $f \in \CinfO
  (M)$
  \begin{equation}\label{eq:Huygenssmooth}
    \int_M u f \;\;d\mu = \int_{\mathcal C} u_0 \rho_1(\Delta f) - u_1
                          \rho_0(\Delta f) \;\;d{\mathcal C},
  \end{equation}
  where $u_0 \equiv \rho_0(u):={u\restriction}_{\mathcal C}$, $u_1 \equiv
  \rho_1(u) := {(n_\mu \nabla^\mu u)\restriction}_{\mathcal C}$ and
  $d{\mathcal C}$ denotes the natural volume element on ${\mathcal C}$
  determined by the induced metric $h_{ab}$ (compare
  subsection~\ref{sec:splittingSpacetime} of this chapter).
\end{Prop}
\begin{pf}
  For the proof, which is merely an application of Greens formula, the
  reader is referred to Dimock's paper~\cite{dimock:80} Lemma~A.1.
\end{pf}
Now let $u$ be a distributional solution of $Pu=0$. Then $u$ admits a
restriction to ${\mathcal C}$ according to
Theorem~\ref{thm:hoermander2.5.11}, since by
Theorem~\ref{Thm:prop_of_sing} its wave front set contains at most
lightlike directions.  Thus Eqn.~(\ref{eq:Huygenssmooth}) can by
extended by continuity and we proved
\begin{Prop}\label{Prop:Huygensdist}
  Let $\hat{\varphi} \colon (M \times {\mathcal C}) \ni (p_1,p_2) \mapsto
  (p_1,p_2) \in (M \times M)$ be the natural embedding of $M \times
  {\mathcal C}$ into $M\times M$. Then the following identity holds for
  all $f\in D(M)$:
  \begin{equation}\label{eq:HuygensDis}
    \begin{split}
      u(f)
      & =
      \hat{\varphi}^* \left\{ ( 1 \otimes u) \cdot \left(
      (1 \otimes n_a \nabla^a)E \right) \right\}
      (f \otimes \identity d{\mathcal C}) \\
      & \phantom{=}
      - \hat{\varphi}^* \left\{ ( 1 \otimes n_a \nabla^a u) \cdot E
                  \right\}
        (f \otimes \identity d{\mathcal C}) \\
        & \mbox{``} \equiv \mbox{''}
        \int_M \int_{\mathcal C} u(z) n_{a'} {\nabla^{a'}}
        E(x,z) f(x)
        - (n_{a'} {\nabla^{a'}} u(z)) E(x,z) f(x)
        d{\mathcal C}_z d\mu_x
    \end{split},
  \end{equation}
  where $ \cdot \otimes \cdot$ is the usual tensor product of
  distributions, $1$ being the function $1$ and $\identity$ denotes
  the restriction of $1$ to ${\mathcal C}$. Note that
  Eqn.~(\ref{eq:HuygensDis}) is well defined, since $\Bigl(
  J^+\bigl(\supp(f)\bigr) \cup J^+\bigl(\supp(f)\bigr) \Bigl) \cap
  {\mathcal C} $ is compact.
\end{Prop}
It is worth noting that the fundamental solution $E$, being a distributional
bisolution of the Klein Gordon equation, can be restricted to ${\mathcal C}$.
For the restriction of $E$ and its normal derivative one finds:
\begin{Prop}\label{Prop:restrictionParamscalar}
  Let $E$ be the (unique) fundamental solution of the scalar Klein
  Gordon equation.
  \begin{itemize}
  \item The restriction of $E$ to any Cauchy surface ${\mathcal C}$
    vanishes, i.e., let $\tilde{\varphi}$ denote the embedding of ${\mathcal
      C}\times {\mathcal C}$ into $M\times M$, then
    \begin{equation}\label{eq:restrE}
      \tilde{\varphi}^*(E) = 0.
    \end{equation}
  \item The normal derivative of $E$ with respect to the
    first entry restricted to ${\mathcal C}\times {\mathcal C}$ is minus the
    Dirac delta distribution, i.e.,
    \begin{equation}\label{eq:restrEderiv}
      \tilde{\varphi}^*(n^a\nabla_a E)(x,y) = - \bfm{\delta} (x,y).
    \end{equation}
  \end{itemize}
\end{Prop}
\begin{rem}
  These two statements can equivalently be formulated in terms of the
  operator $\Delta$. They state that $\Delta$, when restricted to
  ${\mathcal C}\times {\mathcal C}$ vanishes, whereas the restriction of its
  normal derivative with respect to the first entry is minus the
  identity on $\CinfO({\mathcal C})$.
\end{rem}
\begin{pf}
  For $y \in {\mathcal C}$ let $u_y \in D'(M)$ be the unique
  distributional solution of $(\Box+m^2) u_y =0$ with initial
  data
  \[
  (u_y)\restriction_{\mathcal C} =0 \qquad (n^a\nabla_a u_y)
  \restriction_{\mathcal C} = - \bfm{\delta}(\cdot,y),
  \]
  where $\bfm{\delta}(\cdot,y)$ denotes Dirac delta distribution at $y
  \in {\mathcal C}$, i.e., for $\bfm{g}d{\mathcal C}\in D({\mathcal C})$, $
  \bfm{\delta}(\bfm{g}d{\mathcal C},y) = \bfm{g}(y)$. Inserting these initial
  datas in Eqn.~(\ref{eq:HuygensDis}) we obtain for all $f\in
  D(M)$ --using the notation of Proposition~\ref{Prop:Huygensdist}--
  \begin{equation*}
    \begin{split}
      u_y(f)
      & =
      0 - \hat{\varphi}^*\{1 \otimes n^a\nabla_a u_y\}
          \hat{\varphi}^*\{E\} (f \otimes
        \identity d{\mathcal C})\\
      & = ( 1 \otimes \bfm{\delta}(\cdot,y)) \cdot \hat{\varphi}^*\{E\} (f
          \otimes \identity d {\mathcal C})\\
      & = + \hat{\varphi}^* \{E\} ( f \otimes y )\\
      & = E(f,y).
    \end{split}
  \end{equation*}
  This proves $u_y = E(\cdot,y)$ for $y \in {\mathcal C}$. Thus
  $E(\cdot,y)\restriction_{\mathcal C} =0$ together with $(n^a \nabla_a
  E(\cdot,y))\restriction_{\mathcal C} = - \bfm{\delta}(\cdot,y)$ for all $y\
 \in
  {\mathcal C}$, which was claimed in the Corollary.
\end{pf}
Note that $E(f,y)$ is a smooth function on $M$ for all $f \in
\CinfO(M)$, since $\Delta$ applied to a smooth function with compact
support is smooth.  Hence $E\restriction_{{\mathcal C}\times{\mathcal
    C}}(\cdot,y) \equiv E(\cdot,y)\restriction_{\mathcal C}$ if\
 $y\in{\mathcal
  C}$.  The next Lemma states that derivatives of distributions in
direction tangential to ${\mathcal C}$ commute with the restriction to
${\mathcal C}$.
\begin{Lemma}\label{Lemma:derivTangC}
  Let ${\mathcal C}$ be a spacelike hypersurface and let $u \in D' (M)$
  be a distribution on $M$ such that $\varphi^*(u)$, the restriction of
  $u$ to ${\mathcal C}$, exists. Let $v^a$ be a smooth vector field
  tangential to ${\mathcal C}$ and let $h_{ab}$, $D_a$ denote the induced
  metric and Levi-Civita connection respectively on ${\mathcal C}$ (compare
  Definition~\ref{Dfn:inducedMetric} and
  Lemma~\ref{Lemma:inducedLeviCiv}). Then
  \[
  \varphi^*(v^a \nabla_a u ) = v^a D_a \varphi^*(u).
  \]
\end{Lemma}
\begin{pf}
  Suppose $u$ is given by a smooth function, then
  \begin{equation*}
    \begin{split}
      \varphi^*(v^a\nabla_a u)(\bfm{f})
      & :=
       \int_{\mathcal C} ( v^a \nabla_a u(x) ) \bfm{f}(x) d{\mathcal C} \\
      & =
      \int_{\mathcal C} ( v^a D_a u(x) ) \bfm{f}(x) d{\mathcal C}, \\
      \intertext{since $v^a$ is tangential on ${\mathcal C}$}
      & =
      - \int_{\mathcal C} u(x) D_a(v_a \bfm{f}) (x) d{\mathcal C},\\
      \intertext{by Stokes Theorem, i.e., integration by parts, which
        is valid, since $D_a$ is the induced Levi-Civita connection on
        ${\mathcal C}$}
      & =: \bigl( v^a D_a \varphi^*(u)\bigr) (\bfm{f})
    \end{split}
  \end{equation*}
  By continuity this extends to $D'(M)$.
\end{pf}
\begin{Cor}\label{Cor:restrictionParamscalarDeriv}
  Let $E$ be the fundamental solution of the Klein Gordon equation, $v^a$
  denote a smooth vector field and $\tilde{\varphi}$ is the embedding of
  ${\mathcal C}\times{\mathcal C}$ into $M\times M$. Then
  \[
  \tilde{\varphi}^*(v^a \nabla_a E) = - v^a n_a \bfm{\delta}.
  \]
\end{Cor}
\begin{pf}
  Split the vectorfield $v^a$ in its normal and tangential components
  and use the linearity of $\tilde{\varphi}^*$.
  \[
  \tilde{\varphi}^*(v^a \nabla_a E) = \tilde{\varphi}^*(v^a {h_{a}}^b
  \nabla_b E) + \tilde{\varphi}^*(v^a n_a n^b \nabla_b E)
  \]
  By Lemma~\ref{Lemma:derivTangC} the first summand is equal to $v^a
  {h_a}^b D_b \tilde{\varphi}^*(E)$, which vanishes by
  Proposition~\ref{Prop:restrictionParamscalar}. By
  Eqn.~(\ref{eq:restrEderiv}) the second summand is equal to $ - v^a
  n_a \bfm{\delta}$, which finishes the proof.
\end{pf}
We now establish analogous properties for the
fundamental solution of the Dirac equation.
%
%%%%%%%%%%%%
\subsection{The fundamental solution of the Dirac equation}
\label{sec:fundamSolDirac}
%%%%%%%%%%%%
%
Let $(M,g_{ab})$ be a globally hyperbolic Lorentz manifold with
Majorana spinor (cospinor) bundle $DM$ $(D^*M)$. Let ${\mathcal C}$ be a
Cauchy surface in $M$. We remind the reader that functions and
densities on $M$ and ${\mathcal C}$ can be identified using the
appropriate metric.  The Dirac operator $\notD := (i\notnabla - m)$
fulfils the following important identity (See
e.g.~\cite{benn:spinors})
\[
(i\notnabla - m) (- i\notnabla - m)
= \left( \Box + ( m^2 + \frac{1}{4} R ) \right) \frakI,
\]
{\sloppy
where $\frakI$ denotes the identity on the spinor bundle. Note that
$\Dsquare$ is the SUSY coupled Klein Gordon operator, which is
strictly hyperbolic in the sense of Leray. It follows that
\[
\Dsquare u =0
\]
possesses unique advanced and retarded fundamental solutions
viewed either as distributions in \\
$D'( D^*M \times DM)$, denoted by
${E^\pm_A}^{A'}$, or as linear operators from $\Gamma_0(DM)$ to
$\Gamma(DM)$ , denoted by ${\Delta^\pm_A}^{A'}$. Their difference will
be written as ${E_A}^{A'} := {{E^+}_A}^{A'} - {{E^-}_A}^{A'}$ and
${\Delta_A}^{A'} := {{\Delta^+}_A}^{A'} - {{\Delta^-}_A}^{A'}$
respectively. The existence of a unique fundamental solution for the
Dirac equation $\notD u =0$ is a consequence of the following Theorem
of Dimock~\cite{dimock:82}.
}
\begin{Prop}\label{Prop:Dimock2_2.1}
  $\phantom{~}$\par
  \begin{itemize}
  \item The operator $\notD$ on $\Gamma_0(DM)$ has a unique
    fundamental solutions
    \[{{\Xi^\pm}} \colon \Gamma_0(DM) \rightarrow
    \Gamma (DM)
    \] given by
    \[
    {{\Xi^\pm}} := {{\miDm \Delta^\pm}}
    \]
  \item Its Dirac adjoint, ${\notD}^+$, which is an operator on
    $\Gamma_0(D^*M)$, has  unique fundamental solutions
    \[
    {{\hat{\Xi}^\pm}} \colon \Gamma_0(D^*M) \rightarrow \Gamma(D^*M)
    \]
    given by
    \[
    {{\hat{\Xi}^\pm}} := {{\iDm \Delta^\pm}},
    \]
  \end{itemize}
  where $\Delta^\pm$ are the fundamental solutions of $\Dsquare
  \frakI$ uniquely determined by their support properties.  $\frakI$
  denotes the identity on the spinor (cospinor) bundle and   the spinor
  indices have been suppressed.
\end{Prop}
{\sloppy
We may identify
$\Xi^\pm$ ($\hat{\Xi}^\pm$) with distributions denoted
by \\
$S^\pm\in D'( D^*M \times DM)$  ($\hat{S}^\pm \in
D'(DM D^*M)$).
Their difference is written as $\Xi$ ($\hat{\Xi}$) and $S$ ($\hat{S}$)
respectively.
}
For the formulation of Huygens principle it is necessary to define
products of distributions in these bundles. Let us call $D'(DM)$
($D'(D^*M)$) the space of spinor (cospinor) valued distributions.
Using a similar notation as in the previous subsection, we denote the
testdensities for $D'(DM)$ ($D'(D^*M)$) by $D(D^*M)$ ($D(DM)$).  It is
worth noting that the smooth sections in the corresponding bundles are
dense in these spaces. Moreover, on a globally hyperbolic manifold
things are greatly simplified by the fact that every spin structure
$SM$ over $M$ necessarily is the trivial bundle~\cite{geroch:68}. It
follows that all calculation can be performed in `coordinates': Pick a
global spin frame $E \colon M \rightarrow SM$. We denote the
corresponding moving frames by $(E_1,\ldots,E_4)$ and their duals by
$(E^1\ldots,E^4)$ respectively. Then every element $u^A \in D'(DM)$,
$v_A\in D'(D^*M)$, $f^A\in D(D^*M)$, $g_A \in D(DM)$ can uniquely be
written as
\begin{alignat*}{2}
  u & = \sum u^\Lambda E_\Lambda & \qquad
  v & = \sum v_\Lambda E^\Lambda\\
  f & = \sum f^\Sigma E_\Sigma & \qquad
  g & = \sum g_\Sigma E^\Sigma,
\end{alignat*}
where $u^\Lambda$ and $v_\Lambda$, $\Lambda=1,\ldots,4$ are
distributions on $M$ and $ f^\Sigma$, $g_\Sigma$, $\Sigma=1,\cdots,4$
are densities with compact support. Note that we use an upper index
for $f^A\in D(D^*M)$, since this density can be identified with a {\em
  spinor\/} field.
A {\em distributional product\/} of $u^A$ and $v_A$ is well defined if
and only if $\sum_\Lambda u^\Lambda \cdot v_\Lambda$ exists, where
each term in the sum is the usual product of distributions. It is
obvious that in the special case $u^A\in \Gamma(DM)\subset D'(DM)$ and
$v_A \in \Gamma(D^*M) \subset D'(D^*M)$ this definition is frame
independent. Due to the continuity of the product and since
$\Gamma(DM)$ and $\Gamma(D^*M)$ are dense in $D'(DM)$ and $D'(D^*M)$
respectively, this independence can be lifted and $u^A \cdot v_A\in
D'(M)$ defines a distribution on $M$ if the product exists.
We state now the distributional version of Huygens principle for the
`squared' Dirac operator on the spinor bundle.  Recall that all
distributional solutions of $\Dsquare u = 0$ can be restricted to
${\mathcal C}$, since by Theorem~\ref{Thm:prop_of_sing} their wave front
sets contain at most lightlike directions.
\begin{Prop}[Huygens principle (distributional form)]
  \label{Prop:HuygensDisfermi}
  Let $\hat{\varphi}$ denote the natural embedding of $M \times {\mathcal C}$
  into $M \times M$.  For any distributional solution $u_A \in
  D'(D^*M)$ of $\Dsquare u=0$, the following identity holds for all
  $f^A \in D(D^*M)$:
  \begin{equation}\label{eq:HuygensFermi}
    \begin{split}{}
      u_A(f^A) & =
      \hat{\varphi}^*
      \left\{ ({\frakI_{A'}}^{B'} \otimes u_A) \cdot
        \left( ({\frakI_{B'}}^{C'}\otimes n_a \nabla_a) {E_{C'}}^A \right)
      \right\}
      ( f^{A'} \otimes  \bfm{1} d{\mathcal C})\\
     & \phantom{=\hat{\varphi}^* \biggl\{ }
     -
     \hat{\varphi}^*
     \left\{
       ({\frakI_{A'}}^{B'} \otimes n_a \nabla^a u_A) \cdot
       {E_{B'}}^A
     \right\}
     ( f^{A'} \otimes \bfm{1} d{\mathcal C}),
   \end{split}
 \end{equation}
 where $\cdot \otimes \cdot$ is the usual tensor product of
 distributions, ${\frakI_{A'}}^{B'}$ denotes the identity on $DM$,
 whereas $\bfm{1}$ denotes the restriction of the function $1$ to
 ${\mathcal C}$. Note that Eqn.~(\ref{eq:HuygensFermi}) is well defined,
 since $\Bigl( J^+\bigl(\supp(f^A)\bigr) \cup J^+\bigl(\supp(f)\bigr)
 \Bigl) \cap {\mathcal C} $ is compact.
\end{Prop}
The fundamental solution $S$ of the Dirac equation can be restricted
to ${\mathcal C}$, too, since its wave front set is contained in the wave
front set of $E$. The result is:
\begin{Prop}\label{Prop:DiracParam}
  Let ${S_A}^{A'} \in D'(D^*M \times DM)$ be the
  unique fundamental solution of the Dirac operator. The restriction
  of ${S_A}^{A'}$ to the Cauchy surface ${\mathcal C}$ with
  normal covectorfield $n_a$ and embedding $\tilde{\varphi} \colon {\mathcal
    C} \times {\mathcal C} \rightarrow M \times M$ is well defined and
  given by:\\
  For all $\bfm{f}^A d{\mathcal C} \in D(D_{\mathcal C}^*M)$, $\bfm{g}_B\
 d{\mathcal C} \in
  D(D_{\mathcal C}M)$
  \begin{equation}
    \tilde{\varphi}^*
    \left(
      {S_A}^{B'}
    \right) ( \bfm{f}^A,\bfm{g}_{B'} )
    = \int_{{\mathcal C}} \bfm{f}^A {{( i \notn)}_A}^B \bfm{g}_B d{\mathcal\
 C},
  \end{equation}
  where $D_{\mathcal C}M$ ($D^*_{\mathcal C}M$) denotes the restriction of
  $DM$ ($D^*M$) to ${\mathcal C}$.
\end{Prop}
\begin{rem}
  The result of the Proposition is equivalent to the statement that
  $\Xi$ restricted to ${\mathcal C}$ is $( i \notn )$ times the identity
  on $\Gamma_0(D_{\mathcal C}M)$.
\end{rem}
For the proof of the Proposition we need the following two Lemmata.
\begin{Lemma}\label{Lem:restr_fermi}
  Let ${E_A}^{A'} \in D'(D^*M \times DM)$ be the
  unique fundamental solution of the `squared' Dirac operator
  $\Dsquare{}$.
  \begin{itemize}
  \item The
  restriction of ${E_A}^{A'}$ to any smooth Cauchy surface ${\mathcal C}$
  vanishes, i.e., let $\tilde{\varphi}$ be the embedding of ${\mathcal C}
  \times {\mathcal C}$ into $M\times M$, then
  \begin{equation}
    \label{eq:restrEFermi}
    \tilde{\varphi}^*({E_A}^{A'}) = 0
  \end{equation}
  \item The restriction of its normal derivative with
  respect to the first entry gives $ - {{\mathcal J}_A}^{A'}$ times the
  Dirac delta distribution, i.e.,
  \begin{equation}
    \label{eq:restrEderivFermi}
    \tilde{\varphi}^*(n^a \nabla_a {E_A}^{A'}) (x,y) = - {{\mathcal\
 J}_A}^{A'}
                                                 \bfm{\delta}(x,y).
  \end{equation}
  Recall that ${\mathcal J}$ denotes the bi-spinor of parallel
  transport and $\bfm{\delta}$ is defined as in
  Proposition~\ref{Prop:restrictionParamscalar}.
\end{itemize}
\end{Lemma}
\begin{pf}
  Let $ {{u_y}_{A}}^{B'} \in D'(D^*M)$ be the
  unique solution of $\Dsquare {{u_y}_{A}}^{A'} = 0$
  with initial data
  \[
  ({{u_y}_A}^{A'})\restriction_{\mathcal C}(x) =0 \qquad ({(n^a \nabla_a
      u_y)}_{A}^{A'})\restriction_{\mathcal C}(x) = - {{\mathcal\
 J}_A}^{A'}(x,y)
    \bfm{\delta}(x,y),
  \]
  where $\bfm{\delta}(x,y)$ denotes the Dirac delta distribution at $y
  \in {\mathcal C}$.  Using the distributional form of Huygens principle,
  we obtain for all $f^A \in D(D^*M)$ --using the notation of
  Proposition~\ref{Prop:HuygensDisfermi}--
  \begin{equation}
    \begin{split}
      {{u_y}_A}^{B'} (f^A)
      & =
      0 - \varphi^* \left\{
                 ({{\frakI}_C}^A \otimes
                   n^a\nabla_a {{u_y}_{B''}}^{B'}) \cdot
                 {E_A}^{B''}
              \right\}
          ( f^C \otimes \identity d{\mathcal C})  \\
      & =
       \left(
         ({\frakI_C}^A \otimes {{\mathcal J}_{B''}}^{B'}\
 \bfm{\delta}(\cdot,y))
         \varphi^*({E_A}^{B''})
       \right)
       (f^C \otimes \identity d{\mathcal C})\\
      & = {E_A}^{B'}(f^A,y)
    \end{split}
  \end{equation}
  This proves ${{u_y}_A}^{B'} = {E_A}^{B'}(\cdot,y)$ for all points
  $y\in {\mathcal C}$. Note that ${E_A}^{B'}(f^A,y)$ is smooth in $y$,
  since we are dealing with the fundamental solution of the `squared'
  Dirac equation (recall the results for the scalar case). We conclude:
  ${E_A}^{B'}(\cdot,y)\restriction_{\mathcal C} =0$ and $(n^a
  \nabla_a {E_A}^{A'} (\cdot,y))\restriction_{\mathcal C} = - {{\mathcal
      J}_A}^{A'}\bfm{\delta}(\cdot,y)$ for all $y \in {\mathcal C}$, which\
 was
  claimed.
\end{pf}
The second Lemma is the analogon of  Lemma~\ref{Lemma:derivTangC}
above.
\begin{Lemma}\label{Lemma:derivTangCFermi}
  Let ${\mathcal C}$ be a spacelike hypersurface and let $u^A \in D'
  (DM)$ be a spinor valued distribution on $M$ such that
  $\varphi^*(u^A)$, the restriction of $u^A$ to ${\mathcal C}$, exists. Let
  $v^a$ be a smooth vector field tangential to ${\mathcal C}$ and let
  $h_{ab}$, $D_a$ denote the induced metric and Levi-Civita connection
  respectively\footnote{We extend Eqn.~(\ref{eq:inducedLeviCiv})
    without modification to spinors, i.e., for a smooth section $f^A$
    in the spinor bundle
    \[
    D_a f^A \equiv {h_a}^b \nabla_b f^A.
    \]
    Note that we do {\em not\/} project on the spinor indices.} (See
  Definition~\ref{Dfn:inducedMetric} and
  Lemma~\ref{Lemma:inducedLeviCiv}). Then
  \[
  \varphi^*(v^a \nabla_a u^A ) = v^a D_a \varphi^*(u^A).
  \]
\end{Lemma}
\begin{pf}
  {\sloppy
  Suppose $u^A$ is given by a smooth section with compact support in
  $DM$, then for any smooth section $f_A$ in the cospinor bundle $v^a
  u^A f_A$ is a smooth {\em vector field\/} and the integral of
  $D_a(v^a u^A f_A)$ over ${\mathcal C}$ vanishes by Stokes Theorem. We
  can apply the latter, since $D_a$ is the Levi Civita connection on
  ${\mathcal C}$. We obtain the following chain of equation:
  }
  \begin{equation*}
    \begin{split}
      0 & =
      \int_{\mathcal C} D_a(v^a u^A f_A) d{\mathcal C}\\
      & = \int_{\mathcal C}
      (D_a v^a) u^A f_A +v^a \bigl( (\nabla_a u^A) f_A + u^A (\nabla_a
      f_A) \bigr) d{\mathcal C}\\
      \intertext{since $D_a$ is a derivation and on scalars $v^aD_a
        \equiv v^a \nabla_a$}
      & = \int_{\mathcal C}
      u^A D_a(v^a f_A) + v^a (\nabla_a u^A) f_A d{\mathcal C}\\
      \intertext{since $D_a$ is a derivation and $v^a D_a \equiv v^a
        \nabla_a$ for pure (co)spinors}
      & =: \bigl( - v^a D_a \varphi^*(u^A)\bigr) (f_A)  + \varphi^*(
      v^a \nabla_a u^A) (f_A)
    \end{split}
  \end{equation*}
  Note that the last equality is obtained by using the definition of
  the covariant derivative for distributions. Moreover we used the
  metric to identify functions and densities.  By continuity this
  extends to all of $D'(M)$, which proves the Lemma.
\end{pf}
\begin{rem}
  The previous Lemma is not a complete analogon of
  Lemma~\ref{Lemma:derivTangC}. In the former everything is
  independent of the fact that ${\mathcal C}$ is embedded into $M$, e.g., all
  tensorial objects are tangential to ${\mathcal C}$ and covariant
  differentiation does not change this fact.  On the other hand in the
  spinorial case, the embedding enters via the spinor bundle, i.e.,
  spinors on ${\mathcal C}$ are elements in the bundle $D_{\mathcal C}M$
  consisting of all spinors in $DM$, whose base points are elements of
  ${\mathcal C}$.  It follows that $D_{\mathcal C}M$ can {\em not\/} be
  constructed without specifying the embedding ${\mathcal C}
  \stackrel{\text{i}}{\hookrightarrow} M$, i.e., the elements of
  $D_{\mathcal C}M$ do not lie in some subbundle `tangential' to ${\mathcal
    C}$. It is worth noting that such a subbundle actually can be
  defined~(See~\cite{Najmi:84}). However such a treatment yields
  Spin(3)-spinors, which is undesirable from our point of view.
\end{rem}
\begin{pf*}{Proof of Proposition~\ref{Prop:DiracParam}}
  Using Proposition~\ref{Prop:Dimock2_2.1} above we rewrite the
  fundamental solution $S$ (suppressing all spinor indices)
  \begin{equation} \label{eq:diracprop}
    S = \miDm E =
    \left( -i \gamma_a h^{ab} \nabla_b - i
      \notn n^a \nabla_a) -m
    \right) E
  \end{equation}
  For the restriction of $S$ to ${\mathcal C}\times{\mathcal C}$ we obtain by
  Lemma~\ref{Lem:restr_fermi}
  \begin{equation}
    \tilde{\varphi}^*({S})(x,y)
    = \tilde{\varphi}^*\bigl( - (i \gamma_a {h^{ab} \nabla_b E)} \bigr)(x,y)
    + i \notn {\mathcal J} \bfm{\delta}(x,y) -
    m \tilde{\varphi}^*(E)(x,y)
  \end{equation}
  The first summand vanishes by Lemma~\ref{Lemma:derivTangCFermi},
  since $\gamma_a h^{ab}$ is tangential to ${\mathcal C}$, the third term
  is zero by Lemma~\ref{Lem:restr_fermi}. The remaining term in
  the middle is the result that was claimed.
\end{pf*} % Proposition
\begin{pf*}{Proof of Theorem~\ref{Thm:commutQCurvedCx}}
  To prove the Theorem, we calculate the (anti) commutators
  explicitly. It was already shown in
  Proposition~\ref{prop:commutQCurv} that the restrictions of all six
  operator valued distributions $\sideset{^{\cdot}}{_{x_0}}{V}$ yield
  operator valued distributions on ${\mathcal C}$. Consider
  $\tilde{\varphi}^*(\sideset{^A}{_{x_0}}{V}) \in D'({\mathcal
    C}\times{\mathcal C})$. Inserting the definitions and writing all
  distributions as integrals we obtain:
\begin{equation}\label{eq:commutatorA}
  \begin{split}
    & \tilde{\varphi}^* \bigl(
      {{\left[ \sideset{^\cdot}{^+_{A''}}{{Q}}(x_0),A(\cdot) \right]
          }_-} \bigr)  (\bfm{\chi}d{\mathcal C},\bfm{h} d{\mathcal C}) =\
 {\left[
      \sideset{^{\bfm{\chi}d{\mathcal C}}}{_{A''}}{{Q^+}}(x_0),
      A(\bfm{h}d{\mathcal C}) \right]}_-\\
    & =
    \iint_{{\mathcal C}\times{\mathcal C}} {\left[ {k^+}^a_A (x) {{\mathcal
          J}^A}_{A''}(x,x_0) n_a (x),A(y) \right] }_- \bfm{\chi}(x) \,
          \bfm{h} (y) \,
    d{\mathcal C}_x d{\mathcal C}_y \\
    & = \iint_{{\mathcal C}\times{\mathcal C}} (-i)
    \bigl( {\Psi^+} \gamma^a ( - i \underset{x}{\notnabla} -m )
    {\bigr)}_{A} {[A(x),A(y)]}_- {{\mathcal J}^A}_{A''}(x,x_0) n_a(x)\
 \bfm{\chi}(x)
  \, \bfm{h} (y) \, d{\mathcal C}_x d{\mathcal C}_y \\
  \intertext{since $A$
    commutes with $B$ and $\Psi$}
  & = \iint_{{\mathcal C}\times{\mathcal C}}
  (-1) {\left( {\Psi^+}(x) \gamma^a \gamma^b \right)}_A {{\mathcal
      J}^A}_{A''} n_a {\nabla}_b {[A(x),A(y)]}_- (\bfm{\chi}(x)
  \bfm{h}(y))\\
  &
  \phantom{\iint_{{\mathcal C}\times{\mathcal C}}} + i m {\left( {\Psi^+}(x)
    \gamma^a \right)}_A {{\mathcal J}^A}_{A''} n_a {[A(x),A(y)]}_-\
 \bfm{\chi}(x)
  \bfm{h}(y) \, d{\mathcal C}_x d{\mathcal C}_y
  \end{split}
\end{equation}
We are dealing with free fields, hence on the GNS Hilbertspace we have
the identity:
\[ {[A(x),A(y)]}_- = i E(x,y) \identity, \]
where $\identity$ denotes the identity on the Hilbertspace. It was
shown in Proposition~\ref{Prop:restrictionParamscalar} that the
restriction of $E$ to ${\mathcal C}$ vanishes, whereas the restriction of
its derivative in direction $v^a$ is given by $-v^an_a$ times the
Dirac delta distribution by
Corollary~\ref{Cor:restrictionParamscalarDeriv}.  Therefore our result
is
\begin{equation*}
  \begin{split}
    (\ref{eq:commutatorA})
    & = i \int_{{\mathcal C}}
    {{\Psi^+}}_A {{\mathcal J}^A}_{A''}(x,x_0) \bfm{h}(x) d{\mathcal C}_x,
  \end{split}
\end{equation*}
since $\gamma^a \gamma^b n_a n_b = g^{a b} n_a n_a = +1$ and the
$y$-integration can be carried out explicitly. Note that the result
is independent of $\bfm{\chi}$ since we assumed
${\bfm{\chi}\restriction}_{\text{supp}h} \equiv 1$.
The pseudoscalar case (replace $\sideset{^A}{_{x_0}}{V}$ by
$\sideset{^B}{_{x_0}}{V}$) is --up to a factor $(-i\gamma^5)$--
identical to the previous one:
\begin{equation} \label{eq:commutorB}{}
  \begin{split}
    \left[ \sideset{^{\bfm{\chi}d{\mathcal C}}}{^+_{A''}}{Q}(x_0),
           B(\bfm{h}d{\mathcal C}) \right]_-
    & =
     \iint_{{\mathcal C}\times{\mathcal C}}
    (-i)
       \bigl( {\Psi^+} \gamma^a
         ( - i \underset{x}{\notnabla} -m )(-i \gamma^5)
       {\bigr)}_{A}
     {[A(x),A(y)]}_- \\
     & \phantom{= \iint_{{\mathcal C}\times{\mathcal C}}}
     {{\mathcal J}^A}_{A''}(x,x_0) n_a(x)
     \bfm{\chi}(x) \, \bfm{h} (y) \,  d{\mathcal C}_x d{\mathcal C}_y \\
    & = \int_{\mathcal C}
     {{\Psi^+}}_A(x) \gamma^5 {{\mathcal J}^A}_{A''}(x,x_0) \bfm{h}(x)
     \, d{\mathcal C}_x\\
   \end{split}
\end{equation}
Finally for the spinorial case we get:
\begin{equation}  \label{eq:commutatorPsi}
  \begin{split}
      \{ \Psi^{A'}(\bfm{h}_{A'}d{\mathcal C}),&
        \sideset{^{\bfm{\chi}d{\mathcal C}}}{^+_{A''}}{{Q}}(x_0) \}_+
     = \iint_{{\mathcal C}\times{\mathcal C}}
    (-i) \{ {\Psi^{A'}(y),{{\Psi^+}}_B (x) \} }_+
      \bigl(
          \notn ( - i \underset{x}{\notnabla} -m ) \AiB (x)
        {{\bigr)}^B}_A\\
    & \phantom{= \iint_{{\mathcal C}\times{\mathcal C}} }
      {{\mathcal J}^A}_{A''}(x,x_0) \bfm{h}_{A'}(y) \bfm{\chi}(x)
    d{\mathcal C}_x d{\mathcal C}_y \\
\intertext{Since the anti-commutator of $\Psi$ and
  ${\Psi^+}$ is given by ({Note the minus sign!})
  \[ { \{ \Psi^{A'}(y),{{\Psi^+}}_B(x) \} }_+ = -i {S^{A'}}_B(y,x) \identity
  \]
    one obtains
  }
    & = \iint_{{\mathcal C} \times {\mathcal C}}
   (-1) \bfm{h}_{A'}(y) {{S^{A'}}_B(y,x)}\\
    & \phantom{= \iint_{{\mathcal C} \times {\mathcal C}}}
    {{\bigl(
        \notn ( -i {\notnabla} -m ) \AiB (x)
      \bigr)}^B}_A {{\mathcal J}^A}_{A''} (x,x_0) \bfm{\chi}(x)
    \, d{\mathcal C}_y d{\mathcal C}_x \\
    & = \int_{\mathcal C} (-1)
    \bfm{h}_A(x) i \bigl(
     \underbrace{\notn \notn}_{=1} (- i {\notnabla} -m )
     \AiB {\mathcal J}
             {{\bigr)}^A}_{A''} \bfm{\chi}(x)
         \, d{\mathcal C}_x\\
\intertext{where the result of Proposition~\ref{Prop:DiracParam} was used.}
    & = \int_{\mathcal C} i
    \bfm{h}_A(x) {{\bigl(
        ( i {\notnabla} +m ) \AiB (x)
             \bigr) }^A}_B {{\mathcal J}^B}_{A''}(x,x_0)
         \, d{\mathcal C}_x\\
   \end{split}
\end{equation}
The last equation is valid since we assume $\bfm{\chi}\equiv 1$ on those\
 points
where $\bfm{h}_A$ does not vanish.
Let us consider now:
\begin{equation}\label{eq:commutatornablaA}
  \begin{split}
      \bigl[ \sideset{^{\bfm{\chi}d{\mathcal C}}}{^+_{A''}}{Q}(x_0),&
        \nabla_a A(\bfm{g}^ad{\mathcal C})
      \bigr]_-\\
     & = \iint_{{\mathcal C}\times{\mathcal C}}
    (-i)
       \bigl( {\Psi^+} \notn
         ( - i \underset{x}{\notnabla} -m )
       {\bigr)}_{A}
      i (\nabla_{a'} E(x,y))  \bfm{\chi}(x) \bfm{g}^{a'}(y)
     {{\mathcal J}^A}_{A''}(x,x_0)
     \,  d{\mathcal C}_x d{\mathcal C}_y \\
    & = \iint_{{\mathcal C}\times{\mathcal C}}
    {{\Psi^+}}_B (x) \notn
    \biggl(
      -i \gamma^b \underbrace{{h_b}^c {h_{a'}}^{c'}
        \nabla_c \nabla_{c'} E }_{\text{I}}
      -i \notn \underbrace{n^c {h_{a'}}^{c'}
        \nabla_c \nabla_{c'} E  }_{\text{II}}
       \\
    & \phantom{= \iint_{{\mathcal C}\times{\mathcal C}} {{\Psi^+}}_B (x)\
 \notn}
      -i \gamma^b n_{a'} \underbrace{{h_b}^{c}
        n^{c'} \nabla_c \nabla_{c'} E }_{\text{III}}
      -i \notn n_{a'} \underbrace{n^c n^{c'}
        \nabla_c \nabla_{c'} E }_{\text{IV}}
       \\
    & \phantom{= \iint_{{\mathcal C}\times{\mathcal C}} {{\Psi^+}}_B (x)\
 \notn}
      -m \underbrace{{h_{a'}}^{c'} \nabla_{c'} E }_{\text{V}}
      -m n_{a'} \underbrace{n^{c'} \nabla_{c'} E }_{\text{VI}}
    {{\biggr)}^B}_A \\
    & \phantom{= \iint_{{\mathcal C}\times{\mathcal C}} }
    \bfm{\chi}(x) \cdot \bfm{g}^{a'}(y)
    {{\mathcal J}^{A}}_{A''}
    d{\mathcal C}_x d{\mathcal C}_y
  \end{split}
\end{equation}
For the six terms (I-VI) we find the following results (As usual
$\tilde{\varphi}$ denotes the natural embedding of ${\mathcal\
 C}\times{\mathcal
  C}$ into $M\times M$):
\begin{equation}
  \begin{split}
    \text{I:}\qquad
    \tilde{\varphi}^*
    \left(
      {h_b}^c {h_{a'}}^{c'} \nabla_c \nabla_{c'} E
    \right)
    \left(
      {\tilde{\bfm{\chi}}}^b, \tilde{\bfm{g}}^{a'}
    \right)
    & = 0 \qquad \forall
          \tilde{\bfm{\chi}}^a,\tilde{\bfm{g}}^a \in D(T{\mathcal C})
  \end{split}
\end{equation}
by Corollary~\ref{Cor:restrictionParamscalarDeriv}.
\begin{equation}
  \begin{split}
    \text{II:}\qquad
    \tilde{\varphi}^*
    \left( n^c {h_{a'}}^{c'} \nabla_c
    \nabla_{c'} E \right)
    (\bfm{\chi}d{\mathcal C}, \bfm{g}^{a'}d{\mathcal C})
    & =
    {h_{a'}}^{c'} D_{c'} \tilde{\varphi}^*(n^c \nabla_c E)
    (\bfm{\chi}d{\mathcal C},\bfm{g}^{a'}d{\mathcal C})\\
    & =
    + \bfm{\chi}( D_{c}({h_{a}}^{c} \bfm{g}^{a}d{\mathcal C}),
  \end{split}
\end{equation}
since $\tilde{\varphi}^*(n^c \nabla_c E)(x,y) = - \bfm{\delta}(x,y)$
by Proposition~\ref{Prop:restrictionParamscalar}.
For the third term let us consider the whole integral expression:
\begin{equation*}
  \begin{split}
    \text{III:}\qquad
    & -i \tilde{\varphi}^*
    \left(
    {\left(
      {\Psi^+} \notn \gamma^b {\mathcal J}
    \right)}_{A''}
    \left(
      n_{a'} {h_b}^{c} n^{c'}  \nabla_c
      \nabla_{c'} E
    \right)
    \right)
    (\bfm{\chi}d{\mathcal C} , \bfm{g}^{a'}d{\mathcal C})\\
    & =
    \left(
      - i
      \tilde{\varphi}^*
      \left(
        {({\Psi^+} \notn \gamma^b {\mathcal J} )}_{A''}
      \right)
      \cdot
      {h_b}^{c} D_c
      \tilde{\varphi}^*
      ( n^{c'} \nabla_{c'} E )
    \right)
      (\bfm{\chi}d{\mathcal C},n_{a'} \bfm{g}^{a'}d{\mathcal C})\\
      & =
      + i
      D_c \left(
       {h_b}^c \tilde{\varphi}^*
        \left(
          {({\Psi^+} \notn \gamma^b
      {\mathcal J})}_{A''}
        \right)
        \right)
      \cdot
      \tilde{\varphi}^*(n^{c'}\nabla_{c'} E)
      (\bfm{\chi}d{\mathcal C},n_{a'} \bfm{g}^{a'}d{\mathcal C})\\
      & =
      +i {h_{b}}^c D_c
       \tilde{\varphi}^*
        \left(
          {({\Psi^+} \notn \gamma^b
            {\mathcal J})}_{A''}
        \right)
      (n_a \bfm{g}^a d{\mathcal C})\\
   \intertext{Note that $D_c$ is the covariant derivative on ${\mathcal
       C}$, hence $D_b h_{cd} = 0$ and we can apply Leibnitz rule on
     the derivative of products. The boundary term $D_c(\cdot)$
     vanishes since we assume $\bfm{\chi} \equiv 1$ on the support of
     $\bfm{g}^{a'}$. We also used that the normal derivative of $E$ with
     respect to the second entry is the Dirac delta distribution
     (compare VI:).}
      & =
      + i \int_{\mathcal C}
      n_a {h_b}^{c} \nabla_c
      {({\Psi^+} \notn \gamma^b {\mathcal J})}_{A''} g^a d{\mathcal C},
    \end{split}\notag
  \end{equation*}
 since ${h_b}^{c} D_c v^b = {h_b}^{c} \nabla_c v^b$ for all vector
  fields $v^b$ on ${\mathcal C}$.
The fourth terms
vanishes by Huygens principle,
since for the (distributional) solution $u_y$ of the Klein Gordon
Equation with the following initial conditions on ${\mathcal C}$
\begin{alignat*}{2}
  u_y\restriction_{\mathcal C} &= \bfm{\delta}(\cdot,y)
  && \quad
  (n^a \nabla_a u_y)\restriction_{\mathcal C} = 0
\end{alignat*}
we obtain by Proposition~\ref{Prop:Huygensdist}:
\[
D'(M) \ni u_y(\cdot) = n^{c'} \nabla_{c'} E (\cdot,y)
\]
\begin{equation}
    \text{IV:}\qquad
      \tilde{\varphi}^*( n^c n^{c'}\nabla_c \nabla_{c'} E )  = 0
\end{equation}
The fifth term vanishes by Proposition~\ref{Prop:restrictionParamscalar}:
\[
\text{V:}\qquad
\tilde{\varphi}^*({h_a}^c \nabla_c E) = 0.
\]
The result of IV: confirms
\begin{equation}
\text{VI:}\qquad
\tilde{\varphi}^* {(n^{c'} \nabla_{c'} E )}(x,y) = + \bfm{\delta}(x,y).
\end{equation}
Inserting these expressions in~(\ref{eq:commutatornablaA}) yields:
\begin{equation}
  \begin{split}
    (\ref{eq:commutatornablaA})
    & = \int_{{\mathcal C}}
    + i {h_a}^c D_c (\Psi^+ {\mathcal J})_{A''} \bfm{g}^a
    + i
    {h_{b}}^c D_c
       \left(
          {({\Psi^+} \notn \gamma^b
            {\mathcal J})}_{A''}
        \right)
      n_a \bfm{g}^a
    - m {({\Psi^+}\notn {\mathcal J})}_{A''} n_a \bfm{g}^a
    \, d{\mathcal C}
  \end{split}
\end{equation}
Since the result for $B$ differs by a factor $(-i\gamma^5)$ only, we
obtain for the corresponding kernel:
\[
[\sideset{^{\bfm{\chi}d{\mathcal C}}}{^+_{A''}}{Q}(x_0),
      \nabla_a B(x) ]_-
=
-i \left(
+ i D_a {({\Psi^+}\gamma^5 {\mathcal J})}_{A''}
+ i D_b {({\Psi^+}\notn \gamma^b \gamma^5 {\mathcal J})}_{A''} n_a
- m {({\Psi^+}\notn \gamma^5 {\mathcal J})}_{A''} n_a
\right)
\]
Finally for the spinorial case consider the map
\begin{equation}\label{eq:anticomderivSpinor}
  \begin{split}{}
    \bigl\{
      \nabla_a
      \Psi^A(\bfm{g}^a_A d{\mathcal C}),&
       \sideset{^{\bfm{\chi}d{\mathcal C}}}{^+_{A''}}{Q}(x_0)
    \bigr\}_+ \\
    & = \iint_{{\mathcal C}\times{\mathcal C}}
    {\bigl\{
      \nabla_a
      \Psi^A(x), -i
      {\bigl( {\Psi^+} \notn
        ( -i {\rvec{\notnabla}} -m )
      \bigr)}_{A'} {{\AiB}^{A'}}_{B'}(y)
    \bigr\} }_+ \\
    & \phantom{= \iint_{{\mathcal C}\times{\mathcal C}}}
    \bfm{g}^a_A(x) \bfm{\chi}(y) {{\mathcal J}^{B'}}_{A''}(y,x_0)
    \, d{\mathcal C}_x d{\mathcal C}_y \\
    & = \iint_{{\mathcal C}\times{\mathcal C}}
    (-i)  {\nabla_a}
    \underbrace{{\{ \Psi^A(x),{{\Psi^+}}_{A'}(y) \}}_+}_{= -i
      {S^A}_{A'}(x,y)}
    \bigl( \notn (-i \underset{y}{\notnabla} -m ) \AiB (y) {\mathcal J}
    {{\bigr)}^{A'}}_{A''}\\
    & \phantom{= \iint_{{\mathcal C}\times{\mathcal C}}}
    \bfm{g}^a_A (x) \bfm{\chi}(y)
    \, d{\mathcal C}_x d{\mathcal C}_y\\
    & =
    - \iint_{{\mathcal C}\times{\mathcal C}}
    \bigl(
    \notn \bigl( (-i \underset{y}{\notnabla} -m ) \AiB \bigr)
    {\mathcal J}
    {{\bigr)}^{A'}}_{A''} \cdot
    \left( \nabla_a {S^A}_{A'} (x,y) \right) \\
    & \phantom{= \iint_{{\mathcal C}\times{\mathcal C}}}
    \bfm{g}^a_A (x) \bfm{\chi} (y)
    \, d{\mathcal C}_x d{\mathcal C}_y \\
    \intertext{Inserting ${S^{A}}_{A'}(x,y) \equiv {E^A}_{B'}(x,y) (
          + i \underset{y}{\lvec{\notnabla}} - m {{)}^{B'}}_{A'}$, which
      is valid since ${S^A}_{A'}$ is the fundamental solution of the Dirac
      equation (Compare Proposition~\ref{Prop:Dimock2_2.1}), we get}\\
    & =
    - \iint_{{\mathcal C}\times{\mathcal C}}
    \bigl(
    \notn \bigl( (-i \underset{y}{\notnabla} -m ) \AiB \bigr)
    {\mathcal J}
    {{\bigr)}^{A'}}_{A''} \\
    & \phantom{= \iint_{{\mathcal C}\times{\mathcal C}}}
    \times \nabla_a {E^A}_{B'}
    (+ i \underset{y}{\lvec{\notnabla}} - m {)^{B'}}_{A'}
    \bfm{g}^a_A (x) \bfm{\chi} (y)
    \, d{\mathcal C}_x d{\mathcal C}_y \\
    \intertext{Similar to the calculation done above, we treat
      derivatives tangential and normal to ${\mathcal C}$ individually:}
    & =
    - \iint_{{\mathcal C}\times{\mathcal C}}
    \bigl(
    \notn \bigl( (-i \underset{y}{\notnabla} -m ) \AiB \bigr)
    {\mathcal J}
    {{\bigr)}^{A'}}_{A''} \\
    & \phantom{=}
    \Biggl(
      \underbrace{%
        {h_a}^c {h_{b'}}^{c'}
        \nabla_c \nabla_{c'} {E^A}_{B'}
        }_{I}
        {{(+i \gamma^{b'})}^{B'}}_{A'}
    +
    \underbrace{%
      {h_a}^c n_{b'} n^{c'}
      \nabla_c \nabla_{c'} {E^A}_{B'}
      }_{II}
      {{(+i\gamma^{b'})}^{B'}}_{A'}\\
    &\phantom{= \Biggl( }
    +
    \underbrace{%
      n_a n^c {h_{b'}}^{c'}
      \nabla_c \nabla_{c'} {E^A}_{B'}
      }_{III}
      {{(+i\gamma^{b'})}^{B'}}_{A'}
    +
    \underbrace{%
        n_a n^c n_{b'} n^{c'}
        \nabla_c \nabla_{c'} {E^A}_{B'}
        }_{IV}
        {{(+i\gamma^{b'})}^{B'}}_{A'}\\
    & \phantom{= \Biggl( }
    - m (
    \underbrace{{h_a}^c \nabla_c {E^A}_{A'}}_{V} +
    \underbrace{n_a n^c \nabla_c {E^A}_{A'}}_{VI}
        )
    \Biggr)
    \bfm{g}^a_A (x) \bfm{\chi}(y)
    \, d{\mathcal C}_x d{\mathcal C}_y
  \end{split}
\end{equation}
The first term (I:) vanishes since we can apply
Lemma~\ref{Lemma:derivTangCFermi}.
\begin{equation*}
  \begin{split}
    \text{I:} \qquad \tilde{\varphi}^*(
           {h_a}^c {h_{a'}}^{c'} \nabla_c \nabla_{c'}
           {E^A}_{A'})(\tilde{\bfm{\chi}}^a_A, \tilde{\bfm{g}}^{a'A'})
           & = \tilde{\varphi}^*({E^A}_{A'})\bigl( D_c({h_a}^c
                 \tilde{\bfm{\chi}}^a_A),
           D_{c'}( {h_{a'}}^{c'} \tilde{\bfm{g}}^{a'A'})\bigr)\\
           & = 0 \qquad \forall \tilde{\bfm{\chi}}^a_A \in D(D_{\mathcal
             C}M \otimes T^*{\mathcal C}), \tilde{\bfm{g}}^{A'a} \in
           D(D^*_{\mathcal C}M \otimes T^*{\mathcal C})
  \end{split}
\end{equation*}
For the second term let us consider the whole integral expression:
\begin{equation*}{}
  \begin{split}{}
    \text{II:}\qquad
    & - \iint_{{\mathcal C}\times{\mathcal C}}
    \bigl(
    i \notn \notn \bigl( (-i \underset{y}{\notnabla} -m ) \AiB \bigr)
    {\mathcal J}
    {{\bigr)}^{A'}}_{A''}
    {h_a}^b n^{c'} \nabla_c\nabla_{c'} {E^A}_{A'}
    \bfm{g}^a_A(x) \bfm{\chi}(y) d{\mathcal C}_x d{\mathcal C}_y\\
    & =
    + \iint_{{\mathcal C}\times{\mathcal C}}
    \bigl(
    i \bigl( (-i \underset{y}{\notnabla} -m ) \AiB \bigr)
    {\mathcal J}
    {{\bigr)}^{A'}}_{A''}
    n^{c'} \nabla_{c'} {E^A}_{A'}
    D_c({h_a}^{c} \bfm{g}^a_A) \bfm{\chi}(y) d{\mathcal C}_x d{\mathcal\
 C}_y\\
    \intertext{by Lemma~\ref{Lemma:derivTangCFermi}}
    & =
    + \int_{{\mathcal C}}
    \bigl(
    i \bigl( (-i \underset{x}{\notnabla} -m ) \AiB \bigr)
    {\mathcal J}
    {{\bigr)}^{A}}_{A''}
    D_c({h_a}^{c} \bfm{g}^a_A)  d{\mathcal C}_x,\\
    \intertext{since $\bfm{\chi}\equiv 1$ on the support of
      $\bfm{g}^a_A$
      and the normal derivative of ${E^A}_{A'}$ with respect to the {\em
        second\/} entry restricted to ${\mathcal C}$ is plus the Dirac
      delta distribution by Lemma~\ref{Lem:restr_fermi}}
    & =: - \int_{\mathcal C}
    {h_a}^c D_c \bigl(
    i \bigl( (-i \underset{x}{\notnabla} -m ) \AiB \bigr)
    {\mathcal J}
    {{\bigr)}^{A}}_{A''}
    \bfm{g}^a_A d{\mathcal C}_x \\
    & =
    \varphi^*\biggl( - i {h_a}^c \nabla_c
    \Bigl(
     \bigl( (-i \underset{x}{\notnabla} -m ) \AiB \bigr)
    {\mathcal J}
    {{\Bigr)}^{A}}_{A''}
    \biggr) (\bfm{g}^a_A)
  \end{split}
\end{equation*}
The third term yields:
\begin{equation*}
  \begin{split}
    \text{III:}\qquad
    &
    - \iint_{{\mathcal C}\times{\mathcal C}}
    \bigl( i \gamma^{b'}
    \notn \bigl( (-i {\notnabla} -m ) \AiB \bigr)
    {\mathcal J}
    {{\bigr)}^{A'}}_{A''}
    n^c {h_{b'}}^{c'} \nabla_c \nabla_{c'} {E^A}_{A'}
    n^a \bfm{g}^a_A(x) \bfm{\chi}(y) d{\mathcal C}_x d{\mathcal C}_y\\
    & =
    + \iint_{{\mathcal C}\times{\mathcal C}}
    D_{c'} \bigl( {h^{c'}}_{b'} \bfm{\chi}(y) \gamma^{b'} i
    \notn \bigl( (-i {\notnabla} -m ) \AiB \bigr)
    {\mathcal J}
    {{\bigr)}^{A'}}_{A''}
    n^c \nabla_c {E^A}_{A'}
    n^a \bfm{g}^a_A(x) d{\mathcal C}_x d{\mathcal C}_y\\
    \intertext{since ${h_b}^{c} \nabla_c \equiv {h_b}^{c} D_c$ for
      cospinors. This allows the necessary integartion by parts. The
      boundary term vanishes, since $\bfm{\chi}$ and $\bfm{g}^a_A$ have\
 compact
      support.}
    & =
    -
    \int_{\mathcal C}
    \gamma^b {h_b}^c D_c
    \bigl( i
    \notn \bigl( (-i {\notnabla} -m ) \AiB \bigr)
    {\mathcal J}
    {{\bigr)}^{A}}_{A''}
    n_a \bfm{g}^a_A d{\mathcal C},\\
    \intertext{since the normal derivative of ${E^A}_{A'}$ with
      respect to the first entry restricted to ${\mathcal C}$ is minus the
      Dirac delta distribution by Lemma~\ref{Lem:restr_fermi}.
      Recall that $\bfm{\chi}$ is one
      on the support of $\bfm{g}^a_A$ and note that the metric $h_{ab}$ is
      covariantly constant for $D_c$. Moreover, the same result holds
      for the Dirac $\gamma$-matrices, i.e., $D_c \gamma^b = {h_c}^d
      {h^b}_d \nabla_d \gamma^d = 0$.}
    & = \varphi^*
    \biggl(
    -i n_a \gamma^b h{_b}^c \nabla_c
    \bigl(
    \notn \bigl( (-i {\notnabla} -m ) \AiB \bigr)
    {\mathcal J}
    {{\bigr)}^{A}}_{A''}
    \biggr)
    (\bfm{g}^a_A)
  \end{split}
\end{equation*}
The forth term is shown to vanish by using the same argument as in the
scalar case: Consider the unique solution ${{u_y}^A}_{B'} : D(DM)
\rightarrow \Gamma(D^*M)$ of the SUSY-coupled Klein Gordon operator
with distributional initial data ${{u_y}^A}_{B'} \restriction_{{\mathcal
    C}}$ is the Dirac delta distribution and $n^a \nabla_a
{{u_y}^At}_{B'}\restriction_{\mathcal C}$ vanishes. Inserting these
initial data into Eqn.~(\ref{eq:HuygensFermi}), we get
\[
{{u_y}^A}_{B'} = {{(n^{c'} \nabla_{c'} E)}^A}_{B'}
\]
which verifies the statement, i.e.,
\[
\text{IV:}\qquad
\tilde{\varphi}^* (n^c n^{c'} \nabla_c \nabla_{c'} {E^A}_{B'}) = 0.
\]
By Lemma~\ref{Lem:restr_fermi} we get for the fifth term:
\[
\text{V:}\qquad \tilde{\varphi}^*( {h_a}^c \nabla_c {{E^A}_{A'}}) = 0
\]
Finally, since the normal derivative of ${E^A}_{A'}$ with respect to
the first entry restricted to ${\mathcal C}$ is minus the Dirac delta
distribution, the sixth term is
\begin{equation}
  \begin{split}
    \text{VI:}\qquad
    & \iint_{{\mathcal C}\times{\mathcal C}}
    + m
    \bigl(
     \notn
     \bigl(
      ( - i \notnabla -m ) \AiB
     \bigr)
     {\mathcal J}
    {\bigr)^{A'}}_{A''}
    n^c \nabla_c {E^A}_{A'} n_a \bfm{g}^a_A \bfm{\chi}(y)
    d{\mathcal C}_x d{\mathcal C}_y\\
    & =
    \int_{\mathcal C}
    - m
    \bigl(
     \notn
     \bigl(
      ( - i \notnabla -m ) \AiB
     \bigr)
     {\mathcal J}
    {\bigr)^{A}}_{A''}
     n_a \bfm{g}^a_A
    d{\mathcal C}_x\\
    & =
    \varphi^*\biggl(
    - m n_a
    \bigl(
     \notn
     \bigl(
      ( - i \notnabla -m ) \AiB
     \bigr)
     {\mathcal J}
    {\bigr)^{A}}_{A''}
    \biggr)
    (\bfm{g}^a_A)
  \end{split}
\end{equation}
To summarize:
\begin{equation}
  \begin{split}
    (\ref{eq:anticomderivSpinor})
    & =
    \varphi^*
    \biggl(
% II:
    - i {h_a}^c \nabla_c
    \Bigl(
     \bigl( (-i \underset{x}{\notnabla} -m ) \AiB \bigr)
    {\mathcal J}
    {{\Bigr)}^{A}}_{A''}\\
    & \phantom{= \varphi^*\biggl(}
% III:
    -i n_a \gamma^b h{_b}^c \nabla_c
    \bigl(
    \notn \bigl( (-i {\notnabla} -m ) \AiB \bigr)
    {\mathcal J}
    {{\bigr)}^{A}}_{A''}\\
    & \phantom{= \varphi^*\biggl(}
% VI:
    - m n_a
    \bigl(
     \notn
     \bigl(
      ( - i \notnabla -m ) \AiB
     \bigr)
     {\mathcal J}
    {\bigr)^{A}}_{A''}
    \biggr)
    (\bfm{g}^a_A)
  \end{split}
\end{equation}
\end{pf*}
% The use of j instead of k in Q should not change to much
% LUECKE
%
%%%%%%%%
\subsection{Traces for the `supercharge'}
\label{sec:traceSuperCharge}
%%%%%%%%
%
In this subsection we calculate the traces of the anti-com\-muta\-tor of
the `supercharge' with various point-separated currents. The results of
the following Proposition will be used in the next section to
calculate the expectation value of our energy-momentum tensor
(Definition~\ref{Dfn:pointsplitEMTcurved} below) in a locally SUSY
Hadamard product state.
We use the results of Proposition~\ref{prop:commutQCurv}  to
calculate the following traces:
\begin{Prop}\label{prop:trCommQCurv}
  Consider two points $x$ and $y$ on ${\mathcal C}$. The corresponding
  vector and spinor indices are denoted by $a,A$ and $a',A'$
  respectively and a semicolon is a shorthand for covariant
  differentiation. Then we have the following three identities in the sense
  of distributional kernels on ${\mathcal C}$.
  \begin{align}
%%%%%%%%%%%%%%%%%%%%
%      1           %
%%%%%%%%%%%%%%%%%%%%
    \begin{split}{}
      & \phantom{=}
      \Tr{}
          \left( {\gamma^{aD}}_{A}
            \left\{ i \bigl( \AiB (i \lvec{\notnabla} -m ) \gamma^b
              {\bigr)^A}_C {{\mathcal J}^C}_{C'} \Psi^{C'},
              \sideset{^{\mathcal C}}{^+_{B}}{G}
            \right\}_+
          \right)\\
      & =
              2 \biggl(
        {g^b}_{b'}
        \bigl(
        \{A^{;a},A^{;b'}\}_+ +
        \{B^{;a},B^{;b'}\}_+
        \bigr)
        +
        {g^a}_{a'}
        \bigl(
        \{A^{;b},A^{;a'}\}_+ +
        \{B^{;b},B^{;a'}\}_+
        \bigr)\\
        & \phantom{=2 \biggl(}
        -
        g^{ab} {g^c}_{c'}
        \bigl(
        \{A_{;c}, A^{;c'}\}_+ + \{B_{;c}, B^{;c'}\}_+
        \bigr)
        + m^2
        g^{ab}
        \bigl(
        \{A,A'\}_+ + \{B,B'\}_+
        \bigr)\\
% eps korrektur term
        & \phantom{=2 \biggl(}
        + 2 {g^{d'}}_d \epsilon^{abcd} ( A^{;c} B^{;d'} - B^{;c}A^{;d'})\\
        & \phantom{=2 \biggl(}
        + i \Tr\bigl( \gamma^b {\mathcal J} [ \Psi', \Psi^{+;a}]_- \bigr)
        + i K \Tr\bigl(
          (\gamma^a \notn\gamma^b {\mathcal J}) [\Psi', \Psi^+]_-
                 \bigr)
        \biggr)
    \end{split}\label{eq:A1}\\
%%%%%%%%%%%%%%%%%%%%
%      2           %
%%%%%%%%%%%%%%%%%%%%
    \intertext{~} % Abstand
    \begin{split}
      & \phantom{=}
      \Tr
      \left( {\gamma^{aD}}_{A}
        \left\{ \bigl( ( \gamma^b \notnabla - \notnabla \gamma^b ) \AiB
        {\bigr)^A}_{C} {{\mathcal J}^C}_{C'} \Psi^{C'} , \sideset{^{\mathcal
            C}}{^+_{B}}{G}
        \right\}_+
      \right)\\
      & =
      + 4
      \biggl({}
      - g^{ab} {g_c}^{c'}
      \bigl( \{A^{;c}, A^{;c'} \}_+ \{B^{;c},B^{;c'} \}_+\bigr)
      + {g^{a}}_{a'} \bigl( A^{;a'} A^{;b} + B^{;a'} B^{;b}\bigr)\\
      & \phantom{=+4\biggl(}
      + {g^{b}}_{b'} \bigl( A^{;a} A^{;b'} + B^{;a} B^{;b'}\bigr)
% eps korrektur term
      + 2 {g^{d'}}_d \epsilon^{abcd} ( A^{;c} B^{;d'} - B^{;c}A^{;d'})\\
      & \phantom{=+4\biggl(}
      - i \Tr
      \bigl(
      \gamma^a G^{[bc]} {\mathcal J} \left[ \Psi', {\Psi{}^+}_{;c} \right]_-
      \bigr)
      \biggr)\\
    \end{split}\label{eq:A2}\\
%%%%%%%%%%%%%%%%%%%%
%      3           %
%%%%%%%%%%%%%%%%%%%%
    \intertext{~} % Abstand
    \begin{split}
      & \phantom{=}
      \Tr
      \left( {\gamma^{aD}}_{A} {{\mathcal J}^A}_{A'}
        \left\{ {{( A' - i \gamma^{5'} B' )}^{A'}}_{C'} {{\mathcal\
 J}^{C'}}_{C}
          \bigl( (\gamma^b \notnabla - \notnabla\gamma^b) \Psi
          \bigr)^C {{\mathcal J}^C}_{C'} , \sideset{^{\mathcal C}}{^+_{B}}{G}
        \right\}_+
      \right)\\
      & =
      4
      \bigl(
      \left\{ A^{;ab} , A'  \right\}_+ + \left\{ B^{;ab} , B'  \right\}_+
      \bigr)
      +
      4 ( g^{ab} m^2 + \xi R n^a n^b)
      (\left\{ A,A'\right\}_+ + \left\{ B,B'\right\}_+ \\
      & \phantom{=}
      - 4 i \Tr
      \bigl(
      {\mathcal J} \gamma^a G^{[bc]} \left[ \nabla_c \Psi, {\Psi'}^+\
 \right]_-
      \bigr)\\
      & \phantom{=}
      - 2 K
      \Tr
      \biggl(
      \gamma^a G^{[bc]}n_c
      \left\{ \notnabla \AiB,  (A' - i \gamma^5 B' )\right\}_+
      \biggr),\\
    \end{split}\label{eq:A3}
\end{align}
where $\Tr$ means the trace over the suppressed spinor indices, $K$
denotes the extrinsic curvature of ${\mathcal C}$, $G^{[AB]} =
\frac{1}{4}(\gamma^a\gamma^b - \gamma^b \gamma^a)$, $\epsilon^{abcd}$
is the totally anti-symmetric $\epsilon$ tensor on $(M,g_{ab})$ and
$\xi$ denotes the coupling of the Klein Gordon equation for the fields
$A$ and $B$.
\end{Prop}
\begin{pf}
  We start with the first equation.  Write the anti-commutator as a
  sum of commutators and anti-commutators in the basic fields with
  $\Gp{B}$.
  \begin{align*}
    \begin{split}{}
      & \phantom{=}
      \Tr{}
      \left( {\gamma^{aD}}_{A}
        \left\{ i \bigl( \AiB (i \lvec{\notnabla} -m ) \gamma^b
          {\bigr)^A}_C {{\mathcal J}^C}_{C'} \Psi^{C'},
          \sideset{^{\mathcal C}}{^+_{B}}{G}
        \right\}_+
      \right)
    \end{split}\\ % \tag{\ref{eq:A1}}
    \begin{split}
      & =
      \frac{1}{2} \Tr{}
      \left(
        {\gamma^{aD}}_{A} i
        \bigl( \AiB ( i \lvec{\notnabla} - m )\gamma^b {\bigr)^A}_C\
 {{\mathcal
            J}^C}_{C'}
        \left\{ \Psi^{C'}, \Gp{B} \right\}
      \right)
      \end{split}
      \tag{\ref{eq:A1} A}\\
      \begin{split}
      & \phantom{=}
      + \frac{1}{2}
      \left[
      \Gp{D} {\gamma^{aD}}_{A},
      i \bigl( \AiB ( i \lvec{\notnabla} - m ) \gamma^b {\bigr)^A}_C
      \right]_-
      {{\mathcal J}^C}_{C'} \Psi^{C'}
      \end{split}
      \tag{\ref{eq:A1} B}\\
      \begin{split}
      & \phantom{=}
      + \frac{1}{2}
      \Tr
      \left(
        {{\mathcal J}^C}_{C'} \left\{ \Psi^{C'}, \Gp{D} \right\}_+
        {\gamma^{aD}}_{B}
        i \bigl( \AiB ( i \lvec{\notnabla} - m ) \gamma^b {\bigr)^B}_A
      \right)
      \end{split}
      \tag{\ref{eq:A1} C}\\
      \begin{split}
      & \phantom{=}
      - \frac{1}{2}
      \Tr
      \left(
        {{\mathcal J}^{C}}_{C'} \Psi^{C'}
        \left[
          (\sideset{^{\mathcal C}}{^+}{G} \gamma^{a})_{A},
          i \bigl( \AiB ( i \lvec{\notnabla} - m ) \gamma^b {\bigr)^A}_B
        \right]_-
      \right)
      \end{split}
      \tag{\ref{eq:A1} D}\\ \nonumber
  \end{align*}
Consider the part~(\ref{eq:A1}A):
\begin{equation*}
  \begin{split}{}
    &\phantom{=}
    \frac{1}{2} \Tr{}
      \left(
        {\gamma^{aD}}_{A} i
        \bigl( \AiB ( i \lvec{\notnabla} - m )\gamma^b {\bigr)^A}_C\
 {{\mathcal
            J}^C}_{C'}
        \left\{ \Psi^{C'}, \Gp{B} \right\}_+
      \right)\\
    & =
    \frac{1}{2} \Tr{}
      \left(
        {\gamma^{aD}}_{A} i
        \bigl( \AiB ( i \lvec{\notnabla} - m )\gamma^b {\bigr)^A}_C
        {{\mathcal J}^C}_{C'}
        (+i) \bigl( ( i \notnabla + m ) \AiB {\bigr)^{C'}}_{B'}
        {{\mathcal J}^{B'}}_{B}
      \right)\\
    & =
    \frac{1}{2}
    \Tr
    \Biggl(
    i {\gamma^{aD}}_{A}
    \biggl( \gamma^c \nabla_c ( A + i \gamma^5 B) \gamma^b {\biggl)^A}_{C}
    {{\mathcal J}^C}_{C'}
    (-i)
    \biggl(
    \gamma^{d'} \nabla_{d'} (A' - i \gamma^{5'} B')
    {\biggl)^{C'}}_{B'} {{\mathcal J}^{B'}}_{B}\\
    & \phantom{= \frac{1}{2} \Tr \Bigl(}
    +
    i {\gamma^{aD}}_{A}
    \biggl( \gamma^c \nabla_c ( A + i \gamma^5 B) \gamma^b {\biggl)^A}_{C}
    {{\mathcal J}^C}_{C'}
    (-m)
    \biggl(
    (A' - i \gamma^{5'} B')
    {\biggl)^{C'}}_{B'}{{\mathcal J}^{B'}}_{B}\\
    & \phantom{= \frac{1}{2} \Tr \Bigl(}
    - m {\gamma^{aD}}_{A}
    \bigl( \AiB \gamma^b {\bigr)^A}_{C}
    {{\mathcal J}^C}_{C'}
    (-i)
    \biggl(
    \gamma^{d'} \nabla_{d'} (A' - i \gamma^{5'} B')
    {\biggl)^{C'}}_{B'}{{\mathcal J}^{B'}}_{B}\\
    & \phantom{= \frac{1}{2} \Tr \Bigl(}
    - m {\gamma^{aD}}_{A}
    \bigl( \AiB \gamma^b {\bigr)^A}_{C}
    {{\mathcal J}^C}_{C'}
    (-m)
    \biggl(
    (A' - i \gamma^{5'} B')
    {\biggl)^{C'}}_{B'}{{\mathcal J}^{B'}}_{B}
    \Biggr)\\
    & =
    \frac{1}{2}
    \Tr
    \bigl( \gamma^a \gamma^c \gamma^b \gamma^d \bigl)
    {g^{d'}}_d
    \bigl(
    \nabla_c A \nabla_{d'} A' + \nabla_c B \nabla_{d'} B'
    \bigr)\\
    & \phantom{=}
    +
    \frac{1}{2} i \Tr
    \bigl( \gamma^a \gamma^c \gamma^b \gamma^d \gamma^5 \bigl)
    {g^{d'}}_d
    \bigl(
    \nabla_c B \nabla_{d'} A' - \nabla_c A \nabla_{d'} B'
    \bigr)
    +
    \frac{1}{2}
    \Tr
    \bigl(
    \gamma^a \gamma^b
    \bigl)
    m^2 (A A' + B B')\\
    \intertext{due to the cyclicity of the trace and  since ${{\mathcal
          J}^{C}}_{C'} {\gamma^{d'C'}}_{A'} {{\mathcal J}^{A'}}_A =
      {g^{d'}}_{d} {\gamma^{dC}}_{A}$ by Eqn.~(\ref{eq:gammaParallel}),
      $(\gamma^5)^2 = 1$ and the trace over an odd number of
      $\gamma$-matices vanishes.}
    & =
    +2
    \biggl(
    {g^b}_{b'}
    (\nabla^a A \nabla^{b'} A'
    +
    \nabla^a B \nabla^{b'} B')
    -
    g^{ab} {g^c}_{c'}
    ( \nabla_c A \nabla^{c'} A'
    + \nabla_c B \nabla^{c'} B')\\
    & \phantom{=+2\biggl(}
    +
    {g^{a}}_{a'}
    ( \nabla^b A \nabla^{a'} A'
    + \nabla^b B \nabla^{a'} B')
    +
    {g^{d'}}_d \epsilon^{abcd}
      ( \nabla_c A \nabla_{d'} B' - \nabla_c B \nabla_{d'} A' )\\
    & \phantom{=+2\biggl(}
    +
    m^2  g^{ab} ( A A' + B B')
    \biggr),
  \end{split}
\end{equation*}
since the parallel transport commutes with contraction. The various
traces for the $\gamma$-matrices can be found in
Appendix~\ref{sec:usefulForm} section~\ref{sec:gammaident}.
For the next part~(\ref{eq:A1}B) one obtains:
\begin{equation*}{}
  \begin{split}
    &\phantom{=}
     \frac{1}{2}
    \left[
      \Gp{D} {\gamma^{aD}}_{A},
      i \bigl( \AiB ( i \lvec{\notnabla} - m ) \gamma^b {\bigr)^A}_C
    \right]_-
    {{\mathcal J}^C}_{C'} \Psi^{C'}\\
    & =
    \frac{1}{2} i
    \biggl({}
    [ \Gp{D}, i \nabla_c A ]_- \left( {\gamma^{aD}}_{A}
    {\gamma^{cA}}_{C} \right)
    - i
    [ \Gp{D}, i \nabla_c B ]_- \left( {\gamma^{aD}}_{A}
    (\gamma^5 \gamma^{c} {)^{A}}_{C} \right)\\
    & \phantom{= \frac{1}{2}i \biggl(}
    - m
    [ \Gp{D}, A ]_- {\gamma^{aD}}_{C}
    + i m
    [ \Gp{D}, B ]_-
    (\gamma^5 \gamma^{a} {)^{D}}_{C}
    \biggr){}
    {{\mathcal J}^C}_{C'} \Psi^{C'}\\
    & =
    -
    \biggl(
    \bigl( i (D_c \Psi^+) + i n_c (D_d \Psi^+) \notn \gamma^d
     - m n_c \Psi^+ \bigr)_{D}
    \bigl( \gamma^{a}  \gamma^c \gamma^b {\mathcal J} \Psi'
    \bigr)^{D}
    - m \Psi^+ \gamma^{a} \gamma^b {\mathcal J} \Psi'
    \biggr)\\
    & \phantom{=}
    -
    i n_c \Psi^+ (D_d \notn) \gamma^d
    \bigl(
    \gamma^{a} \gamma^c \gamma^b {\mathcal J} \Psi'
    \bigr)^{D'},\\
    \intertext{since $({\gamma^5})^2 = 1$. Note that $D_d(\notn)\gamma^d$
     equals $K$, which is the extrinsic curvature scalar.}
    & =
    -
    \biggl(
    i (\nabla_c \Psi^+) \gamma^a \gamma^c \gamma^b {\mathcal J} \Psi'
    - m \Psi^+ \gamma^a \gamma^b {\mathcal J} \Psi'
    \biggr) - i K \Psi^+\gamma^a \notn\gamma^b {\mathcal J} \Psi'\\
    & =
    - 2 i \nabla^a \Psi^+ \gamma^b {\mathcal J} \Psi'
    - i K \Psi^+\gamma^a \notn\gamma^b {\mathcal J} \Psi'\\
    \intertext{where we used the Dirac equation $\Psi^+
      (i \lvec{\notnabla} + m ) = 0$ for $\Psi^+$.}
  \end{split}
\end{equation*}
The result for the third part (\ref{eq:A1}C) is:
\begin{equation*}{}
  \begin{split}{}
    & \phantom{=}
    + \frac{1}{2}
    \Tr
    \left(
      {{\mathcal J}^C}_{C'} \left\{ \Psi^{C'}, \Gp{D} \right\}_+
      {\gamma^{aD}}_{A}
      i \bigl( \AiB ( i \lvec{\notnabla} - m ) \gamma^b {\bigr)^A}_B
    \right)\\
    & =
    + \frac{1}{2}
    \Tr
    \biggl(
    -i {\mathcal J} \gamma^{c'} \nabla_{c'} ( A' - i \gamma^{5'} B' )\
 {\mathcal J}
    \gamma^{a}  (i) \gamma^d \nabla_d ( A + i \gamma^5 B)
    \gamma^b \\
    & \phantom{= + \frac{1}{2} \Tr \biggl(}
    - i {\mathcal J} \gamma^{c'}\nabla_{c'} ( A' - i \gamma^{5'} B'\
 ){\mathcal J}
    \gamma^{a} (-m) \AiB \gamma^b\\
    & \phantom{= + \frac{1}{2} \Tr \biggl(}
    -m {\mathcal J} (A' - i \gamma^{5'} B' ){\mathcal J} \gamma^{a}  (i)
    \gamma^c \nabla_c ( A + i \gamma^5 B) \gamma^b \\
    & \phantom{= + \frac{1}{2} \Tr \biggl(}
    -m {\mathcal J} (A' - i \gamma^{5'} B' ) {\mathcal J}\gamma^{a}  (-m)
    \AiB \gamma^b
    \biggr)\\
    & =
    +2
    \biggl( {g^{a}}_{a'}
    (\nabla^{a'} A' \nabla^{b} A
    +
    \nabla^{a'} B' \nabla^{b} B)
    -
    g^{ab} {g^c}_{c'}
    ( \nabla_{c'} A' \nabla^{c} A
    + \nabla_{c'} B' \nabla^{c} B)\\
    & \phantom{= + 2 \biggl(}
    +
    {g^b}_{b'}
    ( \nabla^{'b} A' \nabla^{a} A
    + \nabla^{b'} B' \nabla^{a} B)\\
    & \phantom{= + 2 \biggl(}
    +
    {g^{d'}}_d \epsilon^{abcd}
    ( \nabla_c A \nabla_{d'} B' - \nabla_c B \nabla_{d'} A' )
    +
    m^2 g^{ab} ( A' A + B' B')
    \biggr),
  \end{split}
\end{equation*}
since up to a commutation of $x$ and $y$ the same terms as in
(\ref{eq:A1}A) occur.
Finally for the term~(\ref{eq:A1}D) one finds:
\begin{equation*}{}
  \begin{split}
    & \phantom{=}
    - \frac{1}{2}
      \Tr
      \left(
        {{\mathcal J}^{C}}_{C'} \Psi^{C'}
        \left[
          (\sideset{^{\mathcal C}}{^+}{G} \gamma^{a})_{A},
          i \bigl( \AiB ( i \lvec{\notnabla} - m ) \gamma^b {\bigr)^A}_B
        \right]_-
      \right)\\
    & =
    - \Tr
    \biggl(
    {\mathcal J} \Psi' (-2i) \nabla^a \Psi^+ \gamma^b
    - i K {\mathcal J} \Psi' \Psi^+ \gamma^a \notn \gamma^b
    \biggr),
  \end{split}
\end{equation*}
since all terms of (\ref{eq:A1}B) can be used again.
To summarize the results:
\begin{equation*}{}
  \begin{split}{}
    (\ref{eq:A1})
    & =
    %A1A
        +2
        \biggl(
        {g^b}_{b'}
        (\nabla^a A \nabla^{b'} A'
        +
        \nabla^a B \nabla^{b'} B')
        -
        g^{ab} {g^c}_{c'}
        ( \nabla_c A \nabla^{c'} A'
        + \nabla_c B \nabla^{c'} B')\\
        & \phantom{=+2\biggl(}
        +
        {g^a}_{a'}
        ( \nabla^b A \nabla^{a'} A'
        + \nabla^b B \nabla^{a'} B')
        +
        {g^{d'}}_d \epsilon^{abcd}
        ( \nabla_c A \nabla_{d'} B' - \nabla_c B \nabla_{d'} A' )\\
        & \phantom{=+2\biggl(}
        +
        m^2 g^{ab} ( A A' + B B')
        \biggr)\\
%A1B
        & \phantom{=}
        -2 i \Tr
        \bigl( \gamma^b  (\nabla^a \Psi^+ {\mathcal J} \Psi'
        \bigr)
        -
        \bigl(
        i K \Psi^+ \gamma^a \notn\gamma^b {\mathcal J} \Psi'
        \bigl)\\
%A1C
        & \phantom{=}
        +2
        \biggl(
        {g^{a}}_{a'}
        (\nabla^{a'} A' \nabla^{b} A
        +
        \nabla^{a'} B' \nabla^{b} B)
        -
        g^{ab} {g^c}_{c'}
        ( \nabla_{c'} A' \nabla^{c} A
        + \nabla_{c'} B' \nabla^{c} B)\\
        & \phantom{= + 2 \biggl(}
        +
        {g^b}_{b'}
        ( \nabla^{'b} A' \nabla^{a} A
        + \nabla^{b'} B' \nabla^{a} B)
        +
        {g^{d'}}_d \epsilon^{abcd}
        ( \nabla_c A \nabla_{d'} B' - \nabla_c B \nabla_{d'} A' )\\
        & \phantom{= + 2 \biggl(}
        +
        m^2 g^{ab} ( A' A + B' B')
        \biggr)\\
% A1D
        & \phantom{=}
        + 2 i  \Tr
        \bigl(
        {\mathcal J} \Psi' \nabla^a \Psi^+ \gamma^b
        \bigr)
        +
        \Tr
        \bigl(
        i K {\mathcal J} \Psi' \Psi^+ \gamma^a \notn\gamma^b
        \bigr)\\
%%%%%%%%%
        & =
        2 \biggl(
        {g^b}_{b'}
        \bigl(
        \{A^{;a},A^{;b'}\}_+ +
        \{B^{;a},B^{;b'}\}_+
        \bigr)
        +
        {g^a}_{a'}
        \bigl(
        \{A^{;b},A^{;a'}\}_+ +
         \{B^{;b},B^{;a'}\}_+
        \bigr)\\
        & \phantom{=2 \biggl(}
        -
        g^{ab} {g^c}_{c'}
        \bigl(
        \{A_{;c}, A^{;c'}\}_+ + \{B_{;c}, B^{;c'}\}_+
        \bigr)
        + m^2
        g^{ab}
        \bigl(
        \{A,A'\}_+ + \{B,B'\}_+
        \bigr)\\
        & \phantom{=2 \biggl(}
        + 2 {g^{d'}}_{d} \epsilon^{abcd}
        ( A^{;c} B^{d'} - B^{;c} A^{d'})
        + i \Tr\bigl( \gamma^b {\mathcal J} [ \Psi', \Psi^{+;a}]_- \bigr)\\
        & \phantom{=2 \biggl(}
        + i K \Tr\bigl(
          (\gamma^a \notn\gamma^b {\mathcal J}) [\Psi', \Psi^+]_-
                 \bigr)
        \biggr)
  \end{split}
\end{equation*}
We continue with the same calculation for the second Equation:
  \begin{align*}
    \begin{split}{}
      & \phantom{=}
      4 \Tr
      \biggl(
      {\gamma^{aD}}_A \left\{
      \bigl(G^{[bc]} \nabla_c \AiB {\mathcal J} \Psi'\bigr)^A , \Gp{B}
               \right\}_+
      \biggr)
    \end{split}\\ %\tag{\ref{eq:A2}}
    \begin{split}
      & =
      {2}
      \Tr
      \biggl(
      \bigl(\gamma^a G^{[bc]} \nabla_c \AiB {\bigr)^D}_C {{\mathcal
          J}^C}_{C'} \left\{ \Psi^{C'}, \Gp{B} \right\}_+
      \biggr)
    \end{split}
      \tag{\ref{eq:A2} A}\\
      \begin{split}
      & \phantom{=}
      + 2
      \left[ \Gp{B}{\gamma^{aB}}_A, \bigl( G^{[bc]} \nabla_c \AiB
        {\bigr)^{A}}_{C}  \right] {{\mathcal J}^C}_{C'} \Psi^{C'}
      \end{split}
      \tag{\ref{eq:A2} B}\\
      \begin{split}
      & \phantom{=}
      + 2 \Tr
      \biggl(
      {{\mathcal J}^C}_{C'} \left\{\Psi^{C'}, \Gp{B}  \right\}_+
      \bigl( \gamma^a G^{[bc]} \nabla_c \AiB {\bigr)^{A}_D}
      \biggr)
      \end{split}
      \tag{\ref{eq:A2} C}\\
      \begin{split}
      & \phantom{=}
      -2 \Tr
      \biggl(
      {{\mathcal J}^C}_{C'} \Psi^{C'}
      \left[ \Gp{B}{\gamma^{B}}_A,
        \bigl(G^{[bc]} \nabla_c \AiB {\bigr)^A}_D \right]
      \biggr)
      \end{split}
      \tag{\ref{eq:A2} D}\\ \nonumber
  \end{align*}
For the part~(\ref{eq:A2}A) one calculates:
\begin{equation*}{}
  \begin{split}
    & \phantom{=}
    {2}
      \Tr
      \biggl(
      \bigl(\gamma^a G^{[bc]} \nabla_c \AiB {\bigr)^D}_C {{\mathcal
          J}^C}_{C'} \left\{ \Psi^{C'}, \Gp{B} \right\}_+
      \biggr)\\
    & =
    2 \Tr
    \biggl(
    \gamma^a G^{[bc]} \nabla_c \AiB {\mathcal J}
    (+i)
    \bigl( (i \notnabla + m ) (A' - i \gamma^{5'} B') \bigr) {\mathcal J}
    \biggr)\\
    & =
    - 2 \Tr \bigl(\gamma^a G^{[bc]} \gamma^d \bigr) {g_d}^{d'}
    \bigl( \nabla_c A \nabla_{d'} A' + \nabla_c B \nabla_{d'} B'
    \bigr)\\
    & \phantom{=}
    + 2 i \Tr \bigl(\gamma^a G^{[bc]} \gamma^d \gamma^5 \bigr) {g_d}^{d'}
    \bigl( \nabla_c A \nabla_{d'} B' - \nabla_c B \nabla_{d'} A'
    \bigr)\\
    & = -4 \biggl( g^{ab} {g^c}_{c'}
      (\nabla_c A \nabla^{c'} A' + \nabla_c B \nabla^{c'} B')
      -
      {g^b}_{b'}
      (\nabla^a A \nabla^{b'} A' + \nabla^a B \nabla^{b'} B' )
      \biggr)\\
    & \phantom{=}
      +4 \epsilon^{abcd} {g^{d'}}_d
       (\nabla_c A \nabla_{d'} B' - \nabla_c B \nabla_{d'} A')
  \end{split}
\end{equation*}
The next part~(\ref{eq:A2}B) gives:
\begin{equation*}{}
  \begin{split}
    & \phantom{=}
    + 2
    \left[ \Gp{B}{\gamma^{aB}}_A, \bigl( G^{[bc]} \nabla_c \AiB
      {\bigr)^{A}}_{C}  \right] {{\mathcal J}^C}_{C'} \Psi^{C'}\\
    & =
    2
    \biggl(
    \left[\Gp{B},\nabla_c A   \right]_- \bigl( \gamma^a G^{[bc]} \bigr)
    -i\left[\Gp{B},\nabla_c B\right]_- \bigl( \gamma^a G^{[bc]}\gamma^5\
 \bigr)
    \biggr) {\mathcal J} \Psi'\\
    & =
    2
    \biggl(
    \bigl( i D_c \Psi^+ + i n_c (D_d \Psi^+) \notn \gamma^d -m n_c
    \Psi^+ \notn + i K \Psi^+ n_c \bigr) \gamma^a G^{[bc]}\\
    & \phantom{=2\biggl(}
    - i \bigl( i D_c \Psi^+ + i n_c (D_d \Psi^+) \notn \gamma^d -m n_c
    \Psi^+ \notn + i K \Psi^+ n_c \bigr) \gamma^5 \gamma^a G^{[bc]}\gamma^5
    \biggl) {\mathcal J} \Psi'\\
    \intertext{Recall the identity $(D_d \notn) \gamma^d = K$.}
    & =
    4 i (\nabla_c \Psi^+) \gamma^a G^{[bc]} {\mathcal J} \Psi',\\
  \end{split}
\end{equation*}
For the last equality we used the fact that $\Psi^+$ is a solution of
the Dirac equation and that $\notn$ anti-commutes with ${h^a}_b
\gamma^b$.
The third term is again relatively easy to compute:
\begin{equation*}{}
  \begin{split}
    &
    2 \Tr
    \biggl(
    {{\mathcal J}^{C}}_{C'}
    \left\{ \Psi^{C'}, \Gp{B} \right\}_+
      \bigl(\gamma^a G^{[bc]} \nabla_c \AiB{\bigr)^B}_{D}
    \biggr)\\
    & =
    2 \Tr
    \biggl(
    {\mathcal J}
    (i) \bigl(( i \notnabla - m ) (A' - i \gamma^{5'} B' ) \bigr) {\mathcal\
 J}
      \bigl(\gamma^a G^{[bc]} \nabla_c \AiB{\bigr)^B}_{D}
    \biggr)\\
    & =
    -2
    \Tr
    \bigl( \gamma^d \gamma^a G^{[bc]}
    \bigr) {g_d}^{d'}
    ( \nabla_{d'} A' \nabla_c A + \nabla_{d'} B' \nabla_c B )\\
    & \phantom{=}
    + 2 i \Tr \bigl(\gamma^a G^{[bc]} \gamma^d \gamma^5 \bigr) {g_d}^{d'}
    \bigl( \nabla_c A \nabla_{d'} B' - \nabla_c B \nabla_{d'} A'
    \bigr),\\
    \intertext{since all other traces vanish}
    & =
    - 4
    \biggl(
    g^{ab} {g_c}^{c'}
    ( \nabla_{c'} A' \nabla^c A + \nabla_{c'} B' \nabla^c B )
    - {g^a}_{a'} ( \nabla^{a'} A' \nabla^b A + \nabla^{a'} B' \nabla^b B )
    \biggr)\\
    & \phantom{=}
    +4 \epsilon^{abcd} {g_d}^{d'}
    (\nabla_c A \nabla_{d'} B' - \nabla_c B \nabla_{d'} A')
  \end{split}
\end{equation*}
The forth term (\ref{eq:A2}D) gives:
\begin{equation*}
  \begin{split}
    & \phantom{=}
    - 2
    \Tr
    \biggl(
    {\mathcal J} \Psi' \left[  \Gp{B}{\gamma^{aB}}_A ,
    \bigl( G^{[bc]} \nabla_c \AiB {\bigr)^A}_{D} \right]_-
    \biggr)\\
    & =
    - 4 i
    \Tr
    \bigl(\gamma^a G^{[bc]} {\mathcal J} \Psi' \nabla_c \Psi^+ \bigr),
  \end{split}
\end{equation*}
as an immediate consequence of the calculations of part~(\ref{eq:A2}B).
To summarize the second equation:
\begin{equation*}{}
  \begin{split}
    (\ref{eq:A2})
    & =
    -4 \biggl( g^{ab} {g^c}_{c'}
      (\nabla_c A \nabla^{c'} A' + \nabla_c B \nabla^{c'} B'
      -
      {g^b}_{b'}
      (\nabla^a A \nabla^{b'} A' + \nabla^a B \nabla^{b'} B' )\\
    & \phantom{= - 4 \biggl(}
     -  \epsilon^{abcd} {g^{d'}}_d
       (\nabla_c A \nabla_{d'} B' - \nabla_c B \nabla_{d'} A')
       \biggr)\\
    & \phantom{=}
    + 4 i (\nabla_c \Psi^+) \gamma^a G^{[bc]} {\mathcal J} \Psi'\\
    & \phantom{=}{}
    - 4
    \biggl(
    g^{ab} {g_c}^{c'}
    ( \nabla_{c'} A' \nabla^c A + \nabla_{c'} B' \nabla^c B )
    - {g^a}_{a'} ( \nabla^{a'} A' \nabla^b A + \nabla^{a'} B' \nabla^b
    B )\\
    & \phantom{= - 4 \biggl(}
     -  \epsilon^{abcd} {g_d}^{d'}
       (\nabla_c A \nabla_{d'} B' - \nabla_c B \nabla_{d'} A')
    \biggr)\\
    & \phantom{=}{}
    - 4 i
    \Tr
    \bigl(\gamma^a G^{[bc]} {\mathcal J} \Psi' \nabla_c \Psi^+ \bigr)\\
    & =
    + 4
    \biggl({}
    - g^{ab} {g_c}^{c'}
     \bigl( \{A^{;c}, A^{;c'} \}_+ \{B^{;c},B^{;c'} \}_+\bigr)
    + {g^{a}}_{a'} \bigl( A^{;a'} A^{;b} + B^{;a'} B^{;b}\bigr)\\
    & \phantom{=+4\biggl(}
    + {g^{b}}_{b'} \bigl( A^{;a} A^{;b'} + B^{;a} B^{;b'}\bigr)
    + 2 \epsilon^{abcd} {g_d}^{d'}
       (\nabla_c A \nabla_{d'} B' - \nabla_c B \nabla_{d'} A')\\
    & \phantom{=+4\biggl(}
    - i \Tr
    \bigl(
    \gamma^a G^{[bc]} {\mathcal J} \left[ \Psi', {\Psi{}^+}_{;c} \right]_-
    \bigr)
    \biggr)
  \end{split}
\end{equation*}
Finally for the third Equation we consider:
\begin{align*}\allowdisplaybreaks{}
  \begin{split}
    & \phantom{=}
    \Tr
    \left( {\gamma^{aD}}_{A} {{\mathcal J}^A}_{A'}
      \bigl\{ {{(A' - i \gamma^{5'} B' )}^{A'}}_{C'} {{\mathcal J}^{C'}}_{C}
        \bigl( (\gamma^b \notnabla - \notnabla\gamma^b) \Psi
        \bigr)^C , \sideset{^{\mathcal C}}{^+_{B}}{G}
      \bigr\}_+
    \right)\\
  \end{split}\\ % \tag{\ref{eq:A1}}
  \begin{split}
    & =
    2
    \Tr
    \biggl(
    \bigl(
    \gamma^a {\mathcal J} ( A' - i \gamma^{5'} B') {\mathcal J}  G^{[bc]}
    {\bigr)^D}_{C} \left\{ \nabla_c \Psi^C, \Gp{B} \right\}_+
    \biggr)
  \end{split}
  \tag{\ref{eq:A3} A}\\
  \begin{split}
    & \phantom{=}
    + 2
    \left[
      \Gp{B}{\gamma^{aB}}_{A},
       \bigl(G^{[bc]} {\mathcal J} (A' -i \gamma^{5'}B') {\mathcal J}\
 {\bigr)^A}_{C}
    \right]_- \nabla_c \Psi^C
  \end{split}
  \tag{\ref{eq:A3} B}\\
  \begin{split}
    & \phantom{=}
    + 2 \Tr
    \biggl(
    \left\{  \nabla_c \Psi^C , \Gp{B}\right\}_+
      \bigl( \gamma^a G^{[bc]} {\mathcal J} (A'-i \gamma^{5'} B') {\mathcal\
 J}
      {\bigr)^B}_D
    \biggr)
  \end{split}
  \tag{\ref{eq:A3} C}\\
  \begin{split}
    & \phantom{=}
    - 2 \Tr
    \biggl(
    \nabla_c \Psi^C
    \left[ \Gp{B} {\gamma^{aB}}_A,
      \bigl( G^{[bc]} {\mathcal J} (A'- i \gamma^{5'} B') {\mathcal J}\
 {\bigr)^A}_{D}
    \right]_-
    \biggr)
  \end{split}
  \tag{\ref{eq:A3} D}\\ \nonumber
\end{align*}
Consider the part~(\ref{eq:A3}A) and note that the bi-spinors of
parallel transport cancel each other.
\begin{equation*}
  \begin{split}{}
    &
    2
    \Tr
    \biggl(
    \bigl(
    \gamma^a ( A' - i \gamma^{5} B')  G^{[bc]}
    {\bigr)^D}_{C} \left\{ \nabla_c \Psi^C, \Gp{B} \right\}_+
    \biggr)\\
    & =
    2 \Tr
    \biggl(
     \gamma^a (A' -i \gamma^5 B' ) G^{[bc]}
    \Bigl(
    - i D_c \bigl( (-i \notnabla -m ) \AiB \bigr)
    + i n_c \notn \gamma^d D_d \bigl( (-i \notnabla -m ) \AiB \bigr)\\
    & \phantom{= 2 \Tr \biggl(\gamma^a (A' -i \gamma^5 B' )
      G^{[bc]}\Bigl( -}
    -m n_c \notn \bigl( (-i \notnabla -m ) \AiB \bigr)
    -i n_c K  (-i \notnabla -m ) \AiB
    \Bigr)
    \biggr),\\
    \intertext{since $\notn\notn=1$ and $\notn$ commutes with
      $\gamma^a {h_a}^b$}
    & =
    - 2 \Tr
    \biggl(
    \gamma^a (A' -i \gamma^5 B' ) G^{[bc]}
    \Bigl(
    D_c \notnabla \AiB - n_c \notn \gamma^d D_d \notnabla \AiB\\
    & \phantom{= 2 \Tr \biggl(\gamma^a (A' -i \gamma^5 B' )
      G^{[bc]}\Bigl( -}
    - m^2 n_c \notn \AiB + K n_c \notnabla \AiB
    \Bigr)
    \biggr),\\
    \intertext{since traces over an odd number of $\gamma$-matrices
      are zero.}
    & =
    - 2 \Tr
    \biggl(
    \gamma^a G^{[bc]} (A' - i \gamma^5 B' )
     \bigl( \nabla_c \notnabla + \xi R n_c \notn \bigr) \AiB
    \biggr)\\
    &\phantom{=}{}
    -2 K \Tr
    \biggl(
    \gamma^a G^{[bc]} (A' - i \gamma^5 B' ) n_c \notnabla \AiB
    \biggr)\\
    \intertext{For the last equality we used the fact that $A$ and $B$
      are solutions of the Klein Gordon equation with coupling $\xi$, i.e.,
      $\bigl(\Box + (m^2 + \xi R ) \bigr) A = 0$ and $\bigl(\Box + (m^2
      + \xi R ) \bigr) B = 0$. It was also used that $\notnabla
      \notnabla = \Box$ for scalars.}
    & =
    4 \bigl( A' \nabla^a \nabla^b A + B' \nabla^a \nabla^b B \bigr)
    + 4 \bigl( m^2 g^{ab}  + \xi R n^a n^b  \bigr)\bigl(A'A + B'B
    \bigr)\\
    & \phantom{=}
    - 2 K \Tr
    \biggl(
    \gamma^a G^{[bc]} (A' - i \gamma^5 B' ) n_c \notnabla \AiB
    \biggr)\\
    \end{split}
\end{equation*}
The last equality follows, since $\Tr(\gamma^a G^{[bc]} \gamma^d) = 2
( g^{ab} g^{cd} - g^{ac}g^{bd})$, whereas the corresponding terms with
a {\em single\/} $\gamma^5$ gives a result totally anti-symmetric in
the indices $abcd$ which vanishes due to the symmetry of
$\nabla_c\nabla_d$ and $n_c n_d$ respectively.
The calculation for the next part~(\ref{eq:A3}B) results in:
\begin{equation*}{}
  \begin{split}
    &
    + 2
    \left[
      \Gp{B}{\gamma^{aB}}_{A},
       \bigl(G^{[bc]} (A' -i \gamma^{5}B') {\bigr)^A}_{C}
    \right]_- \nabla_c \Psi^C\\
    & = 2
    \biggl(
    \left[\Gp{B}, A'\right]_- {(\gamma^a G^{[bc]})^B}_C
      - i \left[\Gp{B}, A'\right]_- {(\gamma^a G^{[bc]}\gamma^5 )^B}_C
    \biggr) \nabla_c \Psi^C\\
    & = 4 i {\Psi'}^+ {\mathcal J} \gamma^a G^{[bc]} \nabla_c \Psi
  \end{split}
\end{equation*}
The third part~(\ref{eq:A3}C) yields as a consequence of our results
for~(\ref{eq:A3}A):
\begin{equation*}
  \begin{split}
    &
    + 2 \Tr
    \biggl(
    \left\{  \nabla_c \Psi^C , \Gp{B}\right\}_+
      \bigl( \gamma^a G^{[bc]} (A'-i \gamma^{5} B')
      {\bigr)^B}_D
    \biggr)\\
    & =
    4 \bigl( (\nabla^a \nabla^b A) A'  + (\nabla^a \nabla^b B) B' \bigr)
    + 4 \bigl( m^2 g^{ab} + \xi R n^a n^b  \bigr)\bigl(AA' + BB'
    \bigr)\\
    & \phantom{=}
    - 2 K \Tr
    \biggl(
    \notnabla \AiB \gamma^a G^{[bc]} (A' - i \gamma^5 B' ) n_c
    \biggr)
  \end{split}
\end{equation*}
The result for the forth term follows immediately from the second:
\begin{equation*}
  \begin{split}
    & \phantom{=}
    -2 \Tr
    \biggl(
    \nabla_c \Psi^C
    \left[ \Gp{B} {\gamma^{aB}}_A,
      \bigl( G^{[bc]} {\mathcal J} (A'- i \gamma^{5'} B') {\mathcal J}\
 {\bigr)^A}_{D}
    \right]_-
    \biggr)\\
    & =
    - 4 i \Tr
    \biggl(
    {\mathcal J} \gamma^a G^{[bc]} \nabla_c \Psi {\Psi'}^+
    \biggr)
  \end{split}
\end{equation*}
To summarize these results:
\begin{equation*}{}
  \begin{split}
    (\ref{eq:A3}) & =
% A3A
    4 \bigl( A' \nabla^a \nabla^b A + B' \nabla^a \nabla^b B \bigr)
    + 4 \bigl( m^2 g^{ab} + \xi R n^a n^b  \bigr)\bigl(A'A + B'B
    \bigr)\\
    & \phantom{=}
    - 2 K \Tr
    \biggl(
    \gamma^a G^{[bc]} (A' - i \gamma^5 B' ) n_c \notnabla \AiB
    \biggr)\\
% A3B
    & \phantom{=}
    + 4 i \Tr
    \biggl(
    {\mathcal J} \gamma^a G^{[bc]} {\Psi'}^+ \nabla_c \Psi
    \biggr)\\
% A3C
    & \phantom{=}
    4 \bigl( (\nabla^a \nabla^b A) A'  + (\nabla^a \nabla^b B) B' \bigr)
    + 4 \bigl( m^2 g^{ab} + \xi R n^a n^b  \bigr)\bigl(AA' + BB'
    \bigr)\\
    & \phantom{=}
    - 2 K \Tr
    \biggl(
    \notnabla \AiB \gamma^a G^{[bc]} (A' - i \gamma^5 B' ) n_c
    \biggr)\\
% A3D
    & \phantom{=}
    - 4 i \Tr
    \biggl(
    {\mathcal J} \gamma^a G^{[bc]} \nabla_c \Psi {\Psi'}^+
    \biggr)\\
% Result
    & =
    4
    \bigl(
    \left\{ A^{;ab} , A'  \right\}_+ + \left\{ B^{;ab} , B'  \right\}_+
    \bigr)
    +
    4 ( g^{ab} m^2 + \xi R n^a n^b)
      (\left\{ A,A'\right\}_+ + \left\{ B,B'\right\}_+ \\
    & \phantom{=}
    - 4 i \Tr
    \bigl(
        {\mathcal J} \gamma^a G^{[bc]} \left[ \nabla_c \Psi, {\Psi'}^+\
 \right]_-
    \bigr)\\
    & \phantom{=}
    - 2 K
    \Tr
    \biggl(
    \gamma^a G^{[bc]}n_c
    \left\{ \notnabla \AiB,  (A' - i \gamma^5 B' )\right\}_+
    \biggr)
  \end{split}
\end{equation*}
\end{pf}
%
%%%%%%%%
\section{The energy momentum tensor in terms of `supercurrents' and
  `supercharges'}
\label{sec:defEMT}
%%%%%%%%
%
In this section we define a candidate for a point-separated energy
momentum tensor of our locally SUSY model by an analogon of
Eqn.~(\ref{eq:formalTraceCurved}).
{\sloppy
Let $({\EuScript H},D,A,B,\Psi,\Psi^+,\Omega)$ denote our locally SUSY\
 quantum
field theoretical model on the spacetime $(M,g_{ab})$. We define two
point separated `supercurrents':
\begin{align}
  \begin{split}
    \sideset{_1}{^{aA}}{j} (x,y) & = \frac{1}{2}
      \biggl(
       i \AiB ( i \lvec{\notnabla} - m) \gamma^a {\mathcal J} \Psi' \\
       & \phantom{= \frac{1}{2} \biggl(}
         - 2\xi \left\{ \bigl( 4 G^{[ab]} \nabla_b \AiB \bigl)
       {\mathcal J} \Psi' + {\mathcal J} (A' -i \gamma^{5'} B' ) {\mathcal J}
       4 G^{[ab]} \nabla_b \Psi \right\}
      \biggr)
  \end{split}\label{eq:pointsplitsupCur1}\\
  \begin{split}
    \sideset{_2}{^{a'A'}}{j} (x,y) & = \frac{1}{2}
      \biggl(
       i (A' -i \gamma^{5'} B' ) ( i \lvec{\notnabla} - m)
       \gamma^{a'} {\mathcal J} \Psi \\
       & \phantom{= \frac{1}{2} \biggl(}
         - 2\xi \left\{ \bigl( 4 G^{[a'b']} \nabla_{b'} (A'-i\gamma^{5'}B')
                        \bigl)
       {\mathcal J} \Psi + {\mathcal J} \AiB {\mathcal J}
       4 G^{[a'b']} \nabla_{b'} \Psi' \right\}
      \biggr)
  \end{split}\label{eq:pointsplitsupCur2}
\end{align}
}
Note that $\sideset{_{1,2}}{}{j}(x,y)$ are two well defined operator
valued distributions on $\EuScript H$.  The constant $\xi$ is
undetermined at this stage and will be fixed below to the SUSY
coupling $1/4$.  Setting
\[
j^{aA}(x,y) := \sideset{_1}{^{aA}}{j}(x,y)
               + {g^{a}}_{a'} {{\mathcal\
 J}^A}_{A'}\sideset{_2}{^{a'A'}}{j}(x,y),
\]
it follows, using Proposition~\ref{prop:wightmanCur} and the fact that
${g^{a}}_{a'}$ and ${{\mathcal J}^A}_{A'}$ are smooth, that the
restriction of $j^{aA}$ to the diagonal exists and equals the improved
supercurrent of Eqn.~(\ref{eq:supercurrentCurved}). Moreover, since
$[\nabla_a {g^a}_{a'}] = [\nabla_a {{\mathcal J}^A}_{A'}] = 0$, where the
square brackets denote the coincidence limit, it easy to see that
$j^{aA}(x,y)$ is `asymptotically conserved' on
spacetimes with vanishing scalar curvature, i.e.,
\[
[\nabla_a j^{aA}(x,y)] = \nabla_a [j^{aA}(x,y)] = \nabla_a j^{aA}(x) = 0
\qquad \text{iff $R=0$}.
\]
Recall that the last equality was shown in
Proposition~\ref{prop:superCurCurvWeakCons} for classical fields
propagating on a spacetime with $R=0$. This result extends without
modification to the quantized version of the theory.
\begin{Dfn}\label{Dfn:pointsplitEMTcurved}
  Let ${\mathcal C}$ be a smooth maximal spacelike hypersurface in the
  vacuum spacetime $(M,g_{ab})$, i.e., ${\mathcal C}$ has extrinsic
  curvature $K=0$. Then the point separated energy momentum tensor
  $\tilde{T}^{ab}(x,y)$ of a locally SUSY quantum field theoretical
  model $({\EuScript H},D,A,B,\Psi,\Psi^+,\Omega)$ on this spacetime
  is defined as the operator valued distribution on
  ${\mathcal C} \times {\mathcal C}$ with kernel (Compare
  Eqn.~(\ref{eq:SUSY_EMT}))
  \begin{equation}
    \label{eq:pointsplitEMTcurvedKern}
    \begin{split}
    \tilde{T}^{ab}(x,y) & :=
    + \frac{1}{8}
    \biggl(
    \Tr \Bigl( \gamma^{(a} \{ \sideset{_1}{^{b)A}}{j}(x,y) , \Gp{B} \}_+\
 \Bigr)
    +
    {g^{a}}_{a'}{g^{b}}_{b'}
    \Tr
    \Bigl( \gamma^{(a'} \{ \sideset{_2}{^{b')A'}}{j}(x,y),\Gp{B'} \}_+ \Bigr)
    \biggr),
    \end{split}
  \end{equation}
  where $\sideset{_{1,2}}{^{aA}}{j}$ are given by
  Eqn.~(\ref{eq:pointsplitsupCur1}) and
    Eqn.~(\ref{eq:pointsplitsupCur2}) respectively with $\xi = 1/4$.
\end{Dfn}
\begin{rem}
  The r.h.s.\ of Eqn.~(\ref{eq:pointsplitEMTcurvedKern}) is a well
  defined operator valued distribution on ${\mathcal C}\times{\mathcal C}$ by
  Proposition~\ref{prop:commutQCurv}, even if our spacetime does not
  have vanishing scalar curvature, $\xi\neq 1/4$ and ${\mathcal C}$ is not
  maximal.
\end{rem}
We will now use the results of Proposition~\ref{prop:trCommQCurv} to
express the r.h.s.\ of Eqn.~(\ref{eq:pointsplitEMTcurvedKern}) as
commutators and anti-commutators respectively of the basic fields. Let
us relax the assumption on the spacetime $(M,g_{ab})$, the
hypersurface ${\mathcal C}$ and the coupling $\xi$ for a moment.
%
%% M A I N   R E S U L T
\begin{Thm}\label{Thm:EMTPointSplitSuperCur}
  The point separated energy momentum tensor $\tilde{T}^{ab}(x,y)$
  defined by Eqn.~(\ref{eq:pointsplitEMTcurvedKern}) is an operator
  valued distribution on ${\mathcal C}\times{\mathcal C}$, even if the
  spacetime does not have vanishing scalar curvature, $\xi \neq 1/4$
  and ${\mathcal C}$ is not maximal. Its kernel can be written in the
  basic fields as:
  \begin{equation}\label{eq:pointsplitEMTcurvedbasicfields}
    \begin{split}
      \tilde{T}^{ab}(x,y)
      & =
% Result geschoent (Umordnung etc.)
      \frac{1-2 \xi}{4} {g^b}_{b'}
      ( \{A^{;a},A^{;b'}\}_+ + \{B^{;a},B^{;b'}\}_+ )
      +
      \frac{1-2\xi}{4} {g^a}_{a'}
      ( \{A^{;b},A^{;a'}\}_+ + \{B^{;b},B^{;a'}\}_+ )\\
      & \phantom{=}
      - \frac{1-4\xi}{4} g^{ab} {g^c}_{c'}
      ( \{A_{;c},A^{;c'}\}_+ + \{B_{;c},B^{;c'}\}_+ )
      +
      \frac{1-4\xi}{4} g^{ab} m^2
      ( \{A,A'\}_+ + \{B,B'\}_+ )\\
% xi Terme mit A und B
       & \phantom{=}
       - \frac{\xi}{2}
       \left( (\{A^{;ab},A'\}_+ + \{B^{;ab},B'\}_+)
         + {g^a}_{a'} {g^b}_{b'}
           (\{A,A^{;a'b'}\}_+ + \{B,B^{;a'b'} \}_+)
       \right)\\
      & \phantom{=}
      - \xi^2 R n^a n^b
           (\{A,A'\}_+ + \{B,B' \}_+)\\
% Ab hier Spinor Terme
      & \phantom{=}
      + \frac{i}{8}
      \Tr\biggl(
      {\mathcal J} \gamma^{(a}
      \bigl(
        {g^{b)}}_{b'} [\Psi,\Psi^{+;b'}]_- -  [\Psi^{;b)},\Psi^{+'}]_-
      \bigr)
      \Bigr)\\
      % Note the symmetry in a and b !
% Korrektur terme
% K Terme mit \xi
      & \phantom{=}
      + \frac{\xi}{16} K
      \Tr\Bigl(\gamma^{(a} G^{[b)c]} n_c
      \{\notnabla \AiB, (A' -i \gamma^5 B') \}_+ \Bigr)\\
      & \phantom{=}
      + \frac{\xi}{16} K {g^{a}}_{a'} {g^{b}}_{b'}
      \Tr\Bigl(\gamma^{(a'} G^{[b')c']} n_{c'}
      \{ (A - i \gamma^{5'} B),{\notnabla}'(A' -i \gamma^{5'} B')\}_+\Bigr)\\
% die 2 K Terme von oben
      & \phantom{=}
      + \frac{i}{8} K
      \Tr\Bigl( \gamma^{(a}\notn\gamma^{b)} {\mathcal J} [\Psi',\Psi^+]_-\
 \Bigr)
      + \frac{i}{8} K {g^b}_{b'} {g^a}_{a'}
      \Tr\Bigl(\gamma^{(a'}\notn\gamma^{b')}{\mathcal
        J}[\Psi,\Psi^{+'}]_-\Bigr)
% Ende K Terme
    \end{split}
  \end{equation}
\end{Thm}
%% M A I N   R E S U L T     end
%
Now set $\xi=1/4$ and assume $K=R=0$. The result is
  \begin{equation}\label{eq:emtExplicit}{}
    \begin{split}
      \tilde{T}^{ab}(x,y)
      & =
% Result geschoent (Umordnung etc.)
      \frac{1}{8} {g^b}_{b'}
      ( \{A^{;a},A^{;b'}\}_+ + \{B^{;a},B^{;b'}\}_+ )
      +
      \frac{1}{8} {g^a}_{a'}
      ( \{A^{;b},A^{;a'}\}_+ + \{B^{;b},B^{;a'}\}_+ )\\
% xi Terme mit A und B
       & \phantom{=}
       - \frac{1}{8}
       \left( (\{A^{;ab},A'\}_+ + \{B^{;ab},B'\}_+)
         + {g^a}_{a'} {g^b}_{b'}
           (\{A,A^{;a'b'}\}_+ + \{B,B^{;a'b'} \}_+)
       \right)\\
% Ab hier Spinor Terme
      & \phantom{=}
      + \frac{i}{8}
      \Tr\biggl(
      {\mathcal J} \gamma^{(a}
      \bigl(
        {g^{b)}}_{b'} [\Psi,\Psi^{+;b'}]_- -  [\Psi^{;b)},\Psi^{+'}]_-
      \bigr)
      \Bigr) \qquad\text{iff $\xi=1/4$ and $K=R=0$}
      % Note the symmetry in a and b !
    \end{split}
  \end{equation}
As an immediate consequence we get
\begin{Cor}\label{Cor:compatEMT}
  The point separated energy momentum tensor $\tilde{T}^{ab}(x,y)$ with
  $\xi=1/4$ agrees with the Definition of the point separated energy
  momentum tensor $T^{ab}$ of section~\ref{sec:analogonWess},
  Eqn.~(\ref{eq:SUSYemtsplit}), provided $K=R^{ab}=R=0$.
\end{Cor}
This compatibility result will allow us in the next section to
calculate the `vacuum' expectation value \mbox{$<\Omega | \tilde{T}^{ab}
\Omega>$} explicitly using the results of
Christensen~\cite{chris:76,chris:78}. Note that we need $R^{ab}$ here,
i.e., vanishing scalar curvature is not sufficient.
We close this section with the proof of the Theorem:
\begin{pf*}{Proof of Theorem~\ref{Thm:EMTPointSplitSuperCur}}
  For the proof insert the Definitions and use
  Proposition~\ref{prop:trCommQCurv}. After some
  lengthly but elementary calculation one obtains:
  \begin{equation*}{}
    \begin{split}
      \tilde{T}^{ab}(x,y)
      & =
  % Result geschoent (Umordnung etc.)
      \frac{1}{4} {g^b}_{b'}
      ( \{A^{;a},A^{;b'}\}_+ + \{B^{;a},B^{;b'}\}_+ )
      +
      \frac{1}{4} {g^a}_{a'}
      ( \{A^{;b},A^{;a'}\}_+ + \{B^{;b},B^{;a'}\}_+ )\\
      & \phantom{=}
      - \frac{1}{4} g^{ab} {g^c}_{c'}
      ( \{A_{;c},A^{;c'}\}_+ + \{B_{;c},B^{;c'}\}_+ )
      +
      \frac{1}{4} g^{ab} m^2
      ( \{A,A'\}_+ + \{B,B'\}_+ )\\
  % Spinor Terme unten
  % die 2 K Terme sind  weiter unten
  % xi Terme mit A und B
       & \phantom{=}
       - \frac{\xi}{2}
       \left( (\{A^{;ab},A'\}_+ + \{B^{;ab},B'\}_+)
         + {g^a}_{a'} {g^b}_{b'}
           (\{A,A^{;a'b'}\}_+ + \{B,B^{;a'b'} \}_+)
       \right)\\
      & \phantom{=}
      - \xi ( g^{ab} m^2 + \xi R n^a n^b)
           (\{A,A'\}_+ + \{B,B' \}_+)\\
  % \xi Terme 1
      & \phantom{=}
      + \xi g^{ab} {g^{c'}}_{c}
       ( \{A_{;c},A^{;c'}\}_+ + \{B_{;c},B^{;c'}\}_+ )
      - \frac{\xi}{2}  {g^a}_{a'}
        ( \{A^{;b},A^{;a'}\}_+ + \{B^{;b},B^{;a'}\}_+ )\\
      & \phantom{=}
      - \frac{\xi}{2}  {g^b}_{b'}
        ( \{A^{;a},A^{;b'}\}_+ + \{B^{;a},B^{;b'}\}_+ )\\
  % Ab hier Spinor Terme
  %Spinor Terme von oben
      & \phantom{=}
      + \frac{i}{8}\Tr\Bigl( \gamma^{(a} {\mathcal J} [\Psi',\Psi^{+;b)}]_-\
 \Bigr)
      + \frac{i}{8} {g^a}_{a'} {g^b}_{b'}
      \Tr\Bigl( \gamma^{(a'} {\mathcal J} [\Psi,\Psi^{+;b')}]_- \Bigr)\\
      % Note the symmetry in a and b !
  % Korrektur terme
  % Spinor xi Terme
      & \phantom{=}
      + \frac{\xi}{2} i
        \Tr\Bigl( \gamma^{(a} G^{[b)c]} {\mathcal J} [\Psi',\Psi^+_{;c}]_-\
 \Bigr)
      +
      \frac{\xi}{2} {g^a}_{a'} {g^b}_{b'} i
        \Tr\Bigl( \gamma^{(a'} G^{[b')c']} {\mathcal J}
            [\Psi,\Psi^{+'}_{;c'}]_- \Bigr)\\
  % A B Terme oben
      & \phantom{=}
      +\frac{\xi}{2} i
        \Tr\Bigl( {\mathcal J}\gamma^{(a} G^{[b)c]}  [\Psi_{;c},\Psi^{+'}]_-\
 \Bigr)
      +
      \frac{\xi}{2} {g^a}_{a'} {g^b}_{b'} i
        \Tr\Bigl( {\mathcal J} \gamma^{(a'} G^{[b')c']}
            [\Psi'_{;c'},\Psi^{+}]_- \Bigr)\\
      & \phantom{=}
  % K Terme mit \xi
      + \frac{\xi}{16} K
      \Tr\Bigl(\gamma^{(a} G^{[b)c]} n_c
      \{\notnabla \AiB, (A' -i \gamma^5 B') \}_+ \Bigr)\\
      & \phantom{=}
      + \frac{\xi}{16} K {g^{a}}_{a'} {g^{b}}_{b'}
      \Tr\Bigl(\gamma^{(a'} G^{[b')c']} n_{c'}
      \{ (A - i \gamma^{5'} B),{\notnabla}'(A' -i \gamma^{5'} B')\}_+\Bigr)\\
  % die 2 K Terme von oben
      & \phantom{=}
      + \frac{i}{8} K
      \Tr\Bigl( \gamma^{(a}\notn\gamma^{b)} {\mathcal J} [\Psi',\Psi^+]_-\
 \Bigr)
      + \frac{i}{8} K {g^b}_{b'} {g^a}_{a'}
      \Tr\Bigl(\gamma^{(a'}\notn\gamma^{b')}{\mathcal
        J}[\Psi,\Psi^{+'}]_-\Bigr)
 % Ende K Terme
    \end{split}
  \end{equation*}
  By applying Lemma~\ref{Lemma:rearrangeTr} of
  Appendix~\ref{sec:usefulForm} one verifies:
  \begin{align*}\allowdisplaybreaks{}
    \begin{split}
      \frac{\xi}{2} i
      \Tr\Bigl( \gamma^{(a} G^{[b)c]} {\mathcal J} [\Psi',\Psi^+_{;c}]_-
      \Bigr)
      & = -
      \frac{\xi}{2} i
      \Tr\Bigl( {\mathcal J}\gamma^{(a} G^{[b)c]}  [\Psi_{;c},\Psi^{+'}]_-
      \Bigr)
    \end{split}\\
    \begin{split}
      \frac{\xi}{2} {g^a}_{a'} {g^b}_{b'} i
      \Tr\Bigl( \gamma^{(a'} G^{[b')c']} {\mathcal J}
      [\Psi,\Psi^{+'}_{;c'}]_- \Bigr)
      & = -
      \frac{\xi}{2} {g^a}_{a'} {g^b}_{b'} i
      \Tr\Bigl( {\mathcal J} \gamma^{(a'} G^{[b')c']}
      [\Psi'_{;c'},\Psi^{+}]_- \Bigr)
    \end{split}\\
    \intertext{and}
    \begin{split}
      + \frac{i}{8}\Tr\Bigl( \gamma^{(a} {\mathcal J} [\Psi',\Psi^{+;b)}]_-\
 \Bigr)
      & =
      - \frac{i}{8}
      \Tr\Bigl( \gamma^{(a} {\mathcal J} [\Psi^{;b)},\Psi^{+'}]_- \Bigr).
    \end{split}
  \end{align*}
  For the first two equalities one should note that $\gamma^{(a}
  G^{[b)c]} = \frac{1}{4}(2g^{ab}\gamma^c - g^{cb}\gamma^a - g^{ac}
  \gamma^b)$.
  Finally use the fact that the $\gamma$-matrices are
  covariantly constant, i.e., ${g^a}_{a'} {\mathcal J} \gamma^{a'} {\mathcal
    J} = \gamma^a$.  This proves the Theorem.
\end{pf*}
%
%%%%%%%%%
\section{The `vacuum' expectation value}
\label{sec:vacExpValue} \label{sec:christResult}
%%%%%%%%%
%
In this section the results of Christensen's covariant geodesic point
separation~\cite{chris:76,chris:78} are used to calculate the `vacuum'
expectation value of our point-splitting energy momentum tensor
$\tilde{T}^{ab}$. It is emphasized that we suppose the assumption of
Definition~\ref{Dfn:pointsplitEMTcurved}, unless otherwise stated,
throughout this section, i.e., a vacuum spacetime ($R^{ab}=0$) with
maximal spacelike hypersurface ${\mathcal C}$ ($K=0$). Let us start with a
short introduction to the covariant geodesic point separation:
Consider the `vacuum' expectation value of $\tilde{T}^{ab}$. Using the
explicit form for $\tilde{T}^{ab}$ (Eqn.~(\ref{eq:emtExplicit})), one
sees that $<\Omega |\tilde{T}^{ab}(x,y)\Omega>$ depends on the
`vacuum' expectation value of the anti-commutator, or commutator
respectively, of the basic fields, together with their derivatives,
only.  Recall that $\Omega$ arises from a quasifree locally SUSY
product state $\omega$ and $<\Omega |\{ A(x),A(y)\}_+\Omega>$,
$<\Omega|\{ B(x),B(y)\}_+\Omega>$ and $<\Omega | [ \Psi(x),\Psi^+(y)
]_- \Omega>$ are the real and imaginary parts respectively of the
Hadamard distributions $\omega^2$ introduced in
chapter~\ref{sec:analogonWess} section~\ref{sec:hadamardStates}
(Eqn.~(\ref{ScalarHadam})). Using the same notation as in that
section, we denote the corresponding singular kernels by
$\sideset{^A}{^{T,P}_{\epsilon}}{G}$,
$\sideset{^B}{^{T,P}_{\epsilon}}{G}$ and
$\sideset{^\Psi}{^{T,P}_{\epsilon}}{{\mathcal G}}$. We remind the reader
of the fact that in the fermionic case
$\sideset{^\Psi}{^{T,P}_{\epsilon}}{{\mathcal G}}$ is the singular kernel
of the auxiliary two point distribution $\tilde{\omega}^2$. Since both
scalar kernels are identical, we may drop the indices and write $G$
for the scalar and ${{\mathcal G}^B}_{A'}$ for the fermionic case.  As a
result $<\tilde{T}^{ab}>_\omega$ can be written as\footnote{Note that
  the auxiliary two-point distribution ${\mathcal G} + H^{(p)}$ in the
  fermionic case is {\em even\/} in the
  $\gamma$-matrices~\cite{DeWitt:group}, which implies that
  $\Tr(\gamma^a ({\mathcal G}+ H^{(p)}))$ vanishes.}:
\begin{equation}{}
  \label{eq:myEmtFull}
  \begin{split}
    <\tilde{T}^{ab}>_{\omega} & =
    \frac{1}{4}
    \left(
     {g^b}_{b'}G^{;ab'} + {g^a}_{a'} G^{;a'b} -{g^a}_{a'}{g^b}_{b'}G^{;a'b'}
   \right)\\
    & \phantom{=}
    - \frac{1}{8} \Tr
    \left( {\mathcal J} \gamma^{(a} \gamma^c
    \left(
      {g^{b)}}_{b'} {{\mathcal G}_{;c}}^{b'} - {{\mathcal G}_{;c}}^{b)}
    \right)
    \right)\\
    & \phantom{=}
    +\frac{1}{8}
    \biggl(
    {g^b}_{b'} ( \sideset{^A}{^{(p)}}{H} + \sideset{^B}{^{(p)}}{H})^{;ab'}
    +
    {g^a}_{a'} ( \sideset{^A}{^{(p)}}{H} +
    \sideset{^B}{^{(p)}}{H})^{;a'b}\\
    & \phantom{= + \frac{1}{8} \biggl( }
    -
    ( \sideset{^A}{^{(p)}}{H} + \sideset{^B}{^{(p)}}{H})^{;ab}
    -
    {g^a}_{a'} {g^b}_{b'}
    ( \sideset{^A}{^{(p)}}{H} + \sideset{^B}{^{(p)}}{H})^{;a'b'}
    \biggr)\\
    & \phantom{=}
    - \frac{1}{8} \Tr
    \left( {\mathcal J} \gamma^{(a} \gamma^c
    \left(
      {g^{b)}}_{b'} {{\sideset{^\Psi}{^{(p)}}{H}}_{;c}}^{b'}
       - {{\sideset{^\Psi}{^{(p)}}{H}}_{;c}}^{b)}
    \right)
    \right)\\
    & =: <\tilde{T}^{ab}>_{\omega,\text{sing}} +
         <\tilde{T}^{ab}>_{\omega,\text{finite}},\\
  \end{split}
\end{equation}
i.e., as a sum of two parts $<\tilde{T}^{ab}>_{\omega,\text{sing}}$
and $<\tilde{T}^{ab}>_{\omega,\text{finite}}$. The first summand is
constructed solely with the singular kernels $G$ and ${\mathcal G}$, while
the second consists of the remaining regular kernels $H^{(p)}$ and
their derivatives only. The label `finite' was chosen for the latter,
since the corresponding distribution on ${\mathcal C} \times {\mathcal C}$\
 can
be restricted to the diagonal and the result is a finite {\em
  function\/} $<\tilde{T}^{ab}>_{\omega,\text{finite}}$. (Recall that
$H^{(p)}$ can be made sufficiently smooth by choosing $p$ large
enough. It follows, that the coincidence limit exist.)  On the other
hand, due to the divergences in both, $G$ and ${\mathcal G}$, the
coincidence limit of $<\tilde{T}^{ab}>_{\omega,\text{sing}}$ does {\em
  not\/} exist in general, i.e., if we use Eqn.~(\ref{eq:myEmtFull}),
but relax our assumptions on the spacetime and the hypersurface.
Renormalization is necessary in this case.  It is worth noting that
all state dependent information is contained in
$<\tilde{T}^{ab}>_{\omega,\text{finite}}$, since $G$ and ${\mathcal G}$
are purely geometric objects.  We digress for a moment to describe the
renormalization prescription according to Adler \& al.~\cite{adler:77}
or Wald~\cite{wald:78}:
Subtract from $G$ the Hadamard solution $G^{0}$, which is defined by
setting the (arbitrary) function $H_0$ in the expansion of $H^{(p)}$
(see Appendix~\ref{sec:hadamRecRel}) equal to zero and plug in the
result into $\tilde{T}^{ab}(x,y)$. Then, since $G$ and $G^{0}$ have
equal singularities, the coincidence limit $x \rightarrow y$ will
exist and, after performing the same operation on ${\mathcal G}$, defines
a finite expression for
$<\tilde{T}^{ab}(x)>_{\omega,\text{sing,ren.}}$.  Unfortunately, as is
stated in Wald loc.cit., this prescription is not unique, since the
length scale, by which $\sigma$ is measured has not been specified. In
the massive case the Compton wave length of the particle can be used
to introduce such a length scale and the ambiguity can be resolved.
However, for a  massless theory the ambiguity remains due to the fact that
these models do not possess a natural length scale. Moreover, it was
pointed out by Wald loc.cit.\ that an unambiguous prescription for the
renormalized stress tensor can not be given without introducing a new
fundamental length scale. Another problem, which one encounters using
this prescription is related to the fact that not all functions $H_0$
lead to {\em symmetric\/} Hadamard bisolutions. Namely,
$<\tilde{T}^{ab}(x)>_{\omega,\text{sing,ren.}}$ fails to be locally
conserved in general.  To obtain a locally conserved object one must
in a last step add the local curvature term $-\frac{1}{32 \pi^2} [v_1]
g^{ab}$, where $[v_1]$ is the coincidence limit of the second
expansion coefficient of $v^{(p)}$ (See Eqn.~(\ref{eq:hadamSol}) in
Appendix~\ref{sec:hadamRecRel}). Apart from these deficiencies the
geodesic point separation method turned out to be a valuable tool for
explicit calculations. For some of the results obtained by this method
the reader is referred to the books of Birell \&
Davis~\cite{birrell:qft} or Wald~\cite{wald:gr} respectively and the
original literature cited therein.
The result of this section is that in our case a renormalization is
not necessary (at least for the massive theory), since the
divergences of the bosonic and fermionic parts of the model already
cancel each other. Unfortunately there remain direction dependent
terms, which must be dealt with by an `elementary averaging'
procedure, in order to obtain a tensorial, i.e., direction
independent, result.  To establish this result we have to isolate the
divergences of the fermionic and bosonic parts of
$<\tilde{T}^{ab}>_\omega$ respectively in a covariant way. This can be
done  using a covariant Taylor expansion for all terms in
$<\tilde{T}^{ab}>_\omega$, i.e., for $G$, ${\mathcal G}$ and their
derivatives.  The latter is a method frequently used in classical
general relativity. See for instance DeWitt's classical
book~\cite{DeWitt:group}. For the convenience of the reader we
describe the basic idea in the next paragraph:
Covariant Taylor expansion, as well as the usual Taylor expansion,
uses a power series to describe a smooth function near some initial
point. For the covariant Taylor expansion one uses the fact that every
smooth bi-scalar $K(x,x')$ on a Lorentz manifold, i.e., a smooth
function depending on $x$ and $x'$, for $x$ sufficiently close to $x'$
can be written as a power series in $\sigma$ and its covariant
derivatives up to order $q$ with coefficients $k(x)$ depending on $x$
only, plus a remainder, which is of order ${\mathcal O}(\sigma^{;(q+1)})$.
The coefficients $t(x)$ are completely determined by the coincidence
limits of $K$ and its derivatives at $x$ and one writes
\begin{equation} \label{eq:covTayExp}
K(x,x') \equiv k(x) + k_a(x) \sigma^{;a}(x,x') + \frac{1}{2!} k_{ab}
         \sigma^{;ab}(x,x') + \cdots {\mathcal O}(\sigma^{(q+1)}).
\end{equation}
The relation between the coefficients and $K$ can be established by
repeatingly differentiating both sides of Eqn.~(\ref{eq:covTayExp})
with respect to $x'$ and performing the coincidence limit afterwards.
The results, as well as the various coincidence limits of $\sigma$,
which are needed, can be found for instance
in~\cite{chris:76,chris:78}.
Unfortunately we can not expand the kernels $G$ and ${\mathcal G}$
directly according to Eqn.~(\ref{eq:covTayExp}), since their
coincidence limits do not exist. On the other hand the van Vleck
Morette determinant $\Delta^{1/2}$, $v^{(p)}$ and $\sigma$ are smooth
in any convex normal neighborhood and thus {\em their} covariant Taylor
expansions exist to any order. (Note that we do not assume the
convergence of the infinite series without remainder). Therefore the
only problematic terms are those like $(\sigma^{-1})^{;ab}(x,x')$,
which diverge as $x'$ approaches $x$. In order to expand them, we carry
out the differentiation, which yields
\begin{equation}\label{eq:covTaylSigmaInv}
(\sigma^{-1})^{;ab} = 2 \sigma^{-3} \sigma^{;a} \sigma^{;b} -
                      \sigma^{-2}\sigma^{;ab}.
\end{equation}
We can now expand $\sigma^{;ab}$ and $\sigma^{;a}$ and plug in the
result into Eqn.~(\ref{eq:covTaylSigmaInv}). The result, which contains
terms proportional to $\sigma^{-2}$ and $\sigma^{-1/2}$ can be found
e.g.\ in~\cite{chris:76}. It should be noted that this expansion is
only valid if the points $x$ and $x'$ are not null related, i.e., for
$\sigma(x,x')\neq 0$. On the other hand, the distribution
$<\tilde{T}^{ab}(x,y)>_{\omega,\text{sing}}$ is defined on ${\mathcal
  C}\times{\mathcal C}$, hence the only singularities occur at the
diagonal, while everywhere else the corresponding kernel is smooth. We
may therefore expand $G$, ${\mathcal G}$ and their derivatives by
substituting for $\Delta^{1/2}$, $v^{(p)}$, $\sigma^{-1}$,
$\ln(\sigma)$ and their derivative the corresponding power series,
form $<\tilde{T}^{ab}>_{\omega,\text{sing}}$ and collect together
terms like powers of $\sigma^{;a}$. This tremendous task, which
involves much algebra and the calculation of many coincidence limits,
has fortunately already been done by Christensen. In his PhD thesis he
performed the calculation for the scalar field and extended the result
later to fermions~(\cite{chris:76} and~\cite{chris:78}).  He did not
use the representation~(\ref{eq:hadamSol}) for the Hadamard
distributions. Instead he considered a special Hadamard distribution,
usually denoted by $G_{DS}$, which was originally obtained by
DeWitt~\cite{DeWitt:group} using Schwinger's proper time formalism. The
kernel of this DeWitt-Schwinger Hadamard distribution has an
asymptotic expansion of the form~\cite{chris:76}:
\begin{equation}
  \label{eq:dsHadamardExp}
  \begin{split}
    & G_{DS}(x,x')\\
    & = \frac{\Delta^{1/2}}{4 \pi^2}
       \biggl\{ a_0 \left[ \frac{1}{\sigma} + m^2 L (1 + \frac{1}{4}
       m^2 \sigma + \cdots ) - \frac{1}{2} m^2 - \frac{5}{16} m^2
       \sigma + \cdots  \right]\\
    & \phantom{= \frac{\Delta^{1/2}}{4 \pi^2}\biggl\{}
      -a_1 \left[ L(1 + \frac{1}{2} m^2 \sigma + \cdots )
      - \frac{1}{2} m^2 \sigma \right]\\
    & \phantom{= \frac{\Delta^{1/2}}{4 \pi^2}\biggl\{}
      + a_2 \sigma \left[ L ( \frac{1}{2} + \frac{1}{8}m^2 \sigma +
      \cdots ) -\frac{1}{4} - \cdots \right] + \cdots \\
    & \phantom{= \frac{\Delta^{1/2}}{4 \pi^2}\biggl\{}
      + \frac{1}{2m^2} [a_2 + \cdots ]
      + \frac{1}{2m^4}[a_3 + \cdots ] + \cdots
    \biggl\},
  \end{split}
\end{equation}
where $L=(\gamma+ \frac{1}{2}\ln| \frac{1}{2}m^2 \sigma|)$, $\gamma$
is Euler's constant, $\Delta^{1/2}$ is the van Vleck Morette
determinant and the $a_n$ are smooth functions satisfying certain
recursion relations with boundary condition $a_0(x,x)=1$. Note that
Eqn.~(\ref{eq:dsHadamardExp}) is equally valid for the scalar and
fermionic case, but in the former $\Delta^{1/2}$ and $a_n$ are
bi-scalars, while in the latter they are bi-spinors. The omitted terms
are either of order $\sigma^2$ or finite of order $1/m^6$.
%BIS HIER
It was pointed out by Adler, Lieberman and Ng~\cite{adler:77},
and also by
Castagnino and Harari~\cite{castagnino_harari:84}
that $G_{DS}$ can be identified with a Hadamard series in the
DeWitt-Brehme form, i.e., the form given in
Appendix~\ref{sec:hadamRecRel} written as an asymptotic series, by
setting the undetermined function $H_0$ to
\begin{equation}
  \label{eq:HODeW}
  \begin{split}
    \sideset{^{DS}}{_0}{H}
    = & \Delta^{-1/2} \Bigl\{ m^2 ( \ln m^2 - \ln 2
    + 2 \gamma -1) \\
    & - a_1 ( 2\gamma - \ln 2 + \ln m^2) +
    \frac{1}{m^2} a_2 + \frac{1}{m^4} a_3 + \cdots.
  \Bigr\}
  \end{split}
\end{equation}
Moreover, the second authors together with Nu{\~n}ez claimed
in~\cite{CastagHarNue:82} that $G_{DS}$ is the unique kernel which
covariantly generalizes to curved spacetimes the singularity structure
of the Minkowski space Hadamard distribution together with its
derivatives with respect to $m^2$.
Christensen continues by inserting into Eqn.~(\ref{eq:dsHadamardExp})
as well as into its derivatives the covariant Taylor expansions for
the derivatives of $\sigma$, for $\Delta^{1/2}$ and $a_n$ and for
their derivatives. He forms the stress energy tensor for a scalar and
a Fermi field using an analogon of Eqn.~(\ref{eq:emtExplicit}),
collects terms like powers of $\sigma^{;a}$ and observes quartic
($\sigma^{-2}$), quadratic ($\sigma^{-1}$), linear ($1/\sigma^{;a}$)
and logarithmic ($\ln(m^2 \sigma)$) divergences for both fields.
Inserting his results into Eqn.~(\ref{eq:emtExplicit}), one sees that
all divergences cancel on a vacuum solution of the Einstein equations.
One ends up with:
\begin{equation}
  \label{eq:emtDS}
  \begin{split}
    <\tilde{T}^{ab}(x,y)>_{DS} = \frac{1}{4\pi^2} \biggl( &
    -\frac{1}{192}
      R^{cdef}R_{cdef} \frac{\sigma^{;a} \sigma^{;b}}{2\sigma}
    +\frac{1}{3} {{R^{(ac}}_d}^e R_{fcge}
      \frac{\sigma^{;b)}\sigma^{;d}\sigma^{;f}\sigma^{;g}}{(2\sigma)^2}\\
    &
    + \frac{1}{24} R^{cde(a} R_{cdef}
      \frac{\sigma^{;b)}\sigma^{;f}}{2\sigma}\\
    &
    - \frac{1}{24}
      \left[
       R^{acbd} R_{cedf} + {R^{acd}}_e {R^b}_{dcf}
       + {R^{acd}}_e {R^b}_{cdf}
      \right] \frac{\sigma^{;e}\sigma^{;f}}{2\sigma}\\
    &
    + \frac{1}{96} {R^{cde}}_f R_{cdeg}
      \frac{\sigma^{;f}\sigma^{;g}}{2\sigma}
      \left(g^{ab} - 2 \frac{\sigma^{;a}\sigma^{;b}}{2\sigma} \right)\biggr),
  \end{split}
\end{equation}
where the subscript $DS$ reminds of the fact that the special Hadamard
distribution $G_{DS}$ was used.
\begin{rem}$\phantom{=}$ \par
  \begin{itemize}
  \item This result is not exact, but of order $1/m^6$. However all
    missing terms have only finite contributions in the coincidence
    limit (See~\cite{chris:76}).
  \item The result does not depend on the length scale chosen. A
    change of length scale ($\sigma \rightarrow \lambda^2 \sigma'$)
    yields simple factors in front of the quartic, quadratic and
    linear divergent terms associated to the single fields, but
    produces a finite ($\ln(\lambda^2 m^2)$) contribution to be added
    to the logarithmic divergent terms.  On the other hand the fact
    that the divergent terms cancel each other in the situation
    considered here, implies that those finite terms cancel as well.
  \end{itemize}
\end{rem}
It is easy to see that Eqn.~(\ref{eq:emtDS}) remains finite if $y$
approaches $x$, but unfortunately the result is direction dependent,
i.e., it depends on how the points were `separated'.  To remedy this
defect, we use an elementary averaging procedure described by Adler et
al.\ in~\cite{adler:77}. It consists in the following
replacements:
\begin{align*}
  \sigma^{;a}\sigma^{;b} & \rightarrow \frac{1}{2} g^{ab} \sigma\\
  \sigma^{;a}\sigma^{;b}\sigma^{;c}\sigma^{;d} & \rightarrow
    \frac{1}{6} \sigma^2 ( g^{ab}g^{cd} + g^{ac}g^{bd} + g^{ad}g^{cd})
\end{align*}
Note that these replacements correspond to a decomposition of a
direction dependent expression, e.g., $\sigma^{;a}\sigma^{;b}$,
according to irreducible representations of the Lorentz-group, keeping
the scalar part,e.g., $\frac{1}{4}g^{ab}$ $ \sigma^{;c} \sigma_{;c}$,
only.  Let us denote the averaged result for $<\tilde{T}^{ab}>_{DS}$
by $<\tilde{T}^{ab}>_{\overline{DS}}$. Then the coincidence limit is
well defined and we obtain as a final result:
\begin{equation}\label{eq:expTabDS}
  \begin{split}
    <\tilde{T}^{ab}>_{\overline{DS}}
    & =  \frac{1}{4\pi^2}
    \biggl(
     + \frac{1}{288} R^{acde} {R^b}_{cde}
     + \frac{1}{288}  R^{acde} {R^b}_{edc}
     - \frac{1}{96}  R^{acde} {R^b}_{dce}
    \biggr).\\
    & = 0 \qquad{} \text{if $R^{ab}=0$},
  \end{split}
\end{equation}
since $R^{acde}({R^b}_{cde} + {R^b}_{dec} + {R^b}_{ecd}) = 0 =
R^{acde}{R^b}_{cde} + 2 R^{acde} {R^b}_{dec}$ (See Adler et
al.~[loc.~cit.]).
It is worth noting that this result already implies the finiteness of
$<\tilde{T}^{ab}>_\omega$ in the coincidence limit\footnote{After
  elementary averaging.} for {\em every\/} locally SUSY Hadamard product
state $\omega$, since all Hadamard distributions differ by a {\em
  smooth\/} function only. Thus for our state $\omega$, which we
assume to be characterized by its smooth part $H^{(p)}$:
\[
<\tilde{T}^{ab}>_\omega = <\tilde{T}^{ab}>_{\overline{DS}}
                          + <\tilde{T}^{ab}>_{(H^{(p)} -
                            \sideset{^{DS}}{^{(p)}}{H})}
                        \equiv
                          <\tilde{T}^{ab}>_{\overline{DS}}
                          + <\tilde{T}^{ab}>_{H^{(p)}}
                          - <\tilde{T}^{ab}>_{\sideset{^{DS}}{^{(p)}}{H}}
\]
where $<\tilde{T}^{ab}>_{H^{(p)}}$ denotes Eqn.~(\ref{eq:myEmtFull})
with $G$ and ${\mathcal G}$ respectively omitted.
\begin{rem}
  DeWitt has shown in~\cite{DeWitt:75} that
  $<\tilde{T}^{ab}>_{\overline{DS}}$ can be obtained by functionally
  differentiating an effective action with respect to the metric.
  Hence this expression --as is the classical stress-energy tensor--
  is covariantly conserved by construction. On the other hand for the
  smooth part $<\tilde{T}^{ab}>_{\sideset{^{DS}}{^{(p)}}{H}}$ one
  might encounter $\nabla_a
  <\tilde{T}^{ab}>_{\sideset{^{DS}}{^{(p)}}{H}}\neq 0$, since this
  part does not contain the finite contributions from
  $<\tilde{T}^{ab}>_{\overline{DS}}$. However being an auxiliary
  expression only, which was introduced to simplify the calculations
  of $<\tilde{T}^{ab}>_\omega$, this property of
  $<\tilde{T}^{ab}>_{\sideset{^{DS}}{^{(p)}}{H}}$ does not affect the
  covariant conservation of $<\tilde{T}^{ab}>_\omega$ as a whole.
\end{rem}
\begin{Lemma}
  Let $H^{(p)} = \sum_{i=0}^p H_i \sigma^i$ denote the
  smooth\footnote{Strictly speaking $H^{(p)}$ is $p$-times
    differentiable.} part of the scalar and fermionic Hadamard series
  respectively. Then
  \begin{equation*}
    \begin{split}
      <\tilde{T}^{ab}(x,x)>_{H^{(p)}} & \equiv
      [<\tilde{T}^{ab}>_{H^{(p)}}]\\
      & =
      \frac{1}{4}
      \left(
        2 [H_0^{;a}]^{;b} + 2  [H_0^{;b}]^{;a}
        - [H_0]^{;ab} - 4 [ H_0^{;ab}] - 4 [H_1] g^{ab}
      \right)\\
      & \phantom{=}
      - \frac{1}{8} \left\{  \gamma^{(a} \gamma^c
      \left(
        [H_{0;c}]^{;b)} - 2 [ {H_{0;c}}^b] - 2 [H_1] {g_c}^b,
      \right)
      \right\},
    \end{split}
  \end{equation*}
  where $[\cdot]$ as usual denotes the coincidence limit.
\end{Lemma}
\begin{pf}
  Write $H^{(p)}$ in its series expansion:
  \[
  H^{(p)}= \sum_{i=1}^{p} H_i \sigma^i
  \]
  and note that $<\tilde{T^{ab}}>_{H^{(p)}}$ contains second
  derivatives only. Differentiating $H^{(p)}$ two times yields:
  \[
  {H^{(p)}}_{;ab} = H_{0;ab} + H_{1;ab}\sigma + H_{1;a}\sigma_{;b} +
     H_{1;a}\sigma_{;b} + H_1 \sigma_{;ab} + O(\sigma).
  \]
  For the Lemma we are only interested in terms which do not vanish in
  the coincidence limit. These terms are $H_{0;ab}$ and $H_1
  \sigma_{;ab}$, since it is one of the defining properties of
  $\sigma$ that its coincidence limit and that of its covariant
  derivative vanishes. Thus for the coincidence limit of
  $<\tilde{T^{ab}}(x,y)>_{H^{(p)}}$ one gets:
  \begin{equation}
    \label{eq:coincEmtSmooth}
    \begin{split}
      [ <\tilde{T^{ab}}>_{H^{(p)}} ] & =
    2 \cdot \frac{1}{8} \biggl(
      [{g^b}_{c'}] \left( [H_0^{ac'}] + [H_1] [\sigma^{;ac'}] \right)
      + [{g^a}_{c'}] \left( [H_0^{c'b}] + [H_1] [\sigma^{;c'b}]
    \right)\\
    & \phantom{= 2 \cdot \frac{1}{8} \biggl(}
    - \left( [H_0^{;ab}] + [H_1] [\sigma^{;ab}] \right)
    - [ {g^a}_{c'}] [ {g^b}_{d'}] \left( [H_0^{;c'd'}] + [H_1]
    [\sigma^{;c'd'}] \right)\\
    & \phantom{=}
    - \frac{1}{8} \Tr
    \left\{  [{\mathcal J}] \gamma^{(a} \gamma^c
      \left( [{g^{b)}}_{d'}] \bigl( [ {H_{0;c}}^{d'} + [H_1] [
          {\sigma_{;c}}^{d'}] \bigl)
      -    \bigl( [ {H_{0;c}}^{b}] + [H_1] [
          {\sigma_{;c}}^{b}] \bigr)
      \right)
    \right\} \\
    & =
    \frac{1}{4}
    \left(
      2 [H_0^{;a}]^{;b} + 2  [H_0^{;b}]^{;a}
      - [H_0]^{;ab} - 4 [ H_0^{;ab}] - 4 [H_1] g^{ab}
    \right) \\
    & \phantom{=}
    - \frac{1}{8} \left\{  \gamma^{(a} \gamma^c
     \left(
       [H_{0;c}]^{;b)} - 2 [ {H_{0;c}}^b] - 2 [H_1] {g_c}^b,
     \right)
    \right\}\\
    \end{split}
  \end{equation}
  since $[\sigma^{;ab}] = [\sigma^{;a'b'}] = g^{ab}$, $[\sigma^{;ab'}]
  = - g^{ab}$ and $[{g^b}_{c'}] = {\delta^b}_{c}$. We used
  Synge's Theorem (\cite{chris:76}) which states that
  \[
  [H_{a_1,\cdots,a_n,{b'}_1,\cdots,{b'}_m;c'}] = -
  [H_{a_1,\cdots,a_n,{b'}_1,\cdots,{b'}_m;c}] +
  [H_{a_1,\cdots,a_n,{b'}_1,\cdots,{b'}_m}]_{;c},
  \]
  for any bi-tensor $H_{a_1,\cdots,a_n,{b'}_1,\cdots,{b'}_m}$ whose
  coincidence limits are known.
\end{pf}
Now we apply this Lemma to the smooth DeWitt-Schwinger term,
$\sideset{^{DS}}{^{(p)}}{H}$. The necessary coincidence limits for
$\sideset{^{DS}}{_0}{H}$ and its derivatives, as well as for
$\sideset{^{DS}}{_1}{H}$, are collected in
Appendix~\ref{sec:hadamRecRel}. The result is:
\begin{equation}{}
  \label{eq:smoothDSTerm}
  \begin{split}
    <\tilde{T^{ab}}(x,x)>_{\sideset{^{DS}}{^{(p)}}{H}}
    & \phantom{=}{}
    (-1)
    \left(
      - \frac{\delta}{90} R^{cdea}{R_{cde}}^b
      + g^{ab} \bigl( \frac{2\delta-3}{720} W^2 - \frac{1}{4} m^4 (\delta
      -1) \bigr)
    \right)\\
    & \phantom{=}
    - \frac{1}{8} \Tr \biggl( \gamma^{(a} \gamma^{c}
    \Bigl\{ + \frac{\delta}{24} G_{[kd]}G_{[ef]} ( {R^{kdg}}_c
    {{R^{ef}}_g}^{b)} + R^{kdgb)} {R^{ef}}_{gc} )
    + \frac{\delta}{45} {R^{kde}}_{c} {R_{kde}}^{b)} \\
    & \phantom{=}
    - 2 {g_c}^{b)}
    \bigl(
     \frac{2\delta-3}{192} G_{[kd]} G_{[ef]} R^{kdgh} {R^{ef}}_{gh}
     + \frac{2 \delta - 3}{720} W^2 - \frac{m^4}{4} (\delta-1)
    \bigr)
    \Bigl\}
    \biggl)\\
    & =
    \frac{\delta}{90} R^{cdea} {R_{cde}}^b
    -  g^{ab} \bigl( \frac{2\delta-3}{720} W^2 - \frac{1}{4} m^4 (\delta
      -1) \bigr)\\
    & \phantom{=}
    +  g^{ab} \bigl( \frac{2\delta-3}{720} W^2 - \frac{1}{4} m^4 (\delta
      -1) \bigr)
    - g^{ab} W^2 \frac{2 \delta -3}{768}
    - \frac{\delta}{90} R^{cdea} {R_{cde}}^b\\
    & \phantom{=}
    - \frac{\delta}{192}
    \Tr\bigl( \gamma^{(a} \gamma^c G_{[kd]} G_{[ef]} \bigr)
    \bigl( {R^{kdg}}_c {{R^{ef}}_g}^{b)}
           + {R^{kdgb)}} {{R^{ef}}_{gc}} \bigr)\\
    & = \frac{2 \delta -3 }{768} W^2 g^{ab} -
    \frac{\delta}{48} ( R^{acde} {R^b}_{ced} + 2 R^{acde}
    {R^b}_{dec}),
  \end{split}
\end{equation}
where $W^2 = R^{abcde} R_{abcde}$, $\delta=\ln m^2 - \ln 2 + 2
\gamma$ and the trace formulas of Appendix~\ref{sec:usefulForm} were used.
As an example let us now consider the Schwarzschild and Kerr solution
of Einstein's equation.
%
%%%%%%%%
\section{Two examples: The Schwarzschild and Kerr solution}
\label{sec:example}
%%%%%%%%
%
%%%%%%%%%%%
\subsection{The Schwarzschild solution}
\label{sec:exampSchwarzschild}
%%%%%%%%%%%
%
One of the simplest relativistic star models is the Schwarzschild
solution of Einstein's equation. It is an exact vacuum solution which
describes the geometry outside a static spherical symmetric
relativistic system. Moreover, it has been shown implicitly by
Bartnik, Chru\'sciel and Murchdha~\cite{bartChrMu:90} that every point
in this spacetime possesses a maximal spacelike hypersurface passing
through it. With these two properties the Schwarzschild solution
provides a good background for an application of our method.
We are going to calculate the `expectation value' of the stress energy
tensor of a locally SUSY model with respect to the DeWitt Schwinger
pseudostate ($<\tilde{T}^{ab}>_{\overline{DS}}$) together with the
corresponding smooth part
$<\tilde{T}^{ab}>_{\sideset{^{DS}}{^{(p)}}{H}}$. The actual
computations are performed using the {\tt Maple~V} Symbolic
Computation Program by Waterloo Maple
Software~\cite{maple:libref,maple:langref}. (A commented listing and
sample run can be found in the Appendix~\ref{sec:maplelist}.
Using the Schwarzschild coordinates $t,r,\theta,\phi$, the line
element of this spacetime is known to be of the form:
\[
ds^2 = ( 1- \frac{2 M}{r^2}) dt^2 - ( 1- \frac{2 M}{r^2})^{-1} dr^2 -
       - r^2 d\theta^2 - r^2 \sin^2 \theta d\phi^2,
\]
where $M$ can be interpreted as the total mass of the system as
measured from infinity. In these coordinates the only non vanishing
components of the Riemann tensor are
\begin{alignat*}{3}
  R_{0101}        & =  \frac{2M}{r^3} \quad
  & R_{0202}      & =  -\frac{(r-2M)M}{r^2} \quad
  & R_{0303}      & = \frac{(r -2M)\sin^2 \theta}{r^2} \\
  R_{1212}        & = \frac{\sin^2 \theta}{r-2M} \quad
  & R_{2323}      & = -2 r M \sin^2 \theta
\end{alignat*}
and those which follow from symmetry. Contracting the Riemann tensor
with itself yields:
\[
 R^{abcd} R_{abcd} = 48 \frac{M^2}{r^6},
\]
the Kretschman invariant. Inserting these results in
Eqn.~(\ref{eq:expTabDS}) we obtain:
\[ < \tilde{T}^{ab}(x)>_{{\overline{DS}}} =
 \left(
{\begin{array}{rrrr}
0 & 0 & 0 & 0 \\
0 & 0 & 0 & 0 \\
0 & 0 & 0 & 0 \\
0 & 0 & 0 & 0
\end{array}}
 \right) ,
\]
a result which clearly is covariantly constant. For the smooth part
$<\tilde{T}^{ab}>_{\sideset{^{DS}}{^{(p)}}{H}}$ the result is more
complicated:
\begin{eqnarray*}{}
\lefteqn{<\tilde{T}^{ab}>_{\sideset{^{DS}}{^{(p)}}{H}} =
  \left( {\vrule height0.87em width0em depth0.87em}
 \right. \! \! \,{\displaystyle \frac {1}{16}}\,{\displaystyle
\frac {{M}^{2}\,(\,10\,{r}\,{\delta} - 3\,{r} - 16\,{\delta}\,{
M}\,)}{{r}^{5}\,(\,{r} - 2\,{M}\,)^{2}}}\,,  - \,{\displaystyle
\frac {1}{16}}\,{\displaystyle \frac {(\,2\,{\delta} - 3\,)\,{M}
^{2}}{{r}^{6}}}\,, } \\
 & &  - \,{\displaystyle \frac {1}{16}}\,{\displaystyle \frac {(
\,2\,{\delta} - 3\,)\,{M}^{2}}{{r}^{7}\,(\,{r} - 2\,{M}\,)}}\,,
{\displaystyle \frac {1}{16}}\,{\displaystyle \frac {(\,2\,{%
\delta} - 3\,)\,{M}^{2}}{{r}^{7}\, \left( \! \, - {r} + {r}\,
{\rm cos}(\,{\theta}\,)^{2} + 2\,{M} - 2\,{M}\,{\rm cos}(\,{%
\theta}\,)^{2}\, \!  \right) }}\, \! \! \left. {\vrule
height0.87em width0em depth0.87em} \right)  \\
 & &  \left( {\vrule height0.87em width0em depth0.87em}
 \right. \! \! \, - \,{\displaystyle \frac {1}{16}}\,
{\displaystyle \frac {(\,2\,{\delta} - 3\,)\,{M}^{2}}{{r}^{6}}}
\,,  - \,{\displaystyle \frac {1}{16}}\,{\displaystyle \frac {(\,
6\,{r}\,{\delta} + 4\,{\delta}\,{M} + 3\,{r} - 6\,{M}\,)\,{M}^{%
2}\,(\,{r} - 2\,{M}\,)}{{r}^{8}}} \\
 & & , {\displaystyle \frac {1}{16}}\,{\displaystyle \frac {(\,{r
} - 2\,{M}\,)\,(\,2\,{\delta} - 3\,)\,{M}^{2}}{{r}^{9}}}\,,  -
\,{\displaystyle \frac {1}{16}}\,{\displaystyle \frac {{M}^{2}\,(
\,2\,{r}\,{\delta} - 4\,{\delta}\,{M} - 3\,{r} + 6\,{M}\,)}{{r}
^{9}\, \left( \! \, - 1 + {\rm cos}(\,{\theta}\,)^{2}\, \!
 \right) }}\, \! \! \left. {\vrule
height0.87em width0em depth0.87em} \right)  \\
 & &  \left( {\vrule height0.87em width0em depth0.87em}
 \right. \! \! \, - \,{\displaystyle \frac {1}{16}}\,
{\displaystyle \frac {(\,2\,{\delta} - 3\,)\,{M}^{2}}{{r}^{7}\,(
\,{r} - 2\,{M}\,)}}\,, {\displaystyle \frac {1}{16}}\,
{\displaystyle \frac {(\,{r} - 2\,{M}\,)\,(\,2\,{\delta} - 3\,)
\,{M}^{2}}{{r}^{9}}}\,,  \\
 & &  - \,{\displaystyle \frac {1}{16}}\,{\displaystyle \frac {{M
}^{2}\,(\, - 2\,{\delta} + 3 + 8\,{\delta}\,{r}^{2}\,)}{{r}^{10
}}}\,,  - \,{\displaystyle \frac {1}{16}}\,{\displaystyle \frac {%
(\,2\,{\delta} - 3\,)\,{M}^{2}}{{r}^{10}\, \left( \! \, - 1 +
{\rm cos}(\,{\theta}\,)^{2}\, \!  \right) }}\, \! \! \left.
{\vrule height0.87em width0em depth0.87em} \right)  \\
 & &  \left( {\vrule height0.95em width0em depth0.95em}
 \right. \! \! \,{\displaystyle \frac {1}{16}}\,{\displaystyle
\frac {(\,2\,{\delta} - 3\,)\,{M}^{2}}{{r}^{7}\, \left( \! \, -
{r} + {r}\,{\rm cos}(\,{\theta}\,)^{2} + 2\,{M} - 2\,{M}\,{\rm
cos}(\,{ \theta}\,)^{2}\, \!  \right) }}\,,  \\
 & &  - \,{\displaystyle \frac {1}{16}}\,{\displaystyle \frac {{M
}^{2}\,(\,2\,{r}\,{\delta} - 4\,{\delta}\,{M} - 3\,{r} + 6\,{M}
\,)}{{r}^{9}\, \left( \! \, - 1 + {\rm cos}(\,{\theta}\,)^{2}\,
 \!  \right) }}\,,  - \,{\displaystyle \frac {1}{16}}\,
{\displaystyle \frac {(\,2\,{\delta} - 3\,)\,{M}^{2}}{{r}^{10}\,
 \left( \! \, - 1 + {\rm cos}(\,{ \theta}\,)^{2}\, \!  \right) }}
\,,  \\
 & & {\displaystyle \frac {1}{16}}\,{\displaystyle \frac {{M}^{2}
\, \left( \! \,2\,{ \delta} - 3 - 8\,{ \delta}\,{r}^{2} + 8\,{
\delta}\,{r}^{2}\,{\rm cos}(\,{ \theta}\,)^{2}\, \!  \right) }{{r
}^{10}\, \left( \! \,1 - 2\,{\rm cos}(\,{ \theta}\,)^{2} + {\rm
cos}(\,{ \theta}\,)^{4}\, \!  \right) }}\, \! \! \left. {\vrule
height0.95em width0em depth0.95em} \right)
\end{eqnarray*}
\begin{rem}
  It is explicitly shown in the Appendix that this expression is {\em
    not\/} covariantly constant. This only expresses
  Wald's observation~\cite{wald:78} that the divergent part in the
  Adler, Lieberman and Ng renormalization prescription is not
  covariantly constant in general.
\end{rem}
%
%%%%%%%%%%%
\subsection{The Kerr solution}
\label{sec:exampKerr}
%%%%%%%%%%%
%
The Kerr solution of the Einstein equation describes the geometry near
a rotating mass. The results of Bartnik and al.~loc.~cit.\ apply to
this Lorentz manifold, too, i.e., every point in the Kerr spacetime possesses
a maximal spacelike hypersurface passing through it.
We are going to calculate the `expectation value' of the stress energy
tensor of our model with respect to the DeWitt Schwinger pseudostate
($<\tilde{T}^{ab}>_{\overline{DS}}$), by using a slightly modified
version of the {\tt Maple V} program, which was used above for the
Schwarzschild case.  The reader is referred to
Appendix~\ref{sec:maplelist} for a commented listing and a sample run.
It should be noted that this example is already complex enough so that
the smooth part, $<\tilde{T}^{ab}>_{\sideset{^{DS}}{^{(p)}}{H}}$,
can't be handled by the program in reasonable times.
Using the Schwarzschild coordinates $t,r,\theta,\phi$, the line
element of the Kerr spacetime is known to be of the form:
\[
   ds^2 = dt^2 - (r^2 + a^2) \sin^2 \theta d\theta^2
          - \frac{2Mr}{\rho^2} ( dt + a \sin^2\theta d \phi)^2
          - \rho^2 ( d \theta^2 + \frac{dr^2}{r^2 - 2Mr + a^2},
\]
where $\rho^2 := r^2 + a^2 \cos^2 \theta$. The parameter $a$ corresponds
to the angular momentum of the rotating system. The calculation yields
the manifestly conserved result:
\[ < \tilde{T}^{ab}(x)>_{{\overline{DS}}} =
 \left(
{\begin{array}{rrrr}
0 & 0 & 0 & 0 \\
0 & 0 & 0 & 0 \\
0 & 0 & 0 & 0 \\
0 & 0 & 0 & 0
\end{array}}
 \right) .
\]
%
%%%%%%%%%%%%%%%%%%%%%%%%%%%%%%%%%%%%%%%%%%%%%%%%%%%%%%%%%%%%%%%%%%%%%%%%%
\chapter*{Discussion and outlook}
\label{sec:disAndOut}
%%%%%%%%%%%%%%%%%%%%%%%%%%%%%%%%%%%%%%%%%%%%%%%%%%%%%%%%%%%%%%%%%%%%%%%%%
%
For an analogon of the free Wess-Zumino model on a vacuum solution of
the Einstein equations, a relation between conserved `supercurrents'
and the point-separated stress energy tensor, similar to the one on
Minkowski space, was established. The assumption on the scalar
curvature of the underlying manifold, i.e., $R^{ab}=R=0$, is necessary
to yield a {\em conserved\/} current. This in turn ensures --at least
in the flat case-- that the corresponding stress tensor is covariantly
constant. On the other hand if one relaxes this assumption, our
definition of the stress tensor,
Eqn.~(\ref{eq:pointsplitEMTcurvedKern}), would still make sense,
yielding Eqn.~(\ref{eq:pointsplitEMTcurvedbasicfields}).
However, the result in terms of the basic fields would {\em not\/}
reproduce the true improved stress tensor, Eqn.(\ref{eq:SUSYemtsplit}),
i.e., Eqn.~(\ref{eq:pointsplitEMTcurvedbasicfields}) and
Eqn.~(\ref{eq:SUSYemtsplit}) differ by terms which vanish on vacuum
solutions of the Einstein equations only. These difference terms
prevent Eqn.~(\ref{eq:pointsplitEMTcurvedbasicfields}) from being
covariantly conserved. This means that
Eqn.~(\ref{eq:pointsplitEMTcurvedbasicfields}) can not be interpreted
as a physically sensible stress tensor in this case. Moreover, the
divergent parts of the true improved stress energy tensor, according
to Christensen's calculations, do not cancel if $R^{ab} \neq
0$.\footnote{This result can also be verified using the Hadamard
  condition of Kay and Wald instead of Christensen's DeWitt-Schwinger
  ansatz~\cite{koehler:93}.} Both facts considerable limit the use of
the Bose-Fermi cancellation mechanism in the framework of quantum field
theory on curved spacetime, since a `fine tuning' of the underlying
geometry is necessary. Nevertheless the examples of
chapter~\ref{sec:emtCurved} section~\ref{sec:example} show that the
Schwarzschild and Kerr solutions together with the DeWitt-Schwinger
pseudostate\footnote{It is not clear whether the DeWitt-Schwinger
  prescription leads to a {\em state\/} in our case, i.e., whether
  positivity holds.} of a locally SUSY model are  self consistent
solutions of the semiclassical Einstein equation.
A more realistic state on the first manifold is the Hartle-Hawking
vacuum~\cite{hartleHawking:76}, which represents a black hole of mass
$M$ in unstable thermal equilibrium with a bath of black body
radiation of local temperature $T_{\text{loc}} = (8\pi M)^{-1} ( 1-
2M/r)^{-1/2}$. In a future work this state and its generalization to
Fermi fields could be used to calculate the corresponding expectation
value of our stress tensor. One could further investigate whether the
Hawking radiation associated to this state gets modified by the
Bose-Fermi symmetry of the model. This might also answer the question
whether the Hartle Hawking vacuum gives a new self consistent solution
of the semiclassical Einstein equation. On the other hand, independent
of the result, the `fine tuning' problem mentioned above suggests that
such a solution, as well as the one presented in this work, is not
stable, i.e., every small perturbation of the metric which changes the
Ricci tensor immediately yields a divergent result for the stress
tensor, making the semiclassical Einstein equation ill defined. It
should be noted that this statement is qualitatively different
compared to similar statements for the semiclassical Einstein
equation with a {\em renormalized\/} stress tensor on the r.h.s. In
the latter case the equation remains well defined if the metric
changes and the usual phase space methods can be used to check
stability. (See~\cite{castGunNardPaz:86} for partial results on the
stability of the DeSitter universe with a neutral scalar field.)
The second assumption necessary to reproduce the true improved stress
energy tensor, is the setting of the coupling constant $\xi$ to $1/4$.
This setting is motivated by the observation that the `squared' Dirac
operator when applied to spinors equals the SUSY coupled Klein-Gordon
operator times the identity. Moreover, according to Christensen's
calculations, any other setting of $\xi$ would give a divergent result
for the expectation value of the true improved stress energy tensor.
For the physical implications of this setting one should note that the
field equations on a Ricci flat manifold are independent of $\xi$.
Thus on these manifolds SUSY coupling yields --via Dimock's
construction-- the {\em same\/} algebras of local observables as the
usual minimal ($\xi=0$) or conformal ($\xi=1/6$) setting. It follows
that the physics does not depend on the choice of $\xi$, which is
satisfactory.  On the other hand the parameter $\xi$ enters into the
classical definition of the improved stress energy tensor in a non
trivial way, since in order to obtain it, the Lagrangian is variated
with respect to the metric, i.e., the variations are not restricted to
the Ricci flat case. This means that from a Lagrangian point of view,
SUSY coupling may be interpreted as a renormalization: For instance,
in the minimal ($\xi=0$) setting, it consists in adding the
counterterm $+(m^2 + \frac{1}{4}R)^{1/2} - m$ to the scalar masses.
Obviously an analogous statement holds for the conformal setting.
Recall that the latter would be appropriate for a massless model.
Unfortunately for such models, the DeWitt-Schwinger approach can not
be used a priori, since it contains a power series expansion in $1/m$.
On the other hand Christensen showed in~\cite{chris:78} that all his
calculations remain valid in the massless limit, if one sets $m=0$
everywhere except in the logarithmic divergent term. There one has to
substitute the mass $m$ by some `cutoff' mass $\mu$. In our setting
the logarithmic divergent parts cancel each other and we could remove
the cutoff.  Moreover, our final result for the DeWitt-Schwinger
pseudostate (Eqn.~(\ref{eq:expTabDS})) is independent of $m$, which
means that it is valid in the massless limit without any
modifications. The reader should recall that one does not have a
natural lengh scale in the massless case, which would, if
renormalization were necessary, result in an ambiguity of the stress
tensor. It is worth noting that in our model, due to its Bose-Fermi
symmetry, this ambiguity is not present (compare
chapter~\ref{sec:emtCurved} section~\ref{sec:vacExpValue}).
For a successful cancellation of the divergencies, as a third
assumption, equal masses for the three fields contained in the model
were needed. As is well known from experiments, this assumption is
strongly violated in the real world. For supersymmetric models on
Minkowski space, which do face the same problem, the best way to solve
this deficiency seems to be using a mechanism of spontanous symmetry
breaking. However as far as the author knows, such a mechanism has not
yet been implemented into a quantum field theory on a curved
spacetime. This might be due to the fact that it is up to now very
difficult to do perturbation theory on manifolds. The study of more
realistic models should therefore presumable be done on a simple
spacetime first, for instance Schwarzschild, since the corresponding
equations of motion are too complicated otherwise. It is worth noting
that in our model, even for different masses, the worst divergencies
of the point separated stress energy tensor cancel each other, as long
as the metric remains Ricci flat, i.e., $R^{ab}=0$. According to
Christensen, we are left with a logarithmically divergent part only.
Its {\em numerical\/} contribution to the `vacuum expectation value'
of the stress tensor is expected to be small, even if the `separated'
points almost coincide, i.e., if their distance is comparable to the
Planck length.  Recall that this distance is believed to be the border
of validity of our semiclassical approximation, meaning that smaller
distances presumable do not make sense in the framework. Thus there
should exist a regime, where the logarithmic divergent part can be
neglected in explicit calculations.
Up to now we only considered the expectation value of the stress
tensor in some state. For a supersymmetric model on flat space, the
stress tensor itself, due to its relation to the conserved
supercurrent and the supercharge, can be defined as an operator valued
distribution on the vacuum Hilbertspace. Unfortunately one can not
apply the same line of argument to a locally SUSY model on a general
Ricci flat manifold. Although it is true that the latter possesses a
conserved `supercurrent', the definition of a `supercharge' as an {\em
  operator\/} on some GNS-Hilbertspace failed, mainly because on a
general manifold there do not exist covariantly constant classical
spinor fields\footnote{It would be intersting to classify all Ricci
  flat spacetimes, which possesses constant spinors, i.e., which are
  K{\"a}hler manifolds in four dimensions (See~\cite{benn:spinors}).}.
Without such a constant field, the integral of the `supercurrent',
which is necessary for obtaining a charge is not defined. In order to
integrate the current, one parallel transports its spinor index to
some reference point. The result is a bispinor-tensor, which fails to
be conserved. This in turn implies, that no charge {\em operator\/} on
the GNS-Hilbertspace of any Hadamard state exists (see Theorem~6.2.10
of~\cite{lopus:SUSYinQFT}). On the other hand the existence of the
stress energy tensor for a supersymmetric model on Minkowski space as
an operator valued distribution can equally well be established using
Wick products of the basic fields together with the fact that the
vacuum expectation value of the point separated stress tensor vanishes
in the coincidence limit. Namely, using the following definition for
the Wick ordered stress energy tensor:
\begin{equation}
  \label{eq:wickorderedTab}
  :T^{ab}:(x) := \lim{}_{x_1\rightarrow x} \left( T^{ab}(x,x_1) - <0 |
  T^{ab}(x,x_1) 0>  \right),
\end{equation}
where $:\cdot:$ denotes Wick ordering and $|0>$ is the Minkowski
vacuum, shows that
\[T^{ab}(x) := \lim_{x_\rightarrow x} T^{ab}(x,x_1)\]
is a well defined operator valued distribution that differs from
$:T^{ab}:$ by a finite function times the unit operator. Note that
Eqn.~(\ref{eq:wickorderedTab}) is the usual definition for Wick
ordering (compare~e.g.~\cite{strWight:PCT}), since $T^{ab}$ is
quadratic in the basic fields and that the coincidence limit
\[
\lim_{x_1\rightarrow x} <0 |
T^{ab}(x,x_1)0>
\]
exists. Thus showing that Wick products of our basic fields on a
manifold with respect to some Hadamard state can be defined as new
Wightman fields would immediately yield a stress tensor as an operator
valued distribution on the corresponding GNS-Hilbertspace after
elementary averaging. Let us sketch the main idea for the proof of the
first statement. Let $A$ denote a free scalar Wightman field on the
GNS-Hilbertspace $({\EuScript H}, D, A, \Omega)$ of a quasifree
Hadamard state. The kernel
\[
:A^2:(x,y) := A(x)\cdot A(y) - <\Omega|A(x) \cdot A(y) \Omega>
\]
yields a well defined operator valued distribution on ${\EuScript
  H}$. Its restriction to the diagonal is the definition of the Wick
square of $A$, denoted by $:A^2:(x)$. To show that $:A^2:(x,y)$ admits
such a restriction, one proceeds as in the proof of
Proposition~\ref{prop:wightmanCur}, i.e., one considers
\begin{equation}
  \label{eq:normWick}
  \| \cdots A(z_1) \cdots :A^2:(x_1,y_1) \cdots A(z_l) \cdots
  :A^2:(x_m,y_m) \cdots A(z_n) \Omega \|
\end{equation}
and asks, whether the restriction of Eqn.~(\ref{eq:normWick}) to the
diagonal $x_1=y_1,\ldots,x_m=y_m$ exists. Note that
Eqn.~(\ref{eq:normWick}) splits into sums of products of two-point
distributions, since $\Omega$ arises from a quasifree
state. Fortunately all terms containing two-point distributions of the
form $<\Omega|A(x_i)A(y_i) \Omega>$ occur twice thanks to the
definition of $:A^2:(x,y)$ and cancel each other. For the remaining
terms, for instance
\[
\cdots <\Omega|A(x_i)A(x_j) \Omega> \cdots <\Omega|A(y_i)A(y_j)
\Omega> \cdots,
\]
setting $x_i=y_i$ yields products of two-point distributions, which
are well defined, since the state $\Omega$ satisfies the new general
wave front set spectrum condition (Definition~\ref{Dfn:GWFSSC}) by
assumption (compare chapter~\ref{sec:analogonWess}
section~\ref{sec:WFSSC}). We may therefore use the sequential
continuity of the restriction mapping to establish that
Eqn.~(\ref{eq:normWick}) can be restricted to the diagonal. For the
exact proof of this statement and its generalization to arbitrary Wick
powers, we refer the reader to~\cite{brunFredKoe:94}, which is in
preparation. Note that the argument in fact uses only that the state
under consideration is quasifree and satisfies the general wave front
set spectrum condition (GWFSSC). Hence the result extends to the other
basic fields of our local SUSY model. On the other hand the GWFSSC
would be of limited interest without the methods of the microlocal
analysis developed by H{\"o}rmander and
Duistermaat~\cite{Hoermander:71,Hoermander:72} which were extensively
used in this work.  Their mathematical tools together with the wave
front set characterization of globally Hadamard
states~\cite{Radzikowski:92} and our GWFSSC might turn out to be very
powerful for various constructive problems of quantum field theory on
curved spacetime. For example the method developed in
chapter~\ref{sec:emtCurved} section~\ref{sec:superCurCurv} for the
`supercurrent' was successfully used to construct other new Wightman
fields on globally hyperbolic manifolds~\cite{koehler:94}. It was
already mentioned above that its generalization to Wick powers of free
fields is in preparation~\cite{brunFredKoe:94}. Even the formulation
of a perturbation theory on manifolds in terms of Wightman fields
seems possible if one uses the new GWFSSC. Work in this direction is
in progress. As a possible application of the latter one could
consider quantum electrodynamics on globally hyperbolic manifolds and
try to calculate vacuum polarization effects in strong gravitational
fields, e.g., corrections to the life time of excited atoms near
massive objects.  The new GWFSSC might be even useful for a better
understanding of the usual spectrum condition on Minkowski space.
First results in this direction, concerning the compatibility of both
conditions can already be found in Radzikowski's thesis, but the
feeling is that there are still many deep relations to
discover~\cite{Radzikowski:94}. Another thing that should be done is
to clarify the relation between the new GWFSSC and the scaling limit
criterion of Fredenhagen and Haag~\cite{fredenhagenHaag:90}. In
connection with the latter research is done on an {\em algebraic\/}
characterization of wave front sets associated to states.
\begin{ack}
  It is a pleasure to thank my supervisor K.~Fredenhagen for his
  guidance and constant support. I am indebted to him for many
  insights I received during the course of this work. His valuable
  comments and remarks helped me many times to see problems clearer
  and provided substantial motivations. A better way of supervising a
  thesis is hard to imagine.
  This work also profited from helpful and stimulating conversations
  with D.~Buchholz, G.~Mack and H.~Nicolai. It would have been
  impossible without M.~Radzikowski, who generously made the results
  of his PhD thesis available for me prior to publication. My friends
  and colleagues contributed through discussions on topics in general
  relativity and functional analysis. In particular I would like to
  mention W.~Junker, S.~Ohlmeyer and R.~Brunetti. Special thanks are
  due to R.~Verch for many helpful hints and to B.~Rosenow for a
  careful reading of the manuscript.
  The financial support given by the `Studienstiftung des Deutschen
  Volkes' and the `DFG' as part of the `Graduiertenkolleg f\"ur
  theoretische Elementarteilchenphysik' and as part of the project
  `Lokale Quantenphysik und Gravitation' is gratefully acknowledged.
\end{ack}
\appendix
%%%%%%%%%%%%%%%%%%%%%%%%%%%%%%%%%%%%%%%%%%%%%%%%%%%%%%%%%%%%%%%%%%%%%%%%%
%      A P P E N D I X
%%%%%%%%%%%%%%%%%%%%%%%%%%%%%%%%%%%%%%%%%%%%%%%%%%%%%%%%%%%%%%%%%%%%%%%%%
%
%%%%%%%%%%%%%%%%%%%%%%%%%%%%%%%%%%%%%%%%%%%%%%%%%%%%%%%%%%%%%%%%%%%%%%%
\chapter{Algebraic topology}
\label{sec:algTopology}
%%%%%%%%%%%%%%%%%%%%%%%%%%%%%%%%%%%%%%%%%%%%%%%%%%%%%%%%%%%%%%%%%%%%%%%
%
For the convenience of the reader this appendix contains some basic
terminology of algebraic topology. A well written and quite short
introduction into this subject can be found in the book by R. S. Ward
and R. O. Wells~\cite{ward:twistor}. The book by M.
Nakahara~\cite{nak:geometry} is well suited for this purpose, too. We
will restrict ourselves here to the definition and properties of {\em
  homology groups}, since homotopy groups, despite of their close
relation to the former, are not necessary for the understanding of
Theorem~\ref{thm:existenceSpinorStructure}.
Algebraic topology is a branch of mathematics which tries to
characterize topological spaces by their topological invariants.  One
of the oldest and maybe also most useful topological invariant is
the Euler characteristic $\chi(X)$ of polyeders $X$. It is given by
\[
\chi(X) := (\text{number of vertices}) - (\text{number of edges}) +
(\text{number of faces})
\]
Homology groups are roughly speaking generalizations of this
characteristic. To make this statement more precise we need some
basic terminology.
\section{Basic Terminology}
\label{sec:algBasics}
\begin{Dfn}
  The set
  \[ \Delta^k := \{ (x^1, \ldots , x^k) \in \bbbR^k | x^i \ge 0 ,
                   x^1 + \cdots + x^k \leq 1 \} \]
  is called {\em standard simplex}.
\end{Dfn}
Note that $\Delta^k$ is a point for $k=0$, a line for $k=1$, a
triangle for $k=2$, \dots.
\begin{Dfn}
  Let $M$ denote a topological space. A continuous  mapping
  \[ \sigma \colon \Delta^k \rightarrow M \]
  is called a {\em (singular) $k$-simplex\/} in $M$.
\end{Dfn}
{\sloppy
One may think of every $k$-simplex as being generated by a convex
linear combination of $k$ points $P_0, \ldots ,P_{k-1} \in \bbbR^k$. Let
${\mathcal R}$ be a commutative ring.
}
\begin{Dfn}
  The set
  \[ S_k(M,{\mathcal R}) := \{ \sigma |\;\sigma =
                     \sum_{\stackrel{\mu}{ {\rm finite}}
                           } a^\mu
                     \sigma_\mu, \; a^\mu \in {\mathcal R} , \quad
                     \sigma_\mu \: \mbox{singular $\mu$-simplex}\}
  \]
  is called the {\em module\/} of singular $k$-chains of $M$ with
  dimension $k$.
\end{Dfn}
Note that the sum is assumed to denote formal linear combination.
\begin{Dfn}
  The $i$-th face $F^i_k$ of a (standard) $k$-simplex is given by the
  mapping
  \[ F^i_k \colon \Delta ^k \rightarrow \Delta^{k-1} :=
          \text{Span}(P_0,\cdots,\hat{P}_i,\cdots,P_k),\]
  where all symbols under a $\hat{~}$ are omitted.
\end{Dfn}
Let $\sigma$ be a singular $k$-simplex. Its faces are given by the
composition with this mapping.
\begin{Dfn}
  To every $k$-simplex we associate a $(k-1)$-simplex $\partial\sigma$
  called the {\em boundary\/} of $\sigma$ by
  \[
  (\partial \sigma) := \sum_{i=0}^{k} (-1)^i \sigma\circ F^i_k.
  \]
\end{Dfn}
We remark that this boundary operator $\partial$ is null potent and
can be extended canonically to singular $k$-chains.
\begin{Dfn}
  A singular $k$-chain $\sigma\in S_k(M,{\mathcal R})$ is called a {\em
    cycle\/} if and only if $\partial\sigma=0$ and a {\em boundary\/} if
  and only if there exists a singular ($k-1$)-chain $\sigma'\in
  S_k(M,{\mathcal R})$ such that $\sigma=\partial\sigma'$,
\end{Dfn}
In the homology group all cycles which differ by a boundary are
identified. More precisely:
\begin{Dfn}
  Let
   \[ S_k(M,{\mathcal R}) \supset Z_k(M,{\mathcal R})=
                   \{ \mbox{Set of all $k$-cycles} \} \]
 and
 \[ S_k(M,{\mathcal R}) \supset B_k(M,{\mathcal R})=
                   \{ \mbox{Set of all $k$-boundaries} \}, \]
then the $k$-th homology group of $M$ over the ring ${\mathcal R}$ is
given by the quotient of cycles and boundaries
\[ H_k(M,{\mathcal R}) : = Z_k(M,{\mathcal R})/B_k(M,{\mathcal R}). \]
\end{Dfn}
Using objects which are dual to the simplices {\em cohomology groups\/}
are defined analogously.
\begin{Dfn}
  A homomorphism $c$ from $S_k(M,{\mathcal R})$ into $\bbbR$ is called
  {\em (singular) $k$-cochain}. The set of all (formal) linear
  combinations of such cochains will be denoted by $S^k(M,{\mathcal R})$.
\end{Dfn}
\begin{Dfn}
  A {\em coboundary operator\/} $\delta\colon S^k \rightarrow S^{k+1}$
  is given by transposition of $\partial$ with respect of the dual pairing.
\end{Dfn}
\begin{exmp}
  Let $\omega$ be a $k$-form on a manifold $M$, $\sigma\in
  S_k(M,\bbbR)$ a singular $k$-chain. Define
  \[ c(\sigma):= \int_{\sigma} \omega = a^\mu \int_{\sigma_\mu} \omega
             = a^\mu \int_{\Delta^k}\sigma^*_\mu(\omega),\]
  where $a^\mu$ are the coefficients of $\sigma=a^\mu\sigma_\mu$. By
  Stokes Theorem we find on the other hand,
  \[
  c(\sigma) = \int_{\partial\sigma} d\omega = a^\mu\int_{\Delta^{k-1}}
    (\partial \sigma_\mu)^* (d\omega)
  \]
  and conclude that the coboundary operator is given by the exterior
  derivative in this case.
\end{exmp}
Cocycles and coboundaries are defined in analogy to cycles and
boundaries respectively. The corresponding sets are denoted by
$Z^k(M,{\mathcal R})$ and $B^k(M,{\mathcal R})$ respectively.
\begin{Dfn}
  The quotient
  \[
   H^k(M,{\mathcal R}) := Z^k(M,{\mathcal R})/B^q (M,{\mathcal R})
  \]
  is called the {\em $k$-th cohomology group\/} of $M$ over the ring
  ${\mathcal R}$.
\end{Dfn}
For the Definition of \v{C}ech-cohomology groups we need some sheaf
theory.
\section{Sheaf theory}
\label{sec:sheafTheory}
Sheafs were introduced by J. Leray in the late forties of this century
during his research on hyperbolic differential equations. Twenty years
later they became a valuable tool in the algebraic topology and
algebraic geometry. The main reason for this is presumably the fact
that sheafs can be used for non topological problems too: Every sheaf
contains local {\em geometrical\/} information; sheaf cohomology
is roughly speaking a tool that  relates local  sections of
sheafs to global ones.
Let $M$ be a topological space. A `presheaf' $S$ of abelian
groups\footnote{In chapter~\ref{sec:spinorsOnManifolds}
  section~\ref{sec:chech-chomology-groups} these groups were identical
  to $\bbbZ_2$.} on $M$ is an assignment of open subsets $U$ of $M$ to
an abelian group $S(U)$
\[
U \rightarrow S(U),
\]
satisfying the following supplementary condition: For all subsets
$V\subset U \subset M$ there exists a restriction homomorphism
\[
r^U_V \colon S(U) \rightarrow S(V)
\]
with the property that for all $W\subset V \subset U$
\[
r^U_V \circ r^V_W = r^U_W \qquad \text{and}\qquad r^U_U = \bbb1
\]
\begin{rem}
  A bundle for example consists of a family of geometrical objects
  characterized by points (fibers). The corresponding objects in
  a presheaf are characterized by open subsets.
\end{rem}
\begin{Dfn}
  A {\em sheaf\/} is a presheaf satisfying two additional properties:
  \begin{itemize}

  \item[-] Let $(U_\alpha)$ be an open covering of $M$, $s_\alpha \in
    S(U_\alpha)$ and
    \[
     \forall  \alpha,\beta : \;
       r^{U_\alpha}_{U_\alpha \cap U_\beta} (s_\alpha)
    = r^{U_\beta}_{U\alpha \cap U_\beta}(s_\beta) ,\]
    then there exists a $s \in S(U)$, such that all $s_\alpha$
    are restrictions of $s$:
    \[
     \forall\alpha : \;
       r^U_{U_\alpha}(s)=s_\alpha.\]
  \item[-]
     $s$ is unique, i.e., let
     $(U_\alpha)$ and $U$ as above be given and  let $s,s' \in S(U)$,
     such that $r^U_{U_\alpha}(s)=r^U_{U_\beta}(s')$ is satisfied
     then $s \equiv s'$.
 \end{itemize}
\end{Dfn}
\section{Sheaf cohomology}
\label{sec:sheafCohomology}
For the Definition of a sheaf cohomology we will proceed similarly as
in section~\ref{sec:algBasics}. Let $M$ denote a topological space and
let $(U_\alpha)$ be an open covering of $M$. A $k$-simplex is an
ordered set of $(k+1)$ open subsets of $(U_\alpha)$ with non void
intersections.
\begin{Dfn}
  Let $\sigma=(U_0,\ldots,U_k)$ be a $(k+1)$-simplex. Its {\em
    support\/} is defined by
  \[
  |{\sigma}|:= U_0 \cap \cdots\cap U_k.
  \]
\end{Dfn}
\begin{Dfn}
  A {\em $k$-cochain\/} of $(U_\alpha)$ is a continuous mapping
  \[
  f \colon \sigma \mapsto S(|{\sigma}|).
  \]
  The set of all $k$-cochains will be denoted by $S^k(U_\alpha,S)$.
\end{Dfn}
Note that $S^k(U_\alpha,S)$ is an abelian group under pointwise
composition.
\begin{Dfn}\label{dfn:coboundaryOp}
  A {\em coboundary operator\/} $\delta$ is defined by:
  \begin{eqnarray*}
    \delta \colon S^k(U_\alpha,S) & \rightarrow & S^{k+1}(U_\alpha,S) \\
           f             & \mapsto      & \delta f \colon \sigma \mapsto
                \prod_{i=0}^{k+1}
                 {\left( r^{\left| \sigma_i \right|}_\sigma
                 f(\sigma_i) \right)}^{(-1)^i} ,
  \end{eqnarray*}
  where $\sigma_i := (U_0,\ldots,\hat{U}_i,\ldots,U_{k+1})$ and $r$
  denotes the restriction mapping.
\end{Dfn}
It is easy to see that this operator is null potent. We denote by
$Z^k(U_\alpha,S)$ and $B^k(U_\alpha,S)$ the set of all cocycles and
coboundaries respectively.
\begin{rem}
  In chapter~\ref{sec:spinorsOnManifolds}
  section~\ref{sec:chech-chomology-groups} we considered the special
  case $S(U) \equiv S \equiv \bbbZ_2$. The two elements of this
  abelian group are usually denoted by $+1$ and $-1$ and their group
  product is the usual multiplication. Moreover, in this special case
  every element is its own inverse, such that the coboundary operator
  defined in chapter~\ref{sec:spinorsOnManifolds}
  section~\ref{sec:chech-chomology-groups} is just a special case of
  Definition~\ref{dfn:coboundaryOp} above.
\end{rem}
\begin{Dfn}
  The quotient
  \[ H^k(U_\alpha,S) := Z^k(U_\alpha,S)/B^k(U_\alpha,S) \]
  is called {\em $k$-th \v{C}ech-cohomology group\/} of $M$ with
  coefficients in the sheaf $S$ with respect to the covering $(U_\alpha)$.
\end{Dfn}
In a last step the construction will be made independent of the
covering using an inductive limit procedure.
Let $(I,\leq)$ be a partial ordered set, $(S_\alpha)_{\alpha\in I}$ a
family of modules indexed by $I$ and let
\[
f^\alpha_\beta \colon S_\alpha \rightarrow S_\beta \qquad \forall
\alpha,\beta \in I
\]
be a homomorphism with the property
\[ f^\alpha_\alpha  = \bbb1 \]
and
\[ f^\alpha_\beta \circ f^\beta_\gamma
       =f^\alpha_\gamma , \quad  \mbox{ if\ } \alpha \geq \beta \geq
                                                 \gamma. \]
\begin{Dfn}
 The {\em inductive limit\/}
 of $(S_\alpha)_{\alpha\in I}$ denoted by
 \[
 S = \lim_{\alpha\in I} \text{ind} S_\alpha
 \]
 are the equivalence classes in the disjoint union
 \[
  \hat{S} = \bigcup_{\alpha\in I} S_\alpha
 \]
 with respect to the equivalence relation
 \begin{quote}
   $\hat{S} \ni x_\alpha \sim x_\beta \in \hat{S} :\bimply$
   $\exists \gamma$ such that $\alpha \geq \gamma, \beta \geq \gamma$ and
   $f^\alpha_\gamma (x_\alpha)=f^\beta_\gamma (x_\beta)$
 \end{quote}
\end{Dfn}
\begin{Dfn}
  The {\em $k$-th cohomology group\/} of $M$ with coefficients in $S$
  is the inductive limit of $H^k(U_\alpha,S)$, where the partial
  ordering is given by the refinement of coverings.
  \[ H^k(M,S) := \lim_{U_\alpha} \text{ ind}  \:
               \left( Z^k(U_\alpha,S)/B^k(U_\alpha,S) \right) \]
\end{Dfn}
%
%%%%%%%%%%%%%%%%%%%%%%%%%%%%%%%%%%%%%%%%%%%%%%%%%%%%%%%%%%%%%%%%%%%%%%%%%%%
\chapter{The Hadamard recursion relations}
\label{sec:hadamRecRel}
%%%%%%%%%%%%%%%%%%%%%%%%%%%%%%%%%%%%%%%%%%%%%%%%%%%%%%%%%%%%%%%%%%%%%%%%%%%
%
For the purpose of proving existence and uniqueness results for
elliptic and hyperbolic differential operators on manifolds, Hadamard
gave 1923 a prescription for constructing bi-distributional solutions
for the corresponding differential equations~\cite{hadamard:lect}.
For the Klein-Gordon operator with variable mass ($\Box{} + m^2(x)$)
in four dimensions his algorithm is
motivated by writing
\begin{align} \label{eq:hadamSol}
  W^{T,p} (x,x')
   =  \lim{}_{\epsilon\rightarrow 0} \frac{1}{{(2 \pi)}^2}&
  \Biggl( \frac{\Delta^{1/2} (x,x')}
    {\sigma (x,x') + 2 i \epsilon t + \epsilon^2}
   + v^{(p)}(x,x') \ln [\sigma(x,x')+2 i \epsilon t +
    \epsilon^2]\\ \notag
  & + H^{(p)}(x,x')
  \Biggr)\\
  \intertext{with}
  v^{(p)} & = \sum_{i=1}^{p} v_i(x,x') \sigma^i\\
  \intertext{and}
  H^{(p)} & = \sum_{i=1}^{p} H_i(x,x') \sigma^i
\end{align}
where $\sigma(x,x')$ is the square of the geodesic distance,
$\Delta^{1/2}(x,x') := g^{-1/2}(x) \det\bigl(\sigma_{;ab}(x,x')\bigr)
g^{-1/2}(x')$ is the van Vleck Morette determinant and $t = T(x) -
T(x')$ is some global time function. A semicolon denotes covariant
differentiation and $\sigma^{i}$ abbreviates $\sigma^{;i}$. The
normalization factor $(2\pi)^{-1}$ is chosen so as to agree with the
two-point distribution for the scalar field on Minkowski space. One
assumes $v^{(p)}$ to be smooth and $H^{(p)}$ to be at least $p$-times
differentiable. To proceed one substitutes the
ansatz~(\ref{eq:hadamSol}) into the Klein-Gordon equation and equates
the coefficients of the explicitly appearing powers of $\sigma$ and
$\ln (\sigma)$ to zero. This yields to a family of ordinary
differential equations for $v^{(p)}$ and $H^{(p)}$, which can be
written recursively as:
\begin{eqnarray} \label{eq:recrelation1}
0 & = & (n+1) (n+2) v_{n+1} + (n+1) v_{n+1;a}\sigma^{;a}
      - (n+1) v_{n+1} \Delta^{-1/2} {\Delta^{1/2}}_{;a} \sigma^{;a} \\
  &   & + \frac{1}{2} \left( \Box + m^2 + \frac{1}{4}R  \right) v_n \nonumber
\end{eqnarray}
\begin{eqnarray}\label{eq:recrelation2}
0 & = & (n+1) (n+2) H_{n+1} + (n+1) H_{n+1;a}\sigma^{;a}
      - (n+1) H_{n+1} \Delta^{-1/2} {\Delta^{1/2}}_{;a} \sigma^{;a} \\
  &   & + \frac{1}{2} \left( \Box - \frac{1}{4}R + m^2 \right) H_n
        + ( 2 n +3) v_{n+1} + v_{n+1;\mu}\sigma^{;\mu}
        - \Delta^{-1/2} {\Delta^{1/2}}_{;\mu} \sigma^{;\mu} v_{n+1} \nonumber
\end{eqnarray}
together with the boundary condition
\begin{eqnarray}\label{eq:boundScalarHadam}
0 & = & v_{0;a} + \left(
        1 - \Delta^{-1/2} {\Delta^{1/2}}_{;b} \sigma^{;b} \right) v_0
        + \frac{1}{2} \left( \Box + m^2 +  \frac{1}{4}R  \right)\
 \Delta^{1/2},
\end{eqnarray}
provided $m^2(x) \equiv \frac{1}{4} R + m^2 $.  These relations are
called {\em Hadamard recursion relations\/} (See
also \cite{DeWitt:group}). The boundary condition
Eqn.~(\ref{eq:boundScalarHadam}) completely fixes $v^{(p)}$, while
$H^{(p)}$ remains arbitrary. However the latter gets uniquely fixed if
$H_0$ is specified. It should be noted that $\Delta^{1/2}$ and
$v^{(p)}$ are symmetric and hence solve similar equations when
$(\Box+m^2(x))$ is replaced by $(\Box_{x'}+m^2(x'))$ in the
considerations above.
\begin{rem}
  It is explicitly shown in~\cite{wald:78} that $H^{(p)}$ need not to
  be symmetric even if $H_0$ is. Thus in order to obtain a symmetric
  bi-solution (which is necessary for Hadamard bi-distributions
  arising from states) the initial functions $H_0$ are constraint
  (See~\cite{castagnino_harari:84} for the constraint in the scalar
  case). On the other hand the `DeWitt-Schwinger' approach, which was
  used by Christensen, is effectively a choice of $H_0$ which yield to
  a symmetric $H^{(p)}$~\cite{wald:92}, but it is not suitable for
  massless theories. Moreover almost all results in this approach have
  to be understood as being valid asymptotically only.
\end{rem}
For the spinor case the analogous treatment for the auxiliary
bi-distribution ${{\mathcal G}^B}_{A'}$ yields:
\[
{{v^{(p)}}^B}_{A'}= \sum_{i=1}^{p} {{v_i}^B}_{A'}(x,x') \sigma^i
\]
and
\[
{{H^{(p)}}^B}_{A'}= \sum_{i=1}^{p} {{H_i}^B}_{A'} (x,x') \sigma^i.
\]
For the corresponding coefficient functions
the relations (\ref{eq:recrelation1}) and (\ref{eq:recrelation2}) are\
 formally
equally valid --except that now the covariant derivatives
are understood as acting on Spinors~\cite{Najmi:84}. For the boundary
condition one finds
\begin{eqnarray*}
0 & = & v_{0;a} + \left(
        1 - \Delta^{-1/2} {\Delta^{1/2}}_{;b} \sigma^{;b} \right) v_0
        + \frac{1}{2} \left( \Box - \frac{1}{4}R + m^2 \right)
        ( \Delta^{1/2} {\mathcal J}),
\end{eqnarray*}
where ${\mathcal J}$ is the bi-spinor of parallel transport defined by the
relations ${\mathcal J}(x,x)=\frakI$ and $\sigma^{;a} {\mathcal J}_{;a}
(x,x') = 0$~(Compare section~\ref{sec:analogonWess}).  The function
$v^{(p)}$ is again fixed by the geometry, whereas $H_0$ remains
arbitrary. It is worth noting that the Remark written above applies
verbatim to the spinor case, too.  We close this chapter with some
coincidence limits for the smooth DeWitt-Schwinger term
$\sideset{^{DS}}{^{(p)}}{H}$.
%
%%%%%%%%%
\section{The coincidence limit of the smooth DeWitt-Schwinger term}
\label{sec:coincDeWittSchwsmooth}
%%%%%%%%%
%
In this section some coincidence limits of the smooth DeWitt-Schwinger
term $\sideset{^{DS}}{^{(p)}}{H}$ and its derivatives are
collected. The results were used in Eqn.~(\ref{eq:smoothDSTerm}).
The various coincidence limits of the DeWitt-Schwinger coefficients
$a_n$ and ${\mathcal A}_n$ respectively were taken
from~\cite{chris:76,chris:78}.
Consider $\sideset{^{DS}}{_0}{H}$ given by Eqn.~(\ref{eq:HODeW}) up to
order $1/m^2$ and abbreviate $\delta= ( \ln(m^2) - \ln 2 + 2 \gamma)$,
where $\gamma$ is Euler's constant. Then for the scalar case
\[
\sideset{^{DS}}{_0}{H} = (\Delta^{1/2})^{-1} m^2 (\delta-1) -
(\Delta^{1/2})^{-1} \delta a_1,
\]
where $a_1$ is the second DeWitt-Schwinger coefficient defined in
Eqn.~(\ref{eq:dsHadamardExp}) (see also~\cite{castagnino_harari:84}).
We get the following coincidence limits:
\begin{align*}
  [\sideset{^{DS}}{_0}{H}] & = m^2 (\delta -1)\\
  [ \sideset{^{DS}}{^{;a}_0}{H} ]^{;b} & = - \delta [ a_1^{;a}]^{;b}
                                          = 0 \\
  [ \sideset{^{DS}}{^{;ab}_0}{H} ] & = - \delta [ a_1^{;ab} ]
                                     = -\frac{\delta}{90} R^{cdea}
                                       {R_{cde}}^b\\
  [ \sideset{^{DS}}{_0}{H}]^{;ab} & = 0,
\end{align*}
since $[\delta^{1/2}]=1$, $[\delta^{1/2;a}]=[\delta^{1/2}]^{;ab}=0$ on
vacuum spacetimes ($R^{ab}=0$) and the coincidence limits of $a_1$
were used. The coincidence limit of $\sideset{^{DS}}{_1}{H}$ can be
calculated using the Hadamard recursion relation. We obtain:
\begin{equation*}
  \begin{split}
    [ \sideset{^{DS}}{_1}{H} ]
    & =
    - \frac{1}{4} [ \sideset{^{DS}}{^h}{{H_{0;h}}} ]
    - \frac{1}{4} m^2  [\sideset{^{DS}}{_0}{H}] - \frac{3}{4} [ a_2]
    \\
    & = \frac{2\delta - 3}{720} R^{abcd} R_{abcd} - \frac{1}{4} m^4
    (\delta-1)\\
  \end{split}
\end{equation*}
For the fermionic case one finds the following coincidence limits
\begin{align*}
  [\sideset{^{DS}}{_0}{H}] & = m^2 (\delta-1) \\
  [ \sideset{^{DS}}{^{;a}_0}{H}]^{;b} & = - \delta [ {\mathcal
    A}_1^{;a}]^{;b} = 0 \\
    [ \sideset{^{DS}}{^{;ab}_0}{H}] & = - \delta [{\mathcal A}_1^{;ab}] =
      - \frac{\delta}{48} G_{[cd]} G_{[ef]}
         ( R^{cdga} {{R^{ef}}_g}^b + R^{cdgb} {{R^{ef}}_g}^a )
      - \frac{\delta}{90} R^{cdea} {R_{cde}}^b
  \\
   [ \sideset{^{DS}}{^{h}}{{H_{0;h}}} ] & =
      \delta ( -\frac{1}{24} G_{[cd]} G_{[ef]} R^{cdgh} {R^{ef}}_{gh}
               - \frac{1}{90} R^{abcd} R_{abcd}),
\end{align*}
where now ${\mathcal A}_n$ denotes the fermionic  DeWitt-Schwinger
coefficient and the Bianchi identities where used to show that terms
like ${{R^{cdea}}_{;e}}^b$ vanish.
For the coincidence limit of $\sideset{^{DS}}{_1}{H}$ we use the
Hadamard recursion relation again:
\begin{equation*}
  \begin{split}
    [ \sideset{^{DS}}{_1}{H} ]
    & =
    - \frac{1}{4} [ \sideset{^{DS}}{^h}{{H_{0;h}}} ]
    - \frac{1}{4} m^2  [\sideset{^{DS}}{_0}{H}] - \frac{3}{4} [ {\mathcal\
 A}_2]
    \\
    & =
    \frac{2 \delta-3}{192} G_{[cd]} G_{[ef]} R^{cdgh} {R^{ef}}_{gh}
    + \frac{2\delta-3}{720} R^{abcd} R_{abcd}
    - \frac{m^4}{4} (\delta-1)
  \end{split}
\end{equation*}
%
%%%%%%%%%%%%%%%%%%%%%%%%%%%%%%%%%%%%%%%%%%%%%%%%%%%%%%%%%%%%%%%%%%
\chapter{Useful formulas}
\label{sec:usefulForm}
%%%%%%%%%%%%%%%%%%%%%%%%%%%%%%%%%%%%%%%%%%%%%%%%%%%%%%%%%%%%%%%%%%
%
This chapter contains some well known formulas that were frequently
used during the calculations in this work. At the beginning some
identities for the Dirac $\gamma$-matrices are collected. Some of the
Fierz rearrangement formulas valid for classical Majorana spinors are
presented next.  We close this chapter with a proof of a rearrangement
formula for the trace of the commutator of two Majorana spinor fields
which is also valid in a quantized theory.
%
%%%%%%%%
\section{$\gamma$-matrix identities}
\label{sec:gammaident}
%%%%%%%%
In this section some useful and well known properties of the Dirac
$\gamma$-matrices are collected (see e.g.~\cite{itzykson:qft}). We
begin with the basic relation and the definition of the $\gamma^5$
matrix:
\begin{align*}
  \{ \gamma^a , \gamma^b \}_+ & = 2 g^{ab} \\
  \gamma^5 & := \gamma^a \gamma^b \gamma^c \gamma^d \epsilon_{abcd},
\end{align*}
where $g^{ab}$ is the metric, $\epsilon_{abcd}$ denotes the totally
antisymmetric $\epsilon$ tensor. An immediate consequence of this
definition is the fact that $\gamma^5$ anti-commutes with the other
$\gamma$-matrices and has unit square. For every product of the
$\gamma$-matrices the trace of the resulting matrix is well defined
and one finds:
\begin{align*}
  \Tr(\gamma^a \gamma^b) & = 4 g^{ab}\\
  \Tr( \gamma^a \gamma^b \gamma^5 ) & = 0\\
  \Tr( \gamma^a \gamma^b \gamma^c \gamma^d ) & =
  4 ( g^{ab}g^{cd} - g^{ac}g^{bd} + g^{ad} g^{bc})\\
  \Tr( \gamma^a \gamma^b \gamma^c \gamma^d \gamma^5 ) & =
  - 4 i \epsilon^{abcd}
\end{align*}
Moreover, all traces over an odd number of $\gamma$-matrices vanish.
The symbol  $G^{[ab]}$ usually denotes 1/4 times the commutator of
$\gamma^a$ and $\gamma^b$. This implies
\begin{align*}
  \Tr( \gamma^a G^{[bc]} \gamma^d) & =
  2 ( g^{ab} g^{cd} - g^{ac} g^{bd} )\\
  \Tr( \gamma^a G^{[bc]} \gamma^d\gamma^5) & =
  - 2 i \epsilon^{abcd}\\
  \Tr( G^{[kd]} G^{[ef]} ) & = ( g^{de} g^{kf} - g^{df} g^{ke} )\\
    \Tr( \gamma^a \gamma^c G^{[kd]} G^{[ef]} )
    & =
    g^{ac} g^{de} g^{kf} - g^{ac} g^{df} g^{ke} + g^{ka} g^{de} g^{cf}
    - g^{ka} g^{df} g^{ce} \\
    & \phantom{=}
    - g^{ad} g^{ke} g^{cf} + g^{ad} g^{kf} g^{ce} + g^{fa} g^{ck} g^{ed}
    - g^{fa} g^{cd} g^{ek}\\
    & \phantom{=}
    - g^{ea} g^{ck} g^{fd} + g^{ea} g^{cd} g^{fk}
\end{align*}
It is worth noting that on a curved spacetime the $\gamma$-matrices are
covariantly constant, which implies the following identity:
\begin{equation}\label{eq:gammaParallel}
{{\mathcal J}^{A}}_{A'} {\gamma^{a'A'}}_{B'} {{\mathcal J}^{B'}}_{B} =
{g^{a'}}_{a} {\gamma^{aA}}_{B},
\end{equation}
where ${{\mathcal J}^{A}}_{A'}$ and ${g^{a'}}_{a}$ are the bi-spinor and
bi-tensor of parallel transport respectively (Compare
chapter~\ref{sec:analogonWess} and recall that parallel transport
commutes with contraction).
%
%%%%%%%%
\section{Rearrangement of Majorana spinors}
\label{sec:rearrangeTr}
%%%%%%%%
The following Fierz rearrangement formulas for classical Majorana
spinors are valid. The proof of these well known identities can be
found in the literature (see for example~\cite[1.3.4 and
1.4.7]{mueller_kirsten:SUSY} or~\cite[Appendix A]{west:SUSY_Intro}).
It is worth noting that these identities are
equally valid for a metric with signature $(-,+,+,+)$.
\begin{Lemma}[Fierz rearrangement identities]
  For two Majorana spinor fields $\Psi$ and $\chi$ the following
  identities hold.
  \begin{alignat}{2}\label{eq:fierz_indentity}
    \Psi^+\chi                           & = \chi^+ \Psi &  \qquad
                \Psi^+ \gamma^5 \chi & =   \chi^+ \gamma^5 \Psi
                                            \nonumber \tag{\theequation a}\\
    \Psi^+ \gamma^a \chi                 & = - \chi^+ \gamma^a \Psi &
     \qquad
         \Psi^+ \gamma^a \gamma^5 \chi     & = - \chi^+ \gamma^5\gamma^a \Psi
                                             \nonumber \tag{\theequation b}\\
    \Psi^+\gamma^a \gamma^b \chi      & = \chi^+\gamma^b\gamma^a\Psi&
     \qquad
      \Psi^+\gamma^a\gamma^b\gamma^5\chi &=\chi^+\gamma^5\gamma^b\gamma^a\Psi
                                             \nonumber \tag{\theequation c}
  \end{alignat}
  \begin{equation}
    \Psi^+( i \rvec{\notnabla} -m )\chi =
    \nabla_a(- \chi^+ i\gamma^a\Psi)
      +
      \chi^+(i\rvec{\notnabla}-m)\Psi \nonumber \tag{\theequation d}
  \end{equation}
  \addtocounter{equation}{1} % special construct to advance counter
\end{Lemma}
While the identities above do not hold for quantized fields, the
following result for the commutators only uses the Majorana condition
and is hence also valid for quantized fields.
\begin{Lemma}\label{Lemma:rearrangeTr}
  Let $\chi^A$ and ${\phi^+}_B$ denote a Majorana spinor and a Majorana
  cospinor respectively. Then the following identity holds:
  \[
  \Tr\bigl({\gamma^{aA}}_{B} [ \chi^B, {\phi^+}_D ]_- \bigr)
  =
  - \Tr\bigl({\gamma^{aA}}_{B} [ \phi^B, {\chi^+}_D   ]_- \bigr) ,
  \]
  where $[\cdot,\cdot]_-$ denotes the commutator.
\end{Lemma}
\begin{pf}
  Write down the Definition of the trace and the commutator:
  \begin{equation*}
    \begin{split}
      \Tr\bigl({\gamma^{aA}}_{B} [ \chi^B, {\phi^+}_D ]_- \bigr)
      & =
      \chi^A {({\gamma^a}^T)_A}^B {\phi^+}_B
      - {\phi^+}_B {\gamma^{aB}}_{A} \chi^A\\
      & =
      - \chi^A {({\gamma^a}^T)_A}^B C_{BC} \phi^C + \phi^C (C^T)_{CB}
         {\gamma^{aB}}_A \chi^A,\\
      \intertext{where $C$ denotes the charge conjugation
        matrix. Recall that for a {\em Majorana\/}    spinor
        ${\phi^+}_B = - C_{BC} \phi^C$. Using the properties of $C$ we
        obtain}
      & =
      + \chi^A C_{AB} {\gamma^{aB}}_C \phi^C
      + \phi^C {({\gamma^a}^T)_C}^B C_{BA} \chi^A\\
      & = {\chi^+}_B {\gamma^{aB}}_C \phi^C - \phi^C {\gamma^{aB}}_{C}
          {\chi^+}_B \\
      & =
      - \Tr \bigl( {\gamma^{aA}}_B [\phi^A, {\chi^+}_C]_- \bigr)\\
    \end{split}
  \end{equation*}
  % Notation wie in \cite{west:SUSY_Intro}
\end{pf}
%
%%%%%%%%%%%%%%%%%%%%%%%%%%%%%%%%%%%%%%%%%%%%%%%%%%%%%%%%%%%%%%%%%%%%%%%%%%
\chapter{Listing of the {\tt Maple V} programs}
\label{sec:maplelist}
%%%%%%%%%%%%%%%%%%%%%%%%%%%%%%%%%%%%%%%%%%%%%%%%%%%%%%%%%%%%%%%%%%%%%%%%%%
%
This Appendix contains the commented listing and an example run of the
{\tt Maple V} programs that were used for the calculations in
chapter~\ref{sec:emtCurved} section~\ref{sec:example}.
%
%%%%%%%%%
\section{The Schwarzschild spacetime}
\label{sec:mapleSchwarz}
%%%%%%%%%
%
%% Created by Maple V Release 3
\begin{mapleinput}
# Initialization
# Read tensor package
readlib(tensor):
# and linear Algebra
with(linalg):
# Define Variables
x1:=t:
x2:=r:
x3:=theta:
x4:=phi:
# Schwarzschild metric
g11:=1-2*M/r:g22:=-1/g11:g33:=-r^2:g44:=-r^2*sin(theta)^2:
# Compute Riemann etc.
tensor():
# {R_a}^{bcd} = Rlhhh
for a from 1 to 4 do
 for b from 1 to 4 do
  for c from 1 to 4 do
   for d from 1 to 4 do
    Rlhhh.a.b.c.d := 0:
    for e from 1 to 4 do for f from 1 to 4 do for g from 1 to 4 do
     Rlhhh.a.b.c.d := Rlhhh.a.b.c.d +
                      h.b.e * h.c.f * h.d.g * R.a.e.f.g
    od od od
od od od od:
# Kretschman invariant W2 = R^{abcd} R_{abcd}
W2:=0:
for a from 1 to 4 do for b from 1 to 4 do for c from 1 to 4 do
  for d from 1 to 4 do for e from 1 to 4 do
     W2 := W2 + h.e.a * R.a.b.c.d * Rlhhh.e.b.c.d:
    od od od od od:
# SUSY finite Term
# Tsusy = 2 * 4 Pi * T_{ab} =
#                            + 1/288 {R_a}^{cde} R_{bcde}
#                            + 1/288 {R_a}^{cde} R_{bedc}
#                            - 1/96 {R_a}^{cde} R_{bdce}
for a from 1 to 4 do  for b from 1 to 4 do
  Tsusy.a.b :=0:
  for c from 1 to 4 do for d from 1 to 4 do for e from 1 to 4 do
    Tsusy.a.b := Tsusy.a.b
             + 1/288 * Rlhhh.a.c.d.e * R.b.c.d.e
             + 1/288 * Rlhhh.a.c.d.e * R.b.e.d.c
             - 1/96 * Rlhhh.a.c.d.e * R.b.d.c.e :
  od od od
od od:
# Tsusyhh = Tsusy^{ab}
for a from 1 to 4 do for b from 1 to 4 do
 Tsusyhh.a.b := 0:
 for c from 1 to 4 do for d from 1 to 4 do
  Tsusyhh.a.b := Tsusyhh.a.b + h.a.c * h.b.d * Tsusy.c.d
 od od
od od:
# Covariant Derivative
# Tsusyh = T^{ab}_{;a} := T^{ab}_{,a}
#                      + {\gamma_{ad}}^a T^{db}
#                      + {\gamma_{ad}}^b T^{ad}
for b from 1 to 4 do
Tsusyh.b := 0:
 for a from 1 to 4 do
  Tsusyh.b := Tsusyh.b + diff(Tsusyhh.a.b,x.a):
  for d from 1 to 4 do
   Tsusyh.b := Tsusyh.b +
             C.a.d.a * Tsusyhh.d.b + C.a.d.b * Tsusyhh.a.d:
  od
 od
od:
# Smooth terms: Tds = (2*\delta - 3)/768 * W2 -
#                     \delta/48
#                     ( {R_a}^{cde} * {R_bced} + 2 {R_a}^{cde} *
#                       R_{bdec} )
for a from 1 to 4 do for b from 1 to 4 do
 Tds.a.b := 0:
 Tds.a.b := (2 * delta - 3) /768 * W2:
  for c from 1 to 4 do for d from 1 to 4 do for e from 1 to 4 do
   Tds.a.b := Tds.a.b -
              delta / 48 *
              ( Rlhhh.a.c.d.e * R.b.c.e.d
               + 2 * Rlhhh.a.c.d.e * R.b.d.e.c ):
  od od od
od od:
# Tdshh = Tds^{ab}
for a from 1 to 4 do for b from 1 to 4 do
 Tdshh.a.b := 0:
 for c from 1 to 4 do for d from 1 to 4 do
  Tdshh.a.b := Tdshh.a.b + h.a.c * h.b.d * Tds.c.d
 od od
od od:
# Covariant Derivative
# Tdsh = T^{ab}_{;a} := T^{ab}_{,a}
#                      + {\gamma_{ad}}^a T^{db}
#                      + {\gamma_{ad}}^b T^{ad}
for b from 1 to 4 do
Tdsh.b := 0:
 for a from 1 to 4 do
  Tdsh.b := Tdsh.b + diff(Tdshh.a.b,x.a):
  for d from 1 to 4 do
   Tdsh.b := Tdsh.b +
             C.a.d.a * Tdshh.d.b + C.a.d.b * Tdshh.a.d:
  od
 od
od:
# Print results in pretty form
#
# define Matrix
Tsusyhh:= matrix(4,4): # SUSY
Tsusyh:= vector(4):
Tdshh:= matrix(4,4): # Smooth part
Tdsh:= vector(4):
# Set Tsusyh to Tsusyhh
for a to 4 do for b to 4 do
 Tsusyhh[a,b] := simplify(Tsusyhh.a.b):
od od:
# Set Tsusyh to Tsusyh
for a to 4 do
 Tsusyh[a] := simplify(Tsusyh.a):
od:
# Set Tdsh to Tdshh
for a to 4 do for b to 4 do
 Tdshh[a,b] := simplify(Tdshh.a.b):
od od:
# Set Tdsh to Tdsh
for a to 4 do
 Tdsh[a] := simplify(Tdsh.a):
od:
print(`Stress energy Tensor T^{ab}:`);
print(Tsusyhh);
print(`{T^{ab}}_{;a}:`);
print(Tsusyh);
print(`Smooth T^{ab}:`);
print(Tdshh);
print(Divergence);
print(Tdsh);
# Output in LaTeX to files
latex(Tsusyhh,Tsusyhh);
latex(Tsusyh,Tsusyh);
latex(Tdshh,Tdshh);
latex(Tdsh,Tdsh);
quit;
\end{mapleinput}
\begin{maplettyout}
Warning: new definition for   norm
Warning: new definition for   trace
\end{maplettyout}
\begin{maplelatex}
\[
\text{\it Stress energy Tensor $T^{ab}$:}
\]
\end{maplelatex}
\begin{maplelatex}
\[
 \left(
{\begin{array}{rrrr}
0 & 0 & 0 & 0 \\
0 & 0 & 0 & 0 \\
0 & 0 & 0 & 0 \\
0 & 0 & 0 & 0
\end{array}}
 \right)
\]
\end{maplelatex}
\begin{maplelatex}
\[
\text{\it ${T^{ab}}_{;a}$:}
\]
\end{maplelatex}
\begin{maplelatex}
\[
[\,0\,0\,0\,0\,]
\]
\end{maplelatex}
\begin{maplelatex}
\[
\text{\it Smooth $T^{ab}$:}
\]
\end{maplelatex}
\begin{maplelatex}
\begin{eqnarray*}{}
\lefteqn{ \left[ {\vrule height0.87em width0em depth0.87em}
 \right. \! \! \,{\displaystyle \frac {1}{16}}\,{\displaystyle
\frac {{M}^{2}\,(\,10\,{r}\,{ \delta} - 3\,{r} - 16\,{ \delta}\,{
M}\,)}{{r}^{5}\,(\,{r} - 2\,{M}\,)^{2}}}\,,  - \,{\displaystyle
\frac {1}{16}}\,{\displaystyle \frac {(\,2\,{ \delta} - 3\,)\,{M}
^{2}}{{r}^{6}}}\,, } \\
 & &  - \,{\displaystyle \frac {1}{16}}\,{\displaystyle \frac {(
\,2\,{ \delta} - 3\,)\,{M}^{2}}{{r}^{7}\,(\,{r} - 2\,{M}\,)}}\,,
{\displaystyle \frac {1}{16}}\,{\displaystyle \frac {(\,2\,{
\delta} - 3\,)\,{M}^{2}}{{r}^{7}\, \left( \! \, - {r} + {r}\,
{\rm cos}(\,{ \theta}\,)^{2} + 2\,{M} - 2\,{M}\,{\rm cos}(\,{
\theta}\,)^{2}\, \!  \right) }}\, \! \! \left. {\vrule
height0.87em width0em depth0.87em} \right]  \\
 & &  \left[ {\vrule height0.87em width0em depth0.87em}
 \right. \! \! \, - \,{\displaystyle \frac {1}{16}}\,
{\displaystyle \frac {(\,2\,{ \delta} - 3\,)\,{M}^{2}}{{r}^{6}}}
\,,  - \,{\displaystyle \frac {1}{16}}\,{\displaystyle \frac {(\,
6\,{r}\,{ \delta} + 4\,{ \delta}\,{M} + 3\,{r} - 6\,{M}\,)\,{M}^{
2}\,(\,{r} - 2\,{M}\,)}{{r}^{8}}} \\
 & & , {\displaystyle \frac {1}{16}}\,{\displaystyle \frac {(\,{r
} - 2\,{M}\,)\,(\,2\,{ \delta} - 3\,)\,{M}^{2}}{{r}^{9}}}\,,  -
\,{\displaystyle \frac {1}{16}}\,{\displaystyle \frac {{M}^{2}\,(
\,2\,{r}\,{ \delta} - 4\,{ \delta}\,{M} - 3\,{r} + 6\,{M}\,)}{{r}
^{9}\, \left( \! \, - 1 + {\rm cos}(\,{ \theta}\,)^{2}\, \!
 \right) }}\, \! \! \left. {\vrule
height0.87em width0em depth0.87em} \right]  \\
 & &  \left[ {\vrule height0.87em width0em depth0.87em}
 \right. \! \! \, - \,{\displaystyle \frac {1}{16}}\,
{\displaystyle \frac {(\,2\,{ \delta} - 3\,)\,{M}^{2}}{{r}^{7}\,(
\,{r} - 2\,{M}\,)}}\,, {\displaystyle \frac {1}{16}}\,
{\displaystyle \frac {(\,{r} - 2\,{M}\,)\,(\,2\,{ \delta} - 3\,)
\,{M}^{2}}{{r}^{9}}}\,,  \\
 & &  - \,{\displaystyle \frac {1}{16}}\,{\displaystyle \frac {{M
}^{2}\,(\, - 2\,{ \delta} + 3 + 8\,{ \delta}\,{r}^{2}\,)}{{r}^{10
}}}\,,  - \,{\displaystyle \frac {1}{16}}\,{\displaystyle \frac {
(\,2\,{ \delta} - 3\,)\,{M}^{2}}{{r}^{10}\, \left( \! \, - 1 +
{\rm cos}(\,{ \theta}\,)^{2}\, \!  \right) }}\, \! \! \left.
{\vrule height0.87em width0em depth0.87em} \right]  \\
 & &  \left[ {\vrule height0.95em width0em depth0.95em}
 \right. \! \! \,{\displaystyle \frac {1}{16}}\,{\displaystyle
\frac {(\,2\,{ \delta} - 3\,)\,{M}^{2}}{{r}^{7}\, \left( \! \, -
{r} + {r}\,{\rm cos}(\,{ \theta}\,)^{2} + 2\,{M} - 2\,{M}\,{\rm
cos}(\,{ \theta}\,)^{2}\, \!  \right) }}\,,  \\
 & &  - \,{\displaystyle \frac {1}{16}}\,{\displaystyle \frac {{M
}^{2}\,(\,2\,{r}\,{ \delta} - 4\,{ \delta}\,{M} - 3\,{r} + 6\,{M}
\,)}{{r}^{9}\, \left( \! \, - 1 + {\rm cos}(\,{ \theta}\,)^{2}\,
 \!  \right) }}\,,  - \,{\displaystyle \frac {1}{16}}\,
{\displaystyle \frac {(\,2\,{ \delta} - 3\,)\,{M}^{2}}{{r}^{10}\,
 \left( \! \, - 1 + {\rm cos}(\,{ \theta}\,)^{2}\, \!  \right) }}
\,,  \\
 & & {\displaystyle \frac {1}{16}}\,{\displaystyle \frac {{M}^{2}
\, \left( \! \,2\,{ \delta} - 3 - 8\,{ \delta}\,{r}^{2} + 8\,{
\delta}\,{r}^{2}\,{\rm cos}(\,{ \theta}\,)^{2}\, \!  \right) }{{r
}^{10}\, \left( \! \,1 - 2\,{\rm cos}(\,{ \theta}\,)^{2} + {\rm
cos}(\,{ \theta}\,)^{4}\, \!  \right) }}\, \! \! \left. {\vrule
height0.95em width0em depth0.95em} \right]
\end{eqnarray*}
\end{maplelatex}
\begin{maplelatex}
\[
\text{\it Divergence}
\]
\end{maplelatex}
\begin{maplelatex}
\begin{eqnarray*}
\lefteqn{ \left[ {\vrule height0.79em width0em depth0.79em}
 \right. \! \! {\displaystyle \frac {1}{16}}{M}^{2}( - 20\,{
\delta}\,{M}\,{\rm sin}(\,{ \theta}\,) + 8\,{ \delta}\,{\rm sin}(
\,{ \theta}\,)\,{r} - 2\,{ \delta}\,{\rm cos}(\,{ \theta}\,) + 30
\,{M}\,{\rm sin}(\,{ \theta}\,)} \\
 & & \mbox{} - 12\,{\rm sin}(\,{ \theta}\,)\,{r} + 3\,{\rm cos}(
\,{ \theta}\,)) \left/ {\vrule height0.37em width0em depth0.37em}
 \right. \! \! (\,{\rm sin}(\,{ \theta}\,)\,{r}^{7}\,(\,{r} - 2\,
{M}\,)\,){\displaystyle \frac {1}{16}}{M}^{2} \left( {\vrule
height0.44em width0em depth0.44em} \right. \! \!  \\
 & & 180\,{M}^{2}\,{r}^{2}\,{\rm cos}(\,{ \theta}\,)^{2} - 88\,{M
}^{3}\,{r}\,{ \delta} - 132\,{M}^{3}\,{r}\,{\rm cos}(\,{ \theta}
\,)^{2} - 6\,{r}^{2} \\
 & & \mbox{} + 3\,{r}^{2}\,{\rm cos}(\,{ \theta}\,)^{2} + 84\,{M}
\,{r}^{3} + 88\,{M}^{3}\,{r}\,{ \delta}\,{\rm cos}(\,{ \theta}\,)
^{2} \\
 & & \mbox{} + 72\,{M}^{2}\,{ \delta}\,{r}^{2}\,{\rm cos}(\,{
\theta}\,)^{2} - 12\,{r}\,{M}\,{\rm cos}(\,{ \theta}\,)^{2} - 136
\,{M}\,{r}^{3}\,{ \delta}\,{\rm cos}(\,{ \theta}\,)^{2} \\
 & & \mbox{} + 12\,{r}^{4}\,{\rm cos}(\,{ \theta}\,)^{2} + 8\,{M}
\,{r}\,{ \delta}\,{\rm cos}(\,{ \theta}\,)^{2} - 12\,{r}^{4} - 84
\,{M}\,{r}^{3}\,{\rm cos}(\,{ \theta}\,)^{2} \\
 & & \mbox{} - 72\,{M}^{2}\,{ \delta}\,{r}^{2} - 24\,{M}^{2} + 16
\,{M}^{2}\,{ \delta} - 16\,{M}\,{r}\,{ \delta} + 40\,{r}^{4}\,{
\delta}\,{\rm cos}(\,{ \theta}\,)^{2} \\
 & & \mbox{} - 8\,{M}^{2}\,{ \delta}\,{\rm cos}(\,{ \theta}\,)^{2
} - 2\,{ \delta}\,{r}^{2}\,{\rm cos}(\,{ \theta}\,)^{2} + 4\,{
\delta}\,{r}^{2} - 40\,{r}^{4}\,{ \delta} + 136\,{r}^{3}\,{M}\,{
\delta} \\
 & & \mbox{} + 3\,{\rm sin}(\,{ \theta}\,)\,{r}^{3}\,{\rm cos}(\,
{ \theta}\,) + 24\,{r}\,{M} + 12\,{M}^{2}\,{\rm cos}(\,{ \theta}
\,)^{2} \\
 & & \mbox{} - 2\,{\rm sin}(\,{ \theta}\,)\,{r}^{3}\,{\rm cos}(\,
{ \theta}\,)\,{ \delta} + 8\,{\rm sin}(\,{ \theta}\,)\,{r}^{2}\,
{\rm cos}(\,{ \theta}\,)\,{M}\,{ \delta} \\
 & & \mbox{} - 8\,{\rm sin}(\,{ \theta}\,)\,{r}\,{\rm cos}(\,{
\theta}\,)\,{M}^{2}\,{ \delta} - 12\,{\rm sin}(\,{ \theta}\,)\,{r
}^{2}\,{\rm cos}(\,{ \theta}\,)\,{M} \\
 & & \mbox{} + 12\,{\rm sin}(\,{ \theta}\,)\,{r}\,{\rm cos}(\,{
\theta}\,)\,{M}^{2} - 180\,{M}^{2}\,{r}^{2} + 132\,{M}^{3}\,{r}
 \! \! \left. {\vrule height0.44em width0em depth0.44em} \right)
 \left/ {\vrule height0.44em width0em depth0.44em} \right. \! \!
 \left( {\vrule height0.44em width0em depth0.44em} \right. \! \!
\,{r}^{10} \\
 & &  \left( \! \, - {r} + {r}\,{\rm cos}(\,{ \theta}\,)^{2} + 2
\,{M} - 2\,{M}\,{\rm cos}(\,{ \theta}\,)^{2}\, \!  \right) \, \!
\! \left. {\vrule height0.44em width0em depth0.44em} \right)  -
\,{\displaystyle \frac {1}{16}}{M}^{2} \left( {\vrule
height0.44em width0em depth0.44em} \right. \! \!  - 8\,{ \delta}
\,{\rm sin}(\,{ \theta}\,)\,{r} \\
 & & \mbox{} + 8\,{ \delta}\,{r}\,{\rm sin}(\,{ \theta}\,)\,{\rm
cos}(\,{ \theta}\,)^{2} + 12\,{\rm sin}(\,{ \theta}\,)\,{r} - 12
\,{r}\,{\rm sin}(\,{ \theta}\,)\,{\rm cos}(\,{ \theta}\,)^{2} \\
 & & \mbox{} + 20\,{ \delta}\,{M}\,{\rm sin}(\,{ \theta}\,) - 20
\,{M}\,{ \delta}\,{\rm sin}(\,{ \theta}\,)\,{\rm cos}(\,{ \theta}
\,)^{2} - 30\,{M}\,{\rm sin}(\,{ \theta}\,) \\
 & & \mbox{} + 30\,{M}\,{\rm sin}(\,{ \theta}\,)\,{\rm cos}(\,{
\theta}\,)^{2} - 2\,{ \delta}\,{\rm cos}(\,{ \theta}\,)^{3} + 3\,
{\rm cos}(\,{ \theta}\,)^{3} \! \! \left. {\vrule
height0.44em width0em depth0.44em} \right)  \left/ {\vrule
height0.44em width0em depth0.44em} \right. \! \!  \left( {\vrule
height0.44em width0em depth0.44em} \right. \! \! \,{\rm sin}(\,{
\theta}\,)\,{r}^{10} \\
 & &  \left( \! \, - 1 + {\rm cos}(\,{ \theta}\,)^{2}\, \!
 \right) \, \! \! \left. {\vrule
height0.44em width0em depth0.44em} \right) {\displaystyle \frac {
1}{16}}{M}^{2}( - 20\,{ \delta}\,{M}\,{\rm sin}(\,{ \theta}\,) +
8\,{ \delta}\,{\rm sin}(\,{ \theta}\,)\,{r} \\
 & & \mbox{} - 2\,{ \delta}\,{\rm cos}(\,{ \theta}\,) + 30\,{M}\,
{\rm sin}(\,{ \theta}\,) - 12\,{\rm sin}(\,{ \theta}\,)\,{r} + 3
\,{\rm cos}(\,{ \theta}\,)) \left/ {\vrule
height0.37em width0em depth0.37em} \right. \! \!  \left( {\vrule
height0.44em width0em depth0.44em} \right. \! \! \,{\rm sin}(\,{
\theta}\,)\,{r}^{10} \\
 & &  \left( \! \, - 1 + {\rm cos}(\,{ \theta}\,)^{2}\, \!
 \right) \, \! \! \left. {\vrule
height0.44em width0em depth0.44em} \right)  \! \! \left. {\vrule
height0.79em width0em depth0.79em} \right]
\end{eqnarray*}
\end{maplelatex}
\maplesepline
%% End of Maple V Session Output
%
%%%%%%%%%
\section{The Kerr spacetime}
\label{sec:maplekerr}
%%%%%%%%%
%
%% Created by Maple V Release 3
\begin{mapleinput}
# Initialization
# Read tensor package
readlib(tensor):
# and linear Algebra
with(linalg):
# Define Variables
x1:=t:
x2:=r:
x3:=theta:
x4:=phi:
rho:= (r^2 + a^2 * cos(theta)^2):
# Kerr metric
g11:=1- 2*M*r/rho :
g14:= - (2*M*r*a / rho ) * sin(theta)^2 : g41:=g14 :
g22:= - rho / (r^2 + a^2 - 2*M*r) :
g33:= - rho :
g44:= - ( r^2 + a^2) * sin(theta)^2 - (2*M*r*a^2 /rho ) *  sin(theta)^4 :
# Compute Riemann etc.
tensor():
# simplify
# Riemann
print(`Simplify Riemann ...`);
for a from 1 to 4 do
 for b from 1 to 4 do
 print(a,b);
  for c from 1 to 4 do
   for d from 1 to 4 do
     R.a.b.c.d := simplify(R.a.b.c.d)
   od
  od
 od
od:
print(`... done`);
# Metric
for a from 1 to 5 do
 for b from 1 to 4 do
  h.a.b := simplify(h.a.b)
  od
od:
# Christoffel
print(`Simplify Christoffels ...`);
for a from 1 to 4 do
 for b from 1 to 4 do
  print(a,b);
  for c from 1 to 4 do
   C.a.b.c := simplify(C.a.b.c)
  od
 od
od:
print(`... done`);
# Show result
#display(Ricciscalar,Einstein);
# {R_a}^{bcd} = Rlhhh
print(`Calculating {R_a}^{bcd} ...`);
for a from 1 to 4 do
 for b from 1 to 4 do
  print(a,b);
  for c from 1 to 4 do
   for d from 1 to 4 do
    Rlhhh.a.b.c.d := 0:
    for e from 1 to 4 do for f from 1 to 4 do for g from 1 to 4 do
     Rlhhh.a.b.c.d := simplify( Rlhhh.a.b.c.d +
                      h.b.e * h.c.f * h.d.g * R.a.e.f.g)
    od od od
od od od od:
print(`...done`);
# SUSY finite Term
# Tsusy = 2 * 4 Pi * T_{ab} =
#                            + 1/288 {R_a}^{cde} R_{bcde}
#                            + 1/288 {R_a}^{cde} R_{bedc}
#                            - 1/96 {R_a}^{cde} R_{bdce}
print(`Calculating SUSY finite terms ...`);
for a from 1 to 4 do  for b from 1 to 4 do
  Tsusy.a.b :=0: print(a,b);
  for c from 1 to 4 do for d from 1 to 4 do for e from 1 to 4 do
    Tsusy.a.b := (Tsusy.a.b
             + 1/288 * Rlhhh.a.c.d.e * R.b.c.d.e
             + 1/288 * Rlhhh.a.c.d.e * R.b.e.d.c
             - 1/96 * Rlhhh.a.c.d.e * R.b.d.c.e ):
  od od od
od od:
print(`... done`);
# Tsusyhh = Tsusy^{ab}
print(`Calculating Tsusy^{ab} ...`);
for a from 1 to 4 do for b from 1 to 4 do
 Tsusyhh.a.b := 0:
 for c from 1 to 4 do for d from 1 to 4 do
  Tsusyhh.a.b := (Tsusyhh.a.b + h.a.c * h.b.d * Tsusy.c.d)
 od: Tsusyhh.a.b := simplify(Tsusyhh.a.b): od
od od:
print(`... done`);
# Covariant Derivative
# Tsusyh = T^{ab}_{;a} := U^{ab}_{,a}
#                      + {\gamma_{ad}}^a T^{db}
#                      + {\gamma_{ad}}^b T^{ad}
print(`Calculating T^{ab}_{;a} ...`);
for b from 1 to 4 do
Tsusyh.b := 0:
 for a from 1 to 4 do
  print(b,a);
  Tsusyh.b := (Tsusyh.b + diff(Tsusyhh.a.b,x.a)):
  for d from 1 to 4 do
   Tsusyh.b := (Tsusyh.b +
             C.a.d.a * Tsusyhh.d.b + C.a.d.b * Tsusyhh.a.d):
  od
 od
od:
print(`... done`);
# Print results in pretty form
#
# define Matrix
print(`Printing ...`);
Tsusyhh:= matrix(4,4): # SUSY
Tsusyh:= vector(4):
# Set Tsusyhh to Tsusyhh
print(`Tsusyhh ...`);
for a to 4 do for b to 4 do
 print(a,b);
 Tsusyhh[a,b] := simplify(Tsusyhh.a.b):
od od:
print(`... done`);
# Set Tsusyh to Tsusyh
print(`Tsusyh ...`);
for a to 4 do
 print(a);
 Tsusyh[a] := simplify(Tsusyh.a):
od:
print(`... done`);
print(`Stress energy Tensor T^{ab}:`);
print(Tsusyhh);
print(`{T^{ab}}_{;a}:`);
print(Tsusyh);
# Output in LaTeX to files
latex(Tsusyhh,KerrTsusyhh);
latex(Tsusyh,KerrTsusyh);
print(`... done`);
quit;
\end{mapleinput}
\begin{maplettyout}
Warning: new definition for   norm
Warning: new definition for   trace
\end{maplettyout}
\begin{maplelatex}
\[
\text{\it Simplify Riemann ...}
\]
\end{maplelatex}
\begin{maplelatex}
\[
1, 1
\]
\end{maplelatex}
\begin{maplelatex}
\[
1, 2
\]
\end{maplelatex}
\begin{maplelatex}
\[
\vdots
\]
\end{maplelatex}
\begin{maplelatex}
\[
4, 4
\]
\end{maplelatex}
\begin{maplelatex}
\[
\text{\it ... done}
\]
\end{maplelatex}
\begin{maplelatex}
\[
\text{\it Simplify Christoffels ...}
\]
\end{maplelatex}
\begin{maplelatex}
\[
1, 1
\]
\end{maplelatex}
\begin{maplelatex}
\[
\vdots{}
\]
\end{maplelatex}
\begin{maplelatex}
\[
4, 4
\]
\end{maplelatex}
\begin{maplelatex}
\[
\text{\it ... done}
\]
\end{maplelatex}
\begin{maplelatex}
\[
\text{{\it Calculating }} {R_a}^{bcd} ...
\]
\end{maplelatex}
\begin{maplelatex}
\[
1, 1
\]
\end{maplelatex}
\begin{maplelatex}
\[
\vdots{}
\]
\end{maplelatex}
\begin{maplelatex}
\[
4, 4
\]
\end{maplelatex}
\begin{maplelatex}
\[
\text{\it ... done}
\]
\end{maplelatex}
\begin{maplelatex}
\[
\text{\it Calculating SUSY finite terms ...}
\]
\end{maplelatex}
\begin{maplelatex}
\[
1, 1
\]
\end{maplelatex}
\begin{maplelatex}
\[
\vdots{}
\]
\end{maplelatex}
\begin{maplelatex}
\[
4, 4
\]
\end{maplelatex}
\begin{maplelatex}
\[
\text{\it ... done}
\]
\end{maplelatex}
\begin{maplelatex}
\[
\text{\it Calculating } {T_{susy}^{ab} ...}
\]
\end{maplelatex}
\begin{maplelatex}
\[
\text{\it ... done}
\]
\end{maplelatex}
\begin{maplelatex}
\[
\text{\it Calculating } {T^{ab}_{;a} ...}
\]
\end{maplelatex}
\begin{maplelatex}
\[
1, 1
\]
\end{maplelatex}
\begin{maplelatex}
\[
\vdots{}
\]
\end{maplelatex}
\begin{maplelatex}
\[
4, 4
\]
\end{maplelatex}
\begin{maplelatex}
\[
\text{\it ... done}
\]
\end{maplelatex}
\begin{maplelatex}
\[
\text{\it Printing ...}
\]
\end{maplelatex}
\begin{maplelatex}
\[
{\it T_{susyhh} ...}
\]
\end{maplelatex}
\begin{maplelatex}
\[
1, 1
\]
\end{maplelatex}
\begin{maplelatex}
\[
\vdots{}
\]
\end{maplelatex}
\begin{maplelatex}
\[
4, 4
\]
\end{maplelatex}
\begin{maplelatex}
\[
\text{\it ... done}
\]
\end{maplelatex}
\begin{maplelatex}
\[
{\it T_{\text{susyh}} ...}
\]
\end{maplelatex}
\begin{maplelatex}
\[
1
\]
\end{maplelatex}
\begin{maplelatex}
\[
\vdots{}
\]
\end{maplelatex}
\begin{maplelatex}
\[
\text{\it ... done}
\]
\end{maplelatex}
\begin{maplelatex}
\[
\text{\it Stress energy Tensor }{T^{ab}:}
\]
\end{maplelatex}
\begin{maplelatex}
\[
 \left[
{\begin{array}{rrrr}
0 & 0 & 0 & 0 \\
0 & 0 & 0 & 0 \\
0 & 0 & 0 & 0 \\
0 & 0 & 0 & 0
\end{array}}
 \right]
\]
\end{maplelatex}
\begin{maplelatex}
\[
{\it {T^{ab}}_{;a}:}
\]
\end{maplelatex}
\begin{maplelatex}
\[
[\,0\,0\,0\,0\,]
\]
\end{maplelatex}
\begin{maplelatex}
\[
\text{\it ...done}
\]
\end{maplelatex}
\maplesepline
%% End of Maple V Session Output
%
%%%%%%%%%%%%%%%%%%%%%%%%%%%%%%%%%%%%%%%%%%%%%%%%%%%%%%%%%%%%%%%%%%%%%%%%%%
% end of Appendix
%%%%%%%%%%%%%%%%%%%%%%%%%%%%%%%%%%%%%%%%%%%%%%%%%%%%%%%%%%%%%%%%%%%%%%%%%%
%
% Bibliography
%\bibliographystyle{amsalpha}
% amsalpha oder alpha
%\bibliography{bibspin}

\begin{thebibliography}{De{W}75}
\bibitem[ALN77]{adler:77}
S.~L. Adler, J.~Lieberman, and Y.~J. Ng.
\newblock Regularization of the stress-energy tensor for vector and scalar
  particles propagating in a general background metric.
\newblock {\em Ann. Phys.}, 106:279--321, 1977.
\bibitem[B{\etalchar{+}}90]{bog:axiom}
N.~N. Bogolubov et~al.
\newblock {\em General Principles of Quantum Field Theory}.
\newblock Kluwer Academic Publishers, 1990.
\bibitem[Bar54]{barg:54}
V.~Bargmann.
\newblock On unitary ray representations of continuous groups.
\newblock {\em Ann. of Math.}, 59:1, 1954.
\bibitem[Bar84]{bart:84}
R.~Bartnik.
\newblock Existence of maximal surfaces in asymptotically flat spacetimes.
\newblock {\em Comm. Math. Phys.}, 94:155--175, 1984.
\bibitem[BC64]{bishop:geom}
R.~L. Bishop and R.~J. Crittenden.
\newblock {\em Geometry of Manifolds}.
\newblock Academic press, New York, 1964.
\bibitem[BCO90]{bartChrMu:90}
R.~Bartnik, P.~T. Chru{\'s}ciel, and N.~O'{M}urchadha.
\newblock On maximal surfaces in asymptotically flat space-times.
\newblock {\em Comm. Math. Phys.}, 130:95--109, 1990.
\bibitem[BD82]{birrell:qft}
N.~D. Birrell and P.~C.~W. Davis.
\newblock {\em Quantum fields in Curved Space}.
\newblock Cambridge Univ. Press, Cambridge, 1982.
\bibitem[BDLR92]{buchDopLonRob:92}
D.~Buchholz, S.~Doplicher, R.~Longo, and J.~E. Roberts.
\newblock A new look at {G}oldstone's theorem.
\newblock {\em Rev. Math. Phys.,Special Issue}, pages 49--83, 1992.
\bibitem[BF82]{breitenlohner_freedman:82}
P.~Breitenlohner and D.~Z. Freedman.
\newblock Stability in gauged extended supergravity.
\newblock {\em Ann. of Phys.}, 144:249--281, 1982.
\bibitem[BFK94]{brunFredKoe:94}
R.~Brunetti, K.~Fredenhagen, and M.~K{\"o}hler.
\newblock The wave front set spectrum condition and {W}ick ordered products\
 of
  free fields on a manifold.
\newblock In Preparation, 1994.
\bibitem[Bor62]{borchers:62}
H.~J. Borchers.
\newblock On the structure of the algebra of the field operators.
\newblock {\em Nuovo Cimento}, 24:214, 1962.
\bibitem[BT87]{benn:spinors}
I.~M. Benn and R.~W. Tucker.
\newblock {\em An introduction to spinors and geometry with applications in
  physics}.
\newblock Adam Hilger, Bristol and Philadelphia, 1987.
\bibitem[BW92]{baum:causal}
H.~Baumg{\"a}rtel and M.~Wollenberg.
\newblock {\em Causal Nets of Operator Algebras}.
\newblock Akademie Verlag GmbH, Berlin, Germany, 1992.
\bibitem[C{\etalchar{+}}91a]{maple:langref}
B.~W. Char et~al.
\newblock {\em Maple {V} Language reference manual}.
\newblock Springer Verlag, New York, Berlin, 1991.
\bibitem[C{\etalchar{+}}91b]{maple:libref}
B.~W. Char et~al.
\newblock {\em Maple {V} Library reference manual}.
\newblock Springer Verlag, New York, Berlin, 1991.
\bibitem[Car66]{cartan}
E.~Cartan.
\newblock {\em Theory of Spinors}.
\newblock Hermann, Paris, 1966.
\bibitem[CB77]{choquet:77}
Y.~Choquet-Bruhat.
\newblock Hyperbolic differential equations on a manifold.
\newblock In C.~De~Witt and J.~Wheeler, editors, {\em Battelle Rencontres},\
 New
  York, 1977. Benjamin.
\bibitem[CCJ70]{callanEtAl:70}
C.~G.~Jr. Callan, S.~Coleman, and R.~Jackiw.
\newblock A new improved energy-momentum tensor.
\newblock {\em Ann. of Phys.}, 59:42--73, 1970.
\bibitem[CGNP86]{castGunNardPaz:86}
M.~A. Castagnino, E.~Gunzig, P.~Nardone, and J.~P. Paz.
\newblock Hadamard and minimal renormalization.
\newblock {\em Phys. Rev. D}, 34(12):3698--3706, 1986.
\bibitem[CH84]{castagnino_harari:84}
M.~A. Castagnino and D.~D. Harari.
\newblock {H}adamard renormalization in curved space-time.
\newblock {\em Ann. of Phys.}, 152:85--104, 1984.
\bibitem[CHN82]{CastagHarNue:82}
M.~Castagnino, D.~Harari, and C.~Nu{\~n}ez.
\newblock Minimal hypotheses for particle definition in curved space-time.
\newblock In S.~Ferrara and G.~F.~R. Ellis, editors, {\em Proceedings of the
  {E}urophysics Study conference: {U}nification of fundamental interactions
  {II}}, pages 455--466, New York, 1982. Plenum.
\bibitem[Chr76]{chris:76}
S.~M. Christensen.
\newblock Vacuum expectation value of the stress tensor in an arbitrary\
 curved
  background: The covariant point-separation method.
\newblock {\em Phys. Rev. D}, 14(10):2490--2501, 1976.
\bibitem[Chr78]{chris:78}
S.~M. Christensen.
\newblock Regularization, renormalization, and covariant geodesic point
  separation.
\newblock {\em Phys. Rev. D}, 17(4):946--963, 1978.
\bibitem[Col84]{colombeau:NGF}
J.~F. Colombeau.
\newblock {\em New generalized functions and multiplication of\
 distributions},
  volume~84.
\newblock North Holland Math. Studies, 1984.
\bibitem[CW94]{chrusandWald:94}
P.~T Chru{\'s}ciel and R.~M. Wald.
\newblock Maximal hypersurfaces in stationary asymptotically flat spacetimes.
\newblock {\em Comm. Math. Phys.}, 163:561--604, 1994.
\bibitem[DB60]{DeWitt:60}
B.~S. DeWitt and R.~W. Brehme.
\newblock Radiation damping in a gravitational field.
\newblock {\em Ann. Phys.}, 9:220--259, 1960.
\bibitem[DeW65]{DeWitt:group}
B.~S. DeWitt.
\newblock {\em The Dynamical Theory of Groups and Fields}.
\newblock Gordon and Breach, New York, 1965.
\bibitem[De{W}75]{DeWitt:75}
B.~S. De{W}itt.
\newblock Quantum field theory in curved spacetime.
\newblock {\em Phys. Rep.}, 19(6):295--357, 1975.
\bibitem[DH72]{Hoermander:72}
J.~J. Duistermaat and L.~H{\"o}rmander.
\newblock Fourier integral operators {II.}
\newblock {\em Acta Math.}, 128:183, 1972.
\bibitem[Die72]{dieudonne:analysisIII}
J.~Dieudonn{\'e}.
\newblock {\em Treatise on Analysis}, volume {III}.
\newblock Academic Press, New York, 1972.
\bibitem[Dim80]{dimock:80}
J.~Dimock.
\newblock Algebras of local observables on a manifold.
\newblock {\em Comm. Math. Phys.}, 77:219--228, 1980.
\bibitem[Dim82]{dimock:82}
J.~Dimock.
\newblock {D}irac quantum fields on a manifold.
\newblock {\em Trans. Am. Math. Soc.}, 269:133--147, 1982.
\bibitem[Dim92]{dimock:92}
J.~Dimock.
\newblock Quantized electromagnetic field on a manifold.
\newblock {\em Rev. Math. Phys.}, 4:223--233, 1992.
\bibitem[FH87]{fredenhagen:87}
K.~Fredenhagen and R.~Haag.
\newblock Generally covariant quantum field theory and scaling limit.
\newblock {\em Comm. Math. Phys.}, 108:91, 1987.
\bibitem[FH90]{fredenhagenHaag:90}
K.~Fredenhagen and R.~Haag.
\newblock On the derivation of {H}awkings radiation associated with the
  formation of a black hole.
\newblock {\em Comm. Math. Phys.}, 127:273--284, 1990.
\bibitem[Fit81]{fitch}
V.~L. Fitch.
\newblock Update on {CP} violation.
\newblock In {\em On The Unity of the Fundamental Interactions, Erice}, pages
  677--693, 1981.
\bibitem[FNW81]{FullingNarcowichWald:81}
S.~A. Fulling, F.~J. Narcowich, and R.~M. Wald.
\newblock Singularity structure of the two-point function in quantum field
  theory in curved spacetime, {II}.
\newblock {\em Ann. Phys. (NY)}, 136:243--272, 1981.
\bibitem[FSW78]{fullingSweenyWald:78}
S.~A. Fulling, M.~Sweeny, and R.~M. Wald.
\newblock Singularity structure of the two-point function in quantum field
  theory in curved spacetime.
\newblock {\em Com. Math. Phys.}, 63:257--264, 1978.
\bibitem[Ful89]{Fulling:aspects_of_qft}
S.~A. Fulling.
\newblock {\em Aspects of Quantum Field Theory in Curved Space-Time}.
\newblock Cambridge University Press, Cambridge, 1989.
\bibitem[FZ75]{ferraraZumino:75}
S.~Ferrara and B.~Zumino.
\newblock Transformation properties of the supercurrent.
\newblock {\em Nucl. Phys. B}, 87:207--220, 1975.
\bibitem[Ger68]{geroch:68}
R.~Geroch.
\newblock Spinor structure of space-times in general relativity. {I}.
\newblock {\em J. Math. Phys.}, 9:1739--1744, 1968.
\bibitem[Ger70]{geroch:70}
R.~Geroch.
\newblock Spinor structure of space-times in general relativity. {II}.
\newblock {\em J. Math. Phys.}, 11:343--348, 1970.
\bibitem[Haa92]{haag:alg}
R.~Haag.
\newblock {\em Local quantum physics: Fields, particles, algebras.}
\newblock Springer, Berlin, Germany, 1992.
\bibitem[Had23]{hadamard:lect}
J.~Hadamard.
\newblock {\em Lectures on {C}auchy's problem in Linear Partial Differential
  Equations}.
\newblock Yale University, New Haven, 1923.
\bibitem[Haw75]{hawking:75}
S.~W. Hawking.
\newblock Particle creation by black holes.
\newblock {\em Comm. Math. Phys.}, 43:199--220, 1975.
\bibitem[HE73]{hawking:largeScale}
S.~W. Hawking and G.~F.~R. Ellis.
\newblock {\em The large scale structure of space-time}.
\newblock Cambridge University Press, Cambridge, 1973.
\bibitem[HH76]{hartleHawking:76}
J.~B. Hartle and S.~W. Hawking.
\newblock Path integral derivation of black hole radiance.
\newblock {\em Phys. Rev. D}, 21:2188--2203, 1976.
\bibitem[HNS84]{haagnarnhofer:84}
R.~Haag, H.~Narnhofer, and U.~Stein.
\newblock On quantum field theory in gravitational background.
\newblock {\em Comm. Math. Phys.}, 94:219--238, 1984.
\bibitem[H{\"o}r71]{Hoermander:71}
L.~H{\"o}rmander.
\newblock Fourier integral operators {I.}
\newblock {\em Acta Math.}, 127:79, 1971.
\bibitem[H{\"o}r83]{hoermander:analysisI}
L.~H{\"o}rmander.
\newblock {\em The Analysis of Linear Partial Differential Operators~{I}}.
\newblock Springer, Berlin, 1983.
\bibitem[Hor90]{hor:intro}
Sergei~Sergeevich Horuzhy.
\newblock {\em Introduction to Algebraic Quantum Field Theory.}
\newblock Kluwer, Dordrecht, Netherlands, 1990.
\bibitem[IZ80]{itzykson:qft}
C.~Itzykson and J.-B. Zuber.
\newblock {\em Quantum field theory}.
\newblock McGraw-Hill, New York, 1980.
\bibitem[Kay92]{kay:92}
B.~S. Kay.
\newblock The principle of locality and quantum field theory on (non globally
  hyperbolic) curved spacetimes.
\newblock To be published, 1992.
\bibitem[KN63]{kob1:foundation}
S.~Kobayashi and K.~Nomizu.
\newblock {\em Foundations of Differential Geometry {I}}.
\newblock John Wiley \& Sons, 1963.
\bibitem[K{\"o}h91]{koehler:91}
M.~K{\"o}hler.
\newblock {M}ultiplikation von {D}istributionen und {S}kalenlimes.
\newblock {D}iplomarbeit, {TU}-{B}erlin, 1991.
\bibitem[K{\"o}h93]{koehler:93}
M.~K{\"o}hler.
\newblock On the leading singularities of the expectation value of the energy
  momentum tensor of a locally supersymmetric quantum field on a curved space
  time.
\newblock Internal Report, 1993.
\bibitem[K{\"o}h94]{koehler:94}
M.~K{\"o}hler.
\newblock New examples for {W}ightman fields on a manifold.
\newblock {\em DESY--Preprint 94--161; Submitted to Class. and Quant. Grav.},
  1994.
\bibitem[KW91]{kay:91}
B.~S. Kay and R.~M. Wald.
\newblock Theorems on the uniqueness and thermal properties of stationary,
  nonsingular, quasifree states on spacetimes with a bifurcate {K}illing
  horizon.
\newblock {\em Phys. Rep.}, 207(2):49--136, 1991.
\bibitem[Ler63]{leray:diffeqn}
J.~Leray.
\newblock {\em Hyperbolic Differential Equations}.
\newblock Lecture notes. Institute for Advanced Study, Princeton, N.J., 1963.
\bibitem[{\L}op91]{lopus:SUSYinQFT}
J.~{\L}opusz{\'a}nski.
\newblock {\em An Introduction to symmetry and supersymmetry in quantum field
  theory}.
\newblock World Scientific, Singapore, 1991.
\bibitem[LR90]{luedersRoberts:90}
C.~L{\"u}ders and J.~E. Roberts.
\newblock Local quasiequivalence and adiabatic vacuum states.
\newblock {\em Comm. Math. Phys.}, 134:29--63, 1990.
\bibitem[Mil63]{milnor:63}
J.~Milnor.
\newblock Spin structures on a manifold.
\newblock {\em L'Enseig. Math.}, 9:168--173, 1963.
\bibitem[MKW87]{mueller_kirsten:SUSY}
H.~J.~W. M{\"u}ller-Kirsten and A.~Wiedemann.
\newblock {\em Supersymmetry}.
\newblock World Scientific, Singapore, 1987.
\bibitem[MTW73]{misner:grav}
C.~W. Misner, K.~S. Thorne, and J.~A. Wheeler.
\newblock {\em Gravitation}.
\newblock Freeman, San Francisco, USA, 1973.
\bibitem[Nak90]{nak:geometry}
M.~Nakahara.
\newblock {\em Geometry, Topology and Physics}.
\newblock Adam Hilger, Bristol and New York, 1990.
\bibitem[Nie81]{nieuwenhuizen:81}
P.~van Nieuwenhuizen.
\newblock Supergravity.
\newblock {\em Phys. Rep.}, 68(4):189--398, 1981.
\bibitem[NO84]{Najmi:84}
A.-H. Najmi and A.~C. Ottewill.
\newblock Quantum states and the {H}adamard form. {II.} {E}nergy minimization
  for spin-$1/2$ fields.
\newblock {\em Phys. Rev. D}, 30(12):2573, 1984.
\bibitem[Ohl92]{ohlmeyer:94}
S.~Ohlmeyer.
\newblock {R}aumzeiten ohne {S}pinstruktur.
\newblock {D}iplomarbeit, {U}niversit{\"a}t {H}amburg, {II.} {I}nstitut\
 f{\"u}r
  {T}heoretische {P}hysik, 1992.
\bibitem[PR84]{pen:twist1}
R.~Penrose and W.~Rindler.
\newblock {\em Spinors and Spacetime {I}}.
\newblock Cambridge University Press, 1984.
\bibitem[Rad]{Radzikowski:94}
M.~Radzikowski.
\newblock Private communication.
\newblock 1994.
\bibitem[Rad92]{Radzikowski:92}
M.~J. Radzikowski.
\newblock {\em The {H}adamard condition and {K}ay's conjecture in (axiomatic)
  quantum field theory on curved space-time}.
\newblock PhD thesis, Princeton University, October 1992.
\bibitem[Rom60]{roman:TheoryElemPart}
P.~Roman.
\newblock {\em Theory of elementary Particles}.
\newblock North-Holland Publishing Company, Amsterdam, 1960.
\bibitem[RS75]{reed:MethodsII}
M.~Reed and B.~Simon.
\newblock {\em Methods of Modern Mathematical Physics {II}}.
\newblock Academic Press Inc. London, 1975.
\bibitem[Ste51]{steenrod:topology}
N.~Steenrod.
\newblock {\em The Topology of Fibre Bundles}.
\newblock Princeton University Press, 1951.
\bibitem[SW64]{strWight:PCT}
R.~F. Streater and A.~S. Wightman.
\newblock {\em {PCT}, Spin \& Statistics and all that}.
\newblock W. A. Benjamin, Inc., New York, 1964.
\bibitem[Tay81]{Taylor:81}
M.~E. Taylor.
\newblock {\em Pseudodifferential Operators}.
\newblock Princeton University Press, Princeton, New Jersey, 1981.
\bibitem[Uhl62]{uhlmann:62}
A.~Uhlmann.
\newblock {\"U}ber die {D}efinition der {Q}uantenfelder nach {W}ightman und
  {H}aag.
\newblock {\em Wiss. Zeitschrift Karl Marx Univ.}, 11:213, 1962.
\bibitem[Unr76]{unruh:76}
W.~G. Unruh.
\newblock Notes on black hole evaporation.
\newblock {\em Phys. Rev. D.}, 14:870--892, 1976.
\bibitem[Ver93]{Verch:93}
R.~Verch.
\newblock Nuclearity, split property, and duality for the {K}lein-{G}ordon
  field in curved spacetime.
\newblock {\em Lett. Math. Phys.}, 29:297--310, 1993.
\bibitem[Ver94]{Verch:94}
R.~Verch.
\newblock Local definiteness, primarity and quasiequivalence of quasifree
  {H}adamard quantum states in curved spacetime.
\newblock {\em Comm. Math. Phys}, 160:507--536, 1994.
\bibitem[W{\etalchar{+}}57]{wu}
C.~S. Wu et~al.
\newblock Experimental test of parity conservation in beta decay.
\newblock {\em Phys. Rev.}, 105:1413--1415, 1957.
\bibitem[Wal]{wald:92}
R.~M. Wald.
\newblock Quantum field theory in curved spacetime.
\newblock To appear in the Proceedings of the 1992 Les Houches School on
  Gravitation and Quantization.
\bibitem[Wal77]{wald:77}
R.~M. Wald.
\newblock The back reaction effect in particle creation in curved spacetime.
\newblock {\em Comm. Math. Phys.}, 54:1, 1977.
\bibitem[Wal78]{wald:78}
R.~M. Wald.
\newblock Trace anomaly of a conformally invariant quantum field in curved
  spacetime.
\newblock {\em Phys. Rev. D}, 17(6):1477--1484, 1978.
\bibitem[Wal84]{wald:gr}
R.~Wald.
\newblock {\em General Relativity}.
\newblock The University of Chicago Press, 1984.
\bibitem[Wes90]{west:SUSY_Intro}
P.~West.
\newblock {\em Introduction to Supersymmetry and Supergravity}.
\newblock World Scientific, Singapore, 1990.
\bibitem[Wig39]{wig:39}
E.~Wigner.
\newblock On unitary representations of the inhomogeneous {L}orentz group.
\newblock {\em Ann. of Math.}, 40:149, 1939.
\bibitem[Wol92]{wollenberg:92b}
M.~Wollenberg.
\newblock Scaling limits and type of local algebras over curved spacetime.
\newblock In W.~B. Arveson et~al., editors, {\em Operator algebras and
  topology}. Putman Research notes in Mathematics 270, Harlow: Longman, 1992.
\bibitem[WW90]{ward:twistor}
R.~S. Ward and R.~O. Wells.
\newblock {\em Twistor Geometry and Field Theory}.
\newblock Cambridge University Press, 1990.
\bibitem[WZ74]{wess_zumino:74}
J.~Wess and B.~Zumino.
\newblock A lagrangian model invariant under supergauge transformations.
\newblock {\em Phys. Lett.}, 49b:52, 1974.
\end{thebibliography}
\newcommand{\etalchar}[1]{$^{#1}$}

\end{document}